\documentclass[a4paper,12pt]{report}

\usepackage[left=2cm,  right=2cm,  top=2cm,  bottom=2cm]{geometry} 

\usepackage[utf8]{inputenc}	
\usepackage[english]{babel}

\usepackage{blindtext}
\usepackage{amsmath}	
\usepackage{bm}
\usepackage{amssymb}	
\usepackage{setspace}	
\usepackage{hyperref}	
\usepackage{titlesec}

\usepackage{indentfirst} 
\usepackage{booktabs}	
\usepackage{placeins}	
\usepackage{flafter}	

\usepackage[nottoc,numbib]{tocbibind}	

\usepackage{mathtools} 
\usepackage{caption}
\usepackage{subcaption}
\usepackage{placeins}	
\usepackage{flafter}	
\usepackage{xcolor}
\usepackage{verbatim}
\usepackage{tensor}
\usepackage{cancel}
\usepackage{textcomp}

\usepackage{empheq} 
\usepackage{mmacells}
\usepackage{fixltx2e}

\usepackage{pgfornament} 
\usepackage{braket}

\usepackage{bibentry}

\usepackage[citation-order, nobysame]{amsrefs}

\BibSpec{article}{%
    +{}  {\PrintAuthors}                {author}
    +{,} { \textit}                     {title}
    +{.} { }                            {part}
    +{:} { \textit}                     {subtitle}
    +{,} { \PrintContributions}         {contribution}
    +{.} { \PrintPartials}              {partial}
    +{,} { }                            {journal}
    +{}  { \textbf}                     {volume}
    +{,} { \PrintConference}            {conference}
    +{}  { \PrintDatePV}                {date}
    +{,} { \issuetext}                  {number}
    +{,} { \eprintpages}                {pages}
    +{,} { }                            {status}
    +{,} { \PrintDOI}                   {doi}
    +{,} { \eprint}                     {eprint} 
    +{}  { \parenthesize}               {language}
    +{}  { \PrintTranslation}           {translation}
    +{;} { \PrintReprint}               {reprint}
    +{.} { }                            {note}
    +{.} {}                             {transition}
    +{}  {\SentenceSpace \PrintReviews} {review}
}

\renewcommand{\eprint}[1]{\href{https://arxiv.org/abs/#1}{arXiv:#1}}

\mmaDefineMathReplacement[≤]{<=}{\leq}
\mmaDefineMathReplacement[≥]{>=}{\geq}
\mmaDefineMathReplacement[≠]{!=}{\neq}
\mmaDefineMathReplacement[→]{->}{\to}[2]
\mmaDefineMathReplacement[⧴]{:>}{:\hspace{-.2em}\to}[2]
\mmaDefineMathReplacement{∉}{\notin}
\mmaDefineMathReplacement{∞}{\infty}
\mmaDefineMathReplacement{𝕕}{\mathbbm{d}}

\newcommand{\de}{\mathrm{d}}
\newcommand{\be}{\begin{equation}}
\newcommand{\ee}{\end{equation}}

\mmaSet{
  morefv={gobble=2},
  linklocaluri=mma/symbol/definition:#1,
  morecellgraphics={yoffset=1.9ex}
}

\newtheorem{mydef}{Definition}

\begin{document}

\pagenumbering{gobble}

\hypersetup{pageanchor=false}

\newcommand*\varhrulefill[1][0.4pt]{\leavevmode\leaders\hrule height#1\hfill\kern0pt}

\begin{titlepage}
    \begin{center}

    \vspace*{.15\textheight}

    \noindent\varhrulefill[0.7mm] \\[0.9cm]

    {\Huge \bfseries Gravity and its wonders: braneworlds and holography \par}\vspace{0.9cm} 
    
    \noindent\varhrulefill[0.7mm] \\[1.5cm]
    
      \textsc{
      {\Large André Juan Ferreira Martins de Moraes}}\\[4cm]

    \includegraphics[width=0.35\linewidth]{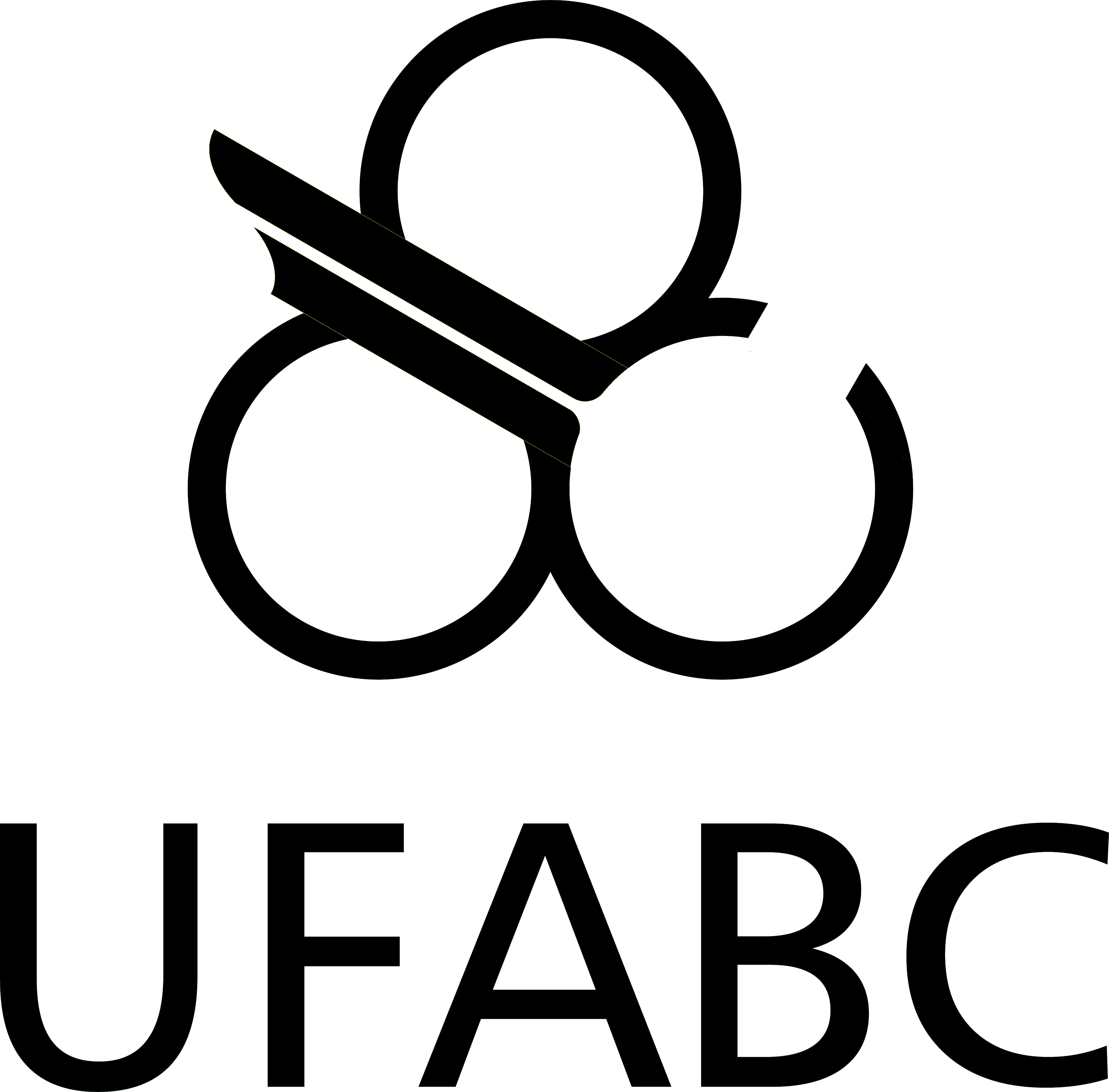}\\[1cm]
    

    {\LARGE 2021}\\[4cm] 

    \vfill
    \end{center}
\end{titlepage}

\begin{titlepage}

    \newpage

    \begin{center}

    {\large Federal University of ABC}\\[6cm]

    \LARGE{\bf Gravity and its wonders: braneworlds and holography} \\[16ex] 

    \Large {\bf Master Thesis}\\[3ex]
    \large {\bf Approved by Dr. Horatiu Nastase and Dr. Ricardo Pazko}\\[2ex]
    \large {\bf Advised by Dr. Roldão da Rocha}\\[14ex]

    {\Large André Juan Ferreira Martins de Moraes }\\[4cm]

    {\large Santo André}\\[0.2cm]
    {\large 2021}
    \end{center}

\end{titlepage}

%
%
%
%
%

\onehalfspacing

%
%
%

\section*{Acknowledgments}

The chance of being able to spend a brief time in the sun is so ridiculously small, that it feels stupefying. I am part of a privileged few, who won the lottery of birth against all odds, and with it the chance to contemplate the wonders of the universe --- that is what I am more grateful for.

Prof. Roldão is more than an advisor to me: above all, he is a huge inspiration and role model as a teacher, researcher and friend. I learned so, so much from him, and not only about Physics and Math, but about science and life itself. Roldão gave me the opportunity and resources to delve deep into our field and actually contribute to science. Without his guidance, none of this would've been possible. Given all that, all I have to express is my deepest gratitude, and expectations of many years to come of collaboration and friendship.

My studies wouldn't have been possible without the support of my family, and people with whom I do not share blood, but who are family just as much. Your support, which has always been present, since my first moments in the world, was absolutely essential for me to be where I am, and I can't thank you enough. You know who you are, and I love you all!

Many thanks go to Universidade Federal do ABC, funding agencies which supported my work at some point (namely Coordenação de Aperfeiçoamento de Pessoal de Nível Superior - Brasil (CAPES) - Finance Code 001), and every Brazilian citizen who indirectly financed my education and academic work. I am working hard to be able to return to society everything that was invested in me, and hopefully I will achieve that through science and education.

I also thank Dr. Horatiu Nastase and Dr. Ricardo Pazko for composing my examination board, and providing important suggestions for the final version of the thesis.

Lastly, I'd like to thank you, who somehow ended up interested in reading these pages. If in any way the words in this thesis serve to guide, inspire or anyhow help your journey through physics, I will feel very grateful and accomplished. So, to you, dear reader, thank you very much, and I hope you have fun!

\newpage


\section*{Abstract}

Gravity is wonderful. The main goal of this thesis is to explore and discuss two aspects of gravity --- two wonders, which illustrate the richness of the theoretical gravitational landscape: braneworlds and holography. Each one of these topics is the core idea of the two parts of which the thesis consists. We begin by presenting gravity as a geometrical theory, then discuss extra dimensions and braneworld scenarios, to motivate the following derivation of the effective Einstein Field Equations on the brane. Afterward, we introduce the Randall--Sundrum model and derive the Minimal Geometric Deformation (MGD) method and its extension (EMGD), which are later on constrained by the classical tests of General Relativity. After this, we discuss black hole thermodynamics and the basic features of the AdS spacetime and black holes in it. We then present the basics of linear response theory and hydrodynamics, and discuss the AdS/CFT duality and its methods, which are employed in the calculation of the shear viscosity-to-entropy density ratio in different gravitational backgrounds, whose result is used to constrain the parameters of generalized 4D and 5D black branes. The relationship between the membrane paradigm and AdS/CFT is also presented. This is followed by a discussion of generalized actions and the violation of the Kovtun--Son--Starinets bound. Afterward, we present the fluid/gravity correspondence as well as an alternative to it in the context of soft hairy horizons. We finish the presentation with a summary of the main results and concluding remarks.\\

\textbf{Keywords:} holographic correspondences, gravitation, black hole physics, quantum mechanics, fluid dynamics.


\newpage


\section*{Resumo}

A Gravidade é maravilhosa. O principal objetivo desta dissertação é explorar e discutir dois aspectos da gravidade --- duas maravilhas, que ilustram a riqueza do panorama da gravitação teórica: mundos-brana e holografia. Cada um destes assuntos é a ideia central das duas partes que compõem esta dissertação. Nós começamos aprensentando a gravidade como uma teoria geométrica, depois passando para a discussão de modelos de dimensão extra e de mundos-brana, de modo a motivar a conseguinte derivação das equações de Einstein efetivas na brana. A seguir, introduzimos o modelo de Randall--Sundrum e derivamos o método Minimal Geometric Deformation (MGD) e sua extensão (EMGD), que são, a seguir, limitados pelos testes clássicos da Relatividade Geral. Depois disso, discutimos termodinâmica de buracos negros e as características básicas do espaço-tempo AdS e buracos negros nele. Então, apresentamos a base da teoria de resposta linear e hidrodinâmica, e discutimos a dualidade AdS/CFT e seus métodos, que são empregados para o cálculo da razão viscosidade-densidade de entropia em diferentes cenários gravitacionais, cujo resultado é utilizado para limitar os parâmetros de branas negras generalizadas em 4D e 5D. A relação entre o paradigma de membranas e AdS/CFT também é apresentada. Depois disso, discutimos ações generalizadas e violações do limite de Kovtun--Son--Starinets. Em seguida, apresentamos a correspondência fluido/gravidade, bem como uma alternativa desta, no contexto de horizontes com soft hair. Terminamos a apresentação com um resumo dos principais resultados e comentários finais.\\

\textbf{Palavras-chave:} correspondências holográficas, gravitação, física de buracos negros, mecânica quântica, fluidodinâmica.

\newpage

\tableofcontents

\newpage

\onehalfspacing

\hypersetup{pageanchor=true}

\cleardoublepage
\phantomsection
\addcontentsline{toc}{chapter}{Publications list}

\begingroup
\def\refname{Publications list}
\def\bibname{Publications list}

\endgroup

\pagenumbering{arabic}

\chapter{Introduction}

Amongst the fundamental interactions, gravity is probably the one with which we are more familiar: we explicitly experience its influence everywhere in our daily life, since the day we were born, and our human intuition grew used to it and its effects.

Now, despite this intuitiveness of gravity, its action and working mechanisms should by no means be taken for granted, as it is absolutely not obvious at all the fact that the same physical phenomenon responsible for making things fall on earth, is also the responsible for the motion of astronomical bodies. And, of course, beyond the ``what'', there is the ``how'' and the ``why'' --- \emph{how} does gravity exert its influence, and \emph{why} does it exist?

Throughout history, the answer to the questions above changed dramatically, and those changes go way beyond the mathematical refinement of the quantitative theory describing gravity and its effects --- the very conceptual basis of our understanding of the gravitational interaction has shifted tremendously at the beginning of the 20\textsuperscript{th} century. Since then, gravity started to be seen in an entirely different way, and this novel view produced quite bizarre concepts, such as black holes.

As time passed, the picture got even more interesting: although the other fundamental interactions of nature found a common description, the familiar and intuitive gravity persistently resisted to this unification of description. Quite the opposite, actually: many efforts of unification were proposed, and until the present days, none of them was successful. With a totally different description and many subtleties, the familiar and intuitive interaction grew more mysterious, and the quest for answers was only responsible for the formulation of more and more questions.

Driven by those questions, we were eventually led to discoveries even more fascinating --- little by little, it became clear that, despite its unique and particular description, gravity is more intimately related to several other physical phenomena than we could have ever thought...

Full of mysteries, unanswered questions and expecting for light to be shed on it, the current gravitational landscape is wide, rich, and very, very beautiful: a \emph{Dark Chest of Wonders!}\footnote{This is the title of a song by the Finnish band Nightwish.}

So, the aim of this thesis is that: to open the chest and explore gravity and its many wonders. We begin by presenting the fundamental new way in which gravity came to be seen in the 20\textsuperscript{th} century, and once the fundamental bases are established, we turn to the study of two particular wonders of gravity, each one at the core of the two main parts of this thesis,

Part I is dedicated to the introduction of the geometrical theory of gravity, General Relativity, and to the introduction of a very interesting concept: the possibility of extra spatial dimensions, in addition to the ones we commonly experience. After the introduction of these so-called extra dimensions, a question that may naturally appear is: what changes in gravity then? By answering this question, the concept of branes and \emph{braneworlds} will be introduced, and we will study some particular examples of such scenarios. That is the first wonder!

Part II is dedicated to the exploration of very intimate relationships between gravity and other theories unrelated to it, at first glance. This set of ideas is called \emph{holography}, the second wonder! We will then learn how to use gravity to study objects of the related theory, and vice-versa, via the establishment of the so-called holographic correspondences that shall be introduced and explored.

Each part of the thesis is composed of several chapters, in which we present and detail existing ideas available in the literature, as well as original results. In the first paragraphs of each chapter, it will be made clear whether its content is original, or based on existing work. In either case, the presentation and discussion will be thoroughly detailed, and calculations will be explicitly provided whenever necessary. For a quick report of which portions of the thesis are original results, refer to Sec. \ref{sec:conclusions}, in which a summary of the main results as well as the conclusions are presented.

The reader will notice that this thesis is quite lengthy. The reason for that is twofold: first, because this thesis consists of two main parts, each one with its central theme, and although those themes are related to each other, each one is quite rich on itself, with its extended context, which must be presented in some detail to provide a clear presentation; secondly, this thesis aims to serve as a guide for future students which adventure themselves into studying the wonders of gravity, particularly those in which we focus. Therefore, as aforementioned, we did not spare words in the conceptual discussions, nor mathematical passages in the explicit calculations. As this is a thesis, not a research paper, we judge it adequate to be as long as necessary in favor of clarity --- and hopefully other students will benefit from that in the future!

Except when otherwise explicitly specified, we will use the following notation throughout the thesis: lowercase Greek indices $\mu = \{0, 1, 2, 3\}$, will be used to label 4-dimensional spacetime coordinates; whilst lowercase Latin indices $i = \{1, 2, 3\}$, will denote purely spatial 3-dimensional coordinates; and uppercase Latin indices $M = \{0, 1, 2, 3, 4\}$, will label 5-dimensional spacetime coordinates. Also, Einstein's summation convention is used, so that identical lower and upper indices are always summed over their possible values.

Once again: gravity is wonderful --- let us now open the chest and start our journey into discovering why!

\newpage

\part{Braneworlds}

\newpage

\chapter{A geometric theory of gravity}
\label{sec:units}

In 1915, Albert Einstein published his General Relativity (GR) \cite{gr}, a geometric theory that generalizes the comprehension of spacetime provided by Special Relativity (SR) \cite{sr} to include gravitation. Within GR, gravity is inherent to spacetime itself --- more specifically, it is the manifestation of its \textit{curvature} ---, in an intricate relationship with matter/energy and momentum. 

Before delving into the main topics of this thesis (extra dimensions, branes and holography), it is important to establish the foundations of everything that will follow. Thus, for completeness, in this chapter, we present GR and its formulation, following the presentation of Carroll \cite{carroll}, Wald \cite{wald} and the lectures of Dr. Frederic Schuller \cites{gravlight, gravlightvd}. Then, it will become clear \textit{why} and \textit{how} GR is a geometric theory, and we will be ready to move on.

Before the establishment of GR, gravity was explained in terms of Newtonian physics, which provided the framework in which rules for the relationship between matter and gravitation were constructed: given two particles of mass $M$ and $m$ separated by a distance $\bm{r} = r \bm{\hat{r}}$, a \emph{gravitational force} $\bm{F}$ is exerted on both bodies, as expressed in the notorious inverse-square law

\begin{equation}
 \bm{F} = - G\frac{M m}{r^2} \hat{\bm{r}} \ ,
 \label{eq:isl}
\end{equation}

\noindent where $G$ is Newton's gravitational constant. Now, given the force $\bm{F}$ exerted in the particle of mass $m$, the acceleration $\bm{a}$ impressed to it is given by Newton's second law, according to,

\begin{equation}
 \bm{a} = \frac{\bm{F}}{m} \ .
  \label{eq:n2l}
\end{equation}

On Eqs. \ref{eq:isl} and \ref{eq:n2l} there is no explicit mention whatsoever to the notion of spacetime or curvature, although the relationship between gravity and matter is rather explicit and fully determines the gravitational interaction as seen within Newtonian physics.

GR completely changes the scenario, by replacing the status of gravity as a force by a completely different notion governed by the Einstein Field Equations (EFE), which will soon be presented and fully derived. For now, it suffices to say that these equations relate the curvature of spacetime with the energy-momentum content in a given spacetime region, so that the geometric relationship between gravity and matter is made explicit as intrinsic to spacetime. 

In this brief introduction, we mentioned rather informally some of the ideas discussed in detail in Appendix \ref{ap:gr}, where the geometric notions are presented with mathematical precision. From now on, we shall freely use all the concepts developed in this appendix, assuming that they are familiar to the reader, and formally establish how, in GR, gravity is seen as the spacetime curvature induced by matter and momentum.

Prior to any further development, it is worth a discussion on the unit system we will adopt in the next few sections in which we discuss the classical formulation and results of 4-dimensional GR: the \emph{geometrized units}. In such system, the speed of light in vacuum $c$ as well as Newton's gravitational $G$ constant are set to dimensionless unit, that is, $c = G = 1$. Such imposition determines a redefinition of all units in terms of a single fundamental one, which will be defined to be that of length $[L]$. Now, imposing $c=1$ implies that $[L][T]^{-1} = 1 \Rightarrow [T] = [L]$, that is, time is measured in units of length. On the other hand, $G = 1$ implies $[L]^3[M]^{-1}[T]^{-2} = [L][M]^{-1} = 1 \Rightarrow [M] = [L]$, i.e., mass also is measured as length. The same occurs with energy $[E] = [M][L]^2[T]^{-2} \Rightarrow [E] = [M] = [L]$. Therefore, in such a system, one has a convenient equation between the units of the $4$ quantities most important to our purposes,

\begin{equation}
 [E] = [T] = [M] = [L] \ .
\end{equation}

Now, apart from the geometrized units, in this thesis we shall also use \emph{natural units}, defined by $c = \hbar = k_B = \epsilon_0 = 1$, where $\hbar$ is the reduced Planck's constant, $k_B$ is Boltzmann's constant and $\epsilon_0$ is the electric constant. Once we start considering the multidimensional formulation of GR, as well as AdS/CFT and related topics in high energy physics, it will be more convenient to adopt natural units, which are used in most of the literature of the field, given its proximity with attempts of formulating a quantum theory of gravity.

With the choice of natural units, the fundamental unit in terms of which other units are defined becomes that of energy $[E]$. Now, $\hbar = 1 \Rightarrow [E][T] = 1 \Rightarrow [T] = [E]^{-1}$; $c = 1 \Rightarrow [L][T]^{-1} = 1 \Rightarrow [L] = [T] = [E]^{-1}$; and $[E] = [M][L]^2[T]^{-2} = [M][E]^2[E]^{-2} \Rightarrow [M] = [E]$. Therefore, in this system, one has

\begin{equation}
[L]^{-1} = [T]^{-1} = [M] = [E] \ .
\end{equation}

Although the use of either of these systems is mostly convenient in some situations, when we deal with quantitative analysis of observational or experimental data, it is useful to recover SI units. When such cases occur, and whenever else it is relevant to make clear which units are used, we will make it clear in the text.

The reason of setting unit systems which differ from SI are deeper than those of mere convenience, though. In fact, setting $c = G = 1$ or $c = \hbar = k_B = \epsilon_0 = 1$ in a way imply an even more fundamental role to such \emph{fundamental natural constants}, since the human-established factors of conversion between units are abandoned, and the units are redefined accordingly. A lively discussion on this topic is presented in \cite{duff}.

\section{The energy-momentum tensor}
\label{sec:em_tensor}

For massive particles, which follow timelike paths, it is convenient to use the proper time $\tau$ as the parameter along the curve, so that the path is given by $x^\mu(\tau)$. The vector tangent to the curve under such a parametrization is then called the \emph{four-velocity} $U^\mu$, defined by,

\begin{equation}
 U^\mu = \frac{d x^\mu}{d \tau} \ .
\end{equation}

A feature of the four-velocity is that it is always normalized, which comes as a consequence of the proper time definition $d \tau^2 = -g_{\mu \nu} dx^\mu dx^\nu \Rightarrow g_{\mu \nu} U^\mu U^\nu = -1$. This fact shows that particles \emph{always move at a fixed velocity through spacetime} (taken as unitary since we set $c = 1$). On a particle rest frame, its four-velocity has components $U^\mu = (1, 0, 0, 0)^T$, which can be thought of as a motion only through time. In a moving frame, the familiar \emph{Lorentz transformations} act in such a way that the unitary norm is always preserved, by splitting the 4-velocity through spatial directions and time. This is what originates time dilation and other similar phenomena known in SR \cite{sr2}. 

From the four-velocity, we define the \emph{four-momentum}, according to $p^\mu = mU^\mu$, where $m$ is the (rest) mass of the ever-moving particle. The particle \emph{energy} is then the temporal component of its four-momentum $E = p^0$. That is completely consistent with the fact that in the particle rest frame, $p^0 = m$, so that the famous $E = mc^2$ equation is manifested (though not explicitly given our choice of units).

In GR it is often necessary to deal with a system of many particles carrying energy and/or momentum, for example to determine the effect of a given energy-momentum distribution in the curvature of spacetime. In such cases, instead of defining a momentum vector for each particle, it is more convenient to characterize the system as a \emph{fluid} with a single four-velocity field throughout the continuum. As it is common in fluid mechanics, it is necessary to define a tensor quantity to fully describe the energy and momentum contents of a fluid \cite{landau_fluids}, which in GR takes the form of the \emph{energy-momentum} or \emph{stress} tensor, $T = T^{\mu \nu} \partial_\mu \partial_\nu$, a rank-$(2, 0)$ symmetric tensor, encoding all information regarding the macroscopic variables of the fluid (such as energy, shear and pressure). 

A single component $T^{\mu \nu}$ may be thought of as \emph{the flux of four-momentum $p^\mu$ through a surface of constant $x^\nu$}. In that sense, for an element of the fluid in its rest frame, $T^{00}$ is the flux of $p^0$ in the direction of $x^0$, i.e., the flux of energy through time, which is the fluid \emph{energy density} $\rho$ in its rest frame. Also, $T^{0i}$ represents the \emph{momentum density}; the spatial components represent the \emph{momentum flux}, such that non-diagonal terms give \emph{shearing terms}, and the diagonals represent the pressure $p_i$ in the $i$-th direction, $p_i = T^{ii}$, all of that in the rest frame and in inertial coordinates. We then realize that the components $T^{\mu \nu}$ encode all macroscopic quantities characterizing a fluid.

A \emph{perfect fluid} is such that, in the local rest frame, it is fully characterized solely by its rest-frame energy density $\rho = T^{00}$ and an \emph{isotropic} rest-frame pressure $p = T^{ii}$. Isotropy implies that $T^{\mu \nu}$ is diagonal on its rest frame, since no flux of a given momentum component occurs in an orthogonal direction. Therefore, in such a frame, one has $T^{\mu \nu} = \mathrm{diag}(\rho, p, p, p)$. To generalize such an expression to any frame in a manifestly tensor way, one defines the general expression for the energy-momentum tensor components of a perfect fluid according to the \emph{constitutive equation}

\begin{equation}
 T^{\mu \nu} = (\rho + p)U^\mu U^\nu + p g^{\mu \nu} \ .
 \label{eq:ene-mom-up}
\end{equation}

Notice that, in fact, Eq. \ref{eq:ene-mom-up} reduces to $T^{\mu \nu} = \mathrm{diag}(\rho, p, p, p)$ in locally inertial coordinates (in which $g_{\hat{\mu} \hat{\nu}} = \eta_{\hat{\mu} \hat{\nu}}$) of the rest frame ($U^\mu = (1, 0, 0, 0)^T$). It is useful to lower the indices of $T^{\mu \nu}$ by contracting it twice with the metric, to put it into the form that will appear in the EFE,

\begin{equation}
 T_{\mu \nu} = (\rho + p)U_\mu U_\nu + p g_{\mu \nu} \ .
 \label{eq:ene-mom-low}
\end{equation}

When pressure is negligible, the energy-momentum tensor becomes $T_{\mu \nu} = \rho U_\mu U_\nu$, an energy-momentum distribution known as \emph{dust}, the simplest case of a perfect fluid.

An important feature of $T^{\mu \nu}$ is that it is conserved, which in curved spacetime is expressed by the covariant conservation equation,

\begin{equation}
 \nabla_\mu T^{\mu \nu} = 0 \ .
 \label{eq:enermomencons}
\end{equation}

This equation (which in flat spacetime is simply given by $\partial_\mu T^{\mu \nu} = 0$) encodes the continuity equation for the energy density, as well as the Euler equation of fluid mechanics \cite{landau_fluids} for the perfect fluid. Thus, Eq. \ref{eq:enermomencons} together with the constitutive equation of Eq. \ref{eq:ene-mom-up} fully describe the fluid dynamics, and also fully characterize the conservation properties of a fluid as a continuum of energy-momentum. 

\section{Einstein Field Equations}
\label{sec:EFEqs}

The central piece of the formal basis of GR is the Einstein Field Equations (EFE), which effectively relate energy-momentum to curvature and its manifestation as gravity. We shall now motivate its derivation with a set of arguments directly related to the idea that Newtonian gravitation must be recovered from GR in the appropriate limits.

In Newtonian gravitation, the gravitational field may be recovered from Poisson's equation given in space by $\nabla^2 \Phi = 4 \pi \rho$, where $\nabla^2 = \delta^{i j} \partial_i \partial_j$ is the Laplacian and $\rho$ is the density of the mass distribution creating the potential $\Phi$. Now, although it must be manifestly covariant, the relativistic field equation is supposed to at least resemble Poisson's equation, in a way that the latter may be seen as a particular limit of the former. In fact, we know the relativistic substitute of the matter density $\rho$: the energy-momentum tensor constructed in the last section, as it encodes all the information concerning the generator of the gravitational field --- in this case, an energy-momentum distribution. 

Also, motivated by the Laplacian, it is expected for the EFE to consist of second-derivatives of a given tensor which would replace the gravitational potential $\Phi$. The question then becomes: which tensor must we pick? Evidently, after the discussion on geometry provided in Appendix \ref{ap:gr}, it becomes clear that, if gravity is a manifestation of spacetime curvature, this tensor must be the \emph{metric}, since it encodes the geometry of spacetime. There is, however, a more precise argument to make it clear that the metric is the tensor we seek.

First, it is necessary to precisely define the \emph{Newtonian limit} in which the Newtonian results for gravitation must be recovered from GR. Such limit is characterized by three demands: that the particle is moving slowly in space $ \left ( d x^i/d \tau \ll d x^0/d\tau \right)$; that the gravitational field is sufficiently weak to be written as a small perturbation $h_{\mu \nu}$ of flat spacetime ($g_{\mu \nu} = \eta_{\mu \nu} + h_{\mu \nu}$, $|h_{\mu \nu}| \ll 1$); and that the metric is also static ($\partial_0 g_{\mu \nu} = \partial_t g_{\mu \nu} = 0$). 

Now, from the definition that the inverse metric must satisfy $g^{\mu \nu}g_{\nu \sigma} = \delta^\mu_\sigma$, one finds that, in first order of $h$, the inverse metric must be given by $g^{\mu \nu} = \eta^{\mu \nu} - h^{\mu \nu}$, since

\begin{equation}
\begin{aligned}
g^{\mu \nu}g_{\nu \sigma} & = (\eta^{\mu \nu} - h^{\mu \nu})(\eta_{\nu \sigma} + h_{\nu \sigma}) \\&
= \eta^{\mu \nu}\eta_{\nu \sigma} + \eta^{\mu \nu}h_{\nu \sigma} - \eta_{\nu \sigma} h^{\mu \nu} + \mathcal{O}(h^2) \\&
= \delta^\mu_\sigma + \eta^{\mu \nu}h_{\nu \sigma} - \eta_{\nu \sigma} ( \eta^{\mu \rho}\eta^{\nu \lambda} h_{\rho \lambda}) + \mathcal{O}(h^2) \\&
= \delta^\mu_\sigma + \eta^{\mu \nu}h_{\nu \sigma} - \delta_\sigma^\lambda \eta^{\mu \rho} h_{\rho \lambda} + \mathcal{O}(h^2) \\&
= \delta^\mu_\sigma + \eta^{\mu \nu}h_{\nu \sigma} -  \eta^{\mu \rho} h_{\rho \sigma} + \mathcal{O}(h^2) \\&
= \delta^\mu_\sigma + \mathcal{O}(h^2) \ ,
\end{aligned}
\end{equation}

\noindent which guarantees that $h_{\mu \nu}$ indeed does represent a small perturbation taken up to first order. Also notice that $\eta^{\mu \nu}$ was directly used to raise the index of $h_{\mu \nu}$ in the third line above, since using $g^{\mu \nu}$ simply produces more negligible terms of order $\mathcal{O}(h^2)$.

Now, using these three conditions altogether, one notices that the geodesic equation (discussed in Sec. \ref{sec:Geodesics}) reduces to

\begin{equation}
\frac{d^2 x^\mu}{d \tau^2} + \Gamma^\mu_{00} \left ( \frac{dx^0}{d \tau}\right)^2 = 0 \ ,
\end{equation}

\noindent since spatial components of the four velocity, $d x^i/d \tau$, are negligible. Given that the metric is static, the Christoffel symbol is simply given by $\Gamma^\mu_{00} = - \frac{1}{2}g^{\mu \nu}\partial_\nu g_{00}$. By using the definition of $g^{\mu \nu}$ constructed above, as well as $g_{00} = \eta_{00} + h_{00}$, one has $g^{\mu \nu}\partial_\nu g_{00} = (\eta^{\mu \nu} - h^{\mu \nu})\partial_\nu(\eta_{00} + h_{00}) = \eta^{\mu \nu}\partial_\nu h_{00} + \mathcal{O}(h^2)$, so that up to first order in $h$, one gets $\Gamma^\mu_{00} = - \frac{1}{2}\eta^{\mu \nu}\partial_\nu h_{00}$, and the geodesic equation then becomes

\begin{equation}
\frac{d^2 x^\mu}{d \tau^2} - \frac{1}{2}\left ( \frac{dx^0}{d \tau}\right)^2 \eta^{\mu \nu}\partial_\nu h_{00} = 0 \ .
\end{equation}

Now, taking the spatial components of the geodesic equations, which corresponds to taking $\mu = i$, so that $\eta^{\mu \nu} \mapsto \delta^{ij} \Rightarrow \eta^{\mu \nu}\partial_\nu \mapsto \delta^{ij} \partial_j = \partial_i$, one has

\begin{equation}
\frac{d^2 x^i}{d \tau^2} = \frac{1}{2} \left ( \frac{dx^0}{d \tau} \right)^2 \partial_i h_{00} \Rightarrow \left ( \frac{d\tau}{d x^0} \right)^2 \frac{d^2 x^i}{d \tau^2} = \frac{1}{2} \partial_i h_{00} \ ,
\end{equation}

\noindent which is equivalent to

\begin{equation}
\frac{d^2 x^i}{d (x^0)^2} = \frac{d^2 x^i}{d t^2} = \frac{1}{2} \partial_i h_{00} \ .
\end{equation}

In fact, $d^2 x^i/d t^2$ corresponds to the components of the spatial acceleration $\bm{a}$, whilst clearly $\partial_i h_{00}$ are the components of the spatial gradient $\nabla h_{00}$. Now, if we define $h_{00} = -2 \Phi$, we promptly recover the Newtonian expression for the acceleration experienced by a particle in a gravitational potential $\Phi$ $\bm{a} =  \frac{1}{2} \nabla (-2 \Phi) \Rightarrow \bm{a} = - \nabla \Phi$. Therefore, we can finally establish

\begin{equation}
 g_{00} = \eta_{00} + h_{00} \Rightarrow g_{00} = -1 - 2 \Phi \ .
 \label{eq:g00}
\end{equation}

What we just did was to establish a relationship between a metric perturbation and the Newtonian gravitational potential in a way that the curvature of spacetime is entirely sufficient to describe gravity in the Newtonian limit, as it should be. Therefore, it is now clear that in the relativistic generalization of Poisson's Equation, the potential $\Phi$ must be replaced by the metric tensor $g_{\mu \nu}$, since Newtonian gravity was successfully recovered from a metric perturbation, i.e., a purely geometrical, genuinely general-relativistic view of gravity.

So, at this point, we know that the EFE must express a covariant relation between second derivatives of the metric and the energy-momentum tensor. The aim now is to precisely establish such a relation. Fortunately, we already know a tensor object which is defined in terms of first and seconds derivatives of the metric: the Riemann Tensor, defined in Eq. \ref{eq:riemtens}. Of course, though, we must construct an equation between symmetric rank-$(0, 2)$ tensors (as $T_{\mu \nu}$ is), so that we take the obvious contraction of the Riemann tensor: the Ricci tensor. The prototype of the field equation becomes then $R_{\mu \nu} = \alpha T_{\mu \nu}$, where $\alpha$ is a proportionality constant. 

There is something wrong with such an equation, though: it implies a non-physical constraint in the energy-momentum distribution of the universe. First of all, notice that the energy-momentum conservation, $\nabla^\mu T_{\mu \nu} = 0$, would imply $\nabla^\mu R_{\mu \nu} = 0$. But, the Bianchi identity ($\nabla^\mu R_{\mu \nu} = \frac{1}{2} \nabla_\nu R$) would lead to $\nabla_\mu R = 0$. Now, taking the trace of the proposed equation, one gets $R = \alpha T$, so that one has $\nabla_\mu R = 0 \Rightarrow \nabla_\mu T = 0 $. As $T = T\indices{^\mu_\mu}$, the last equation is a total derivative --- which, being null, states that $T$ is constant throughout spacetime. This is the absurd constraint, since, for example, inside any star we would have $T \neq 0$, whilst in vacuum $T_{\mu \nu} = 0 \Rightarrow T = 0$. So, this first guess for the field equation of GR (which was postulated by Einstein) must be discarded.

Fortunately, though, just like the Ricci tensor was a natural choice in the first attempt of constructing a field equation, there is yet another natural choice for a rank-$(0, 2)$ symmetric tensor, constructed from the Ricci tensor --- and, therefore, from the metric: the Einstein tensor, $G_{\mu \nu} = R_{\mu \nu} - \frac{1}{2} R g_{\mu \nu}$, which was previously introduced as a conserved tensor, according to the twice-contracted Bianchi identity, $\nabla^\mu G_{\mu \nu} = 0$. This already makes clear that, if the field equation is of the form $G_{\mu \nu} = \alpha T_{\mu \nu}$, the implication $\nabla^\mu T_{\mu \nu} = 0 \Rightarrow \nabla^\mu G_{\mu \nu} = 0 $ is already guaranteed by the Bianchi identity, creating therefore no additional constraints that made us abandon the first attempt of field equation!

Thus, establishing that the field equation will be

\begin{equation}
 R_{\mu \nu} - \frac{1}{2} R g_{\mu \nu} = \alpha T_{\mu \nu} \ ,
\label{eq:quase_efe}
\end{equation}

\noindent Eq. \ref{eq:quase_efe} by construction satisfy all the requirements first established, so that it remains to determine the constant $\alpha$ as well as to prove that the resulting equation does reproduce the Newtonian gravity in the proper limit. To do this, we will once again consider the conditions characterizing the Newtonian limit: slow particles, week gravitational field and time-independent metric. The energy-momentum tensor will be that of dust, $T_{\mu \nu} = \rho U_\mu U_\nu$, since the pressure produced on the continuum by slow particles within the Newtonian limit is negligible.

In the fluid rest frame, $U^\mu = (1, 0, 0, 0)^T \Rightarrow U_\mu = (-1, 0, 0, 0)$, one has that the only non-vanishing component of $T_{\mu \nu}$ is $T_{00} = \rho$, so that $T = g^{00}T_{00} = (\eta^{00} - h^{00})\rho$. Since we are considering a small energy density $\rho$ --- which is, after all, necessary to consider the weak-field approximation ---, the term $h^{00}\rho$ is negligible, and the trace of $T_{\mu \nu}$ is then $T = \eta^{00}\rho = -\rho$.

Now, taking the trace of Eq. \ref{eq:quase_efe}, one gets $R = -\alpha T$, since $g^{\mu \nu}g_{\mu \nu} = \delta^\mu_\mu = 4$. Thus,

\begin{equation}
 R_{\mu \nu} - \frac{1}{2} R g_{\mu \nu} = R_{\mu \nu} - \frac{1}{2} (-\alpha T) g_{\mu \nu} = \alpha T_{\mu \nu} \Rightarrow R_{\mu \nu} = \alpha(T_{\mu \nu} -\frac{1}{2}T g_{\mu \nu}) \ .
\end{equation}

Given that $T_{00} = \rho = -T$, one gets $R_{00} = \alpha[\rho - \frac{1}{2}(-\rho)(-1 + h_{00})] \Rightarrow R_{00} = \frac{1}{2} \alpha \rho$, where again we neglected the term $h_{00}\rho$. This last equation may be written in terms of the metric by explicitly calculating $R_{00} = R\indices{^\mu_{0 \mu 0}}$, according to $R\indices{^{\mu}_{0 \mu 0}} = \partial_\mu \Gamma^\mu_{0 0} - \partial_0 \Gamma^\mu_{\mu 0} + \Gamma^\mu_{\mu \lambda}\Gamma^\lambda_{0 0} - \Gamma^\mu_{0 \lambda}\Gamma^\lambda_{\mu 0}$. Of course, the temporal derivatives vanish, as well as the last two terms in which the metric is quadratic, so that it yields

\begin{equation}
\begin{aligned}
R_{00} & = \partial_{i} \Gamma^i_{00} \\&
= \frac{1}{2} \partial_i \left (  g^{i \mu} (\partial_0 g_{\mu 0} + \partial_0 g_{0 \mu} - \partial_\mu g_{0 0}) \right ) \\&
= -\frac{1}{2} \partial_i g^{i \mu} (\partial_\mu (\eta_{00} + h_{00})) \\&
= -\frac{1}{2} \partial_i g^{ij} (\partial_j h_{00}) \\&
= -\frac{1}{2} \partial_i (\delta^{ij} - h^{ij}) (\partial_j h_{00}) \\&
= -\frac{1}{2} \delta^{ij} \partial_i \partial_j h_{00} \\&
= -\frac{1}{2} \nabla^2 h_{00} \ ,
\end{aligned}
\end{equation}

\noindent where we neglected terms of order $\mathcal{O}(h^2)$; $\nabla^2$ is the Laplacian; and from the third to the fourth line we used the time-independence of the metric to pass to purely spatial indices $\mu \mapsto j$. 

Now, as $R_{00} = \frac{1}{2} \alpha \rho$, Eq. \ref{eq:quase_efe} implies, in the Newtonian limit $\nabla^2 h_{00} = - \alpha \rho $. However, in such a limit, one has shown that to recover the Newtonian acceleration, we must have $h_{00} = -2 \Phi$, so that 

\begin{equation}
 \nabla^2 (-2 \Phi) = -\alpha \rho \Rightarrow \  \nabla^2 \Phi = \frac{1}{2} \alpha \rho \ ,
\end{equation}

\noindent which is exactly Poisson's equation if we set $\alpha = 8 \pi$! With such a choice, we can finally write Eq. \ref{eq:quase_efe} as

\begin{equation}
  R_{\mu \nu} - \frac{1}{2} R g_{\mu \nu} = 8 \pi T_{\mu \nu} \ ,
 \label{eq:efe_no_cosm}
\end{equation}

\noindent which is the EFE --- without the cosmological constant term, that will be introduced in the next section. 

Notice that this well-motivated derivation of the EFE provided us with the central equation of general relativity, in a way that the Newtonian gravity is by construction recovered on its appropriate limit, which is certainly something remarkable!

It is also possible to construct a Lagrangian formulation for GR, which allows us to derive the EFE from the principle of least action applied to the \emph{Einstein-Hilbert action},

\begin{equation}
 S = \int d^4 x \sqrt{-g} R  \ ,
\end{equation}

\noindent where $g = \det(g_{\mu \nu})$ and $R$ is the Ricci scalar. By varying this action with respect to the metric through the usual methods of the principle of least action \cite{landau_fields}, one arrives at the EFE in vacuum (that is, $R_{\mu \nu} - \frac{1}{2}R g_{\mu \nu} = 0$). The full procedure will not be shown here, and it can be found in the main references \cites{carroll, wald}. However, when later on we start presenting multidimensional gravity in braneworld scenarios, we will further discuss the fundamental actions behind the theory.

\section{The cosmological constant in the EFE}
\label{sec:cc_efe}

The possibility that the vacuum possesses an energy density $\rho_v$ has profound implications, as, for example, a way to explain dark energy \cite{darken}. In GR, the absolute value of energy is of great importance --- since energy is, after all, directly linked with the gravitational field. A vacuum energy-momentum tensor must therefore be Lorentz invariant in locally inertial coordinates $x^{\hat{\mu}}$, which is only possible in such coordinates if it is proportional to the metric $\eta_{\hat{\mu} \hat{\nu}}$ \cite{carroll}. 

As a covariant relation, its generalization to arbitrary coordinates is direct $T_{\mu \nu} = -\rho_v g_{\mu \nu}$. Comparing this with the energy-momentum tensor of a perfect fluid (Eq. \ref{eq:ene-mom-low}), it is easy to see that the vacuum then behaves like a perfect fluid with pressure $p_v = -\rho_v$, where $\rho_v$ is everywhere constant.

Now, by writing the energy-momentum tensor like a sum of matter and vacuum energy content, $T_{\mu \nu} = T_{\mu \nu}^M - \rho_v g_{\mu \nu}$, we can rewrite the EFE as presented in Eq. \ref{eq:efe_no_cosm} like,

\begin{equation}
  R_{\mu \nu} - \frac{1}{2} R g_{\mu \nu} = 8 \pi (T_{\mu \nu}^M - \rho_v g_{\mu \nu}) \Rightarrow R_{\mu \nu} - \frac{1}{2} R g_{\mu \nu} + 8 \pi \rho_v g_{\mu \nu} = 8 \pi T_{\mu \nu}^M \ .
\end{equation}

Now, by resetting $T_{\mu \nu}^M = T_{\mu \nu}$, in which case the vacuum energy is regarded as an isolated contribution to spacetime energy content, we may introduce the \emph{cosmological constant}, defined as

\vspace{-5mm}

\begin{equation}
 \Lambda = 8 \pi \rho_v~,
\end{equation}

\noindent and then the EFE can be completely written in all their glory,

\begin{equation}
 R_{\mu \nu} - \frac{1}{2} R g_{\mu \nu} + \Lambda g_{\mu \nu} = 8 \pi T_{\mu \nu} \ .
\end{equation}

Shortly after the first introduction of the EFE in the form of Eq. \ref{eq:efe_no_cosm} in 1915, Einstein added the cosmological constant to the EFE for GR to provide a static universe, which was supported by observations at the time. Not long after, though, observations by Edwin Hubble indicated an expanding universe, consistent with the initial GR without a cosmological constant. The cosmological constant was then accounted for by Einstein as a mistake, although today it is again seriously considered. There is, though, some obscurity around the concept, mainly expressed through the famous ``cosmological constant problem'', which refers to the incredible discrepancy between the theoretical and observed values for $\Lambda$ --- of 54 orders of magnitude! \cite{cosm}. Nonetheless, the cosmological constant will be very important in what follows, so that it will be considered.

We finally reached the point in which the basis of GR are all established, and it is clear how the geometric theory of gravity is governed by the EFE. We are now ready for what is yet to come, which will start with the most immediate, yet vastly interesting, solution to the EFE: the Schwarzschild metric.

\chapter{The Schwarzschild solution}
\label{sec:schwarzschild}

In 1916, shortly after Einstein's publication of his GR, Karl Schwarzschild published an exact solution for the EFE in vacuum \cite{schwarzschild}, which is today known as the \emph{Schwarzschild metric}, an \emph{exterior solution}, i.e., it describes the empty space around a given mass distribution. It is also \emph{spherically symmetric}, which is adequate to describe the gravitational field created by stars, for example. As the statement of \emph{Birkhoff's theorem} \cites{carroll, wald, birk}, the Schwarzschild metric is the \emph{only} spherically symmetric vacuum solution to the EFE. Instead of presenting the full theorem, though, we will present a derivation to the Schwarzschild solution which directly makes use of the arguments used in the construction of the EFE.

Since we will solve the EFE in vacuum, it is useful to write them in a slightly different manner. For now, we will not consider the cosmological constant, so that $\Lambda = 0$. Thus, taking the trace of Eq. \ref{eq:efe_no_cosm}, one has

\begin{equation}
\begin{gathered}
   g^{\mu \nu}R_{\mu \nu} - \frac{1}{2} R g^{\mu \nu} g_{\mu \nu} = 8 \pi g^{\mu \nu} T_{\mu \nu} \\
   \Rightarrow R - \frac{1}{2} 4 R = 8 \pi T \\ 
   \Rightarrow R = - 8\pi T \ , 
\end{gathered}
\end{equation}

\noindent where $g^{\mu \nu}T_{\mu \nu} = T\indices{^\mu_\mu} = T$ and $g^{\mu \mu}g_{\mu \nu} = \delta^\mu_\nu = 4$. Therefore, we can write

\begin{equation}
\begin{gathered}
 R_{\mu \nu} - \frac{1}{2} R g_{\mu \nu} =  R_{\mu \nu} - \frac{1}{2} (- 8 \pi T) g_{\mu \nu}  = 8 \pi T_{\mu \nu} \\ 
 \Rightarrow R_{\mu \nu} = 8 \pi \left ( T_{\mu \nu} - \frac{1}{2} T g_{\mu \nu} \right) \ .
\end{gathered}
\label{eq:efe_style}
\end{equation}

Of course, Eq. \ref{eq:efe_style} is precisely equivalent to Eq. \ref{eq:efe_no_cosm}, only written differently. Now, for the vacuum region, one has $T_{\mu \nu} = 0 \Rightarrow T = 0$, so that Eq. \ref{eq:efe_style} becomes simply

\begin{equation}
 R_{\mu \nu} = 0 \ .
 \label{eq:efe_vac}
\end{equation}

Eq. \ref{eq:efe_vac} is, therefore, a much more convenient way to write the EFE with no cosmological constant and in vacuum, since it suffices, therefore, to find a $g_{\mu \nu}$ such that $R_{\mu \nu}$ vanishes identically. 

We will demand that the solution metric must be \emph{static}, that is, all metric components are independent of $t$ ($\partial_0 g_{\mu \nu} = \partial_t g_{\mu \nu} = 0$), and invariant under time inversion $t \mapsto -t$. This implies that the metric does not present cross spatial-temporal terms ($\mathrm{d}t\mathrm{d}x^i$ and $\mathrm{d}x^i\mathrm{d}t$), and is therefore diagonal, i.e., $g_{\mu \nu} = 0 \ , \forall \mu \neq \nu$. Thus, written in spherical coordinates $\{t, r, \theta, \varphi \}$, the desired metric takes the form

\begin{equation}
 ds^2 = g_{00}\mathrm{d}t^2 + g_{11}\mathrm{d}r^2 + g_{33}\mathrm{d} \theta^2 +  g_{44}\mathrm{d} \varphi^2 \ .
\end{equation}

Now, to guarantee spherical symmetry, the temporal and radial components of the metric must be functions of $r$ only, and the hypersurfaces of constant $t$ and $r$ must have the metric of a $2$-sphere of radius $r$, which, as discussed in \ref{sec:metric}, is given by $r^2 \mathrm{d} \Omega^2 = r^2(\mathrm{d} \theta^2 + \sin^2 \theta \mathrm{d} \varphi^2)$, where $\mathrm{d} \Omega^2$ denotes the metric of $S^2$ (Eq. \ref{eq:s2metric}). Therefore, one has

\begin{equation}
 ds^2 = - e^{2\alpha(r)} \mathrm{d}t^2 + e^{2\beta(r)}\mathrm{d}r^2 + r^2 \mathrm{d} \Omega^2 \ ,
\end{equation}

\noindent where we set $g_{00} \equiv A(r) = -\exp({2\alpha(r)})$ and $g_{11} \equiv B(r) = \exp({2\beta(r)})$ as exponentials to guarantee that the metric signature $(-, +, +, +)$ is conserved. The factor $2$ in the exponents is used for convenience in the following calculations. 

We must now determine the functions $\alpha(r)$ and $\beta(r)$, using the EFE in vacuum, $R_{\mu \nu} = 0$. First, we calculate the Christoffel symbols, from which we calculate the components of the Riemann tensor, which is then contracted to give the Ricci tensor. The explicit calculations are presented in Appendix \ref{ap:cs_rt_sm}. Here we present solely the components of the Ricci tensor, in which we use the labels $(t, r, \theta, \varphi)$ instead of $(0, 1, 2, 3)$,

\begin{equation}
 \begin{aligned}
  R_{tt} &= e^{2(\alpha - \beta)} \left (-(\partial_r \alpha) (\partial_r \beta) + \partial^2_{r^2} \alpha + (\partial_r \alpha)^2 + \frac{2}{r} \partial_r \alpha \right ) ~;
  \\
  R_{rr} &= (\partial_r \alpha) (\partial_r \beta) -\partial^2_{r^2} \alpha - (\partial_r \alpha)^2 + \frac{2}{r} \partial_r \beta ~;
  \\
  R_{\theta \theta} &= e^{-2\beta} (r\partial_r (\beta - \alpha) - 1) + 1 ~;
  \\
  R_{\varphi \varphi} &= \sin^2 \theta \left [e^{-2\beta}(r \partial_r (\beta - \alpha) - 1) + 1\right ] \ ,
 \end{aligned}
\end{equation}

\noindent where we use $\partial_r$ to denote what in fact is the total derivative $\frac{d}{d r}$. Now, according to the EFE, each of these components must vanish, so that

\begin{equation}
\begin{gathered}
  R_{tt} = R_{rr} = 0 \Rightarrow e^{2(\beta-\alpha)}R_{tt} + R_{rr} = 0  \\
 \Rightarrow \frac{2}{r}(\partial_r (\alpha + \beta)) = 0   \\
 \Rightarrow \partial_r (\alpha + \beta) = 0 \\
 \Rightarrow \alpha + \beta = c \ ,
\end{gathered}
\end{equation}

\noindent where $c$ is an arbitrary integration constant. By rescaling the time coordinate as $t \mapsto e^{-c}t$, one gets $\mathrm{d}t \mapsto e^{-c} \mathrm{d}t$, so that $e^{\alpha} \mapsto e^{\alpha - c}$, and therefore $\alpha + \beta = c \mapsto (\alpha -c) + \beta = c \Rightarrow \alpha + \beta = 2c \Rightarrow c = 2c \Rightarrow c = 0$, so that $\alpha = -\beta$. Now, using one more null component of the Ricci tensor, one gets

\begin{equation}
\begin{gathered}
  R_{\theta \theta} = 0 \Rightarrow e^{-2\beta} (r\partial_r (\beta - \alpha) - 1) + 1  = 0  \\
  \Rightarrow e^{-2\beta} (r\partial_r (\beta - (-\beta)) - 1) + 1 = 0 \\
  \Rightarrow e^{-2\beta} (2r\partial_r \beta - 1) = -1  \\
  \Rightarrow \frac{d}{dr} (r e^{-2\beta}) = 1  \\
  \Rightarrow r e^{-2\beta} = r + k  \\
  \Rightarrow e^{2 \beta} = \left ( 1 + \frac{k}{r} \right )^{-1} \ ,
\end{gathered} 
\end{equation}

\noindent where $k$ is a constant of integration, which must now be determined. First of all, notice that

\begin{equation}
 \alpha = -\beta \Rightarrow e^{2 \alpha} =  1 + \frac{k}{r} = -g_{tt} \equiv -g_{00} \ .
\end{equation}

Now, in Sec. \ref{sec:EFEqs}, in which we derived the EFE following the fact that Newtonian gravity should be retrieved in the Newtonian limit, we found, according to Eq. \ref{eq:g00}, that the temporal component of the metric in such a limit must satisfy $g_{00} = -(1 + 2 \Phi)$, where $\Phi = -M/r$ is the gravitational potential created at distance $r$ by a body of mass $M$ (remember, $G = 1$). Therefore, by imposing that at some limit (which is $r \gg M$) the Schwarzschild solution must necessarily retrieve that of Newtonian gravitation, we can establish the equality,

\begin{equation}
\begin{gathered}
g_{00} = - e^{2 \alpha} = - (1 + 2 \Phi)
\\
\Rightarrow 1 + \frac{k}{r} = 1 - \frac{2M}{r} 
\\
\Rightarrow \ k = -2M \ .
\end{gathered}
\end{equation}

In fact, the constant $k$ is related to the \emph{Schwarzschild radius} $-k =2M \equiv r_S$. Its importance will be clear as we move on to the study of black holes. As of now, the determination of such a constant allows us to finally write the Schwarzschild metric in its final form,

\begin{equation}
 ds^2 = - \left ( 1 - \frac{2M}{r} \right ) \mathrm{d}t^2 + \left ( 1 - \frac{2M}{r} \right )^{-1}\mathrm{d}r^2 + r^2 \mathrm{d} \Omega^2 \ .
 \label{eq:sm}
\end{equation}

Notice that in the limit $M \rightarrow 0$, the metric of flat Minkowski space (in spherical coordinates) is retrieved, as expected. Also, such an identification is progressively possible as $r \rightarrow \infty$, which is known as \emph{asymptotic flatness}, in which the Newtonian limit allowed the identification of the Schwarzschild radius.

An important feature of the Schwarzschild metric is that, even though it must necessarily be spherically symmetric, nothing is said about the distribution of energy/momentum which created it, after all, it solely describes the outer vacuum region. In fact, the imposition we made that the metric should be static does not apply to the matter distribution generating it: a Schwarzschild metric may describe the exterior of a collapsing star (which is clearly not static), as long as the collapse is spherically symmetric.

\section{Singularities}
\label{sec:sings}

By inspecting the Schwarzschild metric as presented in Eq. \ref{eq:sm}, it is clear that at $r = 2M$ and $r=0$ the metric is singular, since at $r = 2M$ the temporal component vanishes and the radial component blows up to infinity and the opposite happens for $r=0$. Points where the metric gets singular are known as \emph{singularities}, and they seem to imply that in both these points --- specifically the point $r=0$ or, more precisely, the hypersurface of constant $r=2M$ --- something is not right with the spacetime geometry itself. But, it is important to remember that the metric \emph{components} are coordinate-dependent, so that such singularities may be a mere product of the choice of a problematic coordinate system instead of an ill underlying manifold. 

\emph{Coordinate singularities}, those which emerge as a subtlety of the chosen coordinate system, may be eliminated by a coordinate transformation, which then makes clear that they are not related to the manifold geometry, after all. On the other hand, points of \emph{real singularities} are coordinate-independent and are directly related to the manifold. Although the full characterization of singularities is not a simple task, and goes beyond the scope of this presentation, there is a simple warning for singularities: when curvature becomes infinite. Of course, the Riemann tensor components are also coordinate-dependent, so that we must find actual coordinate-independent quantities directly related to the manifold geometry which become infinite in the singularities --- which guarantees, therefore, that such singularities are real. There is an obvious choice for such invariants: the scalars constructed from the Riemann tensor.

There are many ways to construct scalars from the Riemann tensor, the most trivial of which we already defined: the Ricci scalar $R = g^{\mu \nu}R_{\mu \nu}$. But, if we are, for example, interested in finding the real singularities of the Schwarzschild metric, the Ricci scalar is utterly useless, since it vanishes everywhere for all vacuum solutions. It is then necessary to build higher-order scalars, which can be done in several ways:  $R^{\mu \nu}R_{\mu \nu}$ and $R^{\mu \nu \sigma \rho} R\indices{_{\mu \nu}^{\lambda \kappa}} R_{\lambda \kappa \sigma \rho } $, for example. If any of the many different scalars blows up to infinity in a given point, it is enough to say (but a sufficient condition only) that this point is in fact a real singularity. 

A particularly useful of such invariants is the \emph{Kretschmann scalar}, defined as $\varkappa = R_{\mu \nu \rho \sigma}R^{\mu \nu \rho \sigma}$. It takes some work to calculate it, but for the Schwarzschild metric (SM) it does not vanish, and is given by

\begin{equation}
 \varkappa_{SM} = R_{\mu \nu \rho \sigma}R^{\mu \nu \rho \sigma} = \frac{48M^2}{r^6} ~.
\end{equation}

As $\lim_{r \rightarrow 0} \kappa_{SM} = \infty$, it becomes clear that $r=0$ is a real singularity. To show that $r=2M$ is \emph{not} a real, but only a coordinate singularity, is a bit harder, and will be done through the establishment of the appropriate coordinates in what follows. Nevertheless, although well behaved, the hypersurface defined by the Schwarzschild radius is effectively very interesting, as we shall now see.

The \emph{causal structure} of spacetime is realized through the light cones attached to its points. The events within the cones are causally connected, whilst those in the outside region are not. In terms of the metric, light cones are defined as the radial null curves (of constant $\theta$ and $\varphi$ and null interval $ds^2=0$), which in the Schwarzschild metric in the form of Eq. \ref{eq:sm}, are conditioned by

\begin{equation}
\frac{dt}{dr} = \pm \left ( 1 - \frac{2M}{r} \right )^{-1} \ ,
\end{equation}

\noindent which clearly represents the slope of the light cones in the $t$-$r$ plane. Notice that $r \rightarrow \infty \Rightarrow dt/dr \rightarrow \pm 1$, as expected for the asymptotically flat space, whilst $r \rightarrow 2M \Rightarrow dt/dr \rightarrow  \pm \infty$, so that it is as if the light cones starting closing as $r \rightarrow 2M$. Thus, in the $\{ t, r, \theta, \phi \}$ coordinates, it is as if the light ray never reaches the surface $r=2M$, and, in fact, this is what would be seen by a distant observer: light, or massive particle for that matter, would never be seen to cross the Schwarzschild radius, which does not mean, though, that they effectively do not. In fact, the particles do cross the Schwarzschild radius in a \emph{finite} amount of proper time, which is a fact masked by the $\{ t, r, \theta, \phi \}$ coordinate system. To make it apparent, we must use coordinates where the Schwarzschild radius does not present a singularity.

Such a coordinate system is achieved with the \emph{Eddington-Finkelstein coordinates}, which differs from the original spherical coordinates solely by the redefinition of the temporal coordinate, 

\begin{equation}
t \mapsto v = t + r + 2M \ln \left ( \frac{r}{2M} - 1 \right  ) \ ,
\end{equation}

\noindent so that the metric becomes,

\begin{equation}
ds^2 = - \left ( 1 - \frac{2M}{r} \right  ) \mathrm{d}v^2 + \mathrm{d}v\mathrm{d}r + \mathrm{d}r\mathrm{d}v + r^2 \mathrm{d}\Omega^2 ~.
\end{equation}

Notice that, although $g_{vv}$ vanishes at $r = 2M$, the metric is not singular at this point, which is clear signal that it is not a real singularity. Also, the condition for the radial null curves is given by

\begin{equation}
\begin{aligned}
\frac{dv}{dr} &= \frac{dt}{dr} +  \frac{dr}{dr} + 2M \frac{d}{dr} \left ( \ln \left ( \frac{r}{2M} - 1\right ) \right )  
\\
&= \pm \left ( 1 - \frac{2M}{r}\right )^{-1} + 1 + 2M \left ( \frac{r}{2M} - 1 \right )^{-1} \frac{1}{2M}  
\\
&= \pm \left (\frac{r - 2M}{r}\right )^{-1} + \left ( \frac{r-2M}{2M} \right )^{-1} + 1 
\\
&= \frac{\pm r + 2M}{r - 2M} + 1 
\\
&= \frac{\pm r + 2M + r - 2M}{r - 2M} 
\\
&= \frac{r(1 \pm 1)}{r \left( 1 - \frac{2M}{r} \right )}  
\\ 
\Rightarrow \frac{dv}{dr} &= \frac{1 \pm 1}{\left( 1 - \frac{2M}{r} \right )} \ .
\end{aligned}
\end{equation}

Therefore, by associating $dt/dr =-\left( 1 - \frac{2M}{r}\right )^{-1}$ with an infalling radial null geodesic directed to $r \rightarrow 2M$, and $dt/dr=\left ( 1 - \frac{2M}{r}\right )^{-1}$ with an outgoing, one has, plugging in the respective signal in the last equation above,

\begin{equation}
\left .\frac{dv}{dr} \right\rvert_{in} = 0 \ ,
\end{equation}

\noindent and,

\begin{equation}
\left .\frac{dv}{dr} \right\rvert_{out} = 2 \left( 1 - \frac{2M}{r} \right )^{-1} \ .
\end{equation}

Notice that in such coordinates, the radial infalling null geodesic is such that $v = \mathrm{constant}$, so that in the $r$-$v$ plane, the corresponding null geodesics are always horizontal, whilst the outgoing geodesics are always inclined, positively for $r>2M$, negatively for $r<2M$ and vertical precisely at $r = 2M$, where the light cone has a $90$° opening. Therefore, the light cones only get closed at $r=0$, a real singularity, which is perfectly acceptable. At $r=2M$, though, we see no problems with spacetime geometry: both the metric and the light cones are well-behaved. 

There is a very important point, though: since for $r<2M$ the outgoing radial null geodesics become negatively inclined, light itself becomes directed to $r\rightarrow 0$. Since massive particles must follow timelike paths, it is clear that both light and massive particles are inescapably directed to the real singularity once they cross the hypersurface $r=2M$, which is implied by the causal structure of spacetime itself. After crossing this surface, it is causally impossible to engage a motion in an outgoing radial direction. Therefore, despite being locally regular, the hypersurface $r=2M$ globally acts as the ultimate limit for light and particles, past which there is no return. This hypersurface receives the name \emph{event horizon}.

The event horizon is a hypersurface in spacetime that separates the events connected to infinity by timelike paths from those which are not. Here, ``infinity'' lies in a region sufficiently far from the black hole, where spacetime is flat. Thus, it is clear why after crossing the horizon there is no turning back: it is a consequence of the intrinsic causal structure of spacetime. More precisely, the event horizon is a \emph{null hypersurface}, beyond which timelike paths cannot escape. 

Therefore, although it is not an actual physical singularity, the hypersurface $r=2M$ indeed is of great interest, and behaves very interestingly. Notice, though, that the introduction of the Eddington-Finkelstein coordinates made clear that nothing stops anything, light or massive particles, to cross the event horizon, as the Schwarzschild metric in spherical coordinates seemed to imply. The fact that a distant observer would not be able to watch such fact, though, is absolutely clear in both coordinate systems, which shows that this is what would happen.

\section{The reality of black holes}

Black holes, which may be defined as a region of spacetime separated from infinity by the event horizon, are outstanding objects whose existence mathematically arises from the Schwarzschild solution. Of course, though, there is a considerable gap between mathematics and the real physical world, which is especially important given the high degree of idealization the Schwarzschild metric carries --- namely, its static and spherically symmetric behavior in absolute vacuum.

In fact, for more common astronomical objects like the Sun, whose radius extends up to $r_{Sun} = 10^6 M_{Sun}$, the region limited by the Schwarzschild radius $r=2M$ is way within the star, where the Schwarzschild solutions is no longer valid. In fact, the spacetime within a star must be such that the interior metric must be perfectly smooth at the origin, whilst matching with the exterior Schwarzschild solution, so that, even though the spacetime in the exterior vacuum is Schwarzschild, the highly interesting event horizon and singularity are absent.

Of course, though, a star may evolve in a particular way such that its gravitational pull lead to a collapse in a mass distribution of radius $r<r_s$ and even further to a singularity where all mass is concentrated in $r=0$, and a black hole then emerges. This does not mean, though, that every star will become a black hole at some point. In fact, the maximum mass that can be accommodated in a gravitational stable spherically-symmetric distribution of fixed radius $R$ is $M_{max} = \frac{4}{9}R$ \cite{carroll}. Above this limit, the star will start to contract until it collapses into a black hole in which all matter is concentrated in the singularity.

For stars, the main agent opposing the gravitational collapse comes from the heat pressure as a product of nuclear fusion. When the nuclear fuel is extinguished, the gravitational collapse freely induces the contraction of the star. By force of the \emph{Pauli exclusion principle}, the gravitational collapse of a star may be stopped due to the electron quantum-statistical repulsion: the degeneracy pressure. This may be the final stage of a star, which then becomes a \emph{white dwarf}. 

Now, if a star is sufficiently massive, the gravitational collapse will be so intense that even degeneracy pressure will be overcome, and electrons will combine with protons producing neutrinos, which are expelled, and neutrons, combined in a very dense form of matter. The star becomes then a \emph{neutron star}. Such a process happens to all stars more massive than about $1.4$ solar masses, which is known as the \emph{Chandrasekhar limit}. If the mass is greater than about $4$ solar masses, the \emph{Oppenheimer-Volkoff limit}, the neutron star continues to collapse, eventually becoming a black hole.

Enough astrophysical evidences support the existence of black holes. Indirect observations are possible through $x$-ray bursts \cites{xray1, xray2} originating from objects falling into black holes, apart from the more indirect via of observation through the orbits of celestial bodies around very massive and invisible objects \cites{orbits1, orbits2, orbits3}. Of course, though, in 2016 the direct observation of gravitational waves produced by a binary black hole merger \cite{ligo} provided a shred of utterly strong evidence to the existence of black holes. The direct observation of the phenomena, with experimental data fitting remarkably well the theoretical prevision, is the most recent example of the triumph of Einstein's theory.

Therefore, there is currently little doubt concerning the existence of such incredible astrophysical objects, which were at first mere mathematical curiosities. Later on, in this thesis, we will explore more general metrics and black holes, which will provide even richer possibilities.

\section{Schwarzschild geodesics}

Now that some of the features of the Schwarzschild metric are clear, the natural question is: what are the geodesics of such a spacetime? In fact, the geodesic equation (see Sec. \ref{sec:Geodesics}) correspond to 4 equations, one for each coordinate of the system $\{t, r, \theta, \varphi \}$. They are all coupled, so that their solution is not such a direct task. Fortunately, though, we can rely on the symmetries of the Schwarzschild metric to simplify the solution. We will treat such symmetries through the use of Killing Vectors, as discussed in Sec. \ref{sec:kllvcts}. 

In fact, beforehand we know there are 3 Killing vectors for the spherical symmetry (which imply conservation of the three components of angular momentum) and one for time translations (which imply conservation of energy), each one leading to a constant of motion along the geodesic of a free particle $K_\mu \dfrac{dx^\mu}{d\lambda}$, where $K^\mu$ is the respective Killing vector and $\lambda$ is an affine parameter. Also, for timelike paths, under the choice $\lambda = \tau$, one has the norm of the $4$-velocity vector $\epsilon = -g_{\mu \nu}U^\mu U^\nu = 1$, which is also a conserved quantity. We also have, for massless particles $\epsilon = 0$, as they follow null paths.

We will use the two Killing vectors which imply the conservation of the direction of angular momentum to demand, for a single particle, that its motion happens in the equatorial plane $\theta = \frac{\pi}{2}$, and therefore $d \theta/d \lambda = 0$. Since $\sin^2 \left (\pi/2 \right ) = 1$, the conserved magnitude of the angular momentum comes from $R^\mu = (0, 0, 0, 1)^T$, or $R_\mu = (0, 0, 0, r^2)$, and is given by $L = R_\mu \dfrac{dx^\mu}{d \lambda} = r^2 \dfrac{d\varphi}{d \lambda} $; and the conserved energy comes from $K^\mu = (1, 0, 0, 0)^T$, or $K_\mu= \left( -\left( 1-(2M/r) \right), 0, 0, 0 \right)$, and is given by $E = -K_\mu \dfrac{dx^\mu}{d \lambda} = \left (1 - \dfrac{2M}{r} \right) \dfrac{dt}{d \lambda}$. Of course, though, for massless particles, $E$ and $L$ are respectively the actual energy and angular momentum of the particle, whilst for massive particles these are the same quantities but for unit mass.

Now, from our definition of $\epsilon$ above, one has

\begin{equation}
\epsilon = \left ( 1 - \frac{2M}{r}\right ) \left ( \frac{dt}{d \lambda} \right )^2 - \left ( 1 - \frac{2M}{r}\right )^{-1} \left ( \frac{dr}{d \lambda} \right )^2 - r^2 \left( \frac{d \varphi}{d \lambda} \right )^2 \ , 
\end{equation}

\noindent which, multiplied by $\left ( 1 - \dfrac{2M}{r}\right )$, and substituting $E$ and $L$ as defined above, yields

\begin{equation}
\begin{gathered}
\left ( 1 - \dfrac{2M}{r}\right ) \epsilon = \left ( 1 - \frac{2M}{r}\right )^2 \left ( \frac{dt}{d \lambda} \right )^2 - \left ( \frac{dr}{d \lambda} \right )^2 - r^2 \left ( 1 - \dfrac{2M}{r}\right ) \left( \frac{d \varphi}{d \lambda} \right )^2 
\\
\Rightarrow \left ( 1 - \dfrac{2M}{r}\right ) \epsilon = \left ( 1 - \frac{2M}{r}\right )^2 \left ( E \left ( 1 - \frac{2M}{r}\right )^{-1} \right )^2 - \left ( \frac{dr}{d \lambda} \right )^2 - r^2 \left ( 1 - \dfrac{2M}{r}\right ) \left( \frac{L}{r^2} \right)^2 
\\
\Rightarrow \left ( 1 - \dfrac{2M}{r}\right ) \epsilon = E^2 - \left ( \frac{dr}{d \lambda} \right )^2 - \frac{L^2}{r^2} \left ( 1 - \dfrac{2M}{r}\right ) 
\\
\Rightarrow  E^2 - \left ( \frac{dr}{d \lambda} \right )^2 - \left ( 1 - \dfrac{2M}{r}\right ) \left (\frac{L^2}{r^2} + \epsilon \right) = 0
\\
\Rightarrow \left ( \frac{dr}{d \lambda} \right )^2 = E^2 - \frac{L^2}{r^2} - \epsilon + \frac{2ML^2}{r^3} + \frac{2\epsilon M}{r} 
\\
\Rightarrow \frac{1}{2} \left ( \frac{dr}{d \lambda} \right )^2 = \frac{E^2}{2} - \frac{L^2}{2r^2} - \frac{\epsilon}{2} + \frac{\epsilon M}{r} +  \frac{ML^2}{r^3} ~. 
\end{gathered}
\end{equation}

By defining an effective potential $V(r) = \dfrac{L^2}{2r^2} + \dfrac{\epsilon}{2} - \dfrac{\epsilon M}{r} - \dfrac{ML^2}{r^3} $ and $\mathcal{E} =\dfrac{E^2}{2}$, we arrive at

\begin{equation}
 \frac{1}{2}\left ( \frac{dr}{d \lambda} \right)^2 + V(r) = \mathcal{E} \ .
 \label{eq:orbit}
\end{equation}

Eq. \ref{eq:orbit} gives the radial separation $r(\lambda)$, and is precisely the one found by applying the methods of Newtonian gravitation \cite{marion}. Also, the effective potential $V(r)$ as above defined is exactly the one found for Newtonian orbits, \emph{except} for the last term, which is a genuinely relativistic contribution. 

In fact, the relativistic term $ ML^2/r^3$ makes a very noticeable difference for orbits of small $r$. Specifically, notice that because of this term, $r \rightarrow 0 \Rightarrow V(r) \rightarrow -\infty$, whilst in the Newtonian case $r \rightarrow 0 \Rightarrow V(r) \rightarrow +\infty$. Because of this, in the relativistic potential there must be a $r$ for which $V(r) = 0$. In fact, it is easy to see that this $r$ is exactly the Schwarzschild radius, $r_s = 2M$. 

Depending on the value of $L$, the effective potential $V(r)$ assumes different forms. The orbit a particle will follow will then depend on the relationship between its energy $\mathcal{E}$ and the potential as it moves through: there may be an unbound movement in which exists a \emph{return point} at $r_r$ such that $V(r_r) = \mathcal{E}$, which alters the direction of motion; or the movement may be free, so that such a point does not exist. Also, the particle may be bound between two points (whose orbit, differently from Newtonian orbits, in general do not describe conic sections). A particular case is that in which the particle moves in a circular orbit of radius $r_c$, which happens in $ \left .\dfrac{d}{dr} V(r) \right\rvert_{r_c} = 0$, that is,

\begin{equation}
\begin{gathered}
  \left .\frac{d}{dr} V(r) \right\rvert_{r_c} =  \left .\frac{L^2}{2}\frac{d}{dr} \left( \frac{1}{r^2} \right) - \epsilon M \frac{d}{dr} \left ( \frac{1}{r} \right ) - ML^2 \frac{d}{dr} \left ( \frac{1}{r^3} \right )\right\rvert_{r_c} = 0
 \\
 \Rightarrow -\frac{L^2}{r_c^3} + \epsilon M \frac{1}{r_c^2} + 3ML^2 \frac{1}{r_c^4} = 0
 \\
 \Rightarrow  \epsilon M r_c^2 - L^2 r_c + 3 M L^2 = 0 \ .
\end{gathered}
\end{equation}

For massless particles, for which $\epsilon = 0$, it is easy to see that the circular orbits have radius $r_c = 3M$, which are however unstable, since $d^2 V/dr^2 = -L^2 < 0$. Therefore, a photon with energy $\mathcal{E} = V(3M) = L^2/54M$ would orbit in a circle with this radius, but any perturbation would push it to $r \rightarrow 0$ or $r \rightarrow \infty$.

For massive particles, for which $\epsilon = 1$, the solutions for the circular orbits radii are

\begin{equation}
r_{c_{\pm}} = \frac{L^2}{2M} \left ( 1 \pm \sqrt{1 - \dfrac{12M^2}{L^2}} \right) \ .
\end{equation}

For $L \gg M$, up to first order in $L^{-2}$, one has $\sqrt{1 - \dfrac{12M^2}{L^2}} \approx 1 - \dfrac{6M^2}{L^2}$, so that the possible radii are $r_{s^1} = \dfrac{L^2}{M} - 3M $, which is stable, and $r_{s^2} = 3M$, unstable. For smaller $L$, the radii approximate to each other, until they coincide, which is given for $1 - \dfrac{12M^2}{L^2} = 0 \Rightarrow L = \sqrt{12}M$, for which $r_s = L^2/2M = 12M^2/2M \Rightarrow r_s = 6M$. For $L < \sqrt{12}M$, circular orbits are no longer possible. Therefore, for massive particles, one has unstable circular orbits for $3M < r < 6M$, and stable ones for $r>6M$.

\chapter{Extra dimensions and the EFE in the bulk}
\label{sec:eefe_rs1}

Now that we presented the basis of General Relativity and studied the simplest solution to the EFE, we are almost ready to introduce and discuss the first wonder of gravity, braneworlds. But, before we get to that, it is important to discuss an important modification to GR as considered so far: the introduction of extra spatial dimensions. In this chapter, we shall present a brief overview of extra-dimensional models, thus paving the way to the introduction of braneworlds.

The \emph{dimension} of a given space may be thought of as the number of coordinates demanded to locate any point in it. It is a property intrinsic to the space, and therefore independent of embedding it in a higher dimensional space. As discussed in Appendix \ref{ap:gr}, the dimensionality of manifolds is the same as that of the local Euclidean space. Thus, a circle is one-dimensional --- only one angular coordinate is necessary to specify all its points --- whilst the 2-sphere is two-dimensional.

In the context of standard GR (the one we discussed so far), spacetime is modeled as a 4-dimensional manifold. In quotidian experience, though, it is mostly obvious that the universe we live in has three spatial dimensions. Taking time as an additional dimension is not immediately obvious, but the full concept of spacetime provided by GR along with all its implications make clear that this is the case, even though troubles are found in attempts of visualizing $4$ dimensional constructions, which is a result of our physical intuition being built within what seems pretty confidently to be a universe with three spatial dimensions. 

Still, throughout history many theories were constructed based on the assumption that the universe has more than the 4 dimensions of GR. Gunnar Nordström published in 1914 \cite{nords} a theory aiming at the unification of gravitation and electromagnetism through a single field defined in a $5$-dimensional space. This was the first theory in which gravitation was conceived in geometrical terms, and although it was rapidly made obsolete by GR, the idea of a universe with an extra dimension remained. 

In the 1920's, Kaluza \cite{kaluza} and Klein \cite{klein} developed a theory in which flat spacetime (the Minkowski space, $\mathcal{M}^{1,3}$, in GR) has $S^1$ as an additional spacial dimension, so that it becomes $\mathcal{M}^{1,3} \times S^1$. The theory aimed to unify GR with electromagnetism, within a quantum mechanical interpretation provided by Klein. The basic theory was used in the construction of several unifying-aiming models, and is seen as one of the precursors of string theory, although the lack of perspective of measuring the extra dimension effects, which only arise at Planck scale --- i.e., at distances of the order of $10^{-35} m$ ---, decreased the interest in such models.

After the development of the quantum field theories and the Standard Model (SM), the pursuit of a quantum theory for gravity naturally relighted the interest in extra dimensions, which are necessary for supergravity and string theories. But there is a clear caveat: the inverse-square law of gravity (which is recovered from GR) is a strong evidence for a universe with three spatial dimensions, since its mathematical form implies that the gravitational force decreases according to the inverse of the area of a 2-sphere, which indicates that gravity spreads out in three dimensions. If there were $k$ extra dimensions, the gravitational force should fall of according to $1/r^{2+k}$, as it would then spread through all $3+k$ dimensions. Nevertheless, experiments show that the inverse-square law holds down to distances on the order of $10^{-4} m$ \cite{i-s-l-dev}, which, since we know the law is also valid for astronomical distances, impose a severe constraint on the size of the extra dimensions. The common way to overcome this problem is by considering compact extra dimensions --- like the ones proposed by Kaluza and Klein ---, which, therefore, raises again the problem of their experimental inaccessibility.  

In the 1980's, Rubakov and Shaposhnikov proposed an effective theory with a single extra dimension, in which the SM fields are located in a 4-dimensional submanifold, a domain wall \cite{rubakov} . This work provided the basis for the creation of large extra dimensional models as well as \emph{braneworlds scenarios}, which will soon be further discussed, as the core of Part I. For some other realizations and applications of extra dimensions, see \cites{1_anton, 2_anton, 7_anton, 8_anton}.

In 1998, Arkani-Hamed, Dimopoulos and Dvali (ADD) presented a new framework in which the universe has 6 dimensions, two of which are compactified in a 2-torus $\mathbb{T}^2$ \cite{ADD}. In this model, the SM fields are confined to a 4-dimensional submanifold (the \emph{brane}), whilst gravity is free to propagate also in the extra dimensions (though the full \emph{bulk}). The model allowed the consideration of large extra dimensions, up to a millimeter. The ADD model was proposed to solve the \emph{hierarchy problem}, which lies in the enormous difference between the electroweak scale (which dictates the mass of the SM particles) and the Planck scale (given by the Planck mass, which is inversely proportional to the square root of Newton's gravitational constant, and therefore dictates the strength of gravity)\cite{lisa}. This is one of the major open problems in the SM, which provides no explanation for such a discrepancy. 

Apart from the ADD model, which turned out to solve the hierarchy problem by creating another one, in 1999 a new model was proposed by Randall and Sundrum \cite{rs1}. This model, known as RS1, provides the underlying framework to the introduction of braneworld scenarios, and is therefore of great importance, so that it will be later on discussed in some detail. But, before introducing the model, it is important to make clear what exactly it aims to solve: higher-dimensional EFE.

The RS1 model yields a construction which is known as a \emph{braneworld scenario}. As it will become clear, in this model, the world as we experience it is modeled as a $(3+1)$-dimensional submanifold, the \emph{brane}, embedded in (or bounding a) $5$-dimensional \emph{bulk}. In general terms, a $n$-brane can be modeled as a submanifold of dimension $n$ enclosing a $n+1$ space. In that sense, $n$-branes may be seen as the generalization of \emph{membranes}, which are $2$-branes enclosing $3$-dimensional space. Therefore, the brane corresponding to our $4$-dimensional world --- the braneworld ---, is seen as a $3$-brane, in which the temporal dimension is not explicitly mentioned, although it is implicitly always there. In braneworld scenarios, in which one has a $(n-1)$-submanifold embedded in a $n$-dimensional manifold, we say that the submanifold is of \emph{co-dimension one}.

Now, to study gravitation within these multidimensional braneworld scenarios, it is necessary to establish how GR is set up in higher-dimensional space --- more specifically, how the EFE are changed given the introduction of the extra dimension. 

From now on, since we will work in a field close to the aims of a quantum theory of gravity, we will abandon the geometrized unit system and start using the natural units, as described in Sec. \ref{sec:units}. 
The speed of light will still be unitary, but we will begin to write the gravitational constant, as it is no longer taken as unit.

In Sec. \ref{sec:EFEqs}, we mentioned how it is possible to derive the EFE in vacuum from the Einstein-Hilbert action. In fact, it is also possible to derive the full $4$-dimensional EFE --- in natural units given by $ R^{(4)}_{\mu \nu} - \frac{1}{2} R^{(4)} g^{(4)}_{\mu \nu} + \Lambda^{(4)} g^{(4)}_{\mu \nu} = 8 \pi G^{(4)} T^{(4)}_{\mu \nu}$ ---, from the following action,

\begin{equation}
S^{(4)} = \int d^4 x \sqrt{-g^{(4)}} \left ( \frac{R^{(4)} - 2 \Lambda^{(4)}}{16 \pi G^{(4)}} + \mathcal{L}^{(4)}_M \right ) \ ,
\end{equation}

\noindent where the superscript ${(4)}$ indicate that the objects are $4$-dimensional, and the energy-momentum tensor is retrieved from the matter action $S_M^{(4)} = \int d^4 x \sqrt{-g^{(4)}} \mathcal{L}^{(4)}_M$, according to \cite{carroll},

\begin{equation}
T_{\mu \nu}^{(4)} = \frac{-2}{\sqrt{-g^{(4)}}} \frac{\delta S_M^{(4)}}{\delta g^{\mu \nu}_{(4)}} \ .
\label{eq:e-m-tensor}
\end{equation}

Now, the \emph{Planck mass} (or simply \emph{Planck scale}), as already mentioned, defines the \emph{fundamental natural energy scale} of a gravitational theory. The $4$-dimensional reduced Planck mass is defined, in the natural system ($\hbar = c = 1$) as $\mathcal{M}^{(4)} = \dfrac{1}{\sqrt{8\pi G^{(4)}}} $. 

We can also include the cosmological constant contribution back to the matter Lagrangian, so that it is explicitly seen as the energy density of vacuum. If this is this is the only distribution of energy considered, one has $\mathcal{L}_M^{(4)} = -\Lambda^{(4)}$. Both these facts allow us to write the fundamental $4$-dimensional action as

\begin{equation}
S^{(4)} = \frac{1}{2} \int d^4 x \sqrt{-g^{(4)}} \left (\mathcal{M}^{(4)} \right )^2 R^{(4)} + \int d^4 x \sqrt{-g^{(4)}} (- \Lambda^{(4)}) \ ,
\end{equation}

\noindent in which case, according to Eq. \ref{eq:e-m-tensor}, the energy-momentum tensor will be of the form $T_{\mu \nu} = \Lambda^{(4)} g_{\mu \nu}$, since $\Lambda^{(4)}$ is constant.

Following the same procedure, we can generalize the gravitational action above to arbitrary dimension $n$, as

\begin{equation}
S^{(n)} = \frac{1}{2 \kappa^{(n)}} \int d^n x \sqrt{-g^{(n)}} R^{(n)} + \int d^n x \sqrt{-g^{(n)}} (- \Lambda^{(n)}) \ ,
\end{equation}

\noindent where $\kappa$ is the appropriate Gravitational constant, which may be written in terms of the $n$-dimensional Planck scale. Again, we used the superscript ${(n)}$ to indicate the dimension of the objects. But, since we will mostly be interested in $n = 5$ dimensions in what follows, for $n=5$ we abandon the superscript, as a matter of convenience. Therefore, one has $\Lambda \equiv \Lambda^{(5)}$, $\mathcal{M} \equiv \mathcal{M}^{(5)}$, $g \equiv g^{(5)}$, etc. Now, as $\kappa = \frac{1}{2\mathcal{M}^3}$, the action becomes

\begin{equation}
S = \int d^5x \sqrt{-g} (\mathcal{M}^3 R - \Lambda) \ .
\label{eq:5d-action}
\end{equation}

By varying the action of Eq. \ref{eq:5d-action} with respect to the metric and applying the principle of least action, we arrive at the following $5$-dimensional EFE,

\begin{equation}
G_{MN} = R_{MN} - \frac{1}{2} R g_{M N} = \frac{1}{2\mathcal{M}^3} T_{M N} \ .
\label{eq:efe_5d}
\end{equation}

In five dimensions, we will use upper case Latin indices, $M = (0, 1, 2, 3, 4) \equiv (\mu, 4)$. Notice that, given the reintroduction of the cosmological constant in the matter action, we still have, given that $\Lambda$ is constant $T_{MN} = - \Lambda g_{MN}$ \cites{gabella, goncalves}, if no other form of matter or energy distribution beyond that of vacuum is considered in the matter action.

%
%
%

\chapter{The Randall--Sundrum model}
\label{sec:rs}

In this chapter, we shall present the first Randall--Sundrum model (RS1) \cite{rs1}, given its major conceptual importance for everything that follows regarding braneworlds.

The RS1 model introduces one extra dimension $y$ which is compactified in a $S^1 / \mathbb{Z}_2$ orbifold, where $S^1$ is a circle of radius $R$ and $\mathbb{Z}_2 = \{ 1, -1 \}$ is the multiplicative parity group. In other words, the extra dimension consists of a circle with both sides identified in an equivalence class, so that $y \sim -y$. In the construction, in each of the boundary points $y = 0$ and $y = \pi R = L$ there is one (co-dimension one and $4$-dimensional) $3$-brane extending in the $x^\mu$ directions, bounding the $5$-dimensional bulk. A major point of the construction (specially relevant to the context in which the model was proposed) is that gravity has access to the full 5D bulk, whilst the matter fields are confined to the brane.

Although RS1 cannot be formally recovered from string theory, some of the relations emerging from the model as consequences of the preservation of Poincaré invariance, are the same which arise in the 5D effective theory of the scenario provided by Witten and Hořava, an $11$-dimensional theory on the orbifold $\mathbb{R}^{10} \times S^1 / \mathbb{Z}_2 $ \cite{witten}. 

The metric of the full $5$-dimensional spacetime must be such that it is flat on the branes, since we are not considering any energy-momentum source to curve it, apart from the vacuum energy. Also, we demand that 4-dimensional Poincaré invariance is respected in the $x^\mu$ directions. Therefore, a non-factorizable metric is proposed, such that a \emph{warp factor}, which is a function of the extra dimension, acts on the $4$-dimensional metric. In Gaussian normal coordinates $\{ x^\mu, y \}$  --- discussed in Appendix \ref{ap:whisk} ---  established on the brane at $y=0$, the ansatz metric reads

\begin{equation}
 ds^2 = e^{-2A(y)} \eta_{\mu \nu} \mathrm{d}x^\mu \otimes \mathrm{d}x^\nu + \mathrm{d}y^2 \ ,
\end{equation}

\noindent or, equivalently,

\begin{equation}
 g_{MN} = e^{-2A(y)} \eta_{\mu \nu} + \delta^5_M \delta^5_N \Rightarrow g^{MN} = e^{2A(y)} \eta^{\mu \nu} + \delta^M_5 \delta^N_5 \ .
 \label{eq:rsmetric1}
\end{equation}

The warp factor is written as an exponential $e^{-2A(y)}$ for the same convenient reasons as that of the Schwarzschild solution, and must be a real function. Naturally, the metric must be a solution of the 5D EFE of Eq. \ref{eq:efe_5d}, which will provide us the precise expression of $A(y)$. First, we calculate the Christoffel symbols, from which we calculate the components of the Riemann tensor, which is then contracted to give the Ricci tensor allowing us to construct the Einstein tensor and identify its components with those of the energy-momentum tensor.

From Eq. \ref{eq:rsmetric1}, it is clear that $g_{MN} = g_{MN}(y)$, so that we will have $ \partial_\mu g_{M N} = 0$, i.e., the only non-vanishing derivative is that with respect to the extra dimension, $\partial_5 = \partial_y$ (which we again will use to denote the total derivative $\frac{d}{dy}$), so that the only non-vanishing Christoffel symbols (Eq. \ref{eq:christsymb}) are

\begin{equation}
\begin{gathered}
  \Gamma^5_{\mu \nu} = \frac{1}{2} g^{5L}(-\partial_L g_{\mu \nu}) = -\frac{1}{2} g^{55}(\partial_5 g_{\mu \nu}) = -\frac{1}{2} \partial_y (e^{-2A} \eta_{\mu \nu}) = -\frac{1}{2} \eta_{\mu \nu} (-2 e^{-2A} \partial_y A)   \\
  \Rightarrow  \Gamma^5_{\mu \nu} = g_{\mu \nu} \partial_y A \ ,
\end{gathered}
\end{equation}

\noindent and

\begin{equation}
\begin{gathered}
  \Gamma^\mu_{\nu 5} = \frac{1}{2} g^{\mu L}(\partial_5 g_{L \nu}) = \frac{1}{2} g^{\mu \lambda}(\partial_5 g_{\lambda \nu}) = \frac{1}{2} e^{2A} \eta^{\mu \lambda}\partial_y (e^{-2A} \eta_{\lambda \nu}) = \frac{1}{2} e^{2A} \eta^{\mu \lambda}\eta_{\lambda \nu} (-2 e^{-2A} \partial_y A) 
  \\
  \Rightarrow  \Gamma^\mu_{\nu 5}  =  - \delta^\mu_\nu \partial_y A \ .
\end{gathered}
\end{equation}

From Eq. \ref{eq:ricci_tens}, it is clear that $R_{\mu 5} = R_{5 \mu} = 0$, so that it remains

\vspace{-.4cm}

\begin{equation}
\begin{aligned}
R_{\mu \nu} &= \partial_S \Gamma^S_{\mu \nu} - \cancelto{^0}{\partial_\nu \Gamma^S_{\mu S}} + \Gamma^S_{S L}\Gamma^L_{\mu \nu} - \Gamma^S_{\nu L}\Gamma^L_{\mu S} 
\\
&= \partial_5 \Gamma^5_{\mu \nu} + \Gamma^\sigma_{\sigma 5}\Gamma^5_{\mu \nu} - \Gamma^\sigma_{\nu 5}\Gamma^5_{\mu \sigma} - \Gamma^5_{\nu \lambda}\Gamma^\lambda_{\mu 5} 
\\
&= \partial_y ( g_{\mu \nu} \partial_y A) + (-\delta^\sigma_\sigma \partial_y A)(g_{\mu \nu} \partial_y A) - (-\delta^\sigma_\nu \partial_y A)(g_{\mu \sigma} \partial_y A) - (g_{\nu \lambda}\partial_y A)(-\delta^\lambda_\mu \partial_y A) 
\\
&= \partial_y ( e^{-2A} \eta_{\mu \nu} \partial_y A) -4 g_{\mu \nu} (\partial_y A)^2 + g_{\mu \nu} (\partial_y A)^2 + g_{\mu \nu} (\partial_y A)^2 
\\
&= \eta_{\mu \nu} [(\partial_y A)(-2 e^{-2A}\partial_y A) + e^{-2A}(\partial^2_{y^2}A)] -2 g_{\mu \nu} (\partial_y A)^2 
\\
&= e^{-2A}\eta_{\mu \nu} (-2(\partial_y A)^2 + (\partial^2_{y^2}A)) - 2 g_{\mu \nu} (\partial_y A)^2 
\\
&= g_{\mu \nu}(-2(\partial_y A)^2 + (\partial^2_{y^2}A -2(\partial_y A)^2) 
\\
\Rightarrow R_{\mu \nu} &= g_{\mu \nu} \left( \partial^2_{y^2}A -4(\partial_y A)^2 \right) \ ;
\end{aligned}
\end{equation}

\noindent And

\vspace{-.4cm}

\begin{equation}
\begin{aligned}
R_{55} &= \partial_S \cancelto{^0}{\Gamma^S_{5 5}} -\partial_5 \Gamma^S_{5 S} + \Gamma^S_{S L}\cancelto{^0}{\Gamma^L_{5 5}} - \Gamma^S_{5 L}\Gamma^L_{5 S} 
\\
&= -\partial_5 \Gamma^\sigma_{5 \sigma} - \Gamma^\sigma_{5 \lambda}\Gamma^\lambda_{5 \sigma} = -\partial_y (-\delta^\sigma_\sigma \partial_y A ) - (-\delta^\sigma_\lambda \partial_y A)(-\delta^\lambda_\sigma \partial_y A) 
\\
&= 4 \partial^2_{y^2}A - \delta^\lambda_\lambda (\partial_y A)^2 =  4 \partial^2_{y^2}A - 4 (\partial_y A)^2 
\\
\Rightarrow R_{55} &= 4 \left( \partial^2_{y^2}A -(\partial_y A)^2 \right)  \ .
\end{aligned}
\end{equation}

Therefore, the Ricci scalar reads

\begin{equation}
\begin{aligned}
R &= g^{MN}R_{MN} = g^{\mu \nu} R_{\mu \nu} + g^55 R_55 
\\
&= g^{\mu \nu}[g_{\mu \nu} ((\partial^2_{y^2}A) -4(\partial_y A)^2)] + 4(\partial^2_{y^2}A -(\partial_y A)^2)
\\
&= \delta^\mu_\mu[(\partial^2_{y^2}A) -4(\partial_y A)^2] + 4(\partial^2_{y^2}A -(\partial_y A)^2) 
\\
&= 4 \partial^2_{y^2}A - 16(\partial_y A)^2 + 4\partial^2_{y^2}A - 4(\partial_y A)^2 
\\
\Rightarrow R &= 4 \left(2 \partial^2_{y^2}A - 5(\partial_y A)^2 \right)
\end{aligned}
\end{equation}

We can now calculate the Einstein tensor $G_{MN} = G_{\mu \nu} + G_{55}$, such that

\begin{equation}
\begin{aligned}
G_{\mu \nu} &= R_{\mu \nu} - \frac{1}{2}R g_{\mu \nu} 
\\
\Rightarrow G_{\mu \nu} &= (g_{\mu \nu} (\partial^2_{y^2}A -4(\partial_y A)^2) - \frac{1}{2}( 4 (2 \partial^2_{y^2}A - 5(\partial_y A)^2) g_{\mu \nu})
\\ 
\Rightarrow G_{\mu \nu} &= g_{\mu \nu}((1-4) \partial^2_{y^2}A + (-4+10)(\partial_y A)^2 ) 
\\
\Rightarrow G_{\mu \nu} &= 3g_{\mu \nu} \left(2 (\partial_y A)^2 -\partial^2_{y^2}A \right)\ ,
\end{aligned}
\end{equation}

\noindent and 

\begin{equation}
\begin{aligned}
G_{55} &= R_{55} - \frac{1}{2}R g_{55} 
\\
\Rightarrow G_{55} &= (4(\partial^2_{y^2}A -(\partial_y A)^2) - \frac{1}{2}(4 (2 \partial^2_{y^2}A - 5(\partial_y A)^2))g_{55} 
\\ 
\Rightarrow G_{55} &= (4-4) \partial^2_{y^2}A + (-4+10)(\partial_y A)^2  
\\
\Rightarrow G_{55} &= 6 (\partial_y A)^2 \ .
\end{aligned}
\end{equation}

Now we can use the EFE $G_{MN} = -\dfrac{1}{2\mathcal{M}^3}T_{MN}$, with $T_{MN} = -\Lambda g_{MN}$. For the extra dimension component,  

\begin{equation}
 G_{55} = -\frac{\Lambda g_{55}}{2\mathcal{M}^3} \Rightarrow \frac{d}{dy} A = \pm \sqrt{-\frac{\Lambda g_{55}}{12\mathcal{M}^3}} \ .
 \label{eq:G55}
\end{equation}

Notice that, since $\mathcal{M} > 0$, to guarantee that $A(y)$ is a real function, so that the warp factor represents an exponential decay without oscillations (which is the aim here), we must impose $\Lambda < 0$. In fact, this characterizes an \emph{anti-de Sitter spacetime}, a maximally symmetric Lorentzian manifold with constant negative curvature, expressed through the cosmological constant, as discussed in Sec. \ref{sec:ads}. This means that the bulk between the branes (in which the $5$-dimensional $\Lambda$ is defined), is a $5$-dimensional anti-de Sitter space, $AdS_5$. This particular spacetime will play a major role later in the text, and shall therefore be revisited.

Now, by defining $k \equiv \sqrt{-\Lambda /12\mathcal{M}^3}$, as $g_{55}=1$, the integration of Eq. \ref{eq:G55}, gives $A(y) = \pm k y $. Nevertheless, since the extra dimension carries the orbifold symmetry (i.e., $y \in S^1 / \mathbb{Z}_2 $, so that invariance under $ y \rightarrow -y$ is guaranteed), one has

\begin{equation}
A(y) = k |y| \ .
\end{equation}

With $A(y)$ defined, we already have the full RS1 metric, as desired. But, we only used the extra-dimensional component of the EFE above, so that there may be still useful information in the $\mu \nu$ components. To perform such an analysis, let us first get the first derivative of $A(y)$,

\begin{equation}
 \frac{d}{dy} A(y) = k \frac{d}{dy} |y| = k \mathrm{sgn}(y) \ ,
\end{equation}

\noindent where $\mathrm{sgn}(y)$ is the signal function

\begin{equation}
\mathrm{sgn}(y) = 
\begin{cases}
1 & \text{ if } y > 0 \ ; \\ 
0 & \text{ if } y = 0 \ ;  \\
-1 & \text {if} y < 0 \ .
\end{cases}
\end{equation}

Of course, it is clear that $(\mathrm{sgn}(y))^2 = 1$, $\forall y \in (-L, L)$. Also, one has $\dfrac{d}{dy}\mathrm{sgn}(y) = 2\delta(y)$, but, since $y \in S^1 / \mathbb{Z}_2 $, it is reasonable to express the delta function in both the bounds of the orbifold, $y=0$ and $y=L$, that is,

\begin{equation}
k \frac{d}{dy}\mathrm{sgn}(y) = \frac{d^2}{dy^2}A(y) = 2k (\delta(y) - \delta(y-L)) \ .
\end{equation}

Then, of course, we can check the solution above by using the $\mu \nu$ components of the EFE,

\begin{equation}
\begin{gathered}
G_{\mu \nu} = -\frac{\Lambda g_{\mu \nu}}{2\mathcal{M}^3} 
\\
\Rightarrow 3g_{\mu \nu} \left(2 \left ( \frac{d}{dy} A(y) \right)^2 -\frac{d^2}{dy^2} A(y) \right) = -\frac{\Lambda g_{\mu \nu}}{2\mathcal{M}^3} 
\\
\Rightarrow 3g_{\mu \nu} \left(2 k^2(\mathrm{sgn}(y))^2 - 2k (\delta(y) - \delta(y-L)) \right) =  -\frac{\Lambda g_{\mu \nu}}{2\mathcal{M}^3}
\\
\Rightarrow (6k^2)g_{\mu \nu} - 6k (\delta(y) - \delta(y-L))g_{\mu \nu} = -\frac{\Lambda g_{\mu \nu}}{2\mathcal{M}^3} \ .
\label{eq:Gmunu}
\end{gathered}
\end{equation}

Now, it is straightforward to identify $6k^2 = 6 \left (\sqrt{-\Lambda/12\mathcal{M}^3} \right )^2 = -\Lambda/2\mathcal{M}^3 $. On the other hand, $6k (\delta(y) - \delta(y-L))g_{\mu \nu} \neq 0, \forall y \in  S^1 / \mathbb{Z}_2$, which means that there is something missing in the $\mu \nu$ components of the 5D EFE. To account for such an absence, we consider the \emph{energy density} of each brane, which can be seen as the \emph{brane tension}. Such consideration is achieved by adding to the action of Eq. \ref{eq:5d-action} the matter action of each brane (the first located at $y=0$ and the second at $y=L$) due to their respective tensions $\sigma_1$ and $\sigma_2$, that is,

\begin{equation}
\begin{gathered}
S_1 = \int d^4x dy \sqrt{-g} (-\sigma_1) \delta(y) = \int d^4x \sqrt{-q_1} (-\sigma_1)\ ,
\end{gathered}
\end{equation}

\noindent and

\begin{equation}
\begin{gathered}
S_2 = \int d^4x dy \sqrt{-g} (-\sigma_2) \delta(y-L) = \int d^4x \sqrt{-q_2} (-\sigma_2) \ .
\label{eq:s1,s2}
\end{gathered}
\end{equation}

Notice how, in fact, the second derivative of $A(y)$ when evaluated and integrated in each one of the branes leave only a $4$-dimensional action, as it should be. Moreover, $q_1$ and $q_2$ are respectively the determinants of the induced metric in each brane, as discussed in Appendix \ref{ap:whisk}.

Now, by considering $S_M = \int d^5x \sqrt{-g}\left(-\Lambda -\sigma_1 \delta(y) - \sigma_2 \delta(y-L) \right )$, and given that both brane tensions are constant (as well as the delta functions of $y$ with respect to $x^\mu$), one has that the $\mu \nu$ components of the energy-momentum tensor becomes

\begin{equation}
T_{\mu \nu} = - \left (\Lambda + \sigma_1 \delta(y) + \sigma_2 \delta(y-L) \right)g_{\mu \nu} \ .
\end{equation}

Notice that despite the redefinition of $T_{\mu \nu}$, Eq. \ref{eq:G55} still holds consistent, because in the bulk the brane tensions are not defined, so that $T_{55} = -\Lambda/12\mathcal{M}^3$, i.e., the extra-dimensional component really does not have any contribution from $S_1$ and $S_2$, as it should be. So, in fact, one has

\begin{equation}
\begin{gathered}
  G_{\mu \nu} =  \left ( -\frac{\Lambda}{2\mathcal{M}^3} - \frac{\sigma_1 \delta(y)}{2\mathcal{M}^3} - \frac{\sigma_2 \delta(y-L)}{2\mathcal{M}^3} \right )  g_{\mu \nu} 
\\
\Rightarrow
6k^2g_{\mu \nu} - 6k\delta(y)g_{\mu \nu} +6k\delta(y-L)g_{\mu \nu} =  -\frac{\Lambda}{2\mathcal{M}^3}g_{\mu \nu} - \frac{\sigma_1 \delta(y)}{2\mathcal{M}^3}g_{\mu \nu} - \frac{\sigma_2 \delta(y-L)}{2\mathcal{M}^3}g_{\mu \nu} \ ,
\end{gathered}
\end{equation}

\noindent which yields the identification

\begin{equation}
-6k = \frac{-\sigma_1}{2\mathcal{M}^3} \Rightarrow \sigma_1 = 12k\mathcal{M}^3 \ ,
\end{equation}

\noindent and

\begin{equation}
6k = \frac{-\sigma_2}{2\mathcal{M}^3} \Rightarrow \sigma_2 = -12k\mathcal{M}^3 \ .
\end{equation}

This makes clear that the branes have tensions of equal magnitude, but opposite sign $\sigma_1 = -\sigma_2 \equiv \sigma$. We also have

\begin{equation}
 \sigma^2 = (12)^2\left( \frac{-\Lambda}{12\mathcal{M}^3} \right )(\mathcal{M}^3)^2 \Rightarrow \Lambda = \frac{-\sigma^2}{12\mathcal{M}^3} \ .
 \label{eq:lambda_and_sigma}
\end{equation}

Eq. \ref{eq:lambda_and_sigma} serves as an \emph{fine tuning} in the model, as the brane tension $\sigma$ is a free parameter which can be chosen to determine the high energy scale of the theory \cite{emgd}. It is the negative cosmological constant what prevents gravity from leaking into the extra dimension at low energies \cite{maartens}.

Now that one has shown that he braneworld scenario proposed is consistent with all the non-vanishing components of the 5D EFE, from which insightful conclusions concerning the braneworld may be derived, we can finally write the full metric underlying the RS1 model

\begin{equation}
 ds^2 = \exp\left(-2\sqrt{\frac{-\Lambda}{12\mathcal{M}^3}}|y|\right) \eta_{\mu \nu} \mathrm{d}x^\mu \otimes \mathrm{d}x^\nu + \mathrm{d}y^2  \ .
\end{equation}

This is the underlying framework. Everything that will be presented in the next few sections concerning braneworld gravity will be constructed within the RS1 braneworld, from which, as we shall see, even deeper developments may be derived.

\chapter{Effective EFE on the brane}
\label{sec:effe}

We are considering GR with 5 dimensions, and at this point we already know an explicit example of braneworld: the one described by the RSI model. But, if there is a fifth dimension, only gravity has access to it --- so, trapped on the brane, how could we tell if there really is an extra bulk dimension? Put differently: is there a way to tell the difference between a 4-dimensional brane embedded in a higher dimensional bulk and a 4D universe independent of any embedding described by the standard formulation of GR?

To address this question, in this section we will present the effective EFE in the co-dimension one $3$-brane, through the projection of the 5D EFE defined in the bulk onto the embedded brane. As we shall see, we will arrive at a high-energy correction to the 4D EFE (in a sense which will become clear), as well as in a term containing information regarding the bulk gravitational field, which allows us to answer the questions above, and tell the difference. We will base our presentation in the work of Shiromizu, Sasaki and Maeda (SSM) \cites{ssm1, kuerten}. 

Given the central importance of the effective EFE in some of the following topics of this thesis, throughout this section we will focus on providing a derivation for the effective EFE in great detail, so that not much of the discussion will be directed to the mathematical details behind hypersurfaces and submanifolds. For this formalism, we wrote Appendix \ref{ap:whisk}.

Let $\Sigma_4$ be the $4$-dimensional submanifold of induced metric $q_{\mu \nu}$ (i.e., the $3$-brane) embedded in the $5$-dimensional bulk $\mathcal{M}_5$, of metric $g_{M N}$. In this section, we will use superscripts (or subscripts) $(4)$ or $(5)$ to denote the dimensionality of the objects whenever necessary.

We start with the Gauss equation, which gives the $4$-dimensional Riemann tensor in terms of the $5$-dimensional one,

\begin{equation}
^{(4)}R\indices{_{ABC}^F} =  \  ^{(5)}R\indices{_{DGH}^E} q\indices{_E^F} q\indices{_A^D} q\indices{_B^G} q\indices{_C^H} + K_{CA}K\indices{_B^F} - K_{CB}K\indices{_A^F} \ ,
\label{eq:gauss}
\end{equation}

\noindent where $n^M$ denotes the unit vector normal to $\Sigma_4$, such that $g_{MN} n^M n^N = 1$,  and  $K_{MN}$ is its extrinsic curvature, of trace $K = K\indices{^M_M}$. By contracting Eq. \ref{eq:gauss} in its indices $B$ and $F$, one gets the $4$-dimensional Ricci tensor

\begin{equation}
\begin{aligned}
^{(4)}R_{AC} &= \ ^{(4)}R\indices{_{AFC}^F} =  \  ^{(5)}R\indices{_{DGH}^E} q\indices{_A^D} q\indices{_C^H} \left (q\indices{_E^F} q\indices{_F^G} \right)  + K_{CA}K\indices{_F^F} - K_{CF}K\indices{^F_A} 
\\
&=  \  ^{(5)}R\indices{_{DGH}^E} q\indices{_A^D} q\indices{_C^H} \left(q\indices{_E^G} \right) + K_{CA}K - K_{CF}K\indices{^F_A} 
\\
&=  \  ^{(5)}R\indices{_{DGH}^E} q\indices{_A^D} q\indices{_C^H} \left(g\indices{_E^G} - n_E n^G \right) + K_{CA}K - K_{CF}K\indices{^F_A} 
\\
&=  \  \left( ^{(5)}R\indices{_{DGH}^E} g\indices{_E^G} \right) q\indices{_A^D} q\indices{_C^H} - \ ^{(5)}R\indices{_{DGH}^E} n_E n^G q\indices{_A^D} q\indices{_C^H}  + K_{CA}K - K_{CF}K\indices{^F_A}
\\
\Rightarrow \ ^{(4)}R_{AC} &=  \  ^{(5)}R_{DH} q\indices{_A^D} q\indices{_C^H} - \tilde{\mathcal{E}}_{AC}  + K_{CA}K - K_{CF}K\indices{^F_A}  \ ,
\label{eq:4dricci}
\end{aligned}
\end{equation}

\noindent where we defined $\tilde{\mathcal{E}}_{AC} \equiv \ ^{(5)}R\indices{_{DGH}^E} n_E n^G q\indices{_A^D} q\indices{_C^H}$, and used the definition of the induced metric $q_{AB} = g_{AB} - n_A n_B \Rightarrow q\indices{^A_B} = g\indices{^A_B} - n^An_B$. Notice that only the 5-dimensional metric can be contracted with 5-dimensional objects. By contracting again, one gets the 4-dimensional Ricci scalar

\begin{equation}
\begin{aligned}
^{(4)}R = \ ^{(4)}R\indices{_{AC}} q^{AC} &= \ \indices{^{(5)}}R\indices{_{DH}} \left (q\indices{^D_A} q\indices{^H_C}q^{AC} \right) - \tilde{\mathcal{E}}_{AC} q^{AC} + \left( q^{AC}K_{CA} \right) K - K_{CB} \left (K\indices{^B_A}q^{AC} \right) 
\\
&= \ ^{(5)}R\indices{_{DH}} \left (q\indices{^{DH}} \right) - \tilde{\mathcal{E}}_{AC} q^{AC} + (K)K - K_{CB}\left(K^{CB}\right) 
\\
\Rightarrow  \ ^{(4)}R &= \ ^{(5)}R\indices{_{DH}} q\indices{^{DH}} - \tilde{\mathcal{E}}_{DH} q^{DH} + K^2 - K_{DH}K^{DH} \ .
\end{aligned}
\end{equation}

With these, we can calculate the $4$-dimensional Einstein tensor,

\begin{equation}
\begin{aligned}
^{(4)}G_{AC} &=  \ ^{(4)}R_{AC} - \frac{1}{2} \ ^{(5)}R q_{AC} 
\\
&=  \  ^{(5)}R_{DH} q\indices{_A^D} q\indices{_C^H} - \tilde{\mathcal{E}}_{AC}  + K_{AC}K - K_{CF}K\indices{^F_A} 
\\
&- \frac{1}{2} \left ( \ ^{(5)}R\indices{_{DH}} q\indices{^{DH}} - \tilde{\mathcal{E}}_{DH} q^{DH} + K^2 - K_{DH}K^{DH} \right )q_{AC} 
\\
\Rightarrow ^{(4)}G_{AC} &=  \  ^{(5)}R_{DH} q\indices{_A^D} q\indices{_C^H} - \tilde{\mathcal{E}}_{AC}  + K_{AC}K - K_{CF}K\indices{^F_A} 
\\
&- \frac{q_{AC}}{2}\left (K^2 - K_{DH}K^{DH} \right ) - \frac{1}{2} \left ( \ ^{(5)}R\indices{_{DH}} q\indices{^{DH}} q_{AC} - \tilde{\mathcal{E}}_{DH} q^{DH} q_{AC} \right ) \ .
\label{eq:g1}
\end{aligned}
\end{equation}

The first of the two terms in the last parenthesis of Eq. \ref{eq:g1} can be written as

\begin{equation}
\begin{aligned}
^{(5)}R\indices{_{DH}} q\indices{^{DH}} q_{AC} &= ^{(5)}R\indices{_{DH}} \left( g\indices{^{DH}} - n^D n^H \right ) q_{AC} 
\\
&= \left ( \ ^{(5)}R\indices{_{DH}} g\indices{^{DH}} \right ) \left( q_{AC} \right ) - \ ^{(5)}R\indices{_{DH}} n^D n^H  q_{AC} 
\\ 
&= \ ^{(5)}R \left ( g_{DH}  q\indices{_A^D} q\indices{_C^H} \right ) - \ ^{(5)}R\indices{_{DH}} n^D n^H  q_{AC} 
\\
\Rightarrow \ ^{(5)}R\indices{_{DH}} q\indices{^{DH}} q_{AC} &=  \ ^{(5)}R  g_{DH}  q\indices{_A^D} q\indices{_C^H} - \ ^{(5)}R\indices{_{DH}} n^D n^H  q_{AC} \ ,
\end{aligned}
\end{equation}

\noindent and, similarly,

\begin{equation}
  \tilde{\mathcal{E}}_{DH} q^{DH} q_{AC}  = \  ^{(5)}R_{DH}n^Dn^H q_{AC} \ ,
\end{equation}

\noindent so that, back in Eq. \ref{eq:g1}, one has

\begin{equation}
\begin{aligned}
^{(4)}G_{AC} &=  \  ^{(5)}R_{DH} q\indices{_A^D} q\indices{_C^H} - \tilde{\mathcal{E}}_{AC}  + K_{AC}K - K_{CD}K\indices{^D_A} - \frac{q_{AC}}{2}\left (K^2 - K_{DH}K^{DH} \right ) 
\\
&- \frac{1}{2} \left [ \left( \ ^{(5)}R  g_{DH}  q\indices{_A^D} q\indices{_C^H} - \ ^{(5)}R\indices{_{DH}} n^D n^H  q_{AC}  \right) - \left( \  ^{(5)}R_{DH}n^Dn^H q_{AC}\right) \right ] 
\\
&=  \  ^{(5)}R_{DH} q\indices{_A^D} q\indices{_C^H} - \tilde{\mathcal{E}}_{AC}  + K_{AC}K - K_{CD}K\indices{^D_A} - \frac{q_{AC}}{2}\left (K^2 - K_{DH}K^{DH} \right ) 
\\
&- \frac{1}{2} \left( \ ^{(5)}R  g_{DH}  q\indices{_A^D} q\indices{_C^H} \right ) + \ ^{(5)}R\indices{_{DH}} n^D n^H  q_{AC} 
\\
\Rightarrow \ ^{(4)}G_{AC} &= \left ( \ ^{(5)}R_{DH} - \frac{1}{2} \ ^{(5)}R  g_{DH} \right ) q\indices{_A^D} q\indices{_C^H} + \ ^{(5)}R\indices{_{DH}} n^D n^H  q_{AC} 
\\
&+ K_{AC}K - K_{CD}K\indices{^D_A} - \frac{q_{AC}}{2}\left (K^2 - K_{DH}K^{DH} \right ) - \tilde{\mathcal{E}}_{AC}  \ .
\label{eq:g2}
\end{aligned}
\end{equation}

Notice that in the first parenthesis one has the 5D EFE $\ ^{(5)}R_{DH} - \frac{1}{2} \ ^{(5)}R  g_{DH}  = k_5^2 T_{DH}$, where $T_{DH}$ is the 5-dimensional energy-momentum tensor and $k_5^2 = 8 \pi G_5$, being $G_5$ the 5-dimensional gravitational constant. By contracting the 5D EFE, to rewrite it in a different manner as we did before in Sec. \ref{sec:schwarzschild}, one has

\begin{equation}
 \begin{gathered}
  \left ( \ ^{(5)}R_{DH}g^{DH} \right) - \frac{1}{2} \ ^{(5)}R  \left (g_{DH}g^{DH} \right )= k_5^2 \left( T_{DH}g^{DH} \right)  \\
  \Rightarrow \ ^{(5)}R  - \frac{5}{2} \ ^{(5)}R = k_5^2 T  \\
  \Rightarrow \ ^{(5)}R  = \frac{-2 k_5^2}{3}T \ ,
 \end{gathered}
\end{equation}

\noindent where $T = T_{DH}g^{DH} = T\indices{_D^D}$ and $g_{DH}g^{DH} = \delta^D_D = 5$. Therefore, back in the 5D EFE,

\begin{equation}
\begin{gathered}
^{(5)}R_{DH} - \frac{1}{2} \left ( \frac{-2 k_5^2}{3}T \right )  g_{DH}  = k_5^2 T_{DH}
\\
\Rightarrow  \ ^{(5)}R_{DH} = k_5^2 \left ( T_{DH} - \frac{T}{3} g_{DH} \right ) \ .
\label{eq:r_dh}
\end{gathered}
\end{equation}

Substituting the 5D EFE and Eq. \ref{eq:r_dh} into Eq. \ref{eq:g2}, one has

\begin{equation}
\begin{aligned}
^{(4)}G_{AC} &= \left ( k_5^2 T_{DH} \right ) q\indices{_A^D} q\indices{_C^H} + \left[ k_5^2 \left ( T_{DH} - \frac{T}{3} g_{DH} \right ) \right ] n^D n^H  q_{AC} 
\\
&+ K_{AC}K - K_{CD}K\indices{^D_A} - \frac{q_{AC}}{2}\left (K^2 - K_{DH}K^{DH} \right ) - \tilde{\mathcal{E}}_{AC} 
\\
&= k_5^2 \left ( T_{DH} q\indices{_A^D} q\indices{_C^H} + \ q_{AC} \left [ T_{DH} n^D n^H - \frac{T}{3} \left (g_{DH} n^D n^H \right ) \right ] \right )  
\\
&+ K_{AC}K - K_{CD}K\indices{^D_A} - \frac{q_{AC}}{2}\left (K^2 - K_{DH}K^{DH} \right ) - \tilde{\mathcal{E}}_{AC}  
\\
\Rightarrow ^{(4)}G_{AC} &= k_5^2 \left [ T_{DH} q\indices{_A^D} q\indices{_C^H} + \ q_{AC} \left ( T_{DH} n^D n^H - \frac{T}{3} \right ) \right ]  
\\
&+ K_{AC}K - K_{CD}K\indices{^D_A} - \frac{q_{AC}}{2}\left (K^2 - K_{DH}K^{DH} \right ) - \tilde{\mathcal{E}}_{AC} \ ,
\label{eq:g3}
\end{aligned}
\end{equation}

\noindent where we used $g_{MN} n^M n^N = 1$.

Our aim is to write an equation for $\ ^{(4)}G_{AC}$ only in terms of the energy-momentum tensor, since this is exactly what the EFE is: an equation between the Einstein tensor, containing information about spacetime geometry, and the energy-momentum content which curves spacetime. To achieve this, we must write the extrinsic curvature $K_{AC}$ and $\tilde{\mathcal{E}}_{AC}$ in terms of $T_{AC}$. We will begin with the latter. 

First, remember that we can decompose the Riemann tensor into the Weyl tensor, the Ricci tensor and the Ricci scalar, as given in Eq. \ref{eq:weyl_2} of Appendix \ref{ap:gr}. For $n=5$, one has

\vspace{-.4cm}

\begin{equation}
 ^{(5)}R_{DGHF} = \ ^{(5)}C_{DGHF} +  \frac{1}{3} \left ( g_{D[H}\ ^{(5)}R_{F]G} - g_{G[H}\ ^{(5)}R_{F]D} \right ) - \frac{1}{12} \left ( g_{D [H} g_{F ] G} \ ^{(5)}R \right) \ .
\end{equation}

Therefore, from the definition of $\tilde{\mathcal{E}}_{AC}$,

\vspace{-.4cm}

\begin{equation}
\begin{aligned}
\tilde{\mathcal{E}}_{AC} &= \ ^{(5)}R\indices{_{DGH}^E} n_E n^G q\indices{_A^D} q\indices{_C^H} 
\\ 
&= \left ( \ ^{(5)}R\indices{_{DGHF}}g^{FE} \right) n_E n^G q\indices{_A^D} q\indices{_C^H} 
\\
&= \left ( \ ^{(5)}R\indices{_{DGHF}} \right ) \left (g^{FE} n_E \right) n^G q\indices{_A^D} q\indices{_C^H} 
\\
&=  \ ^{(5)}C_{DGHF} n^F n^G q\indices{_A^D} q\indices{_C^H} +  \frac{1}{3} \left ( g_{D[H}\ ^{(5)}R_{F]G} - g_{G[H}\ ^{(5)}R_{F]D} \right ) n^F n^G q\indices{_A^D} q\indices{_C^H}
\\
&- \frac{1}{12} \left ( g_{D [H} g_{F ] G} \ ^{(5)}R \right) n^F n^G q\indices{_A^D} q\indices{_C^H} 
\\
\Rightarrow \tilde{\mathcal{E}}_{AC}  &=  \mathcal{E}_{AC} +  \frac{1}{3} \left ( g_{D[H}\ ^{(5)}R_{F]G} n^F n^G q\indices{_A^D} q\indices{_C^H}  \right ) - \frac{1}{3} \left ( g_{G[H}\ ^{(5)}R_{F]D} n^F n^G q\indices{_A^D} q\indices{_C^H}  \right )
\\
&- \frac{1}{12} \left ( g_{D [H} g_{F ] G} \ ^{(5)}R  n^F n^G q\indices{_A^D} q\indices{_C^H} \right) \ .
\label{eq:ecursivo}
\end{aligned}
\end{equation}

The term $\mathcal{E}_{AC} \equiv {}^{(5)}C_{DGHF} n^F n^G q\indices{_A^D} q\indices{_C^H} $ is known as the \emph{electrical part of the Weyl tensor}. As one can see, it is the projection of the 5-dimensional Weyl tensor on the brane. We will further discuss this important tensor once one gets the full effective EFE on the brane.

The first term in parenthesis in Eq. \ref{eq:ecursivo} can be written in terms of the energy-momentum tensor, using Eq. \ref{eq:r_dh} in the fourth line of what follows,

\begin{equation}
\begin{aligned}
&\ \ \  g_{D[H}\ ^{(5)}R_{F]G} n^F n^G q\indices{_A^D} q\indices{_C^H}  
\\
&= g_{DH}\ ^{(5)}R_{FG} n^F n^G q\indices{_A^D} q\indices{_C^H} - g_{DF}\ ^{(5)}R_{HG} n^F n^G q\indices{_A^D} q\indices{_C^H} 
\\
&= \ ^{(5)}R_{FG} n^F n^G \left (g_{DH} q\indices{^D_A} q\indices{^H_C} \right) - \ ^{(5)}R_{HG} n^F n^G \left ( q\indices{_A^D} g_{DF} \right ) q\indices{_C^H}  
\\ 
&= \left [ k_5^2  \left ( T_{FG} - \frac{T}{3} g_{FG} \right ) \right] n^F n^G  q_{AC}  - \ ^{(5)}R_{HG} \left ( q_{AF}\right )  n^F n^G q\indices{_C^H}  
\\ 
&= k_5^2  \left ( T_{FG} - \frac{T}{3} g_{FG} \right ) n^F n^G  q_{AC}  - \ ^{(5)}R_{HG} \left ( g_{AF} - n_A n_F\right )  n^F n^G q\indices{_C^H}  
\\
&= k_5^2  \left ( T_{FG} - \frac{T}{3} g_{FG} \right ) n^F n^G  q_{AC}  - \ ^{(5)}R_{HG} \cancelto{^0}{\left( n_A n^G - n_A n^G\right )} q\indices{_C^H} 
\\
\Rightarrow &g_{D[H}\ ^{(5)}R_{F]G} n^F n^G q\indices{_A^D} q\indices{_C^H} = k_5^2  \left ( T_{FG} - \frac{T}{3} g_{FG} \right ) n^F n^G  q_{AC} \ .
\end{aligned}
\end{equation}

Similarly, one has

\begin{equation}
\begin{gathered}
g_{G[H}\ ^{(5)}R_{F]D} n^F n^G q\indices{_A^D} q\indices{_C^H} = -k_5^2 \left (T_{FG} - \frac{T}{3}q_{FG} \right ) q\indices{_C^F}  q\indices{_A^G} \ , 
\end{gathered} 
\end{equation}

\noindent and 

\begin{equation}
  g_{D [H} g_{F ] G} \ ^{(5)}R  n^F n^G q\indices{_A^D} q\indices{_C^H} = \frac{-2T}{3}k_5^2 q_{AC} \ .
\end{equation}

We can then write Eq. \ref{eq:ecursivo} as

\begin{equation}
\begin{aligned}
\Rightarrow \tilde{\mathcal{E}}_{AC}  &=  \frac{1}{3} \left [ \left(  k_5^2  \left ( T_{FG} - \frac{T}{3} g_{FG} \right ) n^F n^G  q_{AC}\right) - \left( -k_5^2 \left (T_{FG} - \frac{T}{3}q_{FG} \right ) q\indices{_C^F}  q\indices{_A^G}  \right) \right ] 
\\
&- \frac{1}{12} \left [ \left( \frac{-2T}{3}k_5^2 q_{AC} \right) \right] +  \mathcal{E}_{AC}  
\\
&=  \frac{k_5^2}{3} \left [   T_{FG} n^F n^G  q_{AC} - \frac{T}{3} \left (g_{FG}n^F n^G \right) q_{AC} + T_{FG} q\indices{_C^F} q\indices{_A^G}  - \frac{T}{3} \left (q_{FG}q\indices{_C^F}q\indices{_A^G} \right ) \right ] 
\\
&+ \left( \frac{2k_5^2}{3} \frac{T}{12} q_{AC} \right) +  \mathcal{E}_{AC}  
\\
&=  \frac{k_5^2}{3} \left [   T_{FG} n^F n^G  q_{AC} - \frac{T}{3}q_{AC} + T_{FG} q\indices{_C^F} q\indices{_A^G}  - \frac{T}{3} \left (q_{AC} \right ) + \frac{T}{6} q_{AC} \right] + \mathcal{E}_{AC} 
\\
\Rightarrow \tilde{\mathcal{E}}_{AC}  &=  \mathcal{E}_{AC} +  \frac{k_5^2}{3} \left (  q_{AC} \left (  T_{FG} n^F n^G - \frac{T}{2} \right ) + T_{FG} q\indices{_C^F} q\indices{_A^G} \right) \ .
\end{aligned}
\end{equation}

Now, plugging this result into Eq. \ref{eq:g3}, one gets

\begin{equation}
\begin{aligned}
^{(4)}G_{AC} &= k_5^2 \left [ T_{DH} q\indices{_A^D} q\indices{_C^H} + \ q_{AC} \left ( T_{DH} n^D n^H - \frac{T}{3} \right ) \right ]  
\\
&+ K_{AC}K - K_{CD}K\indices{^D_A} - \frac{q_{AC}}{2}\left (K^2 - K_{DH}K^{DH} \right ) 
\\
&- \left [ \mathcal{E}_{AC} +  \frac{k_5^2}{3} \left (  q_{AC} \left (  T_{FG} n^F n^G - \frac{T}{2} \right ) + T_{FG} q\indices{_C^F} q\indices{_A^G} \right)   \right ]  
\\
&= k_5^2 \left [ T_{DH} q\indices{_A^D} q\indices{_C^H} + \ q_{AC} \left ( T_{DH} n^D n^H - \frac{T}{3} \right ) \right ] 
\\
&+ k_5^2 \left [ -\frac{1}{3}  T_{DH} q\indices{_A^D}q\indices{_C^H} - \frac{1}{3} q_{AC} \left (  T_{DH} n^D n^H - \frac{T}{2} \right) \right  ] 
\\
&+  K_{AC}K - K_{CD}K\indices{^D_A} - \frac{q_{AC}}{2}\left (K^2 - K_{DH}K^{DH} \right ) -  \mathcal{E}_{AC} 
\\
\Rightarrow \ ^{(4)}G_{AC} &= \frac{2k_5^2}{3} \left [T_{DH} q\indices{_A^D} q\indices{_C^H} + \ q_{AC} \left ( T_{DH} n^D n^H - \frac{T}{4} \right ) \right ] 
\\
&+  K_{AC}K - K_{CD}K\indices{^D_A} - \frac{q_{AC}}{2}\left (K^2 - K_{DH}K^{DH} \right ) - \mathcal{E}_{AC} \ .
\label{eq:g4}
\end{aligned}
\end{equation}

In Sec. \ref{sec:rs}, we introduced the brane tension $\sigma$ and altered the energy-momentum tensor to contain such a contribution. Here we will do the same, but we will also consider the possibility of other fields on the brane, so that it may have a 4-dimensional energy-momentum tensor $\tau_{AC}$. We then define $S_{AC} = \tau_{AC} - \sigma q_{AC}$, which is a tensor containing both the energy-momentum contributions from the fields on the brane and its tension. The cosmological constant of the bulk, $\Lambda_5$, will also contribute to the 5-dimensional energy-momentum tensor, which will be defined as

\begin{equation}
 T_{AC} = S_{AC}\delta(y) - \Lambda_5 g_{AC} = \left (\tau_{AC} - \sigma q_{AC} \right ) \delta(y) - \Lambda_5 g_{AC} \ .
\label{eq:5d_en_mom_tens}
 \end{equation}

Notice that, just like in Sec. \ref{sec:rs}, we employ a delta function of $y$ to localize the brane contribution to $T_{AC}$. The definition of Eq. \ref{eq:5d_en_mom_tens} allows us to write the terms of Eq. \ref{eq:g4} defined from $T_{AC}$ as functions of $\sigma$, $\Lambda_5$ and the metric,

\begin{equation}
T_{DH} q\indices{_A^D} q\indices{_C^H} = S_{DH}q\indices{_A^D} q\indices{_C^H}\delta(y) - \Lambda_5 q_{AC} \ ,
\end{equation}

\noindent and

\begin{equation}
 T_{DH} n^D n^H - \frac{T}{4}  = \left (S_{DH}n^Dn^H - \frac{S}{4} \right ) \delta(y) + \frac{\Lambda_5}{4} \ ,
\end{equation}

\noindent so that one has, for the sum in Eq. \ref{eq:g4} $T_{DH} q\indices{_A^D} q\indices{_C^H} + q_{AC}\left ( T_{DH} n^D n^H - \dfrac{T}{4} \right ) = \dfrac{-3\Lambda_5}{4} q_{AC}$, and Eq. \ref{eq:g4} can be rewritten as

\begin{equation}
\begin{gathered}
^{(4)}G_{AC} = -\frac{k_5^2 \Lambda_5}{2}q_{AC} +  K_{AC}K - K_{CD}K\indices{^D_A} - \frac{q_{AC}}{2}\left (K^2 - K_{DH}K^{DH} \right ) -  \mathcal{E}_{AC} \ .
\label{eq:g_quase}
\end{gathered}
\end{equation}

Now we are almost done. It only remains to determine the extrinsic curvature tensor in terms of the energy-momentum tensor. By contracting the 5D EFE $G_{DH} = k_5^2 \left ( S_{DH} \delta(y) - \Lambda_5 g_{DH} \right )$, one gets

\begin{equation}
\begin{gathered}
\left ( \ ^{(5)}R_{DH}g^{DH} \right) - \frac{1}{2} \ ^{(5)}R  \left (g_{DH}g^{DH} \right )= k_5^2 \left[ \left ( S_{DH} \delta(y) - \Lambda_5 g_{DH} \right )g^{DH} \right]  
\\
\Rightarrow \ ^{(5)}R  - \frac{5}{2} \ ^{(5)}R = k_5^2 \left ( S \delta(y) - 5 \Lambda_5 \right )  
\\
\Rightarrow \ ^{(5)}R  = \frac{-2 k_5^2}{3} \left( S \delta(y) - 5 \Lambda_5 \right ) \ ,
\end{gathered}
\end{equation}

\noindent so that, back in the EFE, one has

\begin{equation}
\begin{gathered}
^{(5)}R_{DH} - \frac{1}{2} \left [ \frac{-2 k_5^2}{3} \left( S \delta(y) - 5 \Lambda_5 \right ) \right ]  g_{DH}  = k_5^2 \left ( S_{DH} \delta(y) - \Lambda_5 g_{DH} \right )  
\\
\Rightarrow  \ ^{(5)}R_{DH} = k_5^2 \left [ \left ( S_{DH} - \frac{S}{3} g_{DH} \right ) \delta(y) + \frac{2 \Lambda_5}{3} g_{DH} \right ] \ .
\label{eq:r_dh2}
\end{gathered}
\end{equation}

Now, notice that in Gaussian coordinates $\{x^\mu, y \}$, the extrinsic curvature is simply given by $K_{AC} = \frac{1}{2} \partial_y q_{AC}$. Also, one has $\partial_y K_{AC} = K_{CB}K\indices{^B_A} - \tilde{\mathcal{E}}_{AB}$ \cite{kuerten}, so that we can write the 4-dimensional Ricci tensor (Eq. \ref{eq:4dricci}) as

\begin{equation}
\begin{aligned}
 ^{(4)}R_{AC} &=  \  ^{(5)}R_{DH} q\indices{_A^D} q\indices{_C^H} + \left (- \tilde{\mathcal{E}}_{AC} \right )  + K_{CA}K - K_{CF}K\indices{^F_A} 
 \\
&=  \  ^{(5)}R_{DH} q\indices{_A^D} q\indices{_C^H} + (\partial_y K_{AC}- K_{CB}K\indices{^B_A})   + K_{CA}K - K_{CB}K\indices{^B_A} 
 \\ 
 \Rightarrow  ^{(5)}R_{DH} q\indices{_A^D} q\indices{_C^H} &=  \  ^{(4)}R_{AC} + 2  K_{CB}K\indices{^B_A} - K_{CA}K - \partial_y K_{AC}  
 \\
  \Rightarrow  ^{(5)}R_{DH} q\indices{_A^D} q\indices{_C^H} &=  P_{AC} - \partial_y K_{AC} \ ,
 \label{eq:seila}
\end{aligned}
\end{equation}

\noindent where $P_{AC} \equiv \  ^{(4)}R_{AC} + 2  K_{CB}K\indices{^B_A} - K_{CA}K$. Plugging now Eq. \ref{eq:r_dh2} into Eq. \ref{eq:seila}, one gets

\begin{equation}
\begin{aligned}
P_{AC} - \partial_y K_{AC} &= k_5^2 \left [ \left ( S_{DH} - \frac{S}{3} g_{DH} \right ) \delta(y) + \frac{2 \Lambda_5}{3} g_{DH} \right ] q\indices{_A^D} q\indices{_C^H}  
\\
&= k_5^2 \left [ \left ( \left(S_{DH}q\indices{_A^D} q\indices{_C^H}\right ) - \frac{S}{3} \left( g_{DH}q\indices{_A^D} q\indices{_C^H} \right ) \right ) \delta(y) + \frac{2 \Lambda_5}{3} \left (g_{DH}q\indices{_A^D} q\indices{_C^H}\right ) \right ] 
\\
\Rightarrow P_{AC} - \partial_y K_{AC} &= k_5^2 \left [ \left ( S_{AC} - \frac{S}{3} q_{AC} \right ) \delta(y) + \frac{2 \Lambda_5}{3} q_{AC} \right ] \ .
\end{aligned}
\end{equation}

To finally determine $K_{AC}$, we must integrate the equation above on the brane ($y=0$). To do this, we integrate in the interval $(-\epsilon, +\epsilon)$, $\epsilon > 0$, in the limit that $\epsilon \rightarrow 0$. That is,

\begin{equation}
\begin{aligned}
  \lim_{\epsilon \rightarrow 0} \int_{-\epsilon}^{+\epsilon} \left ( P_{AC} - \frac{d}{dy} K_{AC} \right ) dy &= k_5^2 \lim_{\epsilon \rightarrow 0} \int_{-\epsilon}^{+\epsilon} \left [ \left ( S_{AC} - \frac{S}{3} q_{AC} \right ) \delta(y) + \frac{2 \Lambda_5}{3} q_{AC} \right ]dy
  \\
    \Rightarrow \  \lim_{\epsilon \rightarrow 0} \left( P_{AC} \left.y \right\rvert_{-\epsilon}^{+\epsilon} - \left.K_{AC} \right\rvert_{-\epsilon}^{+\epsilon} \right ) &= k_5^2 \left( S_{AC} - \frac{S}{3} q_{AC} \right ) + \frac{2 \Lambda_5}{3} q_{AC} \lim_{\epsilon \rightarrow 0} \left ( \left.y \right\rvert_{-\epsilon}^{+\epsilon} \right ) 
    \\
    \Rightarrow \ \lim_{\epsilon \rightarrow 0} \left ( \left.K_{AC} \right\rvert_{-\epsilon}^{+\epsilon}\right ) &= - k_5^2 \left( S_{AC} - \frac{S}{3} q_{AC} \right )
    \\
    \Rightarrow K_{AC}^{+} - K_{AC}^{-}  &=  - \ k_5^2 \left( S_{AC} - \frac{S}{3} q_{AC} \right ) \ , 
\end{aligned}
\end{equation}

\noindent where $K_{AC}^{+}$ ($K_{AC}^{-}$) denotes the extrinsic curvature in the direction of $y>0$ ($y< 0$). Given the definition of the extrinsic curvature in terms of the derivative with respect to $y$, $K_{AC} = \frac{1}{2} \partial_y q_{AC}$, which must be invariant under $y \rightarrow -y$ given the symmetry of the orbifold $S^1 / \mathbb{Z}^2$, one has $K_{AC}^{+} = -K_{AC}^{-} \equiv K_{AC}$. We then have, for the extrinsic curvature and its trace,

\begin{equation}
K_{AC} = -\frac{k_5^2}{2} \left( S_{AC} - \frac{S}{3} q_{AC} \right ) \ ;
\label{eq:extr}
\end{equation}

\noindent and 

\begin{equation}
 K = q^{AC}K_{AC} = -\frac{k_5^2}{2} \left( S - \frac{S}{3} (4) \right ) = \frac{k_5^2 S}{6}  \ ,
\label{eq:extr_tr}
\end{equation}

\noindent where

\begin{equation}
S_{AC} = \tau_{AC} - \sigma q_{AC} \ ;
\end{equation}

\noindent and 

\begin{equation}
  S = q^{AC}S_{AC} = \tau - 4\sigma \ ,
\label{eq:extr_tr2}
\end{equation}

\noindent so that the extrinsic curvature may be rewritten as

\begin{equation}
K_{AC} = -\frac{k_5^2}{2} \left( \tau_{AC} + \frac{1}{3} \left ( \sigma - \tau \right ) q_{AC} \right ) \ .
\label{eq:extr_junc}
\end{equation}

With $K_{AB}$ and $K$ it is possible to find all the remaining terms of Eq. \ref{eq:g_quase} in terms of $\tau_{AC}$, $\sigma$ and $\Lambda_5$. For instance, one has

\begin{equation}
 \begin{aligned}
  K_{AC}K &= \left ( -\frac{k_5^2}{12} \left( S_{AC} - \frac{S}{3} q_{AC} \right ) \right ) \left ( \frac{k_5^2 S}{6}\right ) = -\frac{k_5^4}{2}S \left( S_{AC} - \frac{S}{3} q_{AC} \right )  
  \\
  &= -\frac{k_5^4}{12} \left ( \tau - 4\sigma \right ) \left [ \left ( \tau_{AC} - \sigma q_{AC}\right ) -  \frac{(\tau - 4\sigma)}{3} q_{AC} \right ]  
  \\
  &= -\frac{k_5^4}{12} \left [ -\frac{\left(\tau - 4\sigma \right )^2}{3} q_{AC} + (\tau - 4\sigma)(\tau_{AC} - \sigma q_{AC} ) \right ] 
  \\
  &= -\frac{k_5^4}{12} \left [ \left (\frac{8\sigma \tau - \tau^2 - 16\sigma^2}{3} \right ) q_{AC} + \tau \tau_{AC} -\sigma\tau q_{AC} - 4 \sigma \tau_{AC} + 4 \sigma^2 q_{AC} \right ] 
  \\
  &= \frac{k_5^4}{12} \left ( 4\sigma \tau_{AC} + \frac{4\sigma^2 q_{AC}}{3} - \frac{5 \sigma \tau q_{AC}}{3} -\tau \tau_{AC} + \frac{\tau^3 q_{AC}}{3} \right ) \ .
 \end{aligned}
\end{equation}

Similarly, for the remaining terms,

\begin{equation}
K_{CD}K\indices{^D_A} = \frac{k_5^4}{4} \left (\tau_{CD}\tau\indices{^D_A} + \frac{2 (\sigma - \tau)\tau_{CA}}{3} + \frac{(\sigma - \tau)^2 q_{AC}}{9} \right ) \ ;
\end{equation}

\begin{equation}
K^2 = \frac{k_5^4 (16\sigma^2 + \tau^2 -8\sigma \tau)}{36} \ ;
\end{equation}

\begin{equation}
K_{DH}K^{DH} = \frac{k_5^4}{4} \left (\tau_{DH}\tau^{DH} + \frac{2 (\sigma - \tau)\tau}{3} + \frac{4(\sigma - \tau)^2}{9} \right ) \ .
\end{equation}

By substituting the equations above in Eq. \ref{eq:g_quase} and performing the appropriate algebraic manipulations, one finally arrives at the effective EFE on the brane. As this is a mere question of tedious algebra, we will not present the explicit manipulations here. Instead, we promptly present one of the most important equations of this thesis, the \emph{effective EFE on the brane},

\begin{equation}
^{(4)}G_{\mu \nu} = -\Lambda_4 q_{\mu \nu} + 8\pi G_4 \tau_{\mu \nu} + (8\pi G_5)^4 \pi_{\mu \nu} - \mathcal{E}_{\mu \nu} \ ,
\label{eq:effe_efe}
\end{equation}

\noindent where

\begin{equation}
 \Lambda_4 = \frac{k_5^2}{2} \left ( \Lambda_5 + \frac{k_5^2 \sigma^2}{6} \right ) \ ;
 \label{eq:lambda_45}
\end{equation}

\begin{equation}
 G_4 = \frac{k_5^4 \sigma}{48 \pi} = \frac{4 \pi G_5^2 \sigma}{3}  \Leftrightarrow  k_4^2 = \frac{1}{6}k_5^4 \sigma ~;
 \label{eq:g_45}
\end{equation}

\begin{equation}
 \pi_{\mu \nu} = \frac{1}{4}\left ( \frac{\tau \tau_{\mu \nu}}{3} - \frac{\tau^2 q_{\mu \nu}}{6} + \frac{\tau_{\rho \lambda}\tau^{\rho \lambda} q_{\mu \nu}}{2} - \tau\indices{_\mu^\lambda}\tau_{\lambda \nu}\right ) \ .
\end{equation}

Notice that we explicitly used Greek indices above, since all the objects are defined only on the 4-dimensional brane.

By comparing Eq. \ref{eq:effe_efe} with the 4D EFE from the classical formulation of GR, derived in Sec \ref{sec:EFEqs}: 
$\ ^{(4)}G_{\mu \nu} = -\Lambda_4 q_{\mu \nu} + 8 \pi G_4 \tau_{\mu \nu}$ (where we used the notation of this section for 4-dimensional objects), it is easy to see that in the effective EFE on the brane, two new terms appeared $k_5^4 \pi_{\mu \nu}$ and $\mathcal{E}_{\mu \nu}$.

The term $\pi_{\mu \nu}$ is quadratic in $\tau_{\mu \nu}$, and is therefore negligible in low energies, becoming dominant only in cases in which the energy-matter density codified in $\tau_{\mu \nu}$ is much greater than the brane tension $\sigma$ (for example, in the early universe \cite{ssm1}). This is the sense in which it can be seen as a high-energy correction --- it does not mean that this term emerges as a consequence of higher order derivatives in the action of the theory, since, as made clear in the calculations above, no mention to the underlying action was made.

And, as already mentioned, $\mathcal{E}_{\mu \nu}$ can be seen as the projection of the Weyl tensor on the brane, so that it does contain non-local gravitational information from the bulk. Although $\mathcal{E}_{\mu \nu}$ is also negligible in low-energy regimes, it is larger than the terms quadratics in $\tau_{\mu \nu}$. It can also be seen as a trace-free effective energy-momentum tensor, through the definition \cites{emgd, ovalle1},

\begin{equation}
\mathcal{E}_{\mu \nu} = \frac{6}{\sigma k_4^2} \left [ \mathcal{U} \left ( u_\mu u_\nu + \frac{h_{\mu \nu}}{3} \right ) + \mathcal{P}_{\mu \nu} + \mathcal{Q}_{(\mu}u_{\nu)} \right ] \ ,
\label{eq:weyl_proj}
\end{equation}

\noindent where we assume that the brane matter is a perfect fluid with 4-velocity $u^\mu$, $h_{\mu \nu} = q_{\mu \nu} - u_\mu u_\nu$ is the induce metric orthogonal to the fluid lines, $\mathcal{U}$ is the bulk Weyl scalar, $\mathcal{P}_{\mu \nu}$ is the anisotropic stress tensor and $\mathcal{Q}_\mu$ is the energy flux.

An important point is that the effective EFE of Eq. \ref{eq:effe_efe} are not closed. Therefore, it is in general necessary to solve the 5-dimensional EFE also for the bulk to get the solution on the brane. The complementary equations are obtained through the Bianchi identities \cites{ssm1, kuerten}. The non-closure of the effective EFE happens due to $\mathcal{E}_{\mu \nu}$, so that it could be suggested the imposition $\mathcal{E}_{\mu \nu} = 0$ in the brane. However, such a condition is incompatible with the Bianchi identities, so that other constraints will have to be imposed. One such constraint is realized via the construction of the Minimal Geometric Deformation method, which will be discussed in the next section.

\chapter{Geometric deformation braneworlds}

In this chapter, we will present a particular method, and an extension thereof, to solve the effective EFE of Eq. \ref{eq:effe_efe}. Both the methods will be based in the idea of \emph{geometric deformation}, in a sense which will be explored in the next sections. The methods will allow the construction of deformed braneworlds, which are reduced to standard GR solutions in the appropriate limit. 

The braneworlds constructed via the geometric deformation methods will be extension of standard GR, which, as will be discussed, leads to very interesting and important physical consequences. Later on, such an extension will be constrained by observational data, which narrows the theoretical deformation down to particular ranges allowed by physical observations. Afterward, we shall employ these braneworlds scenarios to model particular physical systems, which will clearly illustrate how braneworld scenarios modify the description of these systems, compared to the standard GR gravitational setup.

\section{Minimal Geometric Deformation}
\label{sec:mgd}

A useful constraint to the effective EFE on the brane is to demand the GR limit to exist in braneworld solutions. Such constraint, defined itself on the brane, physically corresponds to a condition of \emph{minimal geometrical deformation} projected onto the brane. In this section we will show that this condition in fact produces a physically correct low energy limit, allowing the construction of a braneworld version of \emph{any} classical GR solution. The presentation will mainly follow the work by Ovalle \cites{ovalle1, ovalle2}, which is based in the SSM formalism for the effective EFE on the brane, as presented in Sec. \ref{sec:effe}. In this section we will again use geometrized units, namely $G_4=c=1$.

We begin by proposing a general metric ansatz for the solution of the effective EFE (Eq. \ref{eq:effe_efe}), written in Schwarzschild-like coordinates as

\begin{equation}
 ds^2 = -e^{\nu(r)}\mathrm{d}t^2 + e^{\lambda(r)}\mathrm{d}r^2 + r^2\mathrm{d}\Omega^2 \ ,
 \label{eq:metric_general_spher_symm}
\end{equation}

\noindent that is, $q_{tt} = -\exp(\nu(r))$ and $q_{rr} = \exp(\lambda(r))$. We are interested in a solution for both regions within and exterior to a mass distribution --- a star, for example. For the star region we will assume an energy-momentum tensor of a perfect fluid $\tau_{\mu \nu} = (\rho + p)u_\mu u_\nu - p q_{\mu \nu}$. Therefore, $r$ ranges from the centre of the star ($r=0$) until its surface ($r=R$) and then beyond in the exterior vacuum where $\rho = p = 0 \Rightarrow \tau_{\mu \nu} = 0$. We will encode such a condition in the following equation, which is the general standard GR expression for the radial component of the metric,

\begin{equation}
 \mu(r) = 
 \begin{cases}
1 - \frac{2M}{r} & \text{ for } r > R \ ; \\ 
1 - \frac{8 \pi G_4}{r} \int_0^r \tilde{r}^2 \rho d \tilde{r} = 1 - \frac{2m(r)}{r}& \text{ for } r \leq R \ ,
\end{cases}
\end{equation}

\noindent where $m(r)$ denotes the standard GR interior mass function for $r<R$. In fact, since extra-dimensional effects can be perceived in the brane given the introduction of the brane tension $\sigma$, standard GR is recovered in the limit $1/\sigma \rightarrow 0$, so that the constant $M(\sigma)$ will take the standard GR mass $M_0$ value in the appropriate limit $\left .M\right\rvert_{\sigma^{-1} \rightarrow 0} = M_0 = m(R)$.

To determine $e^{\nu(r)}$ and $e^{\lambda(r)}$, we begin by calculating the Christoffel symbols, the Riemann tensor components, and so forth. Of course, though, given that the effective EFE are considerably more complicated, this is not an easy task, accomplished through the work of many pages of algebra which will not be explicitly shown here. The final result for the radial component of the metric is\cite{casadio}

\begin{equation}
 e^{-\lambda(r)} = \mu(r) + f(r) \ ,
 \label{eq:radial_mgd}
\end{equation}

\noindent where the function $f(r)$ is known as the \emph{geometric deformation}, given by

\begin{equation}
 f(r) = e^{-I} \int \frac{ 2 r e^I}{r\partial_r \nu + 4} \left [ H(p,\rho, \nu) + \frac{(\rho^2 + 3 \rho p)}{\sigma} \right ] dr + \beta e^{-I} \ ,
 \label{eq:geo_def}
\end{equation}

\noindent where $\beta = \beta(\sigma)$ is the integration constant and

\begin{equation}
 I(r) = \int_{r_0}^r \frac{ 2\tilde{r}\left (\partial_{\tilde{r}^2}^2 \nu + \frac{ \left (\partial_{\tilde{r}}\nu \right)^2}{2} + \frac{2 \partial_{\tilde{r}}\nu}{\tilde{r}} + \frac{2}{\tilde{r}^2} \right )}{\tilde{r}\partial_{\tilde r} \nu + 4} d\tilde{r} \ ,
 \label{eq:def_i}
\end{equation}

\noindent where $r_0$ is chosen according to the region of interest.

The geometric deformation is what an observer on the brane experiences due to projected five dimensional gravity effects. In fact, notice that $f(r)$ sums to the standard GR solution given by $\mu(r)$ by \emph{distorting} it, therefore it can indeed be seen as a deformation. 

The constant $\beta$ in Eq. \ref{eq:geo_def} must be zero in the standard GR limit $M \rightarrow M_0$ ($\sigma^{-1} \rightarrow 0 \Rightarrow \beta(\sigma) \rightarrow 0$). Also, $\beta$ must vanish in the interior of the star ($r<R$) to guarantee that the metric is smooth at $r=0$. But, at $r>R$, $\beta$ may be non-zero, and a geometrical deformation can be associated to the standard GR Schwarzschild solution. Now, an important factor regarding this parameter is in the assessment of the physical relevance of braneworld solutions, by the constraint of its value. Such constraint may be achieved through the classical tests of GE, which will be presented later on.

The most important contribution in the geometric deformation comes from the function $H(p, \rho, \nu)$,

\begin{equation}
 H(p, \rho, \nu) = - \left [ \partial_r \mu \left ( \frac{\partial_r \nu}{2} + \frac{1}{r} \right ) + \mu \left ( \partial_{r^2}^2 \nu + \frac{(\partial_r \nu)^2}{2} + \frac{2 \partial_r \nu}{r} + \frac{1}{r^2} \right ) - \frac{1}{r^2} \right ] + 24 \pi G_4 p \ .
\end{equation}

In fact, $H$ vanishes for any temporal metric component corresponding to a standard GR solution $e^{\nu} = e^{\nu_{GR}}$. In this case, we will have a minimum value for the geometric deformation (since $H> 0$ \cite{ct_mgd}), in the sense that the geometrical deformation will in this case be given solely by the energy density and pressure of the source $\rho$ and $p$. When this happens, one has the \emph{minimal geometric deformation}, $\tilde{f}(r)$, explicitly given by

\begin{equation}
 \tilde{f}(r) = \frac{e^{-I}}{\sigma}  \int \left (\frac{2re^{I}}{r \partial_r \nu + 4} \right ) (\rho^2 + 3\rho p) dr + \beta e^{-I(r)} \ .
\end{equation}

Therefore, starting from the choice $\nu = \nu_{GR}$, one can find the deformed radial component of the metric by evaluating Eq. \ref{eq:radial_mgd} with the minimal deformation $\tilde{f}(r)$.

For example, one can choose the Schwarzschild metric as the standard GR solution describing the exterior region ($\rho = p = 0$) of a stellar distribution, that is,

\begin{equation}
 e^{\nu_{GR}} = e^{\nu_S} = e^{-\lambda_S} = 1 - \frac{2M}{r} \ , 
\end{equation}

\noindent so that the MGD function becomes

\begin{equation}
  \left .\tilde{f}(r)\right\rvert_{\rho = p = 0} = \beta e^{-I(r)} =  \frac{ b \left (1-\frac{2M}{r} \right )}{r \left ( 1-\frac{3M}{2r} \right )}\beta \ ,
\end{equation}

\noindent where

\begin{equation}
 b = b(M) \equiv \frac{R\left(1-\frac{3M}{2R}\right)}{1-\frac{2M}{R}} \ .
 \label{eq:b(M)}
\end{equation}

Therefore, the deformed exterior temporal and radial metric components are respectively given by

\begin{equation}
 e^\nu = 1 - \frac{2M}{r} \ ,
 \label{eq:mgd_temp}
\end{equation}

\noindent and

\begin{equation}
 e^{-\lambda} = \left (1 - \frac{2M}{r} \right ) \left ( 1 + \frac{b\beta}{r \left (1- \frac{3M}{2r} \right )}\right ) \ .
 \label{eq:mgd_rad}
\end{equation}

There is extensive literature on applications of the MGD method and ramifications thereof (like, for example, the realization of gravitational decoupling via MGD) to the study of several physical systems, which shows the relevance of the method and its broad spectrum of applications. For some examples, see \cites{3_mgd_related_inicio, 5, 7, 9, 22, 28, 38, 40, 41, 44, 46, 75}.

\section{Extended Minimal Geometric Deformation}
\label{sec:emgd}

The exterior region of a stellar distribution may be seen as filled with a Weyl fluid arising from the bulk. As discussed above, MGD allows the study of such a region by the deformation of the standard GR Schwarzschild metric. Now, it is interesting to generalize the MGD for the exterior region, what is done by considering a deformation not only on the radial but also on the temporal metric component. This determines what is called the \emph{extended minimal geometric deformation (EMGD) method}, which will be discussed in this section, following the work of da Rocha, Casadio and Ovalle \cite{emgd}.

The geometric deformation on the temporal metric component $e^{\nu(r)}$ is defined to be given by

\begin{equation}
 \nu(r) = \nu_S + h(r) \ , 
 \label{eq:temp_def}
\end{equation}

\noindent where $\nu_S$ defines the Schwarzschild temporal metric component, 

\begin{equation}
 e^{\nu_S} = 1 - \frac{2 M}{r} \ ,
\end{equation}

\noindent and $h(r)$ is the temporal deformation, proportional to $\sigma^{-1}$, which guarantees the standard GR limit. Using the vacuum effective EFE, the radial geometric deformation $\hat{f}(r)$ is now changed, and can be written in terms of the $h(r)$ as \cite{emgd}

\begin{equation}
 \hat{f}(r) = e^{-I} \left ( \beta - \int_R^r \frac{2\tilde{r}e^{I} F(h)}{\tilde{r}\partial_{\tilde{r}}\nu + 4} d \tilde{r} \right ) \ ,
 \label{eq:geo_ext}
\end{equation}

\noindent where $I$ is given by the same Eq. \ref{eq:def_i}, and

\begin{equation}
 F(h) =  \frac{(\partial_r\mu) (\partial_r h)}{2} + \mu \left ( \partial_{r^2}^2 h + (\partial_r \nu_{S})(\partial_r h) + \frac{(\partial_r h)^2}{2} +  \frac{2\partial_r h}{r}\right ) \ .
 \label{eq:F(h)}
\end{equation}

Therefore, the exterior deformed radial metric component becomes

\begin{equation}
 e^{-\lambda(r)} =  1 - \frac{2M}{r} + \hat{f}(r) \ , 
 \label{eq:radial_emgd}
\end{equation}

\noindent with the extended geometric deformation $\hat{f}(r)$ redefined according to Eq. \ref{eq:geo_ext}.

Notice that a constant $h$ implies $F=0$, which produces an exterior minimal geometrical deformation, as we had before in MGD. On the other hand, it is also possible to achieve minimal geometric deformation with a non-constant $h(r)$, which is given by setting Eq. \ref{eq:F(h)} to zero. The solution of this differential equation is given by

\begin{equation}
 e^{h / 2} = a + \frac{b}{2M} \left ( 1 - \frac{2M}{r}\right )^{-1/2} \ .
\end{equation}

With the imposition of asymptotic flatness, $r \rightarrow \infty \Rightarrow e^{\nu} \rightarrow 1 \Rightarrow h \rightarrow 0$, the integration constants $a$ and $b$, both function of the brane tension, are related as $a = 1 - \frac{b}{2M} $, so that the minimally-deformed temporal metric component gets the form

\begin{equation}
 e^{\nu(r)} = \left ( 1 - \frac{2M}{r}\right ) \left [ \left (1 + \frac{b(\sigma)}{2M} \right ) \left( \left ( 1 - \frac{2M}{r}\right )^{-1/2} - 1 \right ) \right ]^2 \ ,
\end{equation}

\noindent and the minimally-deformed radial metric component becomes, as it should be,

\begin{equation}
 e^{-\lambda(r)} =  1 - \frac{2M}{r} + \beta e^{-I} \ .
\end{equation}

With the equations above it becomes clear that $r=2M$ is a singularity in the temporal metric component. In the case $\beta = 0$, the radial metric has no longer any geometric deformation, as it gets precisely the Schwarzschild form $\lambda = -\nu_S$. Interestingly, in this case it can be proved that $r = 2M$ is a \emph{real singularity} on the braneworld metric $q_{\mu \nu}$, since the Kretschmann scalar diverges at $r=2M$, given the deformation in the temporal metric component.

For $\beta \neq 0$, the Kretschmann does not diverge at $r=2M$ only if $\beta$ satisfies a very specific equation depending on $M$ and $b$ \cite{emgd}. But, given the strong constraints on $\beta$ obtained with the classical tests of GR, which will soon be discussed, as well as other observational constraints on $\beta$, it is fairly unlikely that this very particular expression actually holds for arbitrary masses $M$. Therefore, given the allowed range of $\beta$, it is more likely that the Kretschmann scalar does diverge at $r=2M$, which then remains a real singularity on the brane even for $\beta \neq 0$. 

This is an interesting result, showing that the introduction of the extra dimension and construction of the braneworld scenario with the deformed temporal component makes $r=2M$ indeed a singularity, most likely. Now, since this is a hypersphere singularity on the brane (not a point, like $r=0$), its physical implications are not entirely clear. Kerr black holes, for instance, present ring singularities, which lead to very interesting physical implications on the actual inner geometry of a rotating black hole. Perhaps, some parallels between Kerr and EMGD geometries could be drawn, thus leading to a better understanding of this EMGD singular hypersurface. A study of the causal structure of the spacetime may shed light on these points. However, they are outside the scope of this work, remaining as open points for future investigation.

Now we will derive a more general solution for the exterior radial metric component of Eq. \ref{eq:radial_emgd}, under a geometric deformation such that $F(h) \neq 0$. By choosing

\begin{equation}
 h(r) = k \ln \left ( 1 - \frac{2M}{r}\right ) \ ,
\end{equation}

\noindent one has

\begin{equation}
 \begin{gathered}
  e^\nu = e^{\nu_S + h} = e^{\nu_S} \exp \left [ k \ln \left ( 1 - \frac{2M}{r}\right ) \right] =  \left( 1 - \frac{2M}{r}\right ) \left ( 1 - \frac{2M}{r}\right )^k
  \\
  \Rightarrow e^\nu = \left ( 1 - \frac{2M}{r}\right )^{k+1} \ ,
  \label{eq:temp_emgd}
 \end{gathered}
\end{equation}

\noindent where $k$ is known as the \emph{deformation parameter}. Naturally, $k=0$ gives no temporal geometric deformation, and is directly associated with the Schwarzschild metric. For $k=1$, one has

\begin{equation}
e^{\nu(r)} = 1 - \frac{4M}{r} + \frac{4M^2}{r^2} \ ,
\end{equation}

\noindent which allows the calculation of the radial metric component, through Eq. \ref{eq:radial_emgd}. The necessary algebraic steps are quite long (as it demands the exhausting calculation of $F(h)$ and $I$ with $r_0 = R$) and will therefore not be explicitly shown here. However, the result is fairly simple,

\begin{equation}
 e^{-\lambda(r)} = 1 - \frac{2M - \kappa}{r} + \frac{2M^2 - \kappa M}{r^2} \ ,
\end{equation}

\noindent where $\kappa = \dfrac{M\beta}{1 - (M/R)}$.

Now, to the radial metric component asymptotically approach the Schwarzschild behavior, with ADM mass $\mathcal{M} = 2M$,

\begin{equation}
 e^{-\lambda(r)} \sim 1 - \frac{2\mathcal{M}}{r} + \mathcal{O}(r^{-2}) \ ,
\end{equation}

\noindent we must necessarily have $\kappa = -2M$, in which case the temporal and spatial components of the metric will be inversely equal to each other (as it is the case of the Schwarzschild solution), containing a tidal charge $\mathcal{Q} = 4M^2$, thus reproducing a tidally charged solution \cite{maartens2},

\begin{equation}
 e^\nu = e^{-\lambda} = 1 - \frac{2\mathcal{M}}{r} + \frac{\mathcal{Q}}{r^2} \ .
 \label{eq:k=1}
\end{equation}

The metric of Eq. \ref{eq:k=1} has a degenerate event horizon at $r_h = 2M = \mathcal{M}$. Since the degenerate horizon lies behind the Schwarzschild radius, i.e., $r_h = \mathcal{M} < r_s = 2\mathcal{M}$, extra dimensional effects are responsible for decreasing the gravitational field strength on the brane.

Now we will construct the exterior solution for $k=2$. From Eq. \ref{eq:temp_emgd}, one has

\begin{equation}
 e^{\nu(r)}  = 1 - \frac{2 \mathcal{M}}{r} + \frac{\mathcal{Q}}{r^2} - \frac{2 \mathcal{M} \mathcal{Q}}{9r^3} \ ,
\end{equation}

\noindent in which we defined $\mathcal{M} = 3M$ and $\mathcal{Q} = 12M^2 $ to put the temporal component of the metric in a Schwarzschild-like form. The radial component, on the other hand, is way more complicated than the case $k=1$, and is given by

\begin{equation}
\begin{gathered}
  e^{-\lambda(r)}  = \left ( 1 - \frac{2 \mathcal{M}}{3r}\right )^{-1} \left [ \frac{128 \kappa}{r} \left ( 1 - \frac{\mathcal{M}}{6r} \right )^7 + \frac{5}{224} \left ( \frac{ \mathcal{Q}}{12r^2}\right )^4  + \left ( \frac{5(16r - 2 \mathcal{M})}{96r} \right )\left ( \frac{ \mathcal{Q}}{12r^2}\right )^3 \right ] 
  \\
   + \left ( 1 - \frac{2 \mathcal{M}}{3r}\right )^{-1} \left [ \left ( \frac{25(6r - \mathcal{M})}{12r}\right )\left (  \frac{ \mathcal{Q}}{12r^2} \right )^2 + \left ( \frac{10r - 5\mathcal{M}}{12r}\right )\frac{Q}{r^2} -\frac{4\mathcal{M}}{3r} + 1     \right ] \ ,
\end{gathered}
\end{equation}

\noindent where $\kappa = \dfrac{(1- ( 2M/R)) R^8 \beta}{(2R - M)^7}$. The asymptotically Schwarzschild behavior is guaranteed with the imposition $\kappa = -M/32$.

Once we numerically analyze the radial and temporal metric components, we realize that there are three horizons. They are $r_i \approx 0.09 \mathcal{M}$, $r_e \approx 1.12 \mathcal{M}$ and $r_c = 2/3 \mathcal{M}$. The outer horizon hides the most interior ones, so that it is more convenient to see the solution as describing the exterior region of a stellar distribution of radius $\tilde{R} > r_c$, excluding the singular region $r=r_c$. Notice that here as well the extra-dimensional effects weaken the gravitational field, since the degenerate horizon lies behind the Schwarzschild radius $r_e < r_s$. A deeper analysis of these singularities, and of the causal structure of the spacetime, is outside the scope of this thesis, but is an open idea for future developments.

EMGD also finds several applications in the literature. The extension of the geometrical deformation to the temporal metric component leads to important consequences, which were explored and applied in different contexts. For some examples, see \cites{16, emgd_1, emgd_2, glueballs, quantum_portrait}.

In the next section we will present the classical tests of GR, originally proposed as ways to verify the results of standard GR with observational data of the Solar System. The same tests will then be used to constrain the value of $k$, under which the braneworld EMGD metric can be made consistent with observational data and become physically meaningful. 

\chapter{The classical tests of GR}

As shown in the previous sections, the consideration of extra dimensions and braneworld scenarios leads to major corrections to the standard GR, which is theoretically very remarkable. Nevertheless, in order for these models to be physically viable, they must match with experimental and/or observational data, otherwise no physical significance can be attributed to these rather interesting, but entirely mathematical consequences. One way to put such theories to the observational test is through the \emph{classical tests of GR}, which are performed at the level of the Solar System: the \emph{perihelion precession of Mercury}, the \emph{deflection of light by the Sun} and the \emph{radar echo delay}. 

In this section, we shall present and discuss these tests, following Böhmer \emph{et al.} \cite{bohmer}, and adopting the same strategy therein, which consists in first developing the tests for arbitrary static and spherically symmetric metrics, which is sufficiently general so that we can first apply it to the standard GR Schwarzschild solution, and later on to the MGD and EMGD metrics.

Back in the beginnings of GR, the tests were used to successfully validate the standard Schwarzschild solution, so that we hope to use them to constrain the parameters of the braneworld models. Since we will use actual observational data, it is convenient to use SI units, which will be adopted in this and the next sections.

Static and spherically symmetric metrics, as already discussed in Sec. \ref{sec:schwarzschild}, are adequate to describe the exterior vacuum of stellar distributions. As we will guide our analysis based on the solar system observational data, such a restriction is perfectly adequate. Therefore, we will assume a metric of the same form of Eq. \ref{eq:metric_general_spher_symm}, which is given in the SI units by

\begin{equation}
 ds^2 = q_{\mu \nu} \mathrm{d}x^\mu \otimes \mathrm{d}x^\nu = -c^2e^{\nu(r)}\mathrm{d}t^2 + e^{\lambda(r)}\mathrm{d}r^2 + r^2\mathrm{d}\Omega^2 \ .
\end{equation}

First of all, we shall determine the geodesics --- parameterized under the affine parameter denoted $\ell$, to not be confused with $\lambda(r)$ --- of particles in the spacetime carrying the metric above. We will again consider the conserved quantity defined in Sec. \ref{sec:schwarzschild},

\begin{equation}
\begin{gathered}
 \epsilon = -q_{\mu \nu} \frac{dx^\mu}{d \ell} \frac{dx^\nu}{d \ell} 
 \\
 \Rightarrow  \epsilon = e^\nu c^2 \left ( \frac{dt}{d\ell} \right )^2 - e^\lambda \left ( \frac{dr}{d\ell} \right )^2 - r^2 \left [ \left ( \frac{d\theta}{d\ell} \right )^2 + \sin^2 \theta \left ( \frac{d\varphi}{d\ell} \right )^2\right ] ~.
 \label{eq:eps_ct}
 \end{gathered}
\end{equation}

Remember that for timelike geodesics one has $\epsilon = 1$, whilst null geodesics get $\epsilon = 0$. Now, from the Killing vector implying the conservation of the direction of angular momentum, we will impose without loss of generality that the motion happens in the equatorial plane $\theta = \frac{\pi}{2}$, and therefore $\frac{d \theta}{d \ell} = 0$. With this, Eq. \ref{eq:eps_ct} may be written as

\begin{equation}
\begin{gathered}
 \epsilon = e^\nu c^2 \left ( \frac{dt}{d\ell} \right )^2 - e^\lambda \left ( \frac{dr}{d\ell} \right )^2 - r^2 \left ((1)^2 \left ( \frac{d\varphi}{d\ell} \right )^2\right ) 
 \\
  \Rightarrow   e^{-\lambda} \epsilon = e^\nu e^{-\lambda} c^2 \left ( \frac{dt}{d\ell} \right )^2 - \left ( \frac{dr}{d\ell} \right )^2 - r^2 e^{-\lambda}\left ( \frac{d\varphi}{d\ell} \right )^2 
  \\
    \Rightarrow  \left ( \frac{dr}{d\ell} \right )^2 +  e^{-\lambda} r^2 \left ( \frac{d\varphi}{d\ell} \right )^2 = e^{-\lambda} \left ( e^\nu c^2 \left ( \frac{dt}{d\ell} \right )^2  - \epsilon \right ) \ .
 \label{eq:eps_ct_2}
 \end{gathered}
\end{equation}

Now, the magnitude $L$ of the angular momentum is conserved, which is implied by the killing vector $R^\mu = (0, 0, 0, 1)^T$, or $R_\mu = (0, 0, 0, r^2)$ (since $\sin^2 \left ( \frac{\pi}{2} \right ) = 1$), whilst the Killing vector $K^\mu = (1, 0, 0, 0)^T$, or $K_\mu= \left( -c^2e^\nu, 0, 0, 0 \right)$, implies the conservation of energy $E$. Therefore, one has

\begin{equation}
  L = R_\mu \dfrac{dx^\mu}{d \ell} = r^2 \dfrac{d\varphi}{d \ell} \ ,
\end{equation}

\noindent and

\begin{equation}
   E = -K_\mu \dfrac{dx^\mu}{d \ell} = c^2 e^\nu \dfrac{dt}{d \ell} \ .
\end{equation}

Substituting the conserved quantities above in Eq. \ref{eq:eps_ct_2} then yields

\begin{equation}
  \left ( \frac{dr}{d\ell} \right )^2 +  e^{-\lambda} \frac{L^2}{r^2} = e^{-\lambda} \left ( e^{-\nu} \frac{E^2}{c^2}  - \epsilon \right ) \ .
  \label{eq:motion}
\end{equation}

Now we shall apply the developments presented above to each one of the classical tests.

\section{The perihelion precession}

In this case, we will consider the geodesics of a planet --- originally Mercury --- around the Sun, so that we will have $\epsilon = 1$. Now, by defining $u \equiv \frac{1}{r}$, and substituting $\frac{d \varphi}{d \ell} = Lu^2 \Rightarrow \frac{d}{d\ell} = Lu^2$; and defining a function $f(u)$ such that $e^{-\lambda(r)} = 1 - f(u)$, after a few lines of algebra we can rewrite Eq. \ref{eq:motion} as

\begin{equation}
   \left ( \frac{du}{d\varphi} \right )^2 + u^2 = f(u)u^2 + \frac{e^{-\lambda} \left(E^2 e^{-\nu} - c^2 \right)}{c^2L^2} \ . 
\end{equation}

By defining the right-hand side of the equation above as $G(u)$, and differentiating both sides with respect to $\varphi$, one has

\begin{equation}
\begin{aligned}
\frac{dG(u)}{d \varphi} &= \frac{d}{d \varphi}\left ( \frac{du}{d\varphi} \right )^2 + \frac{du^2}{d \varphi} 
\\
\Rightarrow \frac{dG(u)}{d \varphi} &= 2 \frac{du}{d\varphi}\frac{d^2u}{d\varphi^2} + 2u \frac{du}{d\varphi} 
\\
\Rightarrow \frac{dG(u)}{d \varphi} &= 2 \frac{du}{d\varphi} \left ( \frac{d^2u}{d\varphi^2} + u \right ) 
\\
\Rightarrow \frac{d^2u}{d\varphi^2} + u  &= F(u) \ ,
\end{aligned}
\end{equation}

\noindent where we defined $F(u) \equiv  \frac{1}{2}\frac{d G(u)}{du} $. Now, a circular orbit with radius $r = r_0 \Rightarrow u = u_0$, where $u_0$ is the root of the equation $u_0 = F(u_0)$, may present a small deviation $\delta = u - u_0$, such that

\begin{equation}
\begin{aligned}
\frac{d^2 }{d\varphi^2}(\delta + u_0) + (\delta + u_0)  &= F(\delta + u_0) 
\\
\Rightarrow \left ( \frac{d^2 \delta}{d\varphi^2} +  \delta \right ) + \left ( \frac{d^2 u_0}{d\varphi^2} + u_0 \right ) &= F(u_0) + \delta \left .\frac{dF}{du}\right\rvert_{u = u_0}  + \mathcal{O}(\delta^2) 
\\
\Rightarrow \frac{d^2 \delta}{d\varphi^2} +  \delta + \left ( F(u_0)\right ) &= F(u_0) + \delta \left .\frac{dF}{du}\right\rvert_{u = u_0}  + \mathcal{O}(\delta^2) 
\\
\Rightarrow  \frac{d^2 \delta}{d\varphi^2} +  \left ( 1 -  \left .\frac{dF}{du}\right\rvert_{u = u_0} \right) \delta &= \mathcal{O}(\delta^2) \ .
\end{aligned}
\end{equation}

Therefore, up to first order of the deviation $\delta$, it will obey a harmonic equation,

\begin{equation}
\frac{d^2 \delta}{d\varphi^2} +  k^2 \delta = 0 \ ,
\end{equation}

\noindent where $k^2 \equiv 1 -  \left .\frac{dF}{du}\right\rvert_{u = u_0} $. The solution assumes the well-known oscillatory form

\begin{equation}
 \delta = \delta_0 \cos (k \varphi + \phi_0) \ ,
\end{equation}

\noindent where $\delta_0$ and $\phi_0$ are integration constants. Naturally, $\delta$ will make the orbit slightly non-circular, so that the deviated orbit will present a perihelion, which will happen in points of minimum $r$ and therefore maximum $u$ or $\delta$. Naturally, $\delta$ will be maximum for $k \varphi + \phi_0 = 2\pi$, so that between two perihelia, the initial phase $\phi_0$ is irrelevant, and the orbital angle variation is then

\begin{equation}
k \varphi = 2\pi \Rightarrow \varphi = \frac{2\pi}{k} \equiv\frac{2\pi}{1 - \alpha} \ ,
\end{equation}

\noindent where we defined $k \equiv 1 - \alpha$, with $\alpha$ used to define the \emph{perihelion advance} $\delta \varphi = 2\pi \alpha$. The parameter $\alpha$ represents how much the perihelia deviate with time, since it will advance $\alpha \varphi$ whilst the planet advances $\varphi$ radians in its orbit. We can explicitly write $\alpha$ as

\begin{equation}
\alpha = 1 - \sqrt{1 - \left .\frac{dF}{du}\right\rvert_{u = u_0}} \ ,
\end{equation}

\noindent which, for small $ \left .\frac{dF}{du}\right\rvert_{u = u_0}$, can be approximated to

\begin{equation}
\begin{gathered}
\alpha = 1 - \left ( 1 - \frac{1}{2} \left .\frac{dF}{du}\right\rvert_{u = u_0} \right) 
\\
\Rightarrow \alpha = \frac{1}{2} \left .\frac{dF}{du}\right\rvert_{u = u_0} \ .
\end{gathered}
\end{equation}

Now, to explicitly calculate the perihelion advance $\delta \varphi = 2\pi\alpha = \pi \left .\frac{dF}{du}\right\rvert_{u = u_0}$ from the orbital parameters of a given planet, we can use some results of Kleperian orbits, given that the planet is moving in the small velocity $v$ limit (i.e., $v \ll c$), which yields \cite{marion} 

\begin{equation}
 \frac{r^2}{2}\frac{d\varphi}{dt} = \frac{\pi a^2 \sqrt{1-e^2}}{T} \ ,
 \label{eq:seila_2}
\end{equation}

\noindent where $a$ is the semi-major axis of the elliptical orbit, $e$ is its eccentricity and $T$ is its period, given by Kepler's third law $T^2 = 4\pi^2 a^3/GM $, where $M$ is the mass of the body generating the gravitational field (in our case, the Sun). In the small velocity limit, we can approximate the affine parameter (which for timelike geodesics can be the proper time) to the temporal coordinate itself, $\ell \approx c t$, so that the conserved angular momentum can be written as $L = r^2 \frac{d \varphi}{d \ell} = r^2 \frac{1}{c} \frac{d \varphi}{dt} \Rightarrow r^2 \frac{d\varphi}{dt } = cL $. Eq. \ref{eq:seila_2} then becomes simply

\begin{equation}
\begin{gathered}
  \frac{1}{2} \left (cL \right )= \frac{\pi a^2 \sqrt{1-e^2}}{T} 
  \\
  \Rightarrow L^2 =  \frac{4\pi^2 a^4 (1-e^2)}{c^2T^2} = \frac{4\pi^2 a^4 (1-e^2)}{c^2  \left (\dfrac{4\pi^2 a^3}{GM} \right ) } 
  \\
  \Rightarrow \frac{1}{L^2} = \frac{c^2}{GM a (1-e^2)} \ .
\end{gathered}
\end{equation}

We can then write

\begin{equation}
\begin{aligned}
\delta \varphi &= \pi \left .\frac{dF}{du}\right\rvert_{u = u_0}  
\\
&= \pi  \left .\frac{d}{du}\left (  \frac{1}{2}\frac{dG}{du}  \right )\right\rvert_{u = u_0}  
\\
&= \left .\frac{\pi}{2}\frac{d^2G}{du^2}\right\rvert_{u = u_0} 
\\
&= \frac{\pi}{2} \left .\left [ \frac{d^2}{du^2} \left ( \left (1-e^{-\lambda} \right )u^2 + \frac{e^{-\lambda} \left(E^2 e^{-\nu} - c^2 \right)}{c^2L^2}  \right ) \right ]\right\rvert_{u = u_0} 
\\
\Rightarrow \delta \varphi &=  \frac{\pi}{2} \left .\left [ \frac{d^2}{du^2} \left ( \left (1-e^{-\lambda(u)} \right )u^2 + \frac{e^{-\lambda(u)} \left(E^2 e^{-\nu(u)} - c^2 \right)}{GMa(1-e^2)}  \right ) \right ]\right\rvert_{u = u_0} \ ,
\label{eq:per_pre1}
\end{aligned}
\end{equation}

\noindent where $u_0$ is found through $u_0 = F(u_0)$, i.e.,

\begin{equation}
\begin{aligned}
u_0 = \frac{1}{2} \frac{d}{du} \left .\left ( \left (1-e^{-\lambda(u)} \right )u^2 + \frac{e^{-\lambda(u)} \left(E^2 e^{-\nu(u)} - c^2 \right)}{GMa(1-e^2)}  \right )\right\rvert_{u = u_0} \ .
\label{eq:per_pre2}
\end{aligned}
\end{equation}

We will soon make explicit examples, so that the procedure will become clearer.

\section{The light shift}

Light follow null geodesics, with $\epsilon = 0$ in Eq. \ref{eq:motion}, which yields

\begin{equation}
  \left ( \frac{dr}{d\ell} \right )^2 +  e^{-\lambda} \frac{L^2}{r^2} = e^{-\nu - \lambda} \frac{E^2}{c^2}  \ .
\end{equation}

By using the same change of variables $u = \frac{1}{r}$ and substitution employed in the derivation of the perihelion precession above, we arrive at

\begin{equation}
 \frac{d^2 u}{d \varphi^2} + u = G(u) \ ,
 \label{eq:light_shift}
\end{equation}

\noindent where we now have

\begin{equation}
 G(u) = \frac{1}{2} \frac{d}{du} \left ( \left (1-e^{-\lambda(u)} \right )u^2 + \frac{E^2 e^{-\nu(u)-\lambda(u)}}{c^2 L^2} \right ) \ .
 \label{eq:light_shift1} 
\end{equation}

$L^2$ is now that of the photon, with which we must have no concerns, since, as seen below, this term does not appear in the final expression for the light shift. 

In first approximation, taking $G(u) = 0$ yields the solution $u(\varphi) = \cos\varphi/R$, where the integration constant $u_0^{-1} = r_0 =R$ represents the smallest distance between the light ray and the Sun. Now we can plug this solution into Eq. \ref{eq:light_shift} to iteratively get a better approximation,

\begin{equation}
 \frac{d^2 u}{d \varphi^2} + u = G \left(\frac{\cos\varphi}{R} \right) \ ,
 \label{eq:light_shift2}
\end{equation}

\noindent which has as general solution $u = u(\varphi)$. If there was no deflection, we would have $u(\pi/2) = 0$, with the light ray at $r(\pi/2) \rightarrow \infty$. Now, with a deflection of $\varepsilon$, one has $u(\frac{\pi}{2} + \varepsilon) = 0$. Therefore, to find the total light deflection $\delta = 2\varepsilon$, we only have to solve Eq. \ref{eq:light_shift}, evaluate $\varphi = (\pi + \delta)/2$ and then set $u=0$, which will allow us to find $\delta$. An explicit example will soon be shown, to make the procedure clearer.

\section{The echo delay}

This test consists in measuring the difference between the time it takes for a radar signal to travel between two planets in two different situations: when the signal passes near the Sun, and when it does not.

A signal traveling between two planets distant $l_1$ and $l_2$ from the Sun, so that the signal does not pass near the Sun, makes it in the time

\begin{equation}
 T_1 = \int_{-l_1}^{l_2} \frac{dx}{c} \ ,
\end{equation}

\noindent where $x$ is taken in the direction of the separation between the planets. Notice that we consider the case in which the maximal approach between the sun and the radar signal, $R$, is such that $R \ll l_1$ and $R \ll l_2$, so that we can approximate the distance traveled by the signal to the sum of $l_1$ and $l_2$, as defined in the integral above. Now, if the signal travels even closer to the Sun, its relative speed will change to $v= ce^{(\nu - \lambda)/2}$, so that the time it will take between the planes will be

\begin{equation}
 T_2 = \int_{-l_1}^{l_2} \frac{dx}{v} = \frac{1}{c}\int_{-l_1}^{l_2} \exp \left(\frac{\lambda - \nu}{2} \right ) dx \ .
\end{equation}

Notice that $r = \sqrt{x^2 + R^2}$. Therefore, for the difference $\delta T = T_2 - T_2$, one has

\begin{equation}
 \delta T = \frac{1}{c} \int_{-l_1}^{l_2} \left [\exp \left ( \frac{\lambda (\sqrt{x^2 + R^2}) - \nu (\sqrt{x^2 + R^2})}{2} \right ) - 1 \right ]dx \ .
 \label{eq:echo_delay}
\end{equation}

These are the three classical tests, formulated in a way that the equations describing the deviations are absolutely general for static spherically symmetric metrics, and can be readily used. In the next section we will make explicit calculations of the classical tests for the standard GR Schwarzschild metric, as the most immediate example.

\section{Classical tests for the standard GR Schwarzschild metric}

We will use Eqs. \ref{eq:per_pre1} and \ref{eq:per_pre2} for the perihelion precession; Eqs. \ref{eq:light_shift1} and \ref{eq:light_shift2} for the light shift and Eq. \ref{eq:echo_delay} for the echo delay, applying them to the Schwarzschild metric, which is given by

\begin{equation}
 e^{\nu_S} = e^{-\lambda_S} = 1 - \frac{2GM}{c^2r} =  1 - \frac{2GM u}{c^2}  \ .
\end{equation}

We start by calculating $u_0$, given by Eq. \ref{eq:per_pre2},

\begin{equation}
\begin{aligned}
u_0 &= \frac{1}{2} \frac{d}{du} \left .\left ( \left (1-e^{-\lambda_S(u)} \right )u^2 + \frac{e^{-\lambda_S(u)} \left(E^2 e^{\lambda_S(u)} - c^2 \right)}{GMa(1-e^2)}  \right )\right\rvert_{u = u_0}  
\\
&= \frac{1}{2} \frac{d}{du} \left .\left ( \left (1-e^{-\lambda_S(u)} \right )u^2 + \frac{E^2 - c^2e^{-\lambda_S(u)}}{GMa(1-e^2)}  \right )\right\rvert_{u = u_0}  
\\
&= \frac{1}{2} \frac{d}{du} \left .\left [ \left(1-\left ( 1 - \frac{2GMu}{c^2}\right )\right )u^2 + \frac{E^2}{GMa(1-e^2)} - \frac{c^2\left ( 1 - \frac{2GMu}{c^2}\right )}{GMa(1-e^2)}  \right ]\right\rvert_{u = u_0}  
\\
&= \frac{1}{2} \frac{d}{du} \left .\left [ \frac{2GMu^3}{c^2} + \frac{E^2 - c^2}{GMa(1-e^2)} + \frac{2GMu}{GMa(1-e^2)}  \right ]\right\rvert_{u = u_0}  
\\
&= \frac{1}{2} \left .\left [ \frac{2GM (3u^2)}{c^2} + \frac{2}{a(1-e^2)}  \right ]\right\rvert_{u = u_0} 
\\
\Rightarrow u_0 &= \frac{3GM u_0^2}{c^2} + \frac{1}{a(1-e^2)}  \ ,
\end{aligned}
\end{equation}

\noindent which has the solution

\begin{equation}
\begin{gathered}
u_0 = \frac{1 \pm \sqrt{ 1 - \frac{12GM}{c^2a(1-e^2)}}}{\frac{6GM}{c^2}} \approx \frac{1 \pm \left (1 - \frac{6GM}{c^2a(1-e^2)} \right )}{\frac{6GM}{c^2}} = \frac{6GM}{c^2a(1-e^2)} \frac{c^2}{6GM} 
\\
\Rightarrow u_0 = \frac{1}{a(1-e^2)} \ ,
\end{gathered}
\end{equation}

\noindent in which we pick the negative signal of the discriminant to guarantee a physical solution. Now, for Eq. \ref{eq:per_pre1}, already substituting the first derivative with respect to $u$ calculated above, as well as using the value of $u_0$, one has the following perihelion advance,

\begin{equation}
\begin{gathered}
\delta \varphi =  \frac{\pi}{2} \left .\left [ \frac{d}{du} \left ( \frac{6GMu^2}{c^2} + \frac{2}{a(1-e^2)}  \right ) \right ]\right\rvert_{u = u_0} 
\\
\Rightarrow \delta \varphi =  \frac{\pi}{2} \left .\left [ \frac{6GM (2u)}{c^2} \right ]\right\rvert_{u = u_0} 
\\
\Rightarrow \delta \varphi = \frac{6 \pi GM}{c^2}u_0 
\\
\Rightarrow \delta \varphi =  \frac{6 \pi GM}{c^2a (1-e^2)} \ .
\end{gathered}
\end{equation}

Now for the light shift. Notice that for the Schwarzschild one has $\nu_S + \lambda_S = 0$, so that Eq. \ref{eq:light_shift1} becomes

\begin{equation}
\begin{aligned}
 G(u) &= \frac{1}{2} \frac{d}{du} \left ( \left (1-e^{-\lambda_S(u)} \right )u^2 + \frac{E^2}{GMa (1-e^2)} \right ) 
 \\
 &= \frac{1}{2} \frac{d}{du} \left ( \left (1- \left ( 1 - \frac{2GMu}{c^2}\right ) \right )u^2 + \frac{E^2}{GMa (1-e^2)} \right ) 
\\
 &= \frac{1}{2} \frac{d}{du} \left ( \frac{2GMu^3}{c^2} + \frac{E^2}{GMa (1-e^2)} \right )  
 \\
  &= \frac{1}{2}\frac{2GM(3u^2)}{c^2}
  \\
  \Rightarrow G(u) &= \frac{3GMu^2}{c^2} \ .
\end{aligned}
\end{equation}

Therefore, Eq. \ref{eq:light_shift2} yields

\begin{equation}
 \frac{d^2 u}{d \varphi^2} + u = G \left(\frac{\cos\varphi}{R} \right) = \frac{3GM}{c^2}\frac{\cos^2 \varphi}{R^2} \ ,
\end{equation}

\noindent an ordinary differential equation whose solution is

\begin{equation}
 u = \frac{\cos\varphi}{R} + \frac{3GM}{2c^2R^2} \left ( 1 - \frac{\cos (2\varphi)}{3} \right ) \ .
\end{equation}

By evaluating the equation above at $\phi = \frac{\pi + \delta}{2}$, setting $u=0$, and using the small angle approximations $\sin\left(\frac{\delta}{2}\right) \approx \frac{\delta}{2}$ and $\cos(\delta) \approx 1$ (used below as equalities, since $\delta \ll 1$); as well as the identities $\cos(\pi + \delta) = -\cos(\delta)$ and $\cos\left(\frac{\pi}{2} +\frac{\delta}{2}\right) = -\sin\left(\frac{\delta}{2}\right )$, one gets the deflection angle under which the light ray is shifted,

\begin{equation}
\begin{gathered}
-\cos\left(\frac{\pi}{2} +\frac{\delta}{2}\right) = \frac{3GM}{2c^2R} \left ( 1 - \frac{\cos(\pi + \delta)}{3} \right ) 
\\
\Rightarrow \sin\left(\frac{\delta}{2}\right ) = \frac{3GM}{2c^2R}  \left ( 1 +\frac{\cos(\delta)}{3} \right ) 
\\
\Rightarrow \frac{\delta}{2} = \frac{3GM}{2c^2R}  \left ( 1 +\frac{1}{3} \right ) = \frac{3GM}{2c^2R}  \frac{4}{3} 
\\
\Rightarrow \delta = \frac{4GM}{c^2R} \ .
\end{gathered}
\end{equation}

Now the radar echo delay. First, notice that, for the Schwarzschild metric,

\begin{equation}
\exp\left( \frac{\lambda_S - \nu_S}{2}\right) = \exp\left( \frac{-\nu_S - \nu_S}{2}\right) = e^{-\nu_S} = \left ( 1 - \frac{2GM}{c^2r}\right )^{-1} \approx 1 + \frac{2GM}{c^2r} \ ,
\end{equation}

\noindent where we approximated up to first order in $r^{-1} = u$. We then have, in Eq. \ref{eq:echo_delay},

\begin{equation}
\begin{aligned}
\delta T &= \frac{1}{c} \int_{-l_1}^{l_2}  \left (1 + \frac{2GM}{c^2r} -1 \right )dx 
\\
&= \frac{2GM}{c^3} \int_{-l_1}^{l_2} \frac{1}{r}dx  
\\
&= \frac{2GM}{c^3} \int_{-l_1}^{l_2} \frac{1}{\sqrt{x^2 + R^2}}dx  
\\
\Rightarrow \delta T &= \frac{2GM}{c^3} \ln \left ( \frac{\sqrt{R^2 + l_2^2} + l_2}{\sqrt{R^2 + l_1^2} -l_1} \right ) \ .
\end{aligned}
\end{equation}

Notice that, using the fact that $\left (\frac{R}{l_1}\right )^2 \ll 1$ and $\left (\frac{R}{l_2}\right )^2 \ll 1$, one has

\begin{equation}
\frac{\sqrt{R^2 + l_2^2} + l_2}{\sqrt{R^2 + l_1^2} -l_1} = \frac{\left (\sqrt{R^2 + l_2^2} + l_2\right )\left (\sqrt{R^2 + l_1^2} + l_1\right )}{R^2} \approx \frac{4l_1l_2}{R^2} \ ,
\end{equation}

\noindent so that the radar echo delay is then given by

\begin{equation}
  \delta T = \frac{2GM}{c^3} \ln \left ( \frac{4l_1l_2}{R^2} \right ) \ .
\end{equation}

These were the original results of the classical tests, achieved when first formulated to validate the back then newly developed GR. The results are well known: expressive success, highly contributing at the time to the acceptance of GR as the most precise theory for gravitation, which, despite its quite elaborated mathematical formulation, was able to explain phenomena that Newtonian gravitation could not, as well as to correctly predict quite formidable new phenomena such as black holes.

\section{Classical tests for braneworld scenarios}

We shall now apply the classical tests of GR to the braneworld metrics, namely those derived from the MGD (following the work of da Rocha, Casadio and Ovalle \cite{ct_mgd}) and EMGD methods, to constrain its parameters based on observational data related to the three tests detailed above. In \cite{bohmer}, one can find the classical tests applied to other braneworld scenarios, with the same purpose of ours. In this section we will explicitly need the numerical value of the following constants and parameters,

\begin{equation}
\begin{aligned}
  c &= 2.998 \times 10^8 ms^{-1} ~;
  \\
  M &= M_{Sun} = 1.989 \times 10^{30} kg ~;
  \\
  a &= 5.791 \times 10^{10} m~;
  \\
  R_0 &= 6.955 \times 10^8 m~;
  \\
  e &= 0.205615~;
  \\
  G &= 6.68 \times 10^{-11} m^3kg^{-1}s^{-2} \ ,
\end{aligned}
\end{equation}

\noindent where $a$ and $e$ refer to the orbit of Mercury.

\subsection{MGD metric}

First, we consider a planet orbiting the braneworld Sun, whose exterior spacetime is given by the MGD metrics of Eqs. \ref{eq:mgd_temp} and \ref{eq:mgd_rad} (but we are now working in the SI units, so that the $G$'s and $c$'s will be added accordingly). By performing the same calculations explicitly shown in the previous sections with the new metric components (which is not algebraically complicated, but extremely lengthy, so that it will be omitted here), one discovers that the perihelion advance $\delta \varphi$ is given by

\begin{equation}
\delta \varphi = \delta \varphi_{GR} - f(\beta) = \frac{6 \pi GM}{c^2a (1-e^2)} -673.94\beta \ ,
\label{eq:bw_per_pre}
\end{equation} 

\noindent where we used the result derived above $\delta \varphi_{GR} =  \frac{6 \pi GM}{c^2a (1-e^2)}$ and $f(\beta) \approx 673.94\beta$, which was calculated plugging the numerical values of the constants into Eqs. \ref{eq:per_pre1} and \ref{eq:per_pre2}. Now, by explicitly calculating $\delta \varphi_{GR}$ with the numerical values and their respective uncertainty, one gets the following difference which can be accounted to braneworld effects,

\begin{equation}
 \Delta (\delta \varphi) = \delta\varphi - \delta\varphi_{GR} \approx 0.13 \pm 0.21 \ \mathrm{arcsec/century} \ .
\end{equation}

By using this value in Eq. \ref{eq:bw_per_pre}, one then gets the following constraint on $\beta$ given by the perihelion precession test,

\begin{equation}
 \beta \lesssim (2.80 \pm 3.45) \times 10^{-11} \ .
\end{equation}

Similarly, by using the metric of MGD and following the same procedure as detailed above, in the limit $\left ( \frac{GM}{c^2 R_0} \right )^2 \ll 1$, $\frac{M}{L} \ll 1 $ and $\left (\frac{E}{c} \right )^2 \ll 1 $, one gets the total braneworld light shift by the Sun, $\delta$, given by

\begin{equation}
 \delta = \delta_{GR} + g(\beta) = \frac{4GM}{c^2R} + b_0 \left ( \frac{E^2 R_0}{c^2 L^2} + \frac{18 \pi c^2 R_0}{GM}\right )\beta \ ,
\end{equation}

\noindent where $b_0 = b(M_0)$, calculated according to Eq. \ref{eq:b(M)}, with $M = M_0 = M_{Sun}$ denoting the GR mass in the limit $\sigma^{-1}$, as aforementioned, and $R_0$ the minimum distance between the light ray and the Sun. Now, by substituting all the numerical values, one gets the constraint

\begin{equation}
 \beta \lesssim (1.07 \pm 4.28) \times 10^{-10} \ .
\end{equation}

Finally, the radar echo delay test gives, within the limits $\left (\frac{R}{l_1}\right )^2 \ll 1$ and $\left (\frac{R}{l_2}\right )^2 \ll 1$, a time delay in the braneworld of

\begin{equation}
 \delta T = \delta T_{GR} + h(\beta) \approx  \frac{2GM}{c^3} \ln \left ( \frac{4l_1l_2}{R^2} \right ) + \frac{b_0}{c^3 R} \left [ \ln \left ( \frac{4 l_1 l_2}{R^2} \right ) - \frac{5\pi GM}{2}\right ]  \ .
\end{equation}

By substituting the numerical data, and using more recent measurements of the radio echo delay \cite{viking}, one gets the bound

\begin{equation}
 \beta \lesssim (3.96 \pm 4.30) \times 10^{-5} \ .
\end{equation}

Notice, therefore, that the strongest bound comes from the perihelion precession, which should, therefore, be taken into consideration as the maximum value of $\beta$ in accordance with the classical tests of GR in the Solar System. This quite strong bound is the reason why in Sec. \ref{sec:emgd} we stated that it is highly improbable that $r = 2GM$ is not a real singularity in braneworld black holes.

\subsection{EMGD metric}
\label{sec:emgd_ct}

For the EMGD metric components, expressed in Eqs. \ref{eq:radial_emgd} and \ref{eq:temp_emgd}, it is also possible to run the same classical tests. As one has already presented an explicit calculation for the Schwarzschild metric, and a braneworld application for the MGD metric, we will only present the constraints, this time on the value of the deformation parameter $k$.

Given that the calculation of the classical tests equations are quite long and algebraically complicated, we used a Mathematica\textsuperscript{\textregistered} script to promptly get the following results,

\begin{equation}
k \lesssim 4.5, \ \ \text{for the perihelion precession;}
\end{equation}

\begin{equation}
k \lesssim 4.3, \ \ \text{for the light shift;}
\end{equation}

\begin{equation}
k \lesssim 4.2, \ \ \text{for the echo delay.}
\end{equation}

The most strict bound is that given by the echo delay, but notice that the constraints are not that different, all of them implying, for integer $k$, $k < 5$. These are original results, presented and used in \cite{glueballs} and citations therein. This work will be discussed in greater detail in Sec. \ref{sec:mgd_gb}.

\chapter{Variable tension braneworld}
 
As discussed in Secs. \ref{sec:eefe_rs1} and \ref{sec:effe}, in the braneworld scenarios we must take into consideration the energy density of the brane itself, which is interpreted as the brane tension $\sigma$. As aforementioned, the brane tension must naturally have a quite high value, whilst in the MGD context its limit tending to infinity yields the recovery of standard GR.

If we think of the brane as the higher-dimensional analogue of a classical fluid membrane, it becomes clear why the brane tension must exist as a non-vanishing quantity. In fact, a classical fluid membrane necessarily needs to possess a tension to exist, otherwise it would collapse. Therefore, the 3-brane should also have a tension to remain a hypersurface in the bulk as it evolves. Given this, several phenomenological tests were performed to pose a lower bound to the value of the brane tension \cite{eotvos2}. The strongest bound, derived by combining results from experiments on possible deviations from Newton's law at sub-millimeter scales with the 4D Planck's constant value, gives $\sigma > 138.59 \ \text{TeV}^4$, whilst astrophysical considerations in the context of braneworld neutron stars give the bound $\sigma > 5 \times 10^8 \ \text{MeV}^4$ \cite{eotvos2}.

Now, in the cosmological context, the brane is seen as our observable universe, so that its expansion is realized as the movement of the brane through the extra bulk dimension. We know, however, that in the cosmological time scale the temperature of the brane changes enormously, from a quite hot early universe to the current low temperature (around 2.7K) given by the cosmic microwave background. Thus, considering such dramatic change in temperature, one might ask if it is reasonable to consider a constant brane tension (as one has done this far), especially in the cosmic braneworld scenarios. In this section, following \cites{eotvos1, eotvos2}, we shall discuss in which way we can incorporate a variable tension to our braneworld scenarios, and what changes under this assumption.

To begin our discussion, we go back to the classical fluid membranes, whose non-vanishing tension is not a constant. In fact, its dependence with temperature is given by the \emph{Eötvös law}, an empirical law established by Eötvös in 1886, which relates the membrane tension $\sigma$ to its temperature $T$ according to

\begin{equation}
 \sigma(T) = K \left (T_c - T \right ) \ ,
\end{equation}

\noindent where $K$ is a proportionality constant and $T_c$ is the \emph{critical temperature}, i.e., the highest temperature in which the membrane exists. Now, based on this analogy, one assumes the Eötvös law to hold also for the 3-brane, and derive its gravitational dynamics under this new assumption. The derivation, in a scenario slightly more general than the one we are considering in this work, is presented in details in \cite{eotvos1}. Namely, the effective EFE on the brane are recovered, but manifest contributions due to the varying brane tension appear, yielding \emph{variable} gravitational and cosmological parameters (no longer constants), as we discuss in what follows.

To make an explicit link with the gravitational scenario, we first rewrite Eötvös law as

\begin{equation}
 \sigma = \sigma_0 - \frac{6L}{\kappa_5^4} \frac{1}{a} \ ,
\end{equation}

\noindent where we set $KT_c \equiv \sigma_0$, and used the standard cosmological model relation $T \propto a^{-1}$, where $ a = a(t)$ is the Friedmann--Lemaître--Robertson--Walker (FLRW) expansion factor, setting the proportionality constant such that $6L/(\kappa_5^4) = K$, where $\kappa_5^4$, the 5-dimensional gravitational coupling constant, was introduced to bring about the 4-dimensional gravitational coupling ``constant'' $\kappa_4^2$, according to: 

\begin{equation}
 \kappa_4^2 = \frac{\kappa_5^4 \sigma_0}{6} - \frac{L}{a} \ .
\end{equation}

Notice that as $a \rightarrow \infty$, that is, for late cosmological times, one has $\sigma \rightarrow \sigma_0$ and $\kappa_4^2 \rightarrow \kappa_5^4 \sigma_0/6$ (in agreement with Eq. \ref{eq:g_45}, with $\kappa_4^2 = 8\pi G_4$). Now, notice that, in this context, the brane tension must keep its sign (i.e., remain positive), otherwise the brane would collapse. Also, $\kappa_4^2$ should remain positive (otherwise effects such as anti-gravity would exist on the brane). To guarantee these conditions, a natural scale factor is introduced in terms of $L$ $a_{\text{min}} \equiv 6L/(\kappa_5^4 \sigma_0)$, which is the minimum value $a$ can assume, thus guaranteeing positive $\sigma$ and $\kappa_4^2$. Therefore, in terms of $a_{\text{min}}$, one has

\begin{equation}
 \sigma(t) = \sigma_0 \left (1 - \frac{a_{\text{min}}}{a(t)}  \right ) \ ,
\label{eq:sigmat}
\end{equation}

\begin{equation}
\kappa_4^2(t) = \frac{\kappa_5^4 \sigma_0}{6} \left (1 - \frac{a_{\text{min}}}{a(t)} \right ) \ .
\label{eq:kappat}
\end{equation}

Therefore, we can now see through Eq. \ref{eq:sigmat} an explicit temporal dependence of the brane tension, which has cosmological relevance and will be further explored in what follows. As to Eq. \ref{eq:kappat}, it generalizes Eq. \ref{eq:g_45} (which was derived with the assumption of a fixed brane tension), by incorporating time dependence to the 4-dimensional brane tension, within the same cosmological braneworld context. 

Now, after motivating a variable tension braneworld and giving a particular example to illustrate its functional time-dependence, we shall see the effect of this consideration on the braneworld gravitational solutions.

\section{Black strings in the variable tension braneworld}

In the following discussion, which follows \cites{mgdvar1, mgdvar2}, we shall explicitly see what changes in the braneworld solutions once we consider a variable tension brane, applied to a \emph{black string} in the bulk. A black string can be seen as an extension of a black hole on the brane into the extra dimension. To construct this object, we perform a Taylor expansion in the extra dimension about a black hole solution on the brane. We shall then see that the consideration of a variable (namely, time-dependent) brane tension leads to additional terms in the expansion which carry the time dependence from the brane tension.

Once again we shall use Gaussian normal coordinates $\{x^\mu, y\}$, in which the vector normal to the brane (the hypersurface defined by $y=0$) is given by $n_A\de x^A = \de y$. As before, the bulk metric components $\bar{g}_{\mu \nu}$ are related to the brane metric components $g_{\mu \nu}$ according to $\bar{g}_{\mu \nu} = g_{\mu \nu} + n_\mu n_\nu$, so that the 5-dimensional bulk metric is

\begin{equation}
 \bar{g}_{M N}\de x^M \de x^N = g_{\mu \nu}(x^\alpha,y) \de x^\mu \de x^\nu + \de y^2 \ .
 \label{eq:bs1}
\end{equation}

We start with a 4-dimensional metric $g_{\mu \nu}$ and then extend it into the bulk by performing an expansion near $y=0$, so that as a result one obtains a black string described by a bulk metric defined near the brane. In this context, a \emph{black string} is just that: a bulk metric built as an extension from the brane metric, that is, with an additional component into the bulk, as in Eq. \ref{eq:bs1}.

Near the brane ($y=0$), the bulk metric components can be expressed as a Taylor expansion in the Gaussian coordinate $y$,

\begin{equation}
 g_{\mu \nu}(x^\alpha, y) = \sum_k \left . \mathcal{L}_{\bm{n}}^k (g_{\mu \nu} (x^\alpha, y)) \right \rvert_{y=0}\frac{|y|^k}{k!} \equiv \sum_k g_{\mu \nu}^{(k)} (x^\alpha) \frac{|y|^k}{k!} \ ,
 \label{eq:taylor1}
\end{equation}

\noindent where $\mathcal{L}_{\bm{n}}$ is the Lie derivative with respect to the normal vector $\bm{n}$, which in Gaussian coordinates is simply $\mathcal{L}_{\bm{n}} = \partial/\partial y$, so that $\mathcal{L}_{\bm{n}}^k = \partial^k/\partial y^k$. The extrinsic curvature on the brane, given by $K_{\mu \nu}= \frac{1}{2} \mathcal{L}_{\bm{n}} g_{\mu \nu}$, is fixed 
by the junction condition, which (as detailed in Sec. \ref{sec:effe} and concluded in Eq. \ref{eq:extr_junc}), yields

\begin{equation}
 K_{\mu \nu} = -\frac{k_5^2}{2} \left( T_{\mu \nu} + \frac{1}{3} \left ( \sigma - T \right ) g_{\mu \nu} \right ) \ .
\label{eq:firstorder}
\end{equation}

From the 5-dimensional Weyl tensor (according to Eq. \ref{eq:weyl}), 

\begin{equation}
 C_{\mu\nu\sigma\rho} = R_{\mu\nu\sigma\rho} - \frac{2}{3} (\bar{g}_{[\mu\sigma} R_{\nu]\rho} + \bar{g}_{[\nu\rho} R_{\mu]\sigma}) - \frac{1}{6} R ( \bar{g}_{\mu[\sigma} \bar{g}_{\nu\rho]}) \ ,
\end{equation}

\noindent we build the electric and magnetic Weyl tensor components, respectively given by the symmetric and trace-free components of $C_{\mu\nu\sigma\rho}$, ${\cal E}_{\mu\nu} = C_{\mu\nu\sigma\rho} n^\sigma n^\rho$ and ${\cal B}_{\mu\nu\alpha} = g_\mu^{\;\rho} g_\nu^{\;\sigma} C_{\rho\sigma\alpha\beta}n^\beta$. Together with the Riemann tensor $R_{\mu\nu\sigma\rho}$ and the extrinsic curvature, these components obey a set of effective field equations obtained from the bulk EFE and Bianchi equations which supplement the effective EFE \cite{ssm1},

\begin{equation}
  \mathcal{L}_{\bf n} K_{\mu\nu}= K_{\mu\alpha}K^\alpha
{}_\nu - {\cal E}_{\mu\nu}-\frac{1}{6}\Lambda_5 g_{\mu\nu} \ ;
\label{exp1}
\end{equation}

\begin{equation}
\begin{gathered}
 \mathcal{L}_{\bf n} {\cal E}_{\mu\nu}  = \nabla^\alpha
{\cal B}_{\alpha(\mu\nu)} + \frac{1}{6}
\Lambda_5\left(K_{\mu\nu}-g_{\mu\nu}K\right)
+K^{\alpha\beta}R_{\mu\alpha\nu\beta}-K{\cal E}_{\mu\nu} + 
\\ 
+3K^\alpha{}_{(\mu}{\cal
E}_{\nu)\alpha} + \left(K_{\mu\alpha}K_{\nu\beta}
-K_{\alpha\beta}K_{\mu\nu}\right)K^{\alpha\beta} \ ;
\label{exp2}
\end{gathered}
\end{equation}

\begin{equation}
 \mathcal{L}_{\bf n} {\cal B}_{\mu\nu\alpha} = -
2\nabla_{[\mu}{\cal E}_{\nu]\alpha}+K_\alpha{}^\beta {\cal
B}_{\mu\nu\beta} -2{\cal B}_{\alpha\beta [\mu }K_{\nu]}{}^\beta \ ;
\label{exp3}
\end{equation}

\begin{equation}
 \mathcal{L}_{\bf n} R_{\mu\nu\alpha\beta}
=-2R_{\mu\nu\gamma [\alpha}K_{\beta]}{}^\gamma
-\nabla_{\mu}{\cal B}_{\alpha\beta\nu} + \nabla_{\mu}{\cal
B}_{\beta\alpha\nu} \ . \label{exp4}
\end{equation}

In addition, these equations are subjected to the boundary condition ${\cal B}_{\mu\nu\alpha} = 2\nabla_{[\mu} K_{\nu]\alpha}$ on the brane \cite{maartens} --- and it is important to notice that all the quantities in the expansion are evaluated on the brane, that is, at $y=0$. Thus, Eqs. \ref{exp1} to \ref{exp4} above together with Eq. \ref{eq:firstorder} are employed to the calculation of the terms of the Taylor expansion in Eq. \ref{eq:taylor1},

\begin{equation}
g_{\mu \nu}^{(0)} (x^\alpha) = \left . \mathcal{L}_{\bm{n}}^0 (g_{\mu \nu} (x^\alpha, y)) \right \rvert_{y=0} = g_{\mu \nu} \ ;
\end{equation} 

\begin{equation}
g_{\mu \nu}^{(1)} (x^\alpha) = \left . \mathcal{L}_{\bm{n}} g_{\mu \nu} (x^\alpha, y) \right \rvert_{y=0} =  2 \left . K_{\mu \nu} \right \rvert_{y=0} =  -k_5^2 \left( T_{\mu \nu} + \frac{1}{3} \left ( \sigma - T \right ) g_{\mu \nu} \right )\ ;
\end{equation} 

\begin{equation}
\begin{gathered}
g_{\mu \nu}^{(2)} (x^\alpha) = \left . \mathcal{L}_{\bm{n}}^2 g_{\mu \nu} (x^\alpha, y) \right \rvert_{y=0} =  2 \left . \mathcal{L}_{\bm{n}} K_{\mu \nu} \right \rvert_{y=0} =  2 K_{\mu\alpha}K^\alpha{}_\nu - 2{\cal E}_{\mu\nu}-\frac{1}{3}\Lambda_5 g_{\mu\nu} 
\\
= \frac{1}{2}\kappa_5^4\left(
T_{\mu\alpha}T^\alpha{}_\nu +\frac{2}{3} (\sigma-T)T_{\mu\nu}
\right) -2{\cal E}_{\mu\nu} +\frac{1}{3}\left( \frac{1}{6}
\kappa_5^4(\sigma-T)^2-\Lambda_5
\right)g_{\mu\nu}\ ,
\end{gathered}
\end{equation} 

\noindent and so forth. Up to fourth order in $y$, one has

\begin{equation}
\begin{aligned}
g_{\mu\nu}(x^\alpha,y) &= 
g_{\mu\nu} 
\\
&-\kappa_5^2\left[
T_{\mu\nu}+\frac{1}{3}(\sigma-T)g_{\mu\nu}\right]\,|y| 
\\
&+\left[\frac{1}{2}\kappa_5^4\left(
T_{\mu\alpha}T^\alpha{}_\nu +\frac{2}{3} (\sigma-T)T_{\mu\nu}
\right) -2{\cal E}_{\mu\nu} +\frac{1}{3}\left( \frac{1}{6}
\kappa_5^4(\sigma-T)^2-\Lambda_5
\right)g_{\mu\nu}\right]\, \frac{y^2}{2!} 
\\
&+\left.\Bigg[2K_{\mu\beta}K^{\beta}_{\;\,\alpha}K^{\alpha}_{\;\,\nu} - 
{\cal E}_{(\mu\vert\alpha}K^{\alpha}_{\;\,\vert\nu)}
-\nabla^\rho{\cal B}_{\rho(\mu\nu)} + \frac{1}{6}
\Lambda_5g_{\mu\nu}K-K{\cal E}_{\mu\nu} 
\right.
\\
&\left.
\quad
\qquad\qquad+3K^\alpha{}_{(\mu}{\cal
E}_{\nu)\alpha}+K_{\mu\alpha}K_{\nu\beta}K^{\alpha\beta}
-K^2K_{\mu\nu} + K^{\alpha\beta}R_{\mu\alpha\nu\beta}\Bigg]\;\frac{|y|^3}{3!} 
\right.
\\
&+\left.\Bigg[\frac{\Lambda_5}{6}\left(R-\frac{\Lambda_5}{3} + K^2\right)g_{\mu\nu} + \left(\frac{K^2}{3}
-\Lambda_5\right)K_{\mu\alpha}K^{\alpha}_{\;\,\nu} + (R-\Lambda_5 + 2K^2){\cal E}_{\mu\nu}\right. 
\\
&+
\left. \left(K^{\alpha}_{\;\,\tau}K^{\tau\beta} + {\cal E}^{\alpha\beta}
+KK^{\alpha\beta}\right)\,R_{\mu\alpha\nu\beta} + K^2\,K\,K_{\mu\nu}- K_{\mu\alpha}K_{\nu\beta}{\cal E}^{\alpha\beta} \right. 
\\
&+\left.
 {\cal E}_{\mu\alpha}\left(\frac{1}{2}KK^\alpha_{\;\,\nu}-3K^\alpha_{\;\,\sigma}K^{\sigma}_{\;\,\nu} \right)-\frac{13}{2}K_{\mu\beta}{\cal E}^\beta_{\;\,\alpha}K^{\alpha}_{\;\,\nu} - 4K^{\alpha\beta}R_{\mu\nu\gamma\alpha}K^{\gamma}_{\;\beta} - \frac{1}{6}\Lambda_5R_{\mu\nu}  
\right.
\\
&\left.
+\frac{7}{2}KK^\alpha_{\;\,\mu} 
{\cal E}_{\nu\alpha}
- \frac{7}{6}K^{\sigma\beta}K^{\;\,\alpha}_{\mu}R_{\nu\sigma\alpha\beta} + 2 K_{\mu\beta}K^{\beta}_{\;\,\rho}K^\rho_{\;\,\alpha}K^\alpha_{\;\,\nu}\Bigg]\,\frac{y^4}{4!}
+\mathcal{O}(|y|^5)
\right.
\ ,
\label{eq:tay}
\end{aligned}
\end{equation}

\noindent where we denote $T\equiv T^\mu{}_\mu $ and  $T^2 \equiv T_{\mu \nu}\,T^{\mu \nu}$, for any rank-2
tensor $T_{\mu \nu}$. 

Now, notice that up to $\mathcal{O}(y^2)$, one has no contributions arising from the variable tension, because it has no variation (covariant derivatives) up to this order. However, from $\mathcal{O}(|y|^3)$ it is easy to see that derivatives of the brane tension start to appear, namely in the term $\nabla^\alpha{\cal B}_{\alpha(\mu\nu)}$, at $\mathcal{O}(|y|^3)$. Imposing the boundary condition ${\cal B}_{\mu\nu\alpha} = 2\nabla_{[\mu}K_{\nu]\alpha}$, one has, also employing $K_{\mu\nu}=-\frac{1}{2}\kappa_5^2 \left[T_{\mu\nu}+ \frac{1}{3}
\left(\lambda-T\right)g_{\mu\nu} \right]$, the following extra term,

\begin{equation}
 g_{\mu\nu}^{(3)\,{}^{\rm variable}}
=
\frac{2}{3}\kappa_5^2\left(\nabla_{(\nu}\nabla_{\mu)}\sigma
-
g_{\mu\nu}\,\Box\sigma\right) \ .
\label{addtruey3}
\end{equation}

Analogously, notice that at $\mathcal{O}(y^4)$ we shall also have derivatives of the brane tension, since at this order one has the Lie derivative $\mathcal{L}_{\bf n} {\cal E}_{\mu\nu}$ and $ \mathcal{L}_{\bf n} R_{\mu\nu\alpha\beta}$ leading to the aforementioned term $\nabla^\alpha{\cal B}_{\alpha(\mu\nu)}$, as well as to $\nabla_{\mu}{\cal B}_{\alpha\beta\nu}$ and $\nabla_{\mu}{\cal B}_{\beta\alpha\nu}$. These, together with second derivatives that naturally appear, at fourth order in the extra dimension lead to the following extra term,

\begin{equation}
 \begin{aligned}
g_{\mu\nu}^{(4)\,{}^{\rm variable}} &= - \frac{\kappa_5^2}{3}\left[\Box(\Box\sigma)g_{\mu\nu}-\nabla_{(\nu}\nabla_{\mu)}(\Box\sigma)\right] + \frac{\kappa_5^2}{3}
\left\{(\Box\sigma)R_{\mu\nu}-\nabla^\alpha\left[(\nabla_{(\mu\vert}\sigma)\,R_{\alpha\vert\nu)}\right]\right\}
\\
&+ \left(\frac{1}{3}\kappa_5^2+2 K\right)
\left\{(\Box\sigma){\cal E}_{(\mu\nu)} - \nabla^\alpha\left[(\nabla_{(\mu}\sigma)\, {\cal E}_{\nu)\alpha}\right]\right\}  
\\
&-2 K^{\tau\beta}
\left\{(\Box\sigma)R_{(\mu\vert\tau\vert\nu)\beta}
- \nabla^\alpha\left[(\nabla_{(\mu\vert}\sigma)\, R_{\alpha\tau\vert\nu)\beta}\right]\right\} 
\\
&+ \left(2\,K^2-\frac{1}{3}\Lambda_5 \right)
\left[(\Box\sigma)g_{\mu\nu}-\nabla_{(\nu}\nabla_{\mu)}\sigma\right] 
\\
&+\frac{\kappa_5^2}{3}
\left\{(\Box\sigma)\left(K_{(\mu\vert\rho}K_{\vert\nu)\beta} K^{\rho\beta} - K^2\,K_{\mu\nu}\right)
- \nabla^\alpha
\left[(\nabla_{(\mu\vert}\sigma)\left(K_{\alpha\sigma}K^{\;\,\sigma}_{\vert\nu)} - K\,K_{\alpha\vert\nu)}\right)
\right]
\right\}
\\
&+ 6 \left\{
(\Box\sigma)K_{(\mu\vert\tau}{\cal E}_{\vert\nu)}^\sigma
- \kappa_5^2\nabla^\alpha\left[
(\nabla_{(\mu}\sigma)\, {\cal E}_{\nu)\alpha}\right]
\right\} 
\\
&+ 2\left(K + \frac{7}{3}\kappa_5^2\right)
\left\{(\Box\sigma)K\,K_{\mu\nu}
-\nabla^\alpha\left[(\nabla_{(\mu\vert}\sigma)\, K\,K_{\alpha\vert\nu)}\right]
\right\}
\ .
\label{addtruey4}
 \end{aligned}
\end{equation}

These two extra terms in the expansion then make clear that the metric components themselves change once we consider a variable tension. To illustrate this formalism, in the next section we shall study how such changes are manifested in MGD black strings.

\section{MGD in the variable tension braneworld}

We now proceed into applying the formalism described above to a black string whose 4-dimensional geometry is determined on the brane by the MGD method as means to solve the effective EFE, as detailed in Sec. \ref{sec:mgd}. In particular, to see the effect of the variable tension, we will discuss how the horizon of MGD black strings evolves in time given the temporal dependence of the brane tension (here we shall consider only a time-dependent brane tension, $\sigma = \sigma(t)$, discarding the possibility of anisotropic branes).

First, as discussed before, the MGD 4-dimensional metric --- describing the exterior region surrounding a self-gravitating system on the brane --- is of the form

\begin{equation}
g_{\mu \nu} \de x^\mu \de x^\nu = -e^{\nu(r)}\mathrm{d}t^2 + e^{\lambda(r)}\mathrm{d}r^2 + r^2\mathrm{d}\Omega^2 \ ,
\label{eq:mgd4d}
\end{equation}

\noindent where, according to Eqs. \ref{eq:mgd_temp} and \ref{eq:mgd_rad},

\begin{equation}
 e^\nu = 1 - \frac{2M}{r} \ , \ \ \  e^{-\lambda} = \left (1 - \frac{2M}{r} \right ) \left ( 1 + \frac{b\beta}{r \left (1- \frac{3M}{2r} \right )}\right ) \ .
 \label{eq:mgdsum}
\end{equation}

To realize the time-variation of the five-dimensional horizon given by the variable brane tension, we can, for example, analyze the time-variation of the component $g_{\theta \theta}(x^\mu, y)$, since the horizon area is given by $A = \int \sqrt{g_{\theta \theta} g_{\varphi \varphi}} \de \theta \de \varphi$ (in the $\{t, r, \theta, \varphi\}$ coordinates). 

Naturally, the component $g_{\theta \theta}$ of Eq. \ref{eq:mgd4d} is not deformed under the MGD procedure. However, the component $g_{\theta \theta}(x^\mu, y)$ --- which, notice, is the y-dependent component \emph{of the black string} given by the expansion of Eq. \ref{eq:tay} ---, will also be time-dependent at order $\mathcal{O}(|y|^3)$ and higher, as argued above and given by the extra expansion terms in Eqs. \ref{addtruey3} and \ref{addtruey4}. At $\mathcal{O}(|y|^3)$, we straightforwardly have

\begin{equation}
 g_{\theta\theta}^{(3)\,{}^{\rm extra}}(t,r)
=
-\frac{2}{3}\kappa_5^2r^2 \sigma''
\ ,
\label{y33}
\end{equation}

\noindent where prime denotes differentiation with respect to time, $\sigma' = \de \sigma / \de t$. 

At $\mathcal{O}(y^4)$, we must calculate terms such as $\cal E_{\theta \theta}$, where the deformed MGD metric components $e^\lambda$ and $e^\nu$ are employed. Remind that we are in the exterior vacuum, where $T_{\mu \nu} = 0$, which implies on the brane $R_{\mu \nu} = -\cal E_{\mu \nu}$, $R^\mu {}_\mu={\cal E}^\mu{}_\mu = 0$, $\nabla^\nu {\cal E}_{\mu\nu}=0$ \cite{mgdvar2}. So, for instance,

\begin{equation}
{\cal E}_{\theta\theta}=-R_{\theta\theta}=\frac{r}{2}\,e^{-\lambda}\left(\partial_r\lambda-\partial_r\nu\right)+1-e^{-\lambda} =\frac{\beta(\sigma)}{2}\frac{\left(1-\frac{M}{r}\right)}{\left(1-\frac{3\, {M}}{2\,r}\right)^2} \ ,
\end{equation}

\noindent so that it is expected a $\beta$-dependence at $\mathcal{O}(y^4)$, which is a trace of the MGD procedure even at the component $g_{\theta \theta}$. The full calculation yields the extra term,

\begin{equation}
g_{\theta\theta}^{(4)\,{}^{\rm extra}}(t,r)
=
\kappa_5^2\,r^2
\left[\frac{\kappa_5^2\sigma^2}{9}
-\frac{\sigma^{\prime\prime\prime\prime}}{3} 
+\frac{\sigma^{\prime\prime}}{9}
\left(\frac{197}{72}\sigma^2\kappa_5^2 
- \Lambda_5-6\,\beta(\sigma)\,\frac{r-2M}{2\,r-3M}\right)
\right]
\ .
\label{termse}
\end{equation}

With these extra terms it then becomes clear that the horizon area will depend on time, whose explicit dependence depends only on the explicit functional expression of $\sigma(t)$. As argued before in this section, in the case of a Eötvös brane we can write the temporal dependence according to Eq. \ref{eq:sigmat}. 

Interesting consequences of a variable brane tension are discussed in \cite{quantum_portrait}, which will be discussed in greater detail in Sec. \ref{sec:portrait}.

\chapter{MGD black strings bulk singularities}
\label{sec:mbg_bs_s}

Following the discussion of last section, we shall now continue to study MGD black strings constructed as a Taylor expansion on the extra dimension near the brane ($y=0$), as given by Eq. \ref{eq:tay}. In particular, we shall be interested in determining whether or not there exist any singularities on the black string geometries \emph{at the bulk}. To do that, we shall compute the 5-dimensional MGD black string Kretschmann scalar, which, as discussed in Sec. \ref{sec:sings}, diverges in points which are real singularities. The results shown in this section are original, and shall be further explored to be published.

We are interested in the \emph{exterior} bulk black string, so that the 4-dimensional metric on the brane is determined by the MGD procedure as in Eq. \ref{eq:mgdsum}. Also, as we mentioned above, since we are in vacuum, one has the following field equations holding on the brane,

\begin{equation}
R_{\mu\nu}=-{\cal E}_{\mu\nu}\ , \ R^\mu {}_\mu ={\cal E}^\mu{}_\mu = 0 \ ,\  \nabla^\nu {\cal E}_{\mu\nu}=0 \ ,
\end{equation}

\noindent which then reduces Eq. \ref{eq:tay} to a much simpler expression \cite{mgdvar2}, up to $\mathcal{O}(|y|^3)$,

\begin{equation}
 \begin{aligned}
  g_{\mu\nu}(x^\alpha,y)&= g_{\mu\nu} -\frac{1}{3}\kappa_5^2\sigma g_{\mu\nu}\,|y| 
  \\
  &+\left[-{\cal E}_{\mu\nu} +\left(\frac{1}{36}\kappa_5^4\sigma^2 - \frac{1}{6}\Lambda_5\right)g_{\mu\nu}\right]\, y^2  
  \\
  &-\left(\left(\frac{193}{216}\sigma^3\kappa_5^6 +\frac{5}{18}\Lambda_5\kappa_5^2\sigma\right)g_{\mu\nu}+\frac{1}{6}\kappa_5^2{R}_{\mu\nu}\right)\,\frac{|y|^3}{3!} + \mathcal{O}(y^4) \ ,
  \label{eq:tay_vac}
 \end{aligned}
\end{equation}

\noindent where, just like before,

\begin{equation}
g_{\mu \nu} \de x^\mu \de x^\nu = -e^{\nu(r)}\mathrm{d}t^2 + e^{\lambda(r)}\mathrm{d}r^2 + r^2\mathrm{d}\Omega^2 \ ,
\end{equation}

\noindent with

\begin{equation}
 e^\nu = 1 - \frac{2M}{r} \ , \ \ \  e^{-\lambda} = \left (1 - \frac{2M}{r} \right ) \left ( 1 + \frac{b\beta}{r \left (1- \frac{3M}{2r} \right )}\right ) \ ,
\end{equation}

\noindent defining the 4-dimensional MGD metric on the brane, so that the black string components in the expansion above (Eq. \ref{eq:tay_vac}) constitutes the 5-dimensional MGD black string bulk metric given by,

\begin{equation}
 \bar{g}_{M N}\de x^M \de x^N = g_{\mu \nu}(x^\alpha,y) \de x^\mu \de x^\nu + \de y^2 \ .
\end{equation}

With our metric fixed, we can proceed to the calculation of the Kretschmann scalar $\varkappa = R_{\mu \nu \rho \sigma}R^{\mu \nu \rho \sigma}$. We shall do this order by order in the expansion (which we shall denote with a n-subscript, $\varkappa_n$) and analyze each answer up to $\mathcal{O}(|y|^3)$. Naturally, the computation of $\varkappa$ is extremely lengthy to be done by hand, so that we used the symbolic computation software Mathematica\textsuperscript{\textregistered} to do so.

At zeroth order $\mathcal{O}(y^0)$, the bulk metric has no y-dependence. Therefore, we do not expect to see any singularity at the bulk. However, it is interesting to see the radial singularities as well, which is why we calculated the corresponding Kretschmann scalar,

\begin{equation}
\begin{aligned}
  \varkappa_0 &= \frac{16}{r^6 }\left [\frac{2 b^2 \beta^2 (54 M^4-108 M^3 r+75 M^2 r^2-22 M r^3+3 r^4)}{(3 M-2 r)^4}  \right.
  \\[1.0ex]
  &\left. \qquad \quad +\, 3 M^2 -\frac{6 b \beta M (6 M^2-6 M r+r^2)}{(3 M-2 r)^2} \right ] \ .
\end{aligned}
\end{equation}

By inspecting its denominator, it is easy to see that, at $\mathcal{O}(y^0)$, the bulk metric has two radial singularities,

\begin{equation}
 r_{0,1} = 0\ , \ \ r_{0,2} = \frac{3M}{2}\ .
\end{equation}

Notice that $r_{0,1}$ has multiplicity 6, whilst $r_{0,2}$ has multiplicity 4. As expected, at $\mathcal{O}(y^0)$ we found no singularities at the bulk.

At first order $\mathcal{O}(|y|)$, we start having a y-dependence on the metric components $g_{\mu\nu}(x^\alpha,y) = g_{\mu\nu} -\frac{1}{3}\kappa_5^2\sigma g_{\mu\nu}\,|y|$. So, it is expected that $\varkappa_1$ will depend on $y$, and therefore might lead to bulk singularities. In fact, one has

\begin{equation}
\begin{aligned}
 \varkappa_1 &= \frac{144 \left [3 M^2 (3 M-2 r)^4-6 b \beta M (3 M-2 r)^2 (6 M^2-6 M r+r^2) \right ]}{r^6 (3 M-2 r)^4  (\kappa_5^2 \sigma |y| - 3)^2} 
 \\[1.0ex]
 &+ \frac{144 \left [2 b^2 \beta^2 (54 M^4-108 M^3 r+75 M^2 r^2-22 M r^3+3 r^4) \right ]}{r^6 (3 M-2 r)^4  (\kappa_5^2 \sigma |y| - 3)^2} \ .
 \end{aligned}
\end{equation}

Now, we easily see that $\varkappa_1$ is both a function of $r$ and $y$. Thus, by finding the roots of its denominator with respect to either variables, we can find the points (radial and at the bulk) in which $\varkappa_1$ diverges --- that is, the points of real singularities. It is easy to see that we recover the same radial singularities as at $\mathcal{O}(y^0)$ (and with the same multiplicities),

\begin{equation}
r_{1,1} = 0\ , \ \ r_{1,2} = \frac{3M}{2}\ .
\end{equation}

On the other hand, by finding the roots of the denominator of $\varkappa_1$ with respect to $y$, we easily find two bulk singularities,

\begin{equation}
y_{1,1} = \frac{3}{\kappa_5^2 \sigma} \ , \ \ y_{1,2} = -\frac{3}{\kappa_5^2 \sigma} \ .
\end{equation}

It is interesting to notice that the singularities are ``mirrored'' with respect to the brane at $y=0$, i.e., they are each located at one of its side, which is a natural consequence of the bulk $\mathbb{Z}_2$ symmetry and of considering $|y|$ in the expansion.

Now, at second order $\mathcal{O}(y^2)$, one gets even more interesting results. The metric components are $g_{\mu\nu}(x^\alpha,y)= g_{\mu\nu} -\frac{1}{3}\kappa_5^2\sigma g_{\mu\nu}\,|y| +\left(-{\cal E}_{\mu\nu} +\left(\frac{1}{36}\kappa_5^4\sigma^2 - \frac{1}{6}\Lambda_5\right)g_{\mu\nu}\right)\, y^2 $, but the corresponding Kretschmann scalar $\varkappa_2$ expression is too long to be displayed here entirely. However, luckily its denominator (denoted $\text{DEN}(\varkappa_2)$) is not that long: 

\vspace{-.4cm}

\begin{equation}
\begin{gathered}
  \text{DEN}(\varkappa_2) = r^2 (3 M-2 r)^2  
  \\
  \times  \left [36 (3 M-2 r)^2 r^2+(72 b \beta (M-r)+ (3 M-2 r)^2 r^2 (-6 \Lambda_5+\kappa_5^4 \sigma^2)) y^2 -12 \kappa_5^2 (3 M-2 r)^2 r^2 \sigma |y| \right]^2 
  \\
  \times \left[72 b \beta y^2+r^2 (-3 M+2 r) (36+(-6 \Lambda_5+\kappa_5^4 \sigma^2) y^2)+12 \kappa_5^2 (3 M-2 r) r^2 \sigma |y|\right]^2 \ .
\end{gathered}
\end{equation}

This denominator is null in $8$ different values of $y$, which determines the bulk singularities,

\begin{equation}
 y_{1,1} = \left(\frac{\sqrt{(3 M-2 r) \left [12 b \beta+\Lambda_5 r^2 (3 M-2 r) \right]}}{r \sqrt{6}  (3 M-2 r)}+\frac{\kappa_5^2 \sigma}{6} \right)^{-1} \ ,
\end{equation}

\begin{equation}
 y_{1,2} = - \left(\frac{\sqrt{(3 M-2 r) \left [12 b \beta+\Lambda_5 r^2 (3 M-2 r) \right]}}{r \sqrt{6}  (3 M-2 r)}+\frac{\kappa_5^2 \sigma}{6} \right)^{-1} \ ,
\end{equation}

\begin{equation}
 y_{1,3} = \left(\frac{\sqrt{(3 M-2 r) \left [12 b \beta+\Lambda_5 r^2 (3 M-2 r) \right]}}{r \sqrt{6}  (3 M-2 r)}-\frac{\kappa_5^2 \sigma}{6} \right)^{-1} \ ,
\end{equation}

\begin{equation}
 y_{1,4} = \left(-\frac{\sqrt{(3 M-2 r) \left [12 b \beta+\Lambda_5 r^2 (3 M-2 r) \right]}}{r \sqrt{6}  (3 M-2 r)}+\frac{\kappa_5^2 \sigma}{6} \right)^{-1} \ ,
\end{equation}

\begin{equation}
 y_{1,5} = \frac{6 r^2 (3 M-2 r)^2}{\sqrt{6 r^4 \Lambda_5 (3 M-2 r)^4 +72 b \beta r^2 (r-M) (3 M-2 r)^2}+\kappa_5^2 \sigma r^2 (3 M-2 r)^2} \ ,
\end{equation}

\begin{equation}
 y_{1,6} = \frac{6 r^2 (3 M-2 r)^2}{\sqrt{6 r^4 \Lambda_5 (3 M-2 r)^4 +72 b \beta r^2 (r-M) (3 M-2 r)^2}-\kappa_5^2 \sigma r^2 (3 M-2 r)^2} \ ,
\end{equation}

\begin{equation}
 y_{1,7} = \frac{-6 r^2 (3 M-2 r)^2}{\sqrt{6 r^4 \Lambda_5 (3 M-2 r)^4 +72 b \beta r^2 (r-M) (3 M-2 r)^2}+\kappa_5^2 \sigma r^2 (3 M-2 r)^2} \ ,
\end{equation}

\begin{equation}
 y_{1,8} = \frac{-6 r^2 (3 M-2 r)^2}{\sqrt{6 r^4 \Lambda_5 (3 M-2 r)^4 +72 b \beta r^2 (r-M) (3 M-2 r)^2}-\kappa_5^2 \sigma r^2 (3 M-2 r)^2} \ .
\end{equation}

It is very interesting to see that the bulk singularities depend on $r$ as well as on $\beta$ and $\sigma$. In fact, notice that at the GR limit $\sigma^{-1} \rightarrow 0$, all the bulk singularities at $\mathcal{O}(|y|^3)$ collapse to $y=0$, which may be seen as a confinement of gravity to the brane --- i.e., there is no black string, as there is actually no warped extra dimension or even the notion of a braneworld in GR, as it should be. Also, notice that we also have at third order in $y$ bulk singularities mirrored by the brane.

At $\mathcal{O}(y^2)$ one has 9 radial singularities, each with multiplicity 2 (i.e., $\text{DEN}(\varkappa_2)$ has 18 roots in $r$), which include $r = 0$ and $r = 3M/2$. The remaining 7 different radial singularities have way too long expressions to be shown here, but we stored them for eventual later use.

We were able to calculate the bulk singularities at third order $\mathcal{O}(|y|^3)$, however, the results are also extremely long and complicated to be worthy to be displayed here. Thus, it suffices to say that $12$ \emph{different} bulk singularities were found. The same happens with the radial singularities at $\mathcal{O}(|y|^3)$: one has 9 distinct radial singularities, each with multiplicity 2 including $r = 0$ and $r = 3M/2$, along with quite complicated expressions for the other 7, which we shall not write here.

The meaning of the aforementioned singularities, their expression, as well as their number, are still not yet entirely clear. Some open questions are: what do these bulk singularities represent? Is their number limited? What are their physical implications? A deeper analysis of these singularities, perhaps via the study of the causal structure of the spacetime leading to them, may shed light on these question, and is a interesting direction for future research. but it is outside the scope of this thesis.

\chapter{Braneworld applications}
\label{sec:apps_branes}

Now that we know the concept of braneworlds, and had contact with some explicit examples of such models, we will turn to new applications of the ideas and methods presented in this part of the thesis, thus, going beyond the analysis of features of the braneworld spacetime itself.

Before we present two original applications, it is important to remark how broad and extensive the landscape of applications and braneworlds constructions is --- in the literature, one can find many works making use of this wonder of gravity. For some examples, see \cites{76, 90_mgd_related_fim, 43_braneworlds_inicio, 65_bw, 84_bw, B8_bw, 2_bwrlds, 3_bwrlds, 5_bwrlds, 6_bwrlds, 7_bwrlds, 9_maartens, 10_maartens, 9_bw, 13_bw, 19_bw, 39_bw, 42_bw, 45_bw, 46_bw, 49_bw, 67_bw, 78_bw, 25_bw, 27_bw, 38_bw, 40_bw, new_bw}.

In this chapter, we shall present the main ideas behind two published works done in collaboration with R. da Rocha and A. Fernandes--Silva. These works apply the concept of braneworlds --- in particular, the MGD and EMGD braneworlds --- to the analysis of quite interesting physical systems: dark SU($N$) glueball condensates; and black holes described via the quantum portrait. These works build on existing ideas by extending their gravitational description to a braneworld scenario rather than a 4-dimensional standard GR scenario. Naturally, the main question is: what changes once we consider a braneworld model? Beyond that, each work examines particular phenomena of each system. 

After learning about these applications, it will then become clear how the braneworld scenarios actually can be applied to the study of concrete physical systems, and, more importantly, what changes when we consider these models. After briefly presenting these applications, we will finish this part of the thesis.

\section{MGD and dark SU(\texorpdfstring{$N$}{}) glueball condensates}
\label{sec:mgd_gb}

In the work \cite{glueballs}, we derive corrections to the gravitational wave radiation emitted by mergers of SU($N$) dark (E)MGD glueballs, as their spectra are obtained for two important cases in the MGD and EMGD setups. As a result, the window of observation of gravitational waves probed by eLISA and LIGO experiments becomes wider.

Hidden SU($N$) gauge sectors may be described by the scalar glueball dark matter paradigm. Glueballs can interact by exchanging gravitons, constituting a self-interacting system. When bosons and fermions interact with the SU($N$) scalar glueball, a Bose--Einstein condensation of glueballs can occur, which then originates compact stellar distributions. Also, when the self-gravity unbalances the energy density of the system, the Bose--Einstein condensation induces the glueball system into stellar distributions. The number $N$ of colors driving the SU($N$) gauge sectors and the scalar glueball mass, $m$, are the parameters that model the glueballs self-interaction. The condensation can occur in the ranges $10^3 \lessapprox N \lessapprox 10^6$ and $10$ eV $\lessapprox  m \lessapprox$ 10 KeV \cites{Soni:2016gzf,Forestell:2016qhc}.

The action for the glueball system is given by 

\begin{equation}\label{pot4}
\mathcal{S} =  \int d^4x\,\left(\frac{1}{2} g^{\mu\nu} \partial_\mu \phi \partial_\nu\phi - V(\phi)\right) \ .
\end{equation}

For more details concerning dark SU($N$) glueball condensates, which are beyond the scope of this thesis, see \cite{glueballs}.

In the work, it was proposed that a 5D Weyl fluid, in the membrane paradigm (see Sec. \ref{sec:memb_para}), can induce experimental signatures at the gravitational waves observatories LIGO and eLISA that are amplified by the EMGD procedure and its $\sigma^{-1}\to0$ GR limit. It is shown that observational signatures evinced from EMGD mergers are more probable to be experimentally detectable at these observatories, as the window for detection are wider than for the SU($N$) MGD dark glueball stars. For more details on mergers gravitational waves spectra, which are also beyond the scope of this thesis, see \cite{glueballs}.

The phenomenological upper bound $k\lesssim 4.2$ to the deformation parameter of EMGD (presented in Sec. \ref{sec:emgd_ct}) is presented and used throughout the work. Although the case for $k=1$ resembles a generalized form for the Reissner--Nordstr\"om metric, with tidal charge from a 5D bulk Weyl fluid, for $k=2$ the metric does not take any known form yet. So, it may be interesting to do future work to study the behavior of EMGD solutions in higher orders of $k$, up to the phenomenological bound, and assess the direct consequences of increasing $k$.

\section{EMGD and the quantum portrait of black holes}
\label{sec:portrait}

In the work \cite{quantum_portrait}, the extended quantum portrait of black holes was studied, by establishing an EMGD Bose--Einstein Condensate (BEC) modeling a compact stellar distribution in the membrane paradigm (see Sec. \ref{sec:memb_para}). The scalar invariants computed for the EMGD metric for all integer possible values of $k$ indicate possible coordinate singularities apart from the standard Schwarzschild ones, and no further 
singularities were observed if the brane tension obeys the bound $\sigma\gtrapprox2.83 \times 10^6 {\rm MeV}^4$. This value increases the range of possible values for the brane tension, compared to the previous $\sigma\gtrsim 3.18 \times 10^6 {\rm MeV}^4$, in \cite{Casadio:2016aum}.

The \emph{quantum portrait of black holes} consists in describing a black hole as a BEC of gravitons \cite{muck}, whose quantum mechanical behavior is captured by the solution of the $s$-wave modes of the time-independent Schr\"odinger equation for a particle of mass $m$ in a spherically symmetric P\"oschl-Teller potential, 

\begin{equation}
      \left( -\frac{\hbar^2}{2m} \, \nabla^2 +V(r) \right) \psi(\bm{x}) = E \psi(\bm{x})~,
\label{eq:eds}
\end{equation}

\noindent with

\begin{equation}
      V(r)=-\xi(\xi+1)\frac{\hbar^2\omega^2}{2m}\sec\!{\rm h}^2(\omega r).
\label{eq:posch_pot}
\end{equation}

The potential is self-generated by the condensate, so that its spherical symmetry is only reasonable if there is no angular momentum, which is why we discard the $l>0$ modes. Further, the $l=0$ modes ($s$-waves) are the only components effectively taking part in the scattering process. This being the case, one arrives at the radial Schr\"odinger equation with $l=0$ for $R(r)$,

\begin{equation}
\left( -\frac{\hbar^2}{2 m} \frac{\de^2}{\de \textcolor{black}{r}^2} + V(r) \right) R(r) = E R(r).
\label{eq:eds_radial}
\end{equation}

Its solution is given in terms of the Gauss hypergeometric function ${}_2\mathrm{F}_1$ by \cite{muck} 

\begin{equation}
R(r) =  {}_2\mathrm{F}_1 \left( \alpha_+ +\frac12, \alpha_- + \frac12; \frac32; -\sinh^2(\omega r) \right) a \sinh(\omega r)  \sec\!{\rm h}^\xi(\omega r) \ ,
\label{eq:radial_sol}
\end{equation}

\noindent where $a$ is a normalization constant and

\begin{equation}
\alpha_\pm = -\frac12 \left(\pm \xi + i \frac{{\rm k}}{\omega} \right) \ ,
\end{equation}
\noindent where the wave number $k$ is such that ${\rm k}^2 = 2mE/\hbar^2$.

For the bound states, the normalization condition imposes a discretization to the values of ${\rm k}$, given by \cite{muck} ${\rm k}_n = i \omega \left ( \xi - 1 -2n \right )$, for $n=0,1,\ldots, \left[\frac{\xi-1}2\right]$. The states with $E>0$ form a continuum of scattered states, given by a superposition of incoming and outgoing spherical waves \cite{muck}. 

To avoid superluminal scattering states, consider now the field $\psi$ to be a relativistic complex Klein--Gordon field, $\psi(t, \bm{x}) \equiv \psi(x)$, which is a solution to the Klein--Gordon equation with rest mass $\mu$ and two independent potentials, $S(\bm{x})$ and $V(\bm{x})$: 

\begin{equation}
\left( \hbar^2 \nabla^2 \!+\! \left(i\hbar\partial_t \!-\!V(\bm{x})\right)^2 \!-\! \left(S(\bm{x}) \!+\! \mu\right)^2 \right) \psi(x) =0 \ .
\end{equation}

The potentials are time independent, which allows us to write for the energy $\epsilon$, $\psi(x) = \mathrm{e}^{-i\epsilon t/\hbar} \psi(\bm{x})$. Thus, the particular choice $V=S$ yields
\begin{equation}
\left(-\frac{\hbar^2}{2(\epsilon+\mu)} \nabla^2 + V \right ) \psi(x) = \frac12( \epsilon-\mu) \psi(x)~,
\end{equation}

\noindent which is just Eq. \eqref{eq:eds} with constraints $m= \epsilon+\mu$ and $E= \frac12 \left(\epsilon-\mu\right)$. Hence, by choosing $V=S=V(r)$, the P\"oschl-Teller potential of Eq. \eqref{eq:posch_pot}, we can directly identify the solution of Eq. \eqref{eq:radial_sol} as the radial solution to the relativistic model, subjected to the constraints which represent the relativistic dispersion relation $\epsilon^2 - \mu^2 = \hbar^2 {\rm k}^2$. 

By imposing the so-called marginal binding condition\cites{casadio_bec, muck}, the ground state is fixed as the single bound state, whilst the first excited state is identified with the onset of the continuum. This condition realizes the Hawking radiation in the BEC black hole model through the \emph{quantum depletion process}, in which an excited graviton is emitted from the BEC as a scattering state. 

Now, this condition fixes the graviton effective mass as $\mu = \hbar \omega$. As $\epsilon_{n=1} = \epsilon_{{\rm k}=0} = \mu$, the energy gap between the ground state and the continuum is the graviton effective mass. In addition, for the ground state, one has $\xi = 2$, which fixes the P\"oschl-Teller potential as 

\begin{equation}
V(r) = {-3\hbar \omega}\,{\sec\!{\rm h}^2(\omega r)} \ .
\label{eq:posch_grd}
\end{equation}

For the interior EMGD metric, whose temporal component is given by (see Sec. \ref{sec:emgd}),

\begin{equation}
 e^{\nu(r)} = \left ( 1 - \frac{2m(r)}{r}\right )^{k+1} \ ,
 \label{eq:enu}
\end{equation}

\noindent where 
$m(r) = 4\pi \int_0^r \bar{r}^2 \epsilon(r) \de \bar{r}$ represents the quasilocal Misner--Sharp mass function.

Now, the main point provided by the quantum portrait of black holes comes into play when the fluid energy density $\epsilon$ (describing the BEC macroscopically) is identified to the charge density of the ground state complex Klein--Gordon field $\psi(x)$ (describing the BEC microscopically). Thus, the Klein--Gordon charge is interpreted as a gravitational charge, which allows one to make the connection between a graviton BEC and a black hole. With this identification, the Misner--Sharp mass function reads

\begin{equation}
\begin{aligned}
m(r) &=\frac{M}2 \tanh^3(\omega r) \left( 5 -3 \tanh^2(\omega r) \right) \ .
\label{bh:M.expl}
\end{aligned}
\end{equation} 

This is how we build the bridge between the quantum BEC description and the gravitational description of the EMGD black hole through its metric: by substituting Eq. \eqref{bh:M.expl} into Eq. \eqref{eq:enu} and rewriting it as

\begin{equation}
\begin{gathered}
e^{\nu(\rho)}_\vartheta =  \left (1 - \frac{1}{\rho} \tanh^3 (\vartheta \rho) \left (5 - 3 \tanh^2 (\vartheta \rho) \right ) \right )^{k+1} \ , 
\end{gathered}
\end{equation}

\noindent where $\vartheta \equiv M \omega $ and $\rho \equiv \frac{r}{M}$. The subscript makes clear that the metric component depends now on the parameter $\vartheta$, which is the de Broglie wavelength of the graviton, $\omega$, scaled by the BEC total mass $M$. Given this, a natural question is if there exists any interval for $\vartheta$ in which the existence of the black hole is limited. We address this question by analyzing the sign of the function $e^{\nu(\rho)}_\vartheta$, as a horizon exists if $e^{\nu(\rho)}_\vartheta \leq 0$. Evaluating $e^{\nu(\rho)}_\vartheta$ at its local minimum,
 
\begin{equation}
\begin{aligned}
e^{\nu(\rho_{\text{min}})}_\vartheta &= 1 - \frac{\vartheta}{y_{\text{min}}} \left [ \tanh^3 (y_{\text{min}}) \left (5 - 3 \tanh^2 (y_{\text{min}}) \right ) \right ]
 \\
 &\equiv 1 - \frac{\vartheta}{\vartheta_{\text{min}}},
\end{aligned}
\end{equation} 

\noindent which only yields a horizon if $ \vartheta \geq \vartheta_{\text{min}}$. It is thus straightforward to see that

\begin{equation}
\vartheta_{\text{min}} = \frac{y_{\text{min}}}{\tanh^3 (y_{\text{min}}) \left (5 \!-\! 3 \tanh^2 (y_{\text{min}}) \right )} \approx 0.69372 \ .
\end{equation}

Thus, the black hole exists if $ \vartheta \gtrsim 0.69372$. As argued in Ref. \cite{muck}, the most natural value for $\vartheta$ following the BEC quantum portrait of black holes is $\vartheta = 1 > \vartheta_{\text{min}}$, so that one can have a BEC black hole even in the EMGD setting. 

Beyond the scope of this thesis, Configurational Entropy (CE) was employed in the published work to probe the EMGD BEC. The CE global minima, for each fixed value of $k$, vary with respect to the value of the parameter $\vartheta$ involving the EMGD BEC mass and the characteristic length scale of the BEC. Specifically for the $k=1$ case, the CE indicates that there is a value for the  EMGD BEC mass and its characteristic length scale, encoded by the $\vartheta$ parameter, below which the EMGD BEC does not collapse, being configurationally stable. For more details on CE and its use in the study, see \cite{quantum_portrait}.

\newpage

\vspace*{\fill}
%
%

With this, we finish Part I of the thesis. At this point, we are familiar with the description of gravity in geometrical terms, how to generalize it to arbitrary dimensions other than 4, and how these ideas led us to the first wonder: braneworld scenarios and their physical implications, including some applications to quite interesting systems as described in the last two sections. In Part II, we switch gears and will be interested in a new set of ideas, which we collectively refer to as \emph{holography}. In a nutshell, in the next part of the thesis we will explore how two theories which are apparently unrelated, are actually deeply linked, in a way that allows us to study objects of one theory using tools of the other, and vice-versa. This is the main idea behind the \emph{holographic correspondences} which will be discussed in Part II.

\

\expandafter\pgfornament\expandafter{89}
\vspace*{\fill}

%
%

\part{Holography}

\chapter{Thermodynamics and black holes}
\label{sec:thermo}

We begin Part II of the thesis by discussing some very interesting relationships between black holes and thermodynamics, which will provide us with the first hints towards the correspondences which will be explored in the next sections.

In this section we will discuss, following \cite{natsuume}, the thermodynamic properties of black holes. We begin by noticing that a stationary black hole is characterized only by a few parameters (the \emph{no-hair theorem}) and constrained by only a few initial conditions. This is similar to thermodynamics, the theory which provides a characterization of a system composed by many bodies in terms of a few macroscopic parameters, such as pressure and temperature. This prescription is known as \emph{coarse-graining}, and since a black hole is described by only a few parameters, one may suggest that the standard black hole is a coarse-grained description of a microscopic black hole theory, which, on the other hand, is still unknown. Still, we may trace a quite close parallel between black holes and thermodynamics.

First, let us look at the thermodynamics zeroth law, which states that a thermodynamic system will eventually reach the thermal equilibrium state, in which the temperature is constant everywhere. On the black hole side, we know that a black hole will eventually become spherically symmetric, even if it initially was not, which implies a constant gravitational force over the horizon. In other words, we can say that the \emph{surface gravity} of a black hole will eventually become constant, which can be seen as an equilibrium state for a stationary black hole. 

The surface gravity is understood as the force per unit mass which is necessary to hold a body at the horizon by an asymptotic observer. For a generic static metric of the form $-g_{tt} = 1/g_{rr} = f(r)$, one has for the surface gravity $\kappa$,

\begin{equation}
 \kappa = \frac{1}{2} \left .\frac{\mathrm{d}f(r)}{\mathrm{d}r} \right \rvert_{r=r_0} \ ,
\end{equation}

\noindent where $r_0$ denotes the horizon radius. For the Schwarzschild black hole (SBH), one has $f(r) = 1- 2M/r$ and $r_0 = 2M$, so that its surface gravity is given by

\begin{equation}
 \kappa_{SBH} = \frac{1}{4M} \ .
\end{equation}

If we finally consider that both the surface gravity as well as the temperature are non-negative quantities, the analogy between the zeroth law of thermodynamics and black hole physics becomes even stronger.

The SBH horizon area is given by $A = 4\pi r_0^2 = 16 \pi M^2$, from which it is easy to see that the horizon area increases as the black hole mass increases, that is, as matter falls in. If we take into consideration that nothing can come out from the black hole, it is natural to expect that its mass will not decrease (we are still not considering Hawking radiation effects), which then implies that the area is a non-decreasing quantity. In thermodynamics, we also have a non-decreasing quantity: \emph{entropy}, which then suggests that the horizon area and entropy may be related. We shall soon see that this is precisely the case, but for now, let us keep our analogy-based discussion.

From the SBH horizon area expression, we can see that a change $\mathrm{d}A$ in the horizon area is related to a change $\mathrm{d}M$ in the black hole mass according to: 

\begin{equation}
\begin{gathered}
\mathrm{d}A = 32 \pi M \mathrm{d}M = 8\pi 4 M \mathrm{d}M 
\\
\Rightarrow \mathrm{d}A =  \frac{8\pi}{\kappa} \mathrm{d}M \ ,
\end{gathered}
\end{equation}

\noindent that is,

\begin{equation}
 \mathrm{d}M = \frac{\kappa}{8\pi} \mathrm{d}A \ .
 \label{eq:bh_1st_law}
\end{equation}

If we now remark that in the black hole picture the surface gravity is related to the thermodynamic temperature, and that entropy is related to the horizon area, we can directly compare Eq. \ref{eq:bh_1st_law} to the first law of thermodynamics, $\mathrm{d}E = T \mathrm{d}S$. We refer to Eq. \ref{eq:bh_1st_law} as \emph{the first law of black holes}.

Before we see that the similarity between black hole physics and thermodynamics go beyond mere analogy, let us first notice that the resemblance of the equations really do not imply an exact same physics. 

Namely, it is not straightforward to directly identify black hole physics to thermodynamics because, if nothing comes out from the black hole, it does not emit thermal radiation, in which case the notion of temperature is not well-defined. And there is another problem: in units where the Boltzmann's constant is set to unit, $k_B = 1$, one has a dimensionless entropy, whilst the horizon area has area dimension, so that it should be divided by a length squared to be dimensionless. However, which length that should be is not entirely obvious.

Now, we must take into account that a black hole is not an isolated system, in the sense that matter can enter a black hole and it is, after all, the result of the gravitational collapse of a star. Since the behavior of matter is microscopically described by quantum mechanics, one may consider such quantum effects when dealing with black hole physics. In fact, one can do \emph{quantum field theory in curved spacetime} and take into account the quantum behavior of matter in the context of black hole physics, in a \emph{semiclassical gravitation} approach.

In 1975, by quantizing matter fields in a black hole background \cite{hawk_rad}, Stephen Hawking was able to show that a black hole indeed \emph{emits} black body radiation, which we now know as \emph{Hawking radiation}. This allows us to precisely define a temperature associated to the black hole, the Hawking Temperature, which is given by,

\begin{equation}
T = \frac{\hbar \kappa}{2\pi k_B  c} \ ,
\label{eq:hawk_temp}
\end{equation}

\noindent in which we recovered the constants to make the units explicit. For the SBH, one has

\begin{equation}
T = \frac{\hbar c^3}{8\pi k_B  GM} \ .
\end{equation}

Recovering $G$, one has, for Eq. \ref{eq:bh_1st_law},

\begin{equation}
\begin{gathered}
 \mathrm{d}M = \frac{\kappa}{8\pi G} \mathrm{d}A 
 \\
 \Rightarrow  \mathrm{d}(Mc^2) = \frac{\kappa c^2}{8\pi G} \mathrm{d}A = \frac{\hbar \kappa}{2 \pi k_B c} \frac{k_B c^3}{4G\hbar} \mathrm{d}A 
 \\
 \Rightarrow  \mathrm{d}(E) = T \frac{k_B c^3}{4G\hbar} \mathrm{d}A \ .
\end{gathered}
\end{equation}

If we now compare the result above with the first law of thermodynamics, $\mathrm{d}E = T \mathrm{d}S$, we finally obtain the precise relation between the entropy and the horizon area,

\begin{equation}
 S = \frac{k_B c^3}{4 G \hbar} A = \frac{k_B}{4 l_{pl}^2} A \ ,
\end{equation}

\noindent where it becomes clear that the length necessary to make the horizon area the same units of the entropy is the \emph{Plank length}, defined as $l_{pl} \equiv (G \hbar/c^3)^{-1/2}$, which establishes the scale in which quantum gravity effects are important. If we now go back to units $k_B = c = \hbar = 1$, but keeping the $D$-dimensional newton constant $G_D$, one has

\begin{equation}
 S = \frac{A}{4 G_D} \ ,
 \label{eq:area_law}
\end{equation}

\noindent which is known as the \emph{area law}, which is always valid for a spacetime which is solution to a gravitational action of the Einstein--Hilbert action form, $S = (1/16\pi G_D) \int \mathrm{d}^D x \sqrt{-g} R$, as introduced in Sec. \ref{sec:EFEqs}.

At this point, let us make an important remark. Notice that the black hole entropy is proportional to the horizon area, whilst the statistical entropy is proportional to the volume of a system. Therefore, a 4-dimensional black hole cannot be directly identified to a 4-dimensional statistical system. However, a 5-dimensional area can be seen as a 4-dimensional volume, so that we should rather identify a 5-dimensional black hole to a 4-dimensional statistical system. 

In sum, the black hole entropy suggests that the description of a black hole in terms of a statistical system must be such that this system is in a space with one spatial dimension lower than that of the gravitational theory! This general idea, that the description of a given theory may be encoded in another theory living in a space with one spatial dimension less, is known as the \emph{holographic principle}, and plays a central role in Part II of the thesis, being a very important hint towards the holographic correspondences, which we will discuss in more detail later on. 

\chapter{The AdS spacetime}
\label{sec:ads}

The Anti-de Sitter (AdS) spacetime is a \emph{constant curvature} spacetime, which is a solution to the Einstein Field Equations (EFE) with a constant and negative cosmological constant, as we shall see in this section. This spacetime will play a major role in our future discussion regarding the AdS/CFT duality, so in this section we present some of its properties, following \cite{natsuume}.

To begin the discussion, let us consider an ambient flat spacetime with two timelike directions $Z$ and $X$, that is,

\begin{equation}
 ds^2 = -\mathrm{d}Z^2 - \mathrm{d}X^2 + \mathrm{d}Y^2 \ .
\end{equation}

The 2-dimensional AdS spacetime, AdS$_2$, can be embedded in such a spacetime, and is defined by the surface

\begin{equation}
 -Z^2 -X^2 + Y^2 = -L^2 \ ,
\end{equation}

\noindent where $L$ is the \emph{AdS radius}. Notice that AdS$_2$ has the $SO(2,1)$ invariance of the ambient spacetime in which it is embedded. Now, by the choice of coordinates

\begin{equation}
 Z = L \cosh \rho \cos \tau \ , \ X = L \cosh \rho \sin \tau \ , \ Y = L \sinh \rho \ ,
\end{equation}

\noindent the metric becomes

\begin{equation}
 ds^2 = -L^2 \cosh^2 \rho \mathrm{d}\tau^2 + L^2 \mathrm{d} \rho^2 \ .
\end{equation}

The coordinate system $\{\tau, \rho\}$, where $-\infty < \tau < \infty$ and $0 < \rho < \infty$, is called \emph{global coordinates}, which makes clear that despite the embedding of AdS$_2$ in a flat spacetime with two timelike directions, AdS$_2$ itself has only one. 

From the metric above, it is easy to check that AdS$_2$ has a constant negative curvature, given by the Ricci scalar $R = -2/L^2$.

A convenient coordinate system is the \emph{Poincaré coordinates} $\{t, r \}$, with $-\infty < t < \infty$ and $0 < r < \infty$, which is defined by

\begin{equation}
 Z = \frac{Lr}{2} \left ( -t^2 + \frac{1}{r^2} + 1 \right ) \ , \ X = Lrt \ , \  Y = \frac{Lr}{2} \left ( -t^2 + \frac{1}{r^2} - 1 \right ) \ ,
\end{equation}

\noindent in which chart the metric becomes,

\begin{equation}
 ds^2 = -L^2  r^2 \mathrm{d}t^2 + \frac{L^2}{r^2} \mathrm{d} r^2 \ .
\end{equation}

We will most often use Poincaré coordinates when we start discussing AdS/CFT later on (when we will often set the AdS radius to unit, $L = 1$). An important thing to notice is that the AdS spacetime naturally has the notion of a boundary, the \emph{AdS boundary} \cite{ads_boundary}, which is located, in the Poincaré coordinates, at $r \rightarrow \infty$. The existence of a boundary will also play a role within AdS/CFT, as we shall soon discuss.

It is easy to generalize the results above to higher dimensions. To do this, we shall add $D$ spatial dimensions to the ambient spacetime,

\begin{equation}
ds^2_{D+3} = -\mathrm{d} X_0^2 - \mathrm{d}X_{D+2}^2 + \mathrm{d}X_1^2 + \cdots + \mathrm{d}X_{D+1}^2 \ .
\end{equation}

The $(D+2)$-dimensional version of the AdS spacetime is defined by

\begin{equation}
-X_0^2 - X_{D+2}^2 + X_1^2 + \cdots + X_{D+1}^2 = -L^2 \ ,
\end{equation}

\noindent so that, in the Poincaré coordinates given by

\begin{equation}
\begin{gathered}
X_0 = \frac{Lr}{2} \left (x_i^2 - t^2 + \frac{1}{r^2} + 1 \right ) \ , \ X_{D+2} = Lrt \ , 
\\
X_i = Lrx_i \ , \ X_{D+1} = \frac{Lr}{2} \left (x_i^2 - t^2 + \frac{1}{r^2} - 1 \right ) \ ,
\end{gathered}
\end{equation}

\noindent where $i = 1, ..., D$, one has the metric,

\begin{equation}
ds^2 = -L^2 r^2 \mathrm{d}t^2 + L^2 r^2 \delta_{ij} \mathrm{d}x^i \mathrm{d}x^j + \frac{L^2}{r^2} \mathrm{d}r^2 \ ,
\end{equation}

\noindent where $\delta_{ij}$ here denotes the D-dimensional Kronecker delta. Notice that AdS$_{D+2}$ carries the $SO(D+2, 1)$ invariance from the ambient spacetime, thus being a \emph{maximally symmetric spacetime}, which allows us to write the Riemann tensor for AdS$_{D+2}$ simply as

\begin{equation}
R_{\mu \nu \sigma \lambda} = - \frac{1}{L^2} \left ( g_{\mu \sigma} g_{\nu \lambda} - g_{\mu \lambda} g_{\nu \sigma} \right ) \ ,
\end{equation}

\noindent where we here use Greek indices running from 0 to D+2. The Ricci tensor is

\begin{equation}
R_{\mu \nu} = - \frac{D+1}{L^2} g_{\mu \nu} \ ,
\end{equation}

\noindent so that the Ricci scalar is simply

\begin{equation}
R = - \frac{(D+1)(D+2)}{L^2} \ .
\end{equation}

Thus, one has, for AdS$_{D+2}$,

\begin{equation}
R_{\mu \nu} - \frac{1}{2} R g_{\mu \nu} = -\frac{D (D+1)}{2L^2} g_{\mu \nu} \ ,
\end{equation}

\noindent which is nothing but the vacuum EFE with a negative cosmological constant,

\begin{equation}
\Lambda = -\frac{D (D+1)}{2L^2} \ ,
\label{eq:lambda_def}
\end{equation}

\noindent as we had anticipated.

In the context of AdS/CFT correspondence, we shall be interested in the AdS$_5$ spacetime, so we shall write it explicitly, with the AdS radius set to 1, for later reference,

\begin{equation}
 ds^2_{AdS_5} = -r^2 \mathrm{d}t^2 + r^2 \left (\mathrm{d}x^2 + \mathrm{d}y^2 + \mathrm{d}z^2 \right ) + \frac{1}{r^2} \mathrm{d}r^2 \ .
\label{eq:ads_5_metric}
\end{equation}

\section{The AdS black hole}
\label{sec:sads}

Black holes are also possible in AdS spacetime. The most simple example is the AdS--Schwarzschild black hole, which is a solution to the vacuum EFE with a negative cosmological constant, like the AdS spacetime. Of particular interest is the 5-dimensional AdS--Schwarzschild (SAdS$_5$) black hole with a planar horizon, given by,

\begin{equation}
ds^2_{SAdS_5} = -r^2 f(r) \mathrm{d}t^2 + \frac{1}{r^2 f(r)} \mathrm{d}r^2 + r^2 \left (\mathrm{d}x^2 + \mathrm{d}y^2 + \mathrm{d}z^2 \right ) \ ,
\label{eq:sads_5_r}
\end{equation}

\noindent where

\begin{equation}
f(r) = 1 - \frac{r_0^4}{r^4} \ .
\end{equation}

The horizon is located at $r=r_0$. Notice that it extends indefinitely in the $(x, y, z)$-directions, i.e., it is a planar horizon, so that it is more appropriate to refer to such a black hole as a \emph{black brane}, which we shall do from now on. Also notice that Eq. \ref{eq:sads_5_r} reduces to Eq. \ref{eq:ads_5_metric} for $r_0 = 0$.

After we discuss the AdS/CFT correspondence and its methods, we shall see how the SAdS$_5$ black brane is related to the strongly-coupled $\mathcal{N}=4$ plasma, and how we can use such a relation to perform non-equilibrium physics calculations. Before one gets to that, it is important to first discuss the import methods of linear response theory, which will be done in the next section.

\chapter{Linear response theory and Kubo formula} 
\label{sec:lrt}

In this section we will shortly describe linear response theory and its methods, namely the Kubo formula, in a presentation which follows \cites{son_hydro, natsuume}.

Let us consider a theory described by an action $S$. It is often of interest to determine what is the response of a given operator $\mathcal{O}$, which we denote $\delta \langle \mathcal{O} \rangle$, when one adds an external source, which we denote $\varphi^{(0)}$, coupled to it. The adding of the external source can be seen as a perturbation of the original theory, that is,

\begin{equation}
 S \mapsto S + \int \mathrm{d}^4x \varphi^{(0)}(t, \bm{x}) \mathcal{O}(t, \bm{x}) \ .
\end{equation}

To analyze such a response, we can employ \emph{linear response theory}, which is \emph{linear} because the response is studied up to linear order in the external source. The method is implemented by applying the time-dependent perturbation theory of quantum mechanics. The goal is to obtain the response according to,

\begin{equation}
\delta \left \langle \mathcal{O} (t, \bm{x})\right \rangle := \left \langle \mathcal{O} (t, \bm{x})\right \rangle_S - \left \langle \mathcal{O} (t, \bm{x})\right \rangle \ ,
\end{equation}

\noindent where $\left \langle \mathcal{O} (t, \bm{x})\right \rangle_S$ denotes the ensemble average of the operator $\mathcal{O}$ in the presence of the external source; and $\left \langle \mathcal{O} (t, \bm{x})\right \rangle$ when it is absent. By considering that the unperturbed system is in equilibrium at $t = t_0 \rightarrow -\infty$, and that the external source $\varphi^{(0)}$ is turned on at $t=t_0$, we can then use the fact that the operator $\mathcal{O}$ evolves in time with the time-evolution operator constructed with the unperturbed Hamiltonian $H_0$, that is,

\begin{equation}
\mathcal{O}(t, t_0) = U_0^{-1} \mathcal{O} U_0 \ , \  U_0 = \exp \left ( -i H_0 (t-t_0)\right ) \ ,
\end{equation}

\noindent which allows us to find \cite{natsuume},

\begin{equation}
\delta \left \langle \mathcal{O} (t, \bm{x})\right \rangle = i \int^{\infty}_{-\infty} \mathrm{d}^4 x \theta (t-t^{\prime}) \left \langle \left [ \mathcal{O} (t, \bm{x}), \mathcal{O}(t^{\prime}, \bm{x^{\prime}}) \right ] \right \rangle \varphi^{(0)}(t^{\prime}, \bm{x^{\prime}}) \ ,
\end{equation}

\noindent where $\theta$ is the step function. If we now define the retarded Green's function associated to $\mathcal{O}$ as

\begin{equation}
 G_R^{\mathcal{O}, \mathcal{O}}(t-t^{\prime}, \bm{x} - \bm{x^{\prime}}) = -i \theta (t-t^{\prime}) \left \langle \left [ \mathcal{O} (t, \bm{x}), \mathcal{O}(t^{\prime}, \bm{x^{\prime}}) \right ] \right \rangle \ ,
\end{equation}

\noindent one gets,

\begin{equation}
\delta \left \langle \mathcal{O} (t, \bm{x})\right \rangle = - \int^{\infty}_{-\infty} \mathrm{d}^4 x  G_R^{\mathcal{O}, \mathcal{O}}(t-t^{\prime}, \bm{x} - \bm{x^{\prime}}) \varphi^{(0)}(t^{\prime}, \bm{x^{\prime}}) \ ,
\end{equation}

\noindent whose Fourier transformation yields, in momentum space, simply,

\begin{equation}
\delta \left \langle \mathcal{O} (\omega, \bm{q})\right \rangle = - G_R^{\mathcal{O}, \mathcal{O}} (\omega, \bm{q}) \varphi^{(0)}(\omega, \bm{q}) \ .
\label{eq:response_def}
\end{equation}

Eq. \ref{eq:response_def} then gives the response to the operator given the introduction of a coupled external source, which is narrowed down to the determination of the retarded Green's function. However, this determination is not always immediate. Ultimately, we shall see that AdS/CFT determines the Green's function in its manner, but we shall not discuss this right now. Rather, we shall explore the relationship between the Green's functions and the transport coefficients, which will become clearer within the formalism of hydrodynamics which we will discuss below. But, before we proceed, we can see the link between the Green's function and the transport coefficients --- the Kubo formula --- right away.

Consider Ohm's law, which can be expressed as the response to the conserved current $J^\mu$ as an external electrical field $\bm{E^{(0)}}$ is added. Particularly, in the x-direction,

\begin{equation}
 \delta \left \langle J^x \right \rangle = \sigma E_x^{(0)} \ ,
\end{equation}

\noindent where $\sigma = \sigma(\omega)$ is the (frequency-dependent) electrical conductivity, a transport coefficient. In the gauge $A_0^{(0)} = 0$, where $A_0^{(0)}$ is the potential conjugate to the charge density $\rho = J^0$, the external electric field is given by $E_x^{(0)} = -\partial_t A_x^{(0)}$, which is Fourier-transformed to $E_x^{(0)} = i \omega A_x^{(0)}$, so that one has

\begin{equation}
 \delta \left \langle J^x \right \rangle = i \omega \sigma A_x^{(0)} \ .
\end{equation}

Notice that the actual external source which couples to $J^x$ to perturb the action is $A_x^{(0)}$, so that, from Eq. \ref{eq:response_def}, one has

\begin{equation}
 \delta \left \langle J^x \right \rangle = -G_R^{xx} A_x^{(0)} \ ,
\end{equation}

\noindent so that we can easily obtain the conductivity as

\begin{equation}
\sigma(\omega)= - \frac{G_R^{xx}(\omega, \bm{0})}{i\omega} \ ,
\label{eq:kubo_condu}
\end{equation}

\noindent where we we must have $\bm{q} = \bm{0}$ in the green function, since the conductivity does not depend on the momentum $\bm{q}$, naturally. The frequency-independent electrical conductivity is obtained in the $\omega \rightarrow 0$ limit of Eq. \ref{eq:kubo_condu},

\begin{equation}
\sigma(\omega \rightarrow 0)= - \lim_{\omega \rightarrow 0} \frac{1}{\omega} \Im \left (G_R^{xx}(\omega, \bm{0}) \right ) \ .
\label{eq:kubo_condu_dc}
\end{equation}

Eq. \ref{eq:kubo_condu} is our first example of a \emph{Kubo formula}, which is a relation between a transport coefficient and a retarded Green's function. In what follows, we will be ultimately interested in obtaining the viscosity $\eta$, a transport coefficient of great relevance in fluid dynamics. This will be done with a Kubo formula for $\eta$, which we will soon derive by applying the general ideas of linear response theory discussed so far. But, before one gets to that, it is convenient to introduce the hydrodynamics formalism, in which the transport coefficients such as $\eta$ will naturally appear playing a major role.

\chapter{Hydrodynamics}
\label{sec:hydro}

In this section, we shall present an overview of hydrodynamics, in a presentation following that of \cite{natsuume}.

The formalism of hydrodynamics provides a description of the macroscopic behavior of a given system. More precisely, it refers to the dynamics of macroscopic variables, among which we are mainly interested in conserved quantities, which are guaranteed to survive in the \emph{hydrodynamic limit}, characterized by a low-energy and long-wavelength regime. This limit will play a major role when we discuss fluid-gravity correspondence later on.

Notice that hydrodynamics is not limited to the description of water or generic fluids, but to any system on which a macroscopic description may be of interest. Thus, from the field theoretical point of view, hydrodynamics is an \emph{effective theory}, and, as such, we cannot expect it to carry the details of a microscopic theory, which are encoded in the \emph{transport coefficients}, that are necessary if we want to know responses, but cannot be determined by the formalism of the effective theory alone. 

To see how one actually determines a transport coefficient, and its relation to the Green's function, let us consider the diffusion problem, which illustrates many of the core ideas of the hydrodynamics formalism.

In the diffusion problem, we consider a current $J^\mu$, such that $J^0 = \rho$ is a conserved charge density, along with a conservation law,

\begin{equation}
 J^\mu = \left ( \rho, J^i \right )^T \ , \ \partial_\mu J^\mu = 0 \ .
\end{equation}

One has, in $(3+1)$ dimensions, 4 variables and a single relation provided by the conservation equation, so that the problem is not closed. To close the equation of motion, we must introduce an additional equation, known as \emph{constitutive equation}, which is of phenomenological origin, and determined to satisfy the second law of thermodynamics. In the diffusion problem, the constitutive equation is Fick's law, which states that a charge gradient produces a current,

\begin{equation}
 J^i = -D \partial^i \rho \ ,
\end{equation}

\noindent where $D$ is the diffusion constant, an (undetermined) transport coefficient which is introduced by the constitutive equation. Notice that the constitutive equation gives us the additional equations (3, in this case) necessary to close the equation of motion, and if we combine it with the conservation law, one obtains an equation for the charge density $\rho$, the \emph{diffusion equation},

\begin{equation}
 \partial_0 \rho - D \partial^2 \rho = 0 \ ,
\end{equation}

\noindent where we denote $\partial_i \partial^i \equiv \partial^2$. Now, to fully solve the problem and be able to determine responses, we must determine the transport coefficient, in this case the diffusion constant $D$, that was introduced. As we saw in Sec. \ref{sec:lrt} when we determined the electrical conductivity, it is possible to to obtain a transport coefficient from linear response theory, which provides a microscopic approach, through the Kubo formula.

But, in the diffusion problem, we are interested in the behavior of the inhomogeneous charge density fluctuation, which is a statistical problem that cannot be expressed as a perturbed Hamiltonian \cite{natsuume}, so that the linear response theory is no longer useful. That is also the case for the fluid dynamics problem which we will soon address. 

To be able to apply linear response theory, we must then introduce an inhomogeneous chemical potential $\mu$, which produces the inhomogeneity of $\rho$ as coming from an external source. That is, we shall rewrite Fick's law as,

\begin{equation}
 J^i = -D \partial^i \rho = -D \left ( \frac{\partial \rho}{\partial \mu} \right ) \partial^i \mu \ .
\end{equation}

If we now define the thermodynamic susceptibility $\chi_T := \partial \rho / \partial \mu$ and interpret the term $-\partial^i \mu$ as an electric field for the current, $E^i := -\partial^i \mu$, one obtains,

\begin{equation}
 J^i = (D \chi_T)E^i \ ,
\end{equation}

\noindent which is precisely Ohm's law, $J^i = \sigma E^i$. Therefore, one has $\sigma = D \chi_T$, and, from Eq. \ref{eq:kubo_condu_dc}, we promptly obtain the Kubo formula for the diffusion constant,

\begin{equation}
D = - \frac{1}{\chi_T} \lim_{\omega \rightarrow 0} \frac{1}{\omega} \Im \left (G_R^{xx}(\omega, \bm{0}) \right ) \ .
\end{equation}

Notice that the formalism of hydrodynamics provided us with a further simplification: first, we had narrowed down the determination of the responses to the determination of the retarded Green's function, using the linear response theory in Sec. \ref{sec:lrt}. Now, hydrodynamics further narrows down the problem to the determination of transport coefficients --- which, notice, are given in terms of only \emph{part} of the Green's function (namely, its form in the hydrodynamic limit $\omega \rightarrow 0$, $\bm{q} = \bm{0}$). We shall then soon see how AdS/CFT allows us to determine the transport coefficients from a microscopic approach, completely solving the problem of determining responses.

\section{Fluid dynamics}
\label{sec:fluids}

Let us now focus on fluid dynamics, within the hydrodynamics formalism discussed above. Fluids will a play a major role in later sections, but for now, a higher-level view of the basic formalism will be very important. For a simple fluid, we consider as the macroscopic variables the energy-momentum (stress) tensor $T^{\mu \nu}$ (discussed in Sec. \ref{sec:em_tensor}), along with its conservation law,

\begin{equation}
 T^{\mu \nu} \ , \ \partial_\mu T^{\mu \nu} = 0 \ .
\end{equation}

In $(3+1)$ dimensions, $T^{\mu \nu}$, a symmetric rank-$(2,0)$ tensor, has 10 independent components, whilst its conservation law provides us with only 4 equations. So, again, to close the equations of motion, we must introduce a constitutive equation, which is conveniently written in terms of the normalized fluid velocity field $u^\mu(x^\mu)$, such that $u^\mu u_\mu = -1$, and the rest-frame energy density $\rho(x^\mu)$. 

We shall introduce a constitutive equation by determining the form of $T^{\mu \nu}$ in a derivative expansion. As we discussed in Part I of the thesis, the \emph{perfect fluid} is such that it is, in the local rest frame, fully characterized by its energy density $\rho$ and isotropic pressure $p$, thus corresponding to the zeroth order in the derivative expansion, so that its stress tensor is given by

\begin{equation}
 T^{\mu \nu} = (\rho + p) u^\mu u^\nu + P \eta^{\mu \nu} \ .
\end{equation}

Notice that in the local rest frame one has $u^\mu = (1, 0, 0, 0)^T$, so that the stress tensor takes the form $T^{\mu \nu} = \text{diag}(\rho, P, P, P)$, as desired, and $T^{\mu \nu}$ encodes all macroscopic quantities characterizing a fluid, as detailed in Sec. \ref{sec:em_tensor}.

With the constitutive equation, we can then write the closed equations of motion for the perfect fluid, which are the continuity equation and the Euler equation, respectively given by,

\begin{equation}
 \partial_\mu T^{\mu 0} = 0 \ ; \ \partial_\mu T^{\mu i} = 0 \ .
\end{equation}

Notice that the introduction of the constitutive equation for the perfect fluid did not introduce any transport coefficient, which changes when we consider the next order in the derivative expansion, since in first order we can include the effects of dissipation which are absent in the perfect fluid. We then have the \emph{viscous fluid}, whose stress tensor is expressed as,

\begin{equation}
 T^{\mu \nu} = (\rho + p) u^\mu u^\nu + P \eta^{\mu \nu} + \tau^{\mu \nu} \ ,
\end{equation}

\noindent where the term $\tau^{\mu \nu}$ contains the dissipative effects. It is such that, in the local rest frame, one has

\begin{equation}
 \tau_{ij} = -\eta \left ( \partial_i u_j + \partial_j u_i -\frac{2}{3} \delta_{ij} \partial_k u^k \right ) - \zeta \delta_{ij} \partial_k u^k \ .
\end{equation}

Notice that two transport coefficients are introduced with the consideration of dissipative effects in the fluid: the \emph{shear viscosity} $\eta$, and the \emph{bulk viscosity} $\zeta$. Like the perfect fluid, the introduction of the constitutive equation for the viscous fluid closes the equations of motion, from which one obtains the continuity equation and the Navier--Stokes equation as a result. Viscous fluids and Navier--Stokes will appear once again in our discussion in due time.

\section{Kubo formula for the shear viscosity}
\label{sec:kubo_visco}

We shall now focus on the determination of the Kubo formula for the shear viscosity $\eta$. We derive the Kubo formula by coupling fictitious gravity to the fluid and then determining the response of $T^{\mu \nu}$ under gravitational perturbations, which is in agreement with general relativity, which states that spacetime fluctuations bring up fluctuations in the stress tensor.

At this stage, one should see this procedure just as a quick way to derive the Kubo formula, as one does not really curve spacetime in fluid experiments. However, we shall soon see that the derivation presented here has a natural interpretation within the AdS/CFT duality framework.

First, we add a gravitational perturbation to the 4-dimensional spacetime, so that the perturbed metric $g_{\mu \nu}^{(0)}$ is, in the $\{t, x, y, z\}$ coordinate system,

\begin{equation}
 g_{\mu \nu}^{(0)}\mathrm{d}x^\mu \mathrm{d}x^\nu = \eta_{\mu \nu} \mathrm{d}x^\mu \mathrm{d}x^\nu + 2h_{xy}^{(0)}(t) \mathrm{d}x \mathrm{d}y \ .
 \label{eq:perturbed_metric}
\end{equation}

Notice that we considered a time-dependent perturbation $h_{xy}^{(0)} = h_{xy}^{(0)}(t)$, only in the $xy$ component of the metric, which is what we need to the evaluation of the shear viscosity. 

Since the perturbed spacetime is no longer flat, we must extend the constitutive equation for the stress tensor to a curved spacetime, according to

\begin{equation}
T^{\mu \nu} = (\rho + p) u^\mu u^\nu + P g^{\mu \nu(0)} + \tau^{\mu \nu} \ ,
\end{equation}

\noindent where the dissipative part is now given by

\begin{equation}
\tau^{\mu \nu} = -P^{\mu \sigma} P^{\nu \lambda} \left [ \eta \left ( \nabla_\sigma u_\lambda + \nabla_\lambda u_\sigma - \frac{2}{3} g_{\sigma \lambda}^{(0)} \nabla_k u^k \right ) + \zeta g_{\sigma \lambda}^{(0)} \nabla_k u^k \right ] \ ,
\label{eq:tau_covar}
\end{equation}

\noindent where $\nabla_\mu$ represents the covariant derivative with respect to the perturbed metric $g_{\mu \nu}^{(0)}$; and $P^{\mu \nu} \equiv g^{\mu \nu (0)} + u^\mu u^\nu$ is the projection tensor, necessary to covariantly write the constitutive equation.

Notice that the perturbation we are considering is homogeneous, and so must be the fluid velocity field $u_i = u_i(t)$. However, parity invariance forbids motion in any direction, so that the fluid must be in rest, i.e., $u^\mu = (1, 0, 0, 0)^T \Rightarrow u_\mu = (-1, 0, 0, 0)$. Therefore, the covariant derivative of the velocity field is simply,

\begin{equation}
 \nabla_\mu u_\nu = \partial_\mu u_\nu - \Gamma^\sigma_{\mu \nu} u_\sigma = \Gamma^t_{\mu \nu} \ .
 \label{eq:cov_der}
\end{equation}

We are now interested in calculating the response in $\tau^{xy}$ up to linear order in the perturbation. Notice that one has

\begin{equation}
 \nabla_x u_y = \Gamma^t_{xy} = \Gamma^t_{yx} = \nabla_y u_x \ ,
\end{equation}

\noindent which is of linear order in the perturbation,

\begin{equation}
\begin{gathered}
 \Gamma^t_{xy} = \frac{1}{2} g^{tt (0)} \left ( \partial_x g_{ty}^{(0)} + \partial_y g_{tx}^{(0)} - \partial_t g_{xy}^{(0)} \right ) = \frac{1}{2} \eta^{tt} \left ( -\partial_t h_{xy}^{(0)} \right )
 \\
 \Rightarrow \Gamma^t_{xy} = \frac{1}{2} \partial_t h_{xy}^{(0)} \ , 
\end{gathered}
\end{equation}

\noindent so that the terms proportional to $ \nabla_k u^k$ in Eq. \ref{eq:tau_covar} will be second order in the perturbation. Now, since $ \nabla_x u_y$ is already linear in the perturbation, we can use the projection tensor in the rest frame and in flat spacetime, $P^{\mu \nu} = \text{diag}(0, 1, 1, 1)$, because considering the perturbed metric contribution yields in terms an order higher in the perturbation, which are not of interest. Therefore, the response in $\tau^{xy}$ is simply given by

\begin{equation}
\delta \left \langle \tau^{xy} \right \rangle = -2 \eta \Gamma^t_{xy} = -\eta \partial_t h_{xy}^{(0)} \ ,
\end{equation} 

\noindent which is Fourier-transformed to

\begin{equation}
 \delta \left \langle \tau^{xy} (\omega, \bm{0}) \right \rangle = i \omega \eta h_{xy}^{(0)} \ .
\end{equation}

If we now compare the result above for the response with the general result from linear response theory of Eq. \ref{eq:response_def}, in this case,

\begin{equation}
\delta \left \langle \tau^{xy} \right \rangle = - G_R^{xy, xy} h_{xy}^{(0)} \ ,
\end{equation}

\noindent one obtains the Kubo formula for $\eta$,

\begin{equation}
 \eta = - \lim_{\omega \rightarrow 0} \frac{1}{\omega} \Im \left ( G_R^{xy, xy} (\omega, \bm{0}) \right )\ .
\end{equation}

Of course, the problem of finding the retarded Green's function $G_R^{xy, xy}$ remains. Our goal now becomes to see how we can use the AdS/CFT duality for this task. But, before one gets to that, let us first discuss the general basis of the correspondence.

\chapter{The AdS/CFT duality}
\label{sec:ads/cft}

In the previous few sections, we mentioned how AdS/CFT will be useful for the determination of the Green's function, or the transport coefficients, which is our goal, as will soon become clear. To see how this is achieved, in this section we shall present an overview of the AdS/CFT duality, following \cites{fol, natsuume}.

The AdS/CFT duality was conjectured by Juan Maldacena in 1997 \cite{malda}, as the correspondence between a 10-dimensional string theory and a 4-dimensional conformal field theory: more specifically, the correspondence matches Type IIB supergravity on AdS$_5 \times $ S$^5$, and $\mathcal{N} = 4$ supersymmetric Yang-Mills (SYM) theory on $\mathbb{R}^{1,3} $ as \emph{dual theories}. 

Before we discuss the duality in more detail, let us make clear what we mean by a correspondence between two theories. First of all, a theory stands for a mathematical model of a physical system, which basically consists of a description of the particular way in which the degrees of freedom of the system evolve in a given domain according to the chosen laws of physics. We can then identify the states of the system under analysis via the measurement of physical observables, dynamical variables which can be measured and bring out the physical content of a system. In this context, we claim that two theories are correspondent if their observables may be matched on both sides in a consistent way, so that both theories may be used to yield the same physical description.

Now, a correspondence is regarded as a \emph{duality} if it matches two theories in opposed coupling regimes. This is why we refer to AdS/CFT correspondence as a duality, because it is established in such a way that when one side of the correspondence is in a strongly coupled regime, the other side will be weakly coupled, and vice versa. For this reason, AdS/CFT presents itself as a powerful tool in the computations of gauge theories at strong coupling, which may be carried out using a weakly coupled dual gravitational theory.

Such weak/strong coupling relation in a duality suggests why we can see the theories as correspondent: when the gauge theory is at strong coupling regime, it is not appropriate to use its own variables, gauge fields, weakly-coupled to perform computations. Thus, it may be interesting to use different variables, which is established by the duality to be the weakly-coupled variables of the dual theory, the gravitational fields.

As we mentioned above, AdS/CFT originated from superstring theory, by relating a supersymmetric gauge theory to a supergravity theory. One often considers a compactification of S$^5$, so that we can see the supergravity theory as 5-dimensional, in AdS$_5$ spacetime. Therefore, AdS/CFT is a \emph{holographic theory}, following the holographic principle, as it claims that a 4-dimensional gauge theory encodes a 5-dimensional gravity theory. Since the AdS spacetime has a natural notion of a spatial boundary, as we discussed in Sec. \ref{sec:ads}, we consider the gauge theory to be located on this 4-dimensional AdS boundary, which is why we often refer to the gravitational theory as \emph{bulk theory}, and to the gauge theory as \emph{boundary theory}.

At zero temperature, AdS/CFT claims the equivalence between a 4-dimensional strongly coupled gauge theory and a gravitational theory in AdS$_5$. At finite temperature, the duality is established between a 4-dimensional strongly-coupled gauge theory \emph{at finite temperature} and a gravitational theory in an AdS$_5$ \emph{black hole}, since, as we discussed in Sec. \ref{sec:thermo}, a black hole is a thermal system. We are interested in studying non-equilibrium phenomena using finite-temperature AdS/CFT, which is how we shall be able to obtain the desired transport coefficients.

More precisely, the core claim of the AdS/CFT duality is that the generating functionals, or partition functions, of the gauge and gravitational theories are equivalent,

\begin{equation}
 Z_{gauge} = Z_{AdS} \ .
\end{equation}

The meaning of such a relation will become clear when we discuss the GKP--Witten relation and its uses below.

Now, it is important to notice that, as typical of analytical methods, the gauge theories considered in AdS/CFT are not very realistic. Namely, AdS/CFT typically considers a supersymmetric gauge theory at strong coupling, that is, at the so-called \emph{large-$N_c$} limit, which differs from the realistic $SU(3)$ QCD. Nonetheless, AdS/CFT is still regarded as a very powerful tool, as it at least provides us with intuition, and a very interesting relationship between two superficially different theories. Also, how we shall discuss later on, AdS/CFT has been applied to the study of the quark-gluon plasma, which increased the interest in the duality beyond its original string theoretical context.

It is also worth a remark that the use of AdS/CFT is not limited to conformal field theories, as one can choose a field theory lacking of conformal invariance to be matched to an appropriate gravitational theory. Thus, it is common to see other denominations of the duality, such as \emph{gauge/gravity}, \emph{bulk/boundary} or \emph{field/gravity}. In this work, we stick to AdS/CFT. As a final remark, it should be stressed that, despite circumstantial evidences, AdS/CFT has not yet been proven, thus remaining at the status of a conjecture, although it has been widely used in different contexts and applications.

\chapter{Non-equilibrium AdS/CFT: GKP--Witten relation}
\label{sec:gkp_w}

In this section, we shall introduce the GKP--Witten relation, which is in the core of the AdS/CFT applications to non-equilibrium physics. Our presentation will closely follow that of \cite{natsuume}.

The main claim of AdS/CFT duality is that the generating functionals, i.e., the partition functions, of a gauge theory and that of a gravitational theory in AdS spacetime are equivalent. This claim may be expressed in terms of the (Lorentzian) GKP--Witten relation,

\begin{equation}
\left \langle \exp \left ( i \int \varphi^{(0)} \mathcal{O} \right )\right \rangle = \exp\left (i \bar{S}[\varphi^{(0)}]\right ) \ ,
\label{eq:gkp_witten}
\end{equation}

\noindent where $\langle . \rangle$ denotes an ensemble average; $\varphi$ and $\mathcal{O}$ schematically represent a field in the gravitational (bulk) theory and an operator in the gauge (boundary) theory, respectively; and $\bar{S}$ is the on-shell action. Also, $\varphi^{(0)} = \left. \varphi \right \rvert_{u=0}$, in coordinates such that the AdS boundary where the gauge theory lives is located at $u=0$.

The left-hand side of Eq. \ref{eq:gkp_witten} is the generating functional of the 4-dimensional boundary gauge theory, when an external source $\varphi^{(0)}$ is added. We know that the computation of this expression provides us with the transport coefficients of the gauge theory, which is, in general, a very difficult computation at strong coupling regimes. AdS/CFT then comes into play by providing us the right-hand side of Eq. \ref{eq:gkp_witten}, which is the generating functional of a five-dimensional bulk gravitational theory, under the saddle-point approximation (large-$N_c$ limit). 

The on-shell action $\bar{S}$ is obtained using the bulk field $\varphi$ as a solution to the equation of motion and under the boundary condition at the AdS boundary, $\left .\varphi \right \rvert_{u=0} = \varphi^{(0)}$, and substituting it into the action. Therefore, since $\varphi$ is the solution of the equation of motion, $\bar{S}$ reduces to a surface term on the AdS boundary, which allows us to obtain from the 5-dimensional action a 4-dimensional quantity, and is then identified with the generating functional of the boundary theory, according to Eq. \ref{eq:gkp_witten}. This is the sense in which we loosely say that the gauge theory ``lives'' on the boundary of the bulk.

An important thing to notice is that, from the 5-dimensional point of view, $\varphi$ is a field propagating in the bulk, whilst $\varphi^{(0)}$ is an external source from the 4-dimensional point of view. Therefore, in the context of AdS/CFT, we can say that a bulk field acts as an \emph{external source of a boundary operator.}

In practice, the greatest operational advantage provided by the GKP--Witten relation is the possibility of obtaining the generating functional of a gauge theory by the evaluation of the classical action of a gravitational theory. As we are interested in the response of a system in the presence of an external source, we then have from the GKP--Witten relation the following expression for the one-point function,

\begin{equation}
\left \langle \mathcal{O} \right \rangle_S = \frac{\delta \bar{S}[\varphi^{(0)}]}{\delta \varphi^{(0)}} \ .
\label{eq:one_point}
\end{equation}

If we are interested in the one-point function in the absence of the external source, all one has to do is to evaluate the expression above for $\varphi^{(0)} = 0$, that is, $\left \langle \mathcal{O} \right \rangle = \left . \left \langle \mathcal{O} \right \rangle_S \right \rvert_{\varphi^{(0)} = 0}$.

Before we consider an example to compute the quantity above explicitly, it is important to make a few remarks on the gravitational theory. As already discussed, we will be interested in the 5-dimensional general relativity with a negative cosmological constant $\Lambda < 0$. The bulk action for such a theory is simply the familiar 5-dimensional Einstein--Hilbert action,

\begin{equation}
S = \frac{1}{16 \pi G_5} \int \mathrm{d}^5x \sqrt{-g} \left ( R - 2 \Lambda \right ) + S_{matter} \ ,
\end{equation}

\noindent where the independent matter action $S_{matter}$ varies depending on the four-dimensional boundary theory we are interested in. Here we shall be interested only in a massless scalar field,

\begin{equation}
 S_{matter} = \frac{1}{16 \pi G_5} \int \mathrm{d}^5x \sqrt{-g} \left ( -\frac{1}{2} \left (\nabla_M \varphi \right )^2  \right ) \ ,
\end{equation}

\noindent where we denote $\left (\nabla_M \varphi \right )^2 \equiv g^{MN} \nabla_M \varphi \nabla_N \varphi $. 

The solution to the 5-dimensional bulk action will be asymptotically AdS spacetimes. A particular case of interest is the AdS$_5$--Schwarzschild (SAdS$_5$) spacetime,

\begin{equation}
ds^2_{SAdS_5} = -\frac{r_0^2}{u^2} f(u) \mathrm{d}t^2 + \frac{1}{u^2 f(u)} \mathrm{d}u^2 + \frac{r_0^2}{u^2} \delta_{ij} \mathrm{d}x^i \mathrm{d}x^j \ ,
\label{eq:sads_5_u}
\end{equation}

\noindent where $f(u) = 1-u^4$; $r_0$ is the horizon radius; and $\delta_{ij}$ is the 3-dimensional Kronecker delta. Notice that we performed a change in the radial coordinate, $u = r_0/r$, in the metric of Eq. \ref{eq:sads_5_r}, so that now the horizon is located at $u=1$, and the AdS boundary is at $u=0$, as it should be. We also set the AdS radius to unit, $L=1$.  The asymptotic behavior of this spacetime is

\begin{equation}
g_{MN} \mathrm{d}x^M \mathrm{d}x^N \sim \frac{r_0^2}{u^2} \left (-\mathrm{d}t^2 + \frac{1}{r_0^2}\mathrm{d}u^2 + \delta_{ij} \mathrm{d}x^i \mathrm{d}x^j \right ) \ , \ (u\rightarrow 0) \ ,
\label{eq:asym_ads}
\end{equation}

\noindent so that $\sqrt{-g} \sim r_0^4/u^5$. We also have

\begin{equation}
g^{MN} \partial_M \partial_N \sim \frac{u^2}{r_0^2} \left (-\partial_t^2 + r_0^2\partial_u^2 + \delta^{ij}\partial_i \partial_j \right ) \ , \ (u\rightarrow 0) \ .
\label{eq:asym_ads_inv}
\end{equation}

The asymptotic behavior outlined above will be very important in the explicit calculations performed in the next section.

\section{Bulk scalar field}
\label{sec:bsf}

We shall now calculate the one-point function according to Eq. \ref{eq:one_point}, by considering a massless scalar field as the bulk field, which will provide us with important results to be used later on. In units where $16 \pi G_5 = 1$, one has, for the bulk action,

\begin{equation}
S = \int \mathrm{d}^5x \sqrt{-g} \left ( -\frac{1}{2} \left ( \nabla_M \varphi \right )^2 \right ) \ .
\end{equation}

Notice that we did not include the Einstein--Hilbert action above, because, since we are interested in computing Eq. \ref{eq:one_point}, and the Einstein--Hilbert action is independent of the scalar field $\varphi$, we can consider only the matter action term to calculate the on-shell action $\bar{S}$. To do so, we shall consider the asymptotic behavior of the background spacetime, Eq. \ref{eq:asym_ads}, and also a scalar field which is static and homogeneous along the boundary directions, $\varphi = \varphi(u)$. We then have

\begin{equation}
\begin{gathered}
S = \int \mathrm{d}^5x \sqrt{-g} \left ( -\frac{1}{2} \left ( \nabla_M \varphi \right )^2 \right ) =\int \mathrm{d}^5x \sqrt{-g} \left ( -\frac{1}{2} g^{uu} \left (\frac{\mathrm{d}}{\mathrm{d}u} \varphi \right )^2 \right )  
\\
\Rightarrow S \sim \int \mathrm{d}^5x \frac{r_0^4}{u^5}\left ( -\frac{1}{2} u^2 \left ( \varphi^{\prime} \right )^2 \right ) = \int \mathrm{d}^5x \left ( -\frac{1}{2} \frac{r_0^4}{u^3} \left ( \varphi^{\prime} \right )^2 \right ) \ , \ (u\rightarrow0) \ ,
\end{gathered}
\end{equation}

\noindent where the prime denotes differentiation with respect to $u$. Integrating by parts, one obtains

\begin{equation}
\begin{gathered}
S \sim \int \mathrm{d}^5x \left ( -\frac{r_0^4}{2u^3} \left ( \varphi^{\prime} \right )^2 \right ) \ , \ (u\rightarrow0) 
\\
= \int \mathrm{d}^4x \int_0^1 \mathrm{d}u \left ( \left ( -\frac{r_0^4}{2u^3} \varphi \varphi^{\prime} \right )^{\prime} + \left ( \frac{r_0^4}{2u^3} \varphi^{\prime} \right )^{\prime} \varphi \right ) \ , \ (u\rightarrow0) 
\\
\Rightarrow S \sim \int \mathrm{d}^4x \left .\left ( \frac{r_0^4}{2u^3} \varphi \varphi^{\prime} \right )\right \rvert_{u=0} + \int \mathrm{d}^5x \,
r_0^4\left ( \frac{1}{2u^3} \varphi^{\prime} \right )^{\prime} \varphi \ , \ (u\rightarrow0) \ ,
\label{eq:onshell_quase}
\end{gathered}
\end{equation}

\noindent where we used the fact that the scalar field vanishes at the horizon. Notice that the second term is just the equation of motion for the scalar field,

\begin{equation}
 \left ( \frac{1}{2u^3} \varphi^{\prime} \right )^{\prime} \sim 0 \ , \ (u\rightarrow0) \ ,
\end{equation}

\noindent whose asymptotic solution is of the form

\begin{equation}
 \varphi \sim \varphi^{(0)} \left (1 + \varphi^{(1)} u^4 \right ) \ , \ (u\rightarrow0) \ .
\end{equation}

Using the fact that the scalar field satisfies the equation of motion, the second term in Eq. \ref{eq:onshell_quase} vanishes, and the action reduces to the surface term on the AdS boundary,

\begin{equation}
S \sim \int \mathrm{d}^4x \left .\left ( \frac{r_0^4}{2u^3} \varphi \varphi^{\prime} \right )\right \rvert_{u=0} \ , \ (u\rightarrow0) \ .
\end{equation}

If we now substitute the asymptotic form of the scalar field satisfying the equation of motion, we finally obtain the on-shell action as

\begin{equation}
\begin{gathered}
\bar{S} = \int \mathrm{d}^4x \left .\left ( \frac{r_0^4}{2u^3} \left (\varphi^{(0)} \left (1 + \varphi^{(1)} u^4 \right ) \right ) \left ( 4 u^3\varphi^{(0)} \varphi^{(1)}  \right ) \right )\right \rvert_{u=0} 
\\
\Rightarrow \bar{S}[\varphi^{(0)}] = \int \mathrm{d}^4x \left ( 2 r_0^4 \left ( \varphi^{(0)} \right )^2 \varphi^{(1)} \right ) \ .
\end{gathered}
\end{equation}

We now can easily evaluate the one-point function from Eq. \ref{eq:one_point},

\begin{equation}
\left \langle \mathcal{O} \right \rangle_S = 4r_0^4 \varphi^{(1)} \varphi^{(0)} \ .
\end{equation}

Notice that in the absence of the external source, one has

\begin{equation}
\left \langle \mathcal{O} \right \rangle = \left . \left \langle \mathcal{O} \right \rangle_S \right \rvert_{\varphi^{(0)} = 0} = 0 \ ,
\end{equation}

\noindent so that,

\begin{equation}
\begin{gathered}
\delta \left \langle \mathcal{O} \right \rangle = \left \langle \mathcal{O} \right \rangle_S - \left \langle \mathcal{O} \right \rangle = \left \langle \mathcal{O} \right \rangle_S 
\\
\Rightarrow \delta \left \langle \mathcal{O} \right \rangle = 4r_0^4 \varphi^{(1)} \varphi^{(0)} \ ,
\label{eq:response_general_gkp_w}
\end{gathered}
\end{equation}

\noindent where the precise operator $\mathcal{O}$ in the boundary theory will depend on the nature of the scalar field we are considering. We shall soon carry out a specific and very important example.

Finally, notice now that we can relate this result to the linear response relation of Eq. \ref{eq:response_def},

\begin{equation}
\delta \left \langle \mathcal{O} (\omega, \bm{q})\right \rangle = - G_R^{\mathcal{O}, \mathcal{O}} (\omega, \bm{q}) \varphi^{(0)}(\omega, \bm{q}) \ ,
\end{equation}

\noindent which then allows us the determination of the retarded Green's function,

\begin{equation}
 G_R^{\mathcal{O}, \mathcal{O}} (k = 0) = -4 r_0^4 \varphi^{(1)} \ ,
\end{equation}

\noindent where the Green's function does not depend on $\omega$ or $\bm{q}$, since $\varphi^{(1)}$ also does not. We then finally see the sense in which AdS/CFT can be used to determine the retarded Green's function, as we mentioned in Sec. \ref{sec:lrt}.

\section{Scalar field in 4-dimensional bulk}
\label{sec:4d_gkp_w}

We shall also consider a particular 4-dimensional gravitational background to the calculation of the $\eta/s$ ratio, which shall be done in Sec. \ref{sec:eta_s_ads_mgd_4}. In this section, we present the results from AdS/CFT detailed above, but applied to a 4-dimensional bulk scalar field, by following an analogous procedure. To begin with, we consider the 4-dimensional bulk action (again we omit the Einstein--Hilbert term, as it is not relevant to our purposes here),

\begin{equation}
S = \int \mathrm{d}^4x \sqrt{-g} \left ( -\frac{1}{2} \left ( \nabla_\mu \varphi \right )^2 \right ) \ ,
\end{equation}

\noindent where we consider the asymptotic behavior of the 4-dimensional spacetime as,

\begin{equation}
\begin{gathered}
g_{\mu \nu} \mathrm{d}x^\mu \mathrm{d}x^\nu \sim \frac{r_0^2}{u^2} \left (-\mathrm{d}t^2 + \frac{1}{r_0^2}\mathrm{d}u^2 + \mathrm{d}x^2 + \mathrm{d}y^2 \right ) \ , \ (u\rightarrow 0) \ ;
\\
g^{\mu \nu} \partial_\mu \partial_\nu \sim \frac{u^2}{r_0^2} \left (-\partial_t^2 + r_0^2\partial_u^2 + \partial_x^2 + \partial_y^2 \right ) \ , \ (u\rightarrow 0) \ ,
\label{eq:asym_4d}
\end{gathered}
\end{equation}

\noindent so that $\sqrt{-g} \sim r_0^3/u^4$.

If we again consider a static and homogeneous scalar field along the boundary directions $\varphi = \varphi(u)$, one has

\begin{equation}
\begin{gathered}
S = \int \mathrm{d}^4x \sqrt{-g} \left ( -\frac{1}{2} \left ( \nabla_\mu \varphi \right )^2 \right ) =\int \mathrm{d}^4x \sqrt{-g} \left ( -\frac{1}{2} g^{uu} \left (\frac{\mathrm{d}}{\mathrm{d}u} \varphi \right )^2 \right )  
\\
\Rightarrow S \sim \int \mathrm{d}^4x \frac{r_0^3}{u^4}\left ( -\frac{1}{2} u^2 \left ( \varphi^{\prime} \right )^2 \right ) = \int \mathrm{d}^4x\left ( -\frac{1}{2} \frac{r_0^3}{u^2} \left ( \varphi^{\prime} \right )^2 \right ) \ , \ (u\rightarrow0) \ .
\end{gathered}
\end{equation}

Integrating by parts yields

\begin{equation}
\begin{gathered}
S \sim \int \mathrm{d}^4x \left ( -\frac{r_0^3}{2u^2} \left ( \varphi^{\prime} \right )^2 \right ) \ , \ (u\rightarrow0) 
\\
= \int \mathrm{d}^3x \int_0^1 \mathrm{d}u \left ( \left ( -\frac{r_0^3}{2u^2} \varphi \varphi^{\prime} \right )^{\prime} + \left ( \frac{r_0^3}{2u^2} \varphi^{\prime} \right )^{\prime} \varphi \right ) \ , \ (u\rightarrow0) 
\\
\Rightarrow S \sim \int \mathrm{d}^3x \left .\left ( \frac{r_0^3}{2u^2} \varphi \varphi^{\prime} \right )\right \rvert_{u=0} + \int \mathrm{d}^4x \,
r_0^3\left ( \frac{1}{2u^2} \varphi^{\prime} \right )^{\prime} \varphi \ , \ (u\rightarrow0) \ ,
\label{eq:onshell_quase_2}
\end{gathered}
\end{equation}

The second term is the equation of motion for the scalar field,

\begin{equation}
 \left ( \frac{1}{2u^2} \varphi^{\prime} \right )^{\prime} \sim 0 \ , \ (u\rightarrow0) \ ,
\end{equation}

\noindent whose asymptotic solution is of the form

\begin{equation}
 \varphi \sim \varphi^{(0)} \left (1 + \varphi^{(1)} u^3 \right ) \ , \ (u\rightarrow0) \ .
\end{equation}

Since the scalar field satisfies the equation of motion, the second term in Eq. \ref{eq:onshell_quase_2} vanishes, and the action reduces to the surface term on the AdS boundary,

\begin{equation}
S \sim \int \mathrm{d}^3x \left .\left ( \frac{r_0^3}{2u^2} \varphi \varphi^{\prime} \right )\right \rvert_{u=0} \ , \ (u\rightarrow0) \ .
\end{equation}

By substituting the asymptotic form of the scalar field, we then obtain the on-shell action

\begin{equation}
\begin{gathered}
\bar{S} = \int \mathrm{d}^3x \left .\left ( \frac{r_0^3}{2u^2} \left (\varphi^{(0)} \left (1 + \varphi^{(1)} u^3 \right ) \right ) \left ( 3 u^2\varphi^{(0)} \varphi^{(1)}  \right ) \right )\right \rvert_{u=0} 
\\
\Rightarrow \bar{S}[\varphi^{(0)}] = \int \mathrm{d}^3x \left ( \frac{3}{2}r_0^3 \left ( \varphi^{(0)} \right )^2 \varphi^{(1)} \right ) \ .
\end{gathered}
\end{equation}

Therefore, the one-point function from Eq. \ref{eq:one_point} is

\begin{equation}
\left \langle \mathcal{O} \right \rangle_S = 3r_0^3\varphi^{(1)} \varphi^{(0)} \ .
\end{equation}

Notice that in the absence of the external source, one has

\begin{equation}
\left \langle \mathcal{O} \right \rangle = \left . \left \langle \mathcal{O} \right \rangle_S \right \rvert_{\varphi^{(0)} = 0} = 0 \ ,
\end{equation}

\noindent so that, in the 4-dimensional analogue,

\begin{equation}
\begin{gathered}
\delta \left \langle \mathcal{O} \right \rangle = \left \langle \mathcal{O} \right \rangle_S - \left \langle \mathcal{O} \right \rangle = \left \langle \mathcal{O} \right \rangle_S 
\\
\Rightarrow \delta \left \langle \mathcal{O} \right \rangle = 3r_0^3 \varphi^{(1)} \varphi^{(0)} \ ,
\label{eq:response_general_gkp_w_2}
\end{gathered}
\end{equation}

By relating this result to the linear response relation of Eq. \ref{eq:response_def}, one obtains the retarded Green's function in our 4-dimensional analogue,

\begin{equation}
 G_R^{\mathcal{O}, \mathcal{O}} (k = 0) = -3 r_0^3 \varphi^{(1)} \ ,
\end{equation}

These results will be applied when we later on consider the deformed AdS$_4$--RN black brane as the gravitational background.

\chapter{\texorpdfstring{$\eta/s$}{} in AdS\texorpdfstring{\textsubscript{5}}{}--Schwarzschild black brane background}
\label{sec:eta_s_sads_5}

In this section, we shall discuss the computation of the transport coefficient $\eta$ for the $\mathcal{N}=4$ plasma using AdS/CFT, following \cite{natsuume}. First of all, let us remind that the duality conjectures that $\mathcal{N}=4$ SYM is dual to the five-dimensional Einstein gravity with a negative cosmological constant, whose solutions are the AdS$_5$ spacetime (at zero temperature), and the SAdS$_5$ black brane at finite temperature. Since we are interested in the determination of transport coefficients of the $\mathcal{N}=4$ plasma, we shall consider the finite temperature case. Thus, let us consider the SAdS$_5$ black brane, according to Eq. \ref{eq:sads_5_r}, 

\begin{equation}
\begin{gathered}
ds^2_{SAdS_5} = -r^2 f(r) \mathrm{d}t^2 + \frac{1}{r^2 f(r)} \mathrm{d}r^2 + r^2 \left (\mathrm{d}x^2 + \mathrm{d}y^2 + \mathrm{d}z^2 \right ) \ ; \\ f(r) = 1 - \frac{r_0^4}{r^4} \ .
\end{gathered}
\end{equation}

Let us first calculate some of the thermodynamic quantities of the SAdS$_5$ black brane. According to Eq. \ref{eq:hawk_temp}, with $\hbar = c = k_b = 1$, one has the temperature

\begin{equation}
\begin{gathered}
T = \frac{\kappa}{2\pi} = \frac{1}{4\pi}\left .\frac{\mathrm{d} (r^2 f(r))}{\mathrm{d}r} \right \rvert_{r=r_0} = \frac{1}{4\pi}\left. \left (2r + 2 \frac{r_0^4}{r^3}\right )\right \rvert_{r=r_0} 
\\
\Rightarrow T = \frac{r_0}{\pi} \ .
\label{eq:sadstemp}
\end{gathered}
\end{equation}

Now, as discussed in Sec. \ref{sec:thermo}, we can calculate the SAdS$_5$ black brane entropy with the area law. However, since one has for black branes a horizon extending infinitely in the $x, y, z$ directions, the horizon area, and consequently the entropy, diverge. Thus, it is more appropriate to think in the \emph{entropy density}, defined from the area law as,

\begin{equation}
 s = \frac{a}{4 G_5} \ ,
\end{equation}

\noindent where $a$ is the \emph{horizon area density}, which is nothing but the horizon area divided by the dual gauge theory volume,

\begin{equation}
V = \iiint_{\mathbb{R}^3} \mathrm{d}x\, \mathrm{d}y\, \mathrm{d}z \ ,
\end{equation}

\noindent that is,

\begin{equation}
\begin{gathered}
a = \frac{A}{V} = \frac{1}{V} \iiint_{\mathbb{R}^3} \left .\sqrt{g_{xx}g_{yy}g_{zz}} \right \rvert_{r = r_0} \mathrm{d}x\, \mathrm{d}y\, \mathrm{d}z  =  \frac{1}{V} \sqrt{(r_0^2)^3} \iiint_{\mathbb{R}^3} \mathrm{d}x\, \mathrm{d}y\, \mathrm{d}z 
\\
\Rightarrow a = r_0^3 \ ,
\end{gathered}
\end{equation}

\noindent so that the entropy density of the SAdS$_5$ black brane is simply

\begin{equation}
 s = \frac{r_0^3}{4G_5}
 \label{eq:s}
\end{equation}

The $\mathcal{N}=4$ plasma is described as a conformal fluid, since $\mathcal{N}=4$ SYM is a conformal field theory. As a conformal fluid, its stress-tensor must be traceless, which immediately fixes a vanishing bulk viscosity \cites{rangamani, natsuume}, $\zeta = 0$. Therefore, the only nontrivial transport coefficient to be found is the shear viscosity $\eta$.

To obtain the shear viscosity, we shall employ the results from linear response theory and the GKP--Witten relation, as discussed above. The procedure goes as follow: we first consider a gravitational perturbation and then calculate the response to the stress tensor, by solving the perturbation equation within the hydrodynamic limit, as it is discussed in \cites{son_hydro, natsuume}. 

We shall fully present the arguments and a similar calculation in the next section, when we consider the deformed AdS$_5$--Schwarzschild black brane as the gravitational background. The calculation will not be exactly the same, but analogous. So, here we give only the final result for the $\eta/s$ ration in the SAdS$_5$ black brane gravitational background (which here has the direct interpretation of corresponding to the $\eta/s$ ratio for the $\mathcal{N}=4$ plasma itself),

\begin{equation}
 \frac{\eta}{s} = \frac{1}{4\pi} \ .
 \label{eq:eta_s_sads_5}
\end{equation}

Of course, we are working in natural units, so that $\hbar = k_B = 1$, which is why the quantity above is dimensionless. However, it is interesting to recover SI units, so that the value above can be more sensibly examined,

\begin{equation}
 \frac{\eta}{s} = \frac{1}{4\pi} \frac{\hbar}{k_B} \approx 6.08 \times 10^{-13} K\,s \ ,
 \label{eq:eta_s_sads_5_units}
\end{equation}

\noindent which is indeed a relatively small value. In fact, the value of this ratio was conjectured to be a lower bound for all relativistic quantum field theories at finite temperature and zero chemical potential, as argued in \cite{kss}, that is, we would have the universal bound for all substances,

\begin{equation}
 \frac{\eta}{s} \geq \frac{1}{4\pi} \frac{\hbar}{k_B} \ ,
 \label{eq:eta_s_bound_units}
\end{equation}

\noindent or, going back to natural units, 

\begin{equation}
 \frac{\eta}{s} \geq \frac{1}{4\pi} \ ,
 \label{eq:eta_s_bound}
\end{equation}

\noindent which is supposed to be valid for a wide class of systems, suggesting that black hole horizons are dual to a wide range of ideal fluids \cite{kss}. This is known as the \emph{KSS bound}, and we shall analyze violations to this bound later on.

The ratio $\eta/s$ of Eq. \ref{eq:eta_s_sads_5_units}, provided by the AdS/CFT calculation, shows that the strongly-coupled, large-$N_c$, $\mathcal{N}=4$ plasma has a quite small value for this ratio, compared with ordinary materials. However, if large-$N_c$ gauge theories considered by AdS/CFT are good approximations to QCD, one could expect that this result may be applied to the quark-gluon plasma (QGP), which is a natural phenomenon in QCD, achieved when at high enough temperature the quarks and gluons are deconfined from inside the protons and neutrons to form the plasma \cite{qgp}.

In fact, experiments in the Relativistic Heavy Ion Collider (RHIC) have shown  \cite{qgp_exps} that the QGP behaves like a viscous fluid with very small viscosity, which implies that the QGP is strongly-coupled, which discards the possibility of using perturbative QCD to the study of the plasma. Therefore, AdS/CFT may present itself as an alternative to the QGP research, which is further supported by the fact that the shear viscosity obtained in the RHIC experiment is close to that predicted by AdS/CFT \cites{qgpexp1, qgpexp2}.

In a few sections, we shall perform the calculation of $\eta/s$ in gravitational backgrounds other than the SAdS$_5$, so that we shall not have an evident dual gauge theory in the discussion. However, the results will be useful to constrain the parameters of the black branes we shall consider.

\chapter{Braneworld, membrane paradigm and AdS/CFT}
\label{sec:brane_memb_ads_cft}

The generalized deformed AdS black branes which we will soon introduce as the gravitational background to the calculation of $\eta/s$, are constructed in the braneworld scenarios \cites{whisker, mgd1}, which we extensively discussed in Part I of the thesis. To transliterate the braneworld models into the AdS/CFT language, we use the \emph{membrane paradigm}. In this section we shall describe how such a transliteration is achieved, following \cite{ads_memb}.

\section{The membrane paradigm}
\label{sec:memb_para}

The membrane paradigm establishes that for an observer in the near-horizon region, the effective action can be seen as a composition of an action carrying contributions from the whole spacetime outside the horizon, $S_{out}$, and a surface action, $S_{surf}$, which carries the contribution from the horizon, representing the influence of the horizon on the external universe. It is often the case to consider this surface action contribution as being defined on the \emph{stretched horizon}, a timelike hypersurface of constant radius outside the true horizon, which acts as a cutoff for the spacetime outside the black hole.

If one consider a massless bulk scalar field for $S_{out}$, and demands that the effective action must be stationary on a solution to the equations of motion, the surface action $S_{surf}$ is required to be such that an external observer will perceive the horizon as a \emph{charged membrane} \cites{mp, ads_memb}, where the charge may be interpreted as the response of the horizon membrane induced by the bulk scalar field around the black hole. In the context of the linear response theory discussed in Sec. \ref{sec:lrt}, this identification leads to the introduction of a membrane transport coefficient.

In the case in which we take the bulk scalar field to be $\varphi = g^{xx}h_{xy}$, where $h_{xy}$ is an off-diagonal gravitational perturbation, the introduced transport coefficient will be the \emph{shear viscosity}, since a perturbation on the metric leads to a perturbation in the stress tensor --- in this case, on its viscous part ---, as was discussed in Sec. \ref{sec:kubo_visco}. This procedure then allows us to obtain the shear viscosity of the horizon membrane.

In this sense, the influence of the black hole on its surroundings may be taken into account by considering charges on the fictitious fluid on its horizon, which then leads to the determination of transport coefficients. However, it is not immediately clear that these coefficients are related to the transport coefficients of the dual field theory provided by AdS/CFT.

We discussed in Sec. \ref{sec:hydro} how hydrodynamics provide an effective description of interacting quantum field theories at finite temperature at the appropriate low-energy, long-wavelength limits. As further discussed in Sec. \ref{sec:ads/cft}, the gravitational dual of such field theories is analyzed through black holes in the bulk geometry, so that the UV/IR connection suggests that the near-horizon region of the bulk geometry should govern the long-scale physics of the dual field theory \cite{ads_memb}.

As it is shown in \cite{ads_memb}, the linear response of the boundary operators, which lead to the determination of transport coefficients of the boundary field theory, is expressed almost identically to the response of the membrane fluid on the horizon. However, in AdS/CFT, the boundary theory is on the AdS boundary, instead of being on the horizon, so that if the membrane is moved from the horizon to the boundary, the membrane paradigm quantities become concrete field theory observables! Therefore, we can say that the low-frequency limit of the boundary theory fluid linear response is completely captured by the horizon fluid. This is the precise sense in which it is possible to identify the membrane fluid on the horizon in the membrane paradigm context and the low-energy, long-wavelength description of the dual field theory in the AdS/CFT context.

Now, as detailed in Part I, in the braneworld scenarios \emph{our universe} is seen as a 4-dimensional brane with a finite tension embedded in a 5-dimensional bulk, which influences the gravity on the brane as encoded in the additional terms in the effective EFE. We can then promptly identify such scenarios with the membrane paradigm, in which \emph{the horizon} is seen as a membrane with charges induced by exterior sources, that is, a membrane whose parameters are influenced by the exterior region, which is in parallel with the gravity itself on the brane being influenced by the bulk in which it is embedded. 

Therefore, we can place the braneworld scenarios within the membrane paradigm. This is a very important observation, because, as discussed above, there is a precise connection between the membrane paradigm and the AdS/CFT correspondence, so that if we can realize the braneworld scenarios through the membrane paradigm, one obtains a connection between the braneworld scenarios and AdS/CFT, in a transliteration given by the membrane paradigm. Such connection then provides us with the precise sense in which we can apply the AdS/CFT techniques to the braneworld models in AdS backgrounds, as we shall do next.

\chapter{\texorpdfstring{$\eta/s$}{} in deformed AdS\texorpdfstring{\textsubscript{5}}{}--Schwarzschild black brane background}
\label{sec:eta_s_ads_mgd_5}

In this section, we shall consider the deformed AdS$_5$--Schwarzschild ($\beta$-SAdS$_5$) black brane, and then calculate the $\eta/s$ ratio in this particular gravitational background. The results in this section are original, and were published in \cite{ads5}, in collaboration with R. da Rocha and P. Meert.

The $\beta$-SAdS$_5$ black brane metric is

\begin{equation}
 ds^2_{\beta\text{-}SAdS_5} = -r^2 N(r) \mathrm{d}t^2 + \frac{1}{r^2 A(r)} \mathrm{d} r^2 + r^2 \delta_{ij} \mathrm{d}x^i \mathrm{d}x^j \ ,
\end{equation}

\noindent where $\delta_{ij}$ is the 3-dimensional Kronecker delta, and,

\begin{equation}
N(r) = 1 - \frac{r_0^4}{r^4} + \left (\beta - 1 \right ) \frac{r_0^6}{r^6} \ ;
\end{equation}

\begin{equation}
A(r) = \left (1 -\frac{r_0^4}{r^4} \right ) \left ( \frac{1 - \frac{3}{2}\frac{r_0^4}{r^4}}{1- \left(\frac{4\beta-1}{2} \right ) \frac{r_0^4}{r^4}} \right ) \ ,
\end{equation}

\noindent where $r_0$ is the horizon radius; and $\beta$ is a constant, which we will here refer to as ``deformation parameter''. Notice that we set the AdS radius to unit, $L = 1$, and that to preserve the metric signature, $N(r)$ and $A(r)$ are non-negative functions.

In \cite{mgd2}, a generalized black string solution is constructed within the MGD method procedure. From this solution, and following a procedure similar to the ADM formalism, it is possible to emulate a black brane version, which is the deformed AdS$_5$--Schwarzschild black brane presented above. The construction of this spacetime is detailed in \cite{ads5}.

The change of coordinate $u = r_0/r$ yields,

\begin{equation}
\begin{gathered}
g_{tt} = -r^2 N(r) \mapsto g_{tt} = -\frac{r_0^2}{u^2} N(u) \ ;
\\
g_{rr} = \frac{1}{r^2 A(r)} \mapsto g_{uu} = \left (\frac{\partial r}{\partial u} \right )^2 \frac{1}{\frac{r_0^2}{u^2} A(u)} = \left ( - \frac{r_0}{u^2} \right )^2 \frac{u^2}{r_0^2 A(u)} \Rightarrow g_{rr} \mapsto g_{uu} = \frac{1}{u^2 A(u)} \ ;
\\
g_{ij} = r^2 \delta_{ij} \mapsto g_{ij} = \frac{r_0^2}{u^2} \delta_{ij} \ ,
\end{gathered}
\end{equation}

\noindent where

\begin{equation}
N(u) = 1 - u^4 + \left (\beta - 1 \right ) u^6 \ ;
\label{eq:Nu}
\end{equation}

\begin{equation}
A(u) = \left (1 - u^4 \right ) \left ( \frac{2 - 3u^4}{2- \left (4\beta-1\right ) u^4}\right ) .
\label{eq:Au}
\end{equation}

Therefore, one has

\begin{equation}
g_{MN} \mathrm{d}x^M \mathrm{d}x^N = -\frac{r_0^2}{u^2} N(u) \mathrm{d}t^2 + \frac{1}{u^2 A(u)} \mathrm{d}u^2 + \frac{r_0^2}{u^2} \delta_{ij} \mathrm{d}x^i \mathrm{d}x^j \ ,
\end{equation}

\noindent whose determinant $\det g_{MN} \equiv g$ is such that,

\begin{equation}
\begin{gathered}
\sqrt{-g} = \sqrt{-\left (-\frac{r_0^2}{u^2}\right)\left(\frac{r_0^2}{u^2}\right)^3 \frac{N(u)}{u^2 A(u)}} 
\\
\Rightarrow \sqrt{-g} = \frac{r_0^4}{u^5} \sqrt{\frac{N}{A}} \ ,
\end{gathered}
\end{equation}

\noindent where, from now on, $N$ and $A$ (which should not be confused with 5-dimensional indices!) refer respectively to $N(u)$ and $A(u)$, unless otherwise specified.

And, according to the definition $g_{MN}g^{NS} = \delta_M^S$, the inverse metric is

\begin{equation}
g^{MN} \partial_M \partial_N = -\frac{u^2}{r_0^2 N(u)} \partial_t^2 + u^2 A(u) \partial_u^2 + \frac{u^2}{r_0^2} \delta^{ij} \partial_i \partial_j \ .
\end{equation}

If we now consider a bulk perturbation $h_{xy}$, so that

\begin{equation}
 ds^2 = ds^2_{\beta\text{-}SAdS_5} + 2h_{xy} \mathrm{d}x \mathrm{d}y \ ,
\end{equation}

\noindent we can determine the response in the boundary energy-momentum tensor, $\delta \left \langle T^{xy} \right \rangle$, according to the linear response theory discussed in Sec. \ref{sec:lrt}, which is given by

\begin{equation}
 \delta \left \langle \tau^{xy} \right \rangle  = i \omega \eta h_{xy}^{(0)} \ ,
 \label{eq:response_1}
\end{equation}

\noindent where $\tau^{xy}$ is the dissipative part of $T^{xy}$; $\eta$ is the viscosity; and $h_{xy}^{(0)}$ is the perturbation added to the boundary theory, which is asymptotically related to $h_{xy}$, the bulk perturbation, by

\begin{equation}
 g^{xx}h_{xy} \approx h_{xy}^{(0)} \left ( 1 + h_{xy}^{(1)} u^4 \right ) \ , \ (u \rightarrow 0) \ ,
 \label{eq:perturb_asym_def}
\end{equation}

\noindent which is valid because $g^{xx}h_{xy}$ obeys the equation of motion for a massless scalar field \cites{natsuume, son_hydro}, so that we can apply the results discussed in Sec. \ref{sec:bsf}. This is only the case because the deformed AdS$_5$--Schwarzschild black brane has the same asymptotic behavior of the SAdS$_5$ black brane (namely, Eq. \ref{eq:asym_ads}), so that we indeed can directly apply the results of Sec. \ref{sec:bsf}, which were obtained considering the asymptotic behavior of the 5-dimensional gravitational background. Also, notice that the discussion of Sec. \ref{sec:bsf} were carried out independently of the Einstein--Hilbert action, so that even with consider its deformation, which leads to the deformed AdS$_5$--Schwarzschild black brane, we still can use the results derived for the massless scalar field.

In the context of what was discussed in Sec. \ref{sec:gkp_w}, we can then see $g^{xx}h_{xy}$ as the bulk field $\varphi$ which acts as an external source of a boundary operator, which, in this case, is nothing but $\tau^{xy}$. Therefore, we can once again use the results we derived for a scalar field, and directly obtain the response $\delta \left \langle \tau^{xy} \right \rangle$, from Eq. \ref{eq:response_general_gkp_w},

\begin{equation}
\delta \left \langle \tau^{xy} \right \rangle = \frac{r_0^4}{16 \pi G_5}4 h_{xy}^{(1)}h_{xy}^{(0)} \ , 
\label{eq:response_2}
\end{equation}

\noindent where it is convenient here to reintroduce $16 \pi G_5$. Comparing Eqs. \ref{eq:response_1} and \ref{eq:response_2}, one has

\begin{equation}
 i \omega \eta = \frac{r_0^4}{4\pi G_5} h_{xy}^{(1)} \ .
\label{eq:eta_quase}
\end{equation}

Like we discussed in Sec. \ref{sec:eta_s_sads_5} for the SAdS$_5$ black brane, the entropy of the deformed AdS$_5$--Schwarzschild black brane will also diverge due to its planar horizon. Therefore, we should rather calculate its entropy density. 
As shown in detail in \cite{ads5}, the entropy density in the deformed AdS$_5$--Schwarzschild black brane background reads

\begin{equation}
s = \frac{r_0^3}{4G} \left ( 15 \beta - 3\beta^2 - 11 \right ) \left(\frac{3-4\beta}{\beta-2}\right)^{1/2}   \ .
\end{equation}

Plugging this result into Eq. \ref{eq:eta_quase} yields,

\begin{equation}
\frac{\eta}{s} = \frac{r_0}{\pi} \left[ \left ( \frac{1}{15 \beta - 3\beta^2 - 11} \right ) \left ( \frac{\beta - 2}{3-4\beta} \right )^{1/2} \right] \frac{h_{xy}^{\left(1\right)}}{i\omega} \ . 
\label{eq:eta_s_geral}
\end{equation}

Therefore, to obtain the ratio $\eta/s$ in the deformed AdS$_5$--Schwarzschild black brane background, all one has to do now is to find $h_{xy}^{(1)}$, which we shall do by solving the equation of motion for the perturbation $g^{xx}h_{xy} \equiv \varphi$, which is that of a massless scalar field \cites{natsuume, son_hydro},

\begin{equation}
 \partial_M \left ( \sqrt{-g} g^{MN} \partial_N \varphi \right ) = 0
\end{equation}

Considering the perturbation of the form $\varphi = \phi(u) e^{-i\omega t}$, one has

\begin{equation}
\begin{gathered}
\partial_u \varphi = e^{-i\omega t} \frac{\mathrm{d}}{\mathrm{d}u} \phi  \ ;
\\
\partial_t \varphi = -i \omega \varphi \Rightarrow \partial_t \left ( \partial_t \varphi \right ) = - \omega^2 \varphi \ ,
\end{gathered}
\end{equation}

\noindent so that

\begin{equation}
\begin{gathered}
 \partial_t \left ( \sqrt{-g} g^{tt} \partial_t \varphi \right ) = \frac{r_0^4}{u^5}\sqrt{\frac{N}{A}}\left ( -\frac{u^2}{r_0^2 N} \right ) \partial_t \left ( \partial_t \varphi \right ) = \frac{-r_0^2}{\sqrt{NA}} \frac{(-\omega^2 \varphi)}{u^3} 
 \\
 \Rightarrow  \partial_t \left ( \sqrt{-g} g^{tt} \partial_t \varphi \right ) = \frac{1}{\sqrt{NA}} \frac{\omega^2 r_0^2}{u^3} \varphi \ ;
\end{gathered}
\end{equation}

\begin{equation}
\begin{gathered}
 \partial_u \left ( \sqrt{-g} g^{uu} \partial_u \varphi \right ) = \partial_u \left ( \frac{r_0^4}{u^5}\sqrt{\frac{N}{A}} u^2 A \partial_u \varphi \right ) 
 \\
 \Rightarrow  \partial_u \left ( \sqrt{-g} g^{uu} \partial_u \varphi \right ) = r_0^4 \frac{\mathrm{d}}{\mathrm{d}u}\left ( \frac{\sqrt{NA}}{u^3} \frac{\mathrm{d}}{\mathrm{d}u} \phi \right ) e^{-i\omega t} \ .
\end{gathered}
\end{equation}

Thus, since $\partial_i \varphi = 0$, one has

\begin{equation}
\begin{gathered}
 \partial_M \left ( \sqrt{-g} g^{MN} \partial_N \varphi \right ) = \partial_t \left ( \sqrt{-g} g^{tt} \partial_t \varphi \right ) + \partial_u \left ( \sqrt{-g} g^{uu} \partial_u \varphi \right ) =  0 
 \\
 \Rightarrow \frac{1}{\sqrt{NA}} \frac{\omega^2 r_0^2}{u^3} \phi e^{-i \omega t} + r_0^4 \frac{\mathrm{d}}{\mathrm{d}u}\left ( \frac{\sqrt{NA}}{u^3} \frac{\mathrm{d}}{\mathrm{d}u} \phi \right ) e^{-i\omega t} = 0 
 \\
 \Rightarrow
\frac{u^3}{\sqrt{NA}} \frac{\mathrm{d}}{\mathrm{d}u}\left ( \frac{\sqrt{NA}}{u^3} \frac{\mathrm{d}}{\mathrm{d}u} \phi \right ) + \frac{1}{NA} \frac{\omega^2}{r_0^2} \phi = 0 \ .
\end{gathered}
\end{equation}

Therefore, the perturbation equation reduces to the following second-order partial equation only for the $u$-dependent function $\phi(u)$: 

\begin{equation}
\frac{u^3}{\sqrt{NA}} \left ( \frac{\sqrt{NA}}{u^3}\phi^{\prime}\right )^{\prime}  + \frac{1}{NA} \frac{\omega^2}{r_0^2} \phi  = 0 \ ,
\label{eq:pertu_edo}
\end{equation}

\noindent or

\begin{equation}
\phi^{\prime \prime} + \left ( \frac{(NA)^{\prime}}{2 NA} - \frac{3}{u}\right ) \phi^{\prime} + \frac{1}{NA} \frac{\omega^2}{r_0^2} \phi = 0 \ ,
\label{eq:pertu_edo2}
\end{equation}

\noindent where primes denote differentiation with respect to $u$.

To the solution of the partial differential equation we shall impose a set of two boundary conditions: an ``incoming wave'' boundary condition in the near-horizon region ($u \rightarrow 1$); and a Dirichlet boundary condition at the AdS boundary ($u \rightarrow 0$), expressed by $\phi (u\rightarrow 0) = \phi^{(0)}$, where $h_{xy}^{(0)} =  \phi^{(0)} e^{-i \omega t}$. 

To incorporate the near-horizon incoming wave boundary condition, we shall first solve Eq. \ref{eq:pertu_edo2} in the limit $u \simeq 1$. To do so, we expand the coefficients of the equation around $u=1$ and then take the leading order term. First, notice that, according to Eqs. \ref{eq:Nu} and \ref{eq:Au}, one has

\begin{equation}
\begin{aligned}
\frac{(NA)^{\prime}}{2 NA} - \frac{3}{u} &= \frac{-2(4\beta -1) u^3}{(4\beta-1)u^4 -2} + \frac{3(\beta-1)u^5-2u^3}{(\beta-1)u^6 - u^4 + 1} 
\\
&+ \frac{u}{1+u^2} + \frac{6u^3}{3u^4-2} + \frac{1}{2(u-1)} + \frac{1}{2(u+1)} - \frac{3}{u} \ ,
\end{aligned}
\end{equation}

\noindent so that, up to leading order,

\begin{equation}
\frac{(NA)^{\prime}}{2 NA} - \frac{3}{u} \approx -\frac{1}{2(1-u)} \ , \ (u \rightarrow 1) \ .
\end{equation}

We also have

\begin{equation}
\frac{1}{NA} \approx \frac{4\beta - 3}{4(\beta - 1)(1-u)} \ , \ (u \rightarrow 1) \ ,
\end{equation}

\noindent so that Eq. \ref{eq:pertu_edo2} becomes,

\begin{equation}
\phi^{\prime \prime} - \frac{1}{2(1-u)} \phi^{\prime} + \frac{4\beta - 3}{4(\beta - 1)(1-u)} \frac{\omega^2}{r_0^2} \phi \approx 0 \ , \ (u \rightarrow 1) \ .
\end{equation}

The solution the reads

\begin{equation}
 \phi = c_1 \cosh \left ( \frac{\omega}{r_o} \sqrt{\frac{4\beta -3}{\beta - 1}}\sqrt{u-1}\right ) + ic_2 \sinh \left ( \frac{\omega}{r_o} \sqrt{\frac{4\beta -3}{\beta - 1}}\sqrt{u-1}\right ) \ , \ (u \rightarrow 1) \ ,
\end{equation}

\noindent that is,

\begin{equation}
\begin{gathered}
  \phi \propto \exp\left ( \pm \frac{\omega}{r_o} \sqrt{\frac{4\beta -3}{\beta - 1}}\sqrt{u-1} \right ) 
 \\
 \Rightarrow \phi \propto \exp \left (\pm i \frac{\omega}{r_o} \sqrt{\frac{4\beta -3}{\beta - 1}}\sqrt{1-u}\right) \ , \ (u \rightarrow 1) \ .
\end{gathered}
\end{equation}

Now, as discussed in \cite{natsuume}, the solution above has a natural interpretation using tortoise coordinates, which allows us to identify this solution to that of a plane wave, in which the positive exponent represents the wave outgoing from the horizon, whilst the negative exponent represents the wave incoming to the horizon, which, according to the near-horizon boundary condition, allows us to fix

\begin{equation}
\phi \approx \exp \left (- i \frac{\omega}{r_o} \sqrt{\frac{4\beta -3}{\beta - 1}}\sqrt{1-u}\right) \ , \ (u \rightarrow 1) \ .
\label{eq:sol_nh}
\end{equation}

We shall now solve the perturbation equation in Eq. \ref{eq:pertu_edo} for all u ($0\leq u \leq 1$), which we will do in a power series in $\omega$ up to leading order, that is, we seek a solution of the form

\begin{equation}
 \phi(u) = \Phi_0(u) + \omega \Phi_i(u) \ .
 \label{eq:sol_omega_power}
\end{equation}

We are interested in a solution in the hydrodynamic limit, $\omega \rightarrow 0$, so that we shall consider terms only up to $\mathcal{O}(\omega)$. Therefore, the second term in Eq. \ref{eq:pertu_edo}, which is proportional to $\omega^2$, will not be considered, and the perturbation equation becomes simply

\begin{equation}
\left ( \frac{\sqrt{NA}}{u^3}\phi^{\prime}\right )^{\prime} = \left ( \frac{\sqrt{NA}}{u^3}\Phi_0^{\prime}\right )^{\prime} + \omega \left ( \frac{\sqrt{NA}}{u^3}\Phi_1^{\prime}\right )^{\prime} = 0 \ ,
\end{equation}

\noindent so that we must have, separately for $i=0,1$,

\begin{equation}
 \left ( \frac{\sqrt{NA}}{u^3}\Phi_i^{\prime}\right )^{\prime} = 0 \ .
\end{equation}

It is straightforward to find the solution

\begin{equation}
 \Phi_i = C_i + K_i \int \frac{u^3}{\sqrt{NA}}\, \mathrm{d} u \ ,
\end{equation}

\noindent where $C_i$ and $K_i$ are integration constants. Consequently, the general solution up to leading order in $\omega$ is, according to Eq. \ref{eq:sol_omega_power},

\begin{equation}
 \phi = \left (C_0 + \omega C_1 \right ) + \left ( K_0 + \omega K_1 \right ) \int \frac{u^3}{\sqrt{NA}}\, \mathrm{d} u \ .
 \label{eq:sol_omega_power_general}
\end{equation}

Given the definition of $N(u)$ and $A(u)$, respectively given by Eqs. \ref{eq:Nu} and \ref{eq:Au}, we cannot analytically write the integral in the solution in terms of standard mathematical functions. However, since we are interested in the $u \rightarrow 0$ and $u \rightarrow 1$ limits of the solution to impose the boundary conditions, we can expand the integral in $u$ around these extreme values, which yields, up to leading order in the respective expansions,

\begin{equation}
\int \frac{u^3}{\sqrt{NA}}\, \mathrm{d} u  \approx \frac{u^4}{4} \ , \ \left (u \rightarrow 0 \right ) \ ;
\end{equation}

\begin{equation}
\int \frac{u^3}{\sqrt{NA}}\, \mathrm{d} u  \approx \frac{3-4\beta}{\beta-1} \sqrt{\frac{\beta-1}{4\beta-3}}\sqrt{1-u} \ , \ \left (u \rightarrow 1 \right ) \ .
\end{equation}

The boundary condition at the AdS boundary then fixes the first pair of integration constants,

\begin{equation}
\begin{gathered}
  \phi (u\rightarrow0)  \approx \left (C_0 + \omega C_1 \right ) + \left ( K_0 + \omega K_1 \right ) \lim_{u\rightarrow0} \frac{u^4}{4} =  \phi^{(0)}   \\
  \Rightarrow \left (C_0 + \omega C_1 \right ) = \phi^{(0)} \ .
\end{gathered}
\end{equation}

Near the horizon, one has

\begin{equation}
   \phi \approx \phi^{(0)} - \left ( K_0 + \omega K_1 \right ) \frac{4\beta-3}{\beta-1} \sqrt{\frac{\beta-1}{4\beta-3}}\sqrt{1-u} \ , \ (u\rightarrow1) \ .
   \label{eq:sol_general_nh}
\end{equation}

Now, if we expand the near-horizon solution of Eq. \ref{eq:sol_nh} in $\omega$, one gets, up to $\mathcal{O}(\omega)$,

\begin{equation}
   \phi \propto 1 - i \frac{\omega}{r_o}\sqrt{\frac{4\beta-3}{\beta-1}}\sqrt{1-u} \ , \ (u\rightarrow1) \ .
\end{equation}

It is easy to see that Eq. \ref{eq:sol_general_nh} fixes the proportionality according to

\begin{equation}
   \phi \approx \phi^{(0)} - i \phi^{(0)}\frac{\omega}{r_o}\sqrt{\frac{4\beta-3}{\beta-1}}\sqrt{1-u} \ , \ (u\rightarrow1) \ .
   \label{eq:sol_nh_power_omega}
\end{equation}

Comparing now Eqs. \ref{eq:sol_general_nh} and \ref{eq:sol_nh_power_omega} then allows us to fix the second pair of integration constants,

\begin{equation}
\begin{gathered}
 \left ( K_0 + \omega K_1 \right ) \frac{4\beta-3}{\beta-1} \sqrt{\frac{\beta-1}{4\beta-3}} = i \phi^{(0)}\frac{\omega}{r_o}\sqrt{\frac{4\beta-3}{\beta-1}} 
 \\
 \Rightarrow  \left ( K_0 + \omega K_1 \right ) = i \phi^{(0)}\frac{\omega}{r_o} \left (\frac{\beta -1}{4\beta-3} \right ) \frac{|4\beta - 3|}{|\beta-1|} \ ,
\end{gathered}
\end{equation}

\noindent so that the full solution is

\begin{equation}
 \phi = \phi^{(0)} \left ( 1 + i \frac{\omega}{r_0} \left (\frac{\beta -1}{4\beta-3} \right ) \frac{|4\beta - 3|}{|\beta-1|} \int \frac{u^3}{\sqrt{NA}}\, \mathrm{d} u \right ) \ , \ 0 \leq u \leq 1 \ .
\end{equation}

Accordingly, the full time-dependent perturbation is given by

\begin{equation}
\varphi = g^{xx}h_{xy} = \phi^{(0)} e^{-i\omega t} \left ( 1 + i \frac{\omega}{r_0} \left (\frac{\beta -1}{4\beta-3} \right ) \frac{|4\beta - 3|}{|\beta-1|} \int \frac{u^3}{\sqrt{NA}}\, \mathrm{d} u \right ) \ , \ 0 \leq u \leq 1 \ ,
\end{equation}

\noindent which is asymptotically given by

\begin{equation}
g^{xx}h_{xy} \approx \phi^{(0)} e^{-i\omega t} \left ( 1 + i \frac{\omega}{r_0} \left (\frac{\beta -1}{4\beta-3} \right ) \frac{|4\beta - 3|}{|\beta-1|} \frac{u^4}{4}\right ) \ , \  (u \rightarrow 0) \ .
\label{eq:perturb_asym_solution}
\end{equation}

Comparing now Eq. \ref{eq:perturb_asym_solution} to Eq. \ref{eq:perturb_asym_def}, and identifying $h_{xy}^{(0)} =  \phi^{(0)} e^{-i \omega t}$, we promptly obtain,

\begin{equation}
h_{xy}^{(1)} = \frac{i\omega}{4r_0} \left (\frac{\beta -1}{4\beta-3} \right ) \frac{|4\beta - 3|}{|\beta-1|} \  = 
\begin{cases} 
-\frac{i\omega}{4r_{0}} & \frac{3}{4}<\beta<1 \ , \\
\frac{i\omega}{4r_{0}} & \beta<\frac{3}{4}\text{ or }\beta>1 \ . 
\end{cases} 
\end{equation}

Finally, substituting the resulting $h_{xy}^{(1)}$ above in Eq. \ref{eq:eta_s_geral} yields,

\begin{equation} \label{eq:etaSfinal}
\frac{\eta}{s}=\begin{cases}
-\frac{1}{4\pi} \left ( \frac{1}{15 \beta - 3\beta^2 - 11} \right ) \left ( \frac{\beta - 2}{3-4\beta} \right )^{1/2} \ , & \frac{3}{4}<\beta<1
\\
\frac{1}{4\pi} \left ( \frac{1}{15 \beta - 3\beta^2 - 11} \right ) \left ( \frac{\beta - 2}{3-4\beta} \right )^{1/2} \ ,  & \beta<\frac{3}{4}\text{ or }\beta>1
\end{cases} \ .
\end{equation}

Naturally, for $\beta =1$, the deformed black brane $\eta/s$ ratio is exactly $\frac{1}{4\pi}$, recovering the KSS result for the AdS$_5$--Schwarzschild black brane. Also, $\eta/s$ diverges for $\beta \approx 0.9$ and vanishes for $\beta = 2$.

As detailed in \cite{ads5}, the deformed black brane temperature diverges at $\beta\to 3/4$, and is purely imaginary for $\beta < 3/4$ or $\beta > 2$. As the deformed black brane temperature cannot have imaginary or divergent values, the temperature of $\beta$-SAdS$_5$ constrains the $\beta$ parameter to the range $3/4 < \beta < 2$. Therefore, a priori, the deformation parameter $\beta$ is limited to the ranges

\begin{equation}\label{ra1}
3/4< \beta < 0.9\;\;\;\text{and}\quad 1 < \beta < 2.
\end{equation}

The value $\beta < 2$ is seen from \eqref{eq:etaSfinal}, since $\beta=2$ makes $\eta/s$ equal to zero, whereas the range  $0.9\leq\beta < 1$ imply $\eta/s<0$, which has no physical significance, in accordance with the second law of thermodynamics.

Now, by demanding that the deformed SAdS$_5$ black brane event horizon must be real, by analyzing Eq. \ref{eq:Nu}, the parameter $\beta$ is further restricted: from $1 < \beta < 2$ to $1 < \beta \leq 1.384$. Also, the $r_0=\lim_{\beta\to1} r_\beta$ horizon (the standard SAdS$_5$ black brane event horizon) is of Killing type. For it to be a good approximation, i.e., $\left|r_0-r_\beta\right|\lesssim 10^{-2}$, $\beta$ must be restricted a little more, to the range $1 < \beta \lesssim  1.2$. Considering these two observations, the $\beta$ parameter is further restricted into the ranges

\begin{equation}\label{ra2}
3/4< \beta < 0.9\;\;\;\text{and}\quad 1 < \beta \lesssim 1.2.
\end{equation}

Now, as shown in \cite{ads5}, the deformation induced by $\beta$ changes thermodynamics and hydrodynamics by a numerical factor, so that in the range $1< \beta \lesssim 1.2$, there is a violation of the KSS bound. The violation may come from the fact that the deformed AdS$_5$--Schwarzschild black brane is not a solution to the standard GR EFE, but the effective 5D EFE. The existence of a range where the KSS bound is violated, namely $1<\beta\lesssim 1.2$, but no pathologies in causality of space-time or thermodynamic functions can be seen, is one of the main results of this work. On the other hand, in the range $3/4<\beta < 0.9$, the KSS bound is not violated. 

The family of solutions obtained with the allowed values of $\beta$ is an interesting result worthy further investigation, mainly in the AdS/QCD (Quantum Chromodynamics) correspondence, as further discussed in \cite{ads5}.

\chapter{\texorpdfstring{$\eta/s$}{} in AdS\texorpdfstring{\textsubscript{4}}{}--Reissner--Nordstr\"om black brane background}
\label{sec:eta_s_ads_mgd_4}

In this section, we shall consider the deformed AdS$_4$--Reissner--Nordstr\"om ($\beta$-AdS$_4$--RN) black brane, and then calculate the ratio $\eta/s$ in this particular gravitational background. The results in this section are original, and were published in \cite{ads4}, in collaboration with R. da Rocha and P. Meert. 

The $\beta$-AdS$_4$--RN black brane background is given by

\begin{equation}
 ds^2_{\beta\text{-}AdS_4\text{--RN}} = -r^2 n(r) \mathrm{d}t^2 + \frac{1}{r^2 a(r)} \mathrm{d} r^2 + r^2 \left ( \mathrm{d}x^2 + \mathrm{d}y^2 \right ) \ ,
 \label{eq:ads_4-mgd_metric}
\end{equation}

\noindent where

\begin{equation}
n(r) = 1 - \left (1 + Q^2\right )\frac{r_0^3}{r^3} + Q^2 \frac{r_0^4}{r^4} \ ;
\end{equation}

\begin{equation}
a(r) = n(r) \left ( \frac{1 - \left (1 + \frac{1}{3} \left (\beta - 1 \right ) \right )\frac{r_0}{r}}{1-\frac{r_0}{r}}\right ) \ ,
\end{equation}

\noindent where again $r_0$ is the horizon radius; $\beta$ is the deformation parameter constant; and $Q$ is the RN black hole charge. In fact, the black brane in Eq. \ref{eq:ads_4-mgd_metric} is built as a deformation of the AdS$_4$--Reissner--Nordström black brane \cites{mgd1, mgd2}, so that $n(r)$ is fixed as defined above. On the other hand, $a(r)$ is determined from the Hamiltonian constraint in the ADM formalism \cite{adm}. Details on the construction of this black brane are in \cite{ads4}. We still have an unit AdS radius and take $n(r)$ and $a(r)$ to be non-negative functions. 

The change of coordinate $u = r_0/r$, yields

\begin{equation}
g_{\mu \nu} \mathrm{d}x^\mu \mathrm{d}x^\nu = -\frac{r_0^2}{u^2} n(u) \mathrm{d}t^2 + \frac{1}{u^2 a(u)} \mathrm{d}u^2 + \frac{r_0^2}{u^2} \left (\mathrm{d}x^2 + \mathrm{d}y^2 \right ) \ ,
\end{equation}

\noindent and

\begin{equation}
g^{\mu \nu} \partial_\mu \partial_\nu = -\frac{u^2}{r_0^2 n(u)} \partial_t^2 + u^2 a(u) \partial_u^2 + \frac{u^2}{r_0^2} \left (\partial_x^2 + \partial_y^2 \right ) \ ,
\end{equation}

\noindent where $n(u)$ and $a(u)$ are given by

\begin{equation}
n(u) = 1 - \left (1 + Q^2\right )u^3 + Q^2u^4 \ ;
\label{eq:nu}
\end{equation}

\begin{equation}
a(u) = n(u) \left ( \frac{1 - \left (1 + \frac{2}{9} \left (\beta - 1 \right ) \right )u}{1-u}\right ) \ .
\label{eq:au}
\end{equation}

The metric determinant $g \equiv \det g_{\mu \nu}$ is such that

\begin{equation}
\begin{gathered}
\sqrt{-g} = \sqrt{-\left (-\frac{r_0^2}{u^2}\right)\left(\frac{r_0^2}{u^2}\right)^2 \frac{n(u)}{u^2 a(u)}} 
\\
\Rightarrow \sqrt{-g} = \frac{r_0^3}{u^4} \sqrt{\frac{n}{a}} \ ,
\end{gathered}
\end{equation}

\noindent where, from now on, $n$ and $a$ refer respectively to $n(u)$ and $a(u)$.

We shall now again consider a bulk perturbation $h_{xy}$, so that,

\begin{equation}
 ds^2 = ds^2_{\beta\text{-}AdS_4\text{--RN}} + 2h_{xy} \mathrm{d}x \mathrm{d}y \ .
\end{equation}

The response in the boundary energy-momentum tensor, $\delta \left \langle T^{xy} \right \rangle$, according to the linear response theory, is, as before, given by

\begin{equation}
 \delta \left \langle \tau^{xy} \right \rangle  = i \omega \eta h_{xy}^{(0)} \ ,
 \label{eq:response_1_2}
\end{equation}

\noindent where again $\tau^{xy}$ is the dissipative part of $T^{xy}$; $\eta$ is the viscosity; and $h_{xy}^{(0)}$ is the perturbation added to the boundary theory, which is asymptotically related to $h_{xy}$, the bulk perturbation, by

\begin{equation}
 g^{xx}h_{xy} \approx h_{xy}^{(0)} \left ( 1 + h_{xy}^{(1)} u^3 \right ) \ , \ (u \rightarrow 0) \ ,
 \label{eq:perturb_asym_def_2}
\end{equation}

\noindent which is valid because $g^{xx}h_{xy}$ obeys the equation of motion for a massless scalar field \cites{natsuume, son_hydro}. Since the deformed AdS$_4$--RN black brane has the asymptotic behavior considered in Sec. \ref{sec:4d_gkp_w} (Eq. \ref{eq:asym_4d}), we can directly apply the results discussed in this section, and treat $g^{xx}h_{xy}$ as the 4-dimensional bulk scalar field $\varphi$ which acts as an external source of a boundary operator, which, in this case, is nothing but $\tau^{xy}$. Thus, from Eq. \ref{eq:response_general_gkp_w_2}, one has

\begin{equation}
\delta \left \langle \tau^{xy} \right \rangle = \frac{r_0^3}{16 \pi G_4}3 h_{xy}^{(1)}h_{xy}^{(0)} \ , 
\label{eq:response_2_2}
\end{equation}

\noindent where it is convenient here to reintroduce $16 \pi G_4$. Comparing Eqs. \ref{eq:response_1_2} and \ref{eq:response_2_2},

\begin{equation}
 i \omega \eta = \frac{3r_0^3}{16\pi G_4} h_{xy}^{(1)} \ .
\label{eq:eta_quase_2}
\end{equation}

Now, the entropy density of the deformed AdS$_4$--RN black brane reads \cite{ads4}

\begin{equation}
\begin{gathered}
 s = \frac{r_{0}^2}{4G_4} \left ( 1 + \frac{1}{3} \left(\beta-1\right) \right)^2 \ ,
\end{gathered}
\end{equation}

\noindent where $\beta$ is fixed in terms of $Q$ by the Killing Equation, $\beta = \beta(Q)$, in a quite lengthy expression, available in \cite{ads4}. Plugging this result into Eq. \ref{eq:eta_quase_2} yields

\begin{equation}
\begin{gathered}
\frac{\eta}{s} = \frac{3r_0}{4\pi\left(1+\frac{1}{3}\left(\beta-1\right)\right)^{2}} \frac{h_{xy}^{(1)}}{i \omega} \ .
\label{eq:eta_s_geral_2}
\end{gathered}
\end{equation}

We now must find $h_{xy}^{(1)}$, by solving the equation of motion for the perturbation $g^{xx}h_{xy} \equiv \varphi$, which is that of a massless scalar in a 4-dimensional background,

\begin{equation}
 \partial_\mu \left ( \sqrt{-g} g^{\mu \nu} \partial_\nu \varphi \right ) = 0 \ .
\end{equation}

Considering again the perturbation of the form $\varphi = \phi(u) e^{-i\omega t}$, one has

\begin{equation}
\begin{gathered}
 \partial_t \left ( \sqrt{-g} g^{tt} \partial_t \varphi \right ) = \frac{r_0^3}{u^4}\sqrt{\frac{n}{a}}\left ( -\frac{u^2}{r_0^2 n} \right ) \partial_t \left ( \partial_t \varphi \right ) = \frac{-r_0}{\sqrt{na}} \frac{(-\omega^2 \varphi)}{u^2} 
 \\
 \Rightarrow  \partial_t \left ( \sqrt{-g} g^{tt} \partial_t \varphi \right ) = \frac{1}{\sqrt{na}} \frac{\omega^2 r_0}{u^2} \varphi \ ;
\end{gathered}
\end{equation}

\begin{equation}
\begin{gathered}
 \partial_u \left ( \sqrt{-g} g^{uu} \partial_u \varphi \right ) = \partial_u \left ( \frac{r_0^3}{u^4}\sqrt{\frac{n}{a}} u^2 a \partial_u \varphi \right ) 
 \\
 \Rightarrow  \partial_u \left ( \sqrt{-g} g^{uu} \partial_u \varphi \right ) = r_0^3 \frac{\mathrm{d}}{\mathrm{d}u}\left ( \frac{\sqrt{na}}{u^2} \frac{\mathrm{d}}{\mathrm{d}u} \phi \right ) e^{-i\omega t} \ ,
\end{gathered}
\end{equation}

\noindent so that

\begin{equation}
\begin{gathered}
 \partial_\mu \left ( \sqrt{-g} g^{\mu \nu} \partial_\nu \varphi \right ) = \partial_t \left ( \sqrt{-g} g^{tt} \partial_t \varphi \right ) + \partial_u \left ( \sqrt{-g} g^{uu} \partial_u \varphi \right ) =  0 
 \\
 \Rightarrow \frac{1}{\sqrt{na}} \frac{\omega^2 r_0}{u^2} \phi e^{-i \omega t} + r_0^3 \frac{\mathrm{d}}{\mathrm{d}u}\left ( \frac{\sqrt{na}}{u^2} \frac{\mathrm{d}}{\mathrm{d}u} \phi \right ) e^{-i\omega t} = 0 
 \\
 \Rightarrow
\frac{u^2}{\sqrt{na}} \frac{\mathrm{d}}{\mathrm{d}u}\left ( \frac{\sqrt{na}}{u^2} \frac{\mathrm{d}}{\mathrm{d}u} \phi \right ) + \frac{1}{na} \frac{\omega^2}{r_0^2} \phi = 0 \ .
\end{gathered}
\end{equation}

Therefore, the perturbation equation reduces to

\begin{equation}
\frac{u^2}{\sqrt{na}} \left ( \frac{\sqrt{na}}{u^2}\phi^{\prime}\right )^{\prime}  + \frac{1}{na} \frac{\omega^2}{r_0^2} \phi  = 0 \ ,
\label{eq:pertu_edo_2}
\end{equation}

\noindent or

\begin{equation}
\phi^{\prime \prime} + \left ( \frac{(na)^{\prime}}{2 na} - \frac{2}{u}\right ) \phi^{\prime} + \frac{1}{na} \frac{\omega^2}{r_0^2} \phi = 0 \ ,
\label{eq:pertu_edo2_2}
\end{equation}

\noindent where primes denote differentiation with respect to $u$.

We shall again impose an ``incoming wave'' boundary condition in the near-horizon region ($u \rightarrow 1$); and the Dirichlet boundary condition at the AdS boundary ($u \rightarrow 0$) $\phi (u\rightarrow 0) = \phi^{(0)}$. To incorporate the near-horizon incoming wave boundary condition, we shall first solve Eq. \ref{eq:pertu_edo2_2} in the limit $u \simeq 1$, which we do by expanding the coefficients of the equation around $u=1$ and then take the leading order term. According to Eqs. \ref{eq:nu} and \ref{eq:au}, one has, up to leading order,

\begin{equation}
\begin{aligned}
\frac{(na)^{\prime}}{2 na} - \frac{2}{u} &= \frac{2\beta +7}{2((2\beta+7)u -9)} + \frac{3Q^2u^2 -2u -1}{Q^2u^3 - u^2 - u - 1} 
\\
&+ \frac{1}{2(u-1)} - \frac{2}{u} 
\\
\Rightarrow \frac{(na)^{\prime}}{2 na} - \frac{2}{u} &\approx -\frac{1}{2(1-u)} \ , \ (u \rightarrow 1) \ ,
\end{aligned}
\end{equation}

\noindent and

\begin{equation}
\frac{1}{na} \approx \frac{9}{2 (1-\beta)(Q^2-3)^2(1-u)} \ , \ (u \rightarrow 1) \ ,
\end{equation}

\noindent so that Eq. \ref{eq:pertu_edo2_2} becomes

\begin{equation}
\phi^{\prime \prime} - \frac{1}{2(1-u)} \phi^{\prime} + \frac{9}{2 (1-\beta)(Q^2-3)^2(1-u)} \frac{\omega^2}{r_0^2} \phi \approx 0 \ , \ (u \rightarrow 1) \ .
\end{equation}

The solution is,

\begin{equation}
 \phi = c_1 \cosh \left ( \frac{3\omega}{r_0|Q^2-3|} \sqrt{\frac{2(u-1)}{1-\beta}}\right ) + ic_2 \sinh \left ( \frac{3\omega}{r_0|Q^2-3|} \sqrt{\frac{2(u-1)}{1-\beta}}\right ) \ , \ (u \rightarrow 1) \ .
\end{equation}

The solution above is equivalent to

\begin{equation}
\begin{gathered}
  \phi \propto \exp\left ( \pm \frac{3\omega}{r_0|Q^2-3|} \sqrt{\frac{2(u-1)}{1-\beta}} \right ) 
 \\
 \Rightarrow \phi \propto \exp \left (\pm i \frac{3\omega}{r_o|Q^2-3|} \sqrt{\frac{2(u-1)}{\beta-1}}\right) \ , \ (u \rightarrow 1) \ .
\end{gathered}
\end{equation}

Once again, we can identify this solution to that of a plane wave, in which the positive exponent represents the wave outgoing from the horizon, whilst the negative exponent represents the wave incoming to the horizon, which, according to the near-horizon boundary condition, fixes

\begin{equation}
\phi \approx \exp \left (-i \frac{3\omega}{r_0|Q^2-3|} \sqrt{\frac{2(u-1)}{\beta-1}}\right) \ , \ (u \rightarrow 1) \ .
\label{eq:sol_nh_2}
\end{equation}

Now we solve the perturbation equation in Eq. \ref{eq:pertu_edo} for all u in a power series in $\omega$ up to leading order, that is, taking

\begin{equation}
 \phi(u) = \Phi_0(u) + \omega \Phi_i(u) \ ,
 \label{eq:sol_omega_power_2}
\end{equation}

\noindent so that, considering terms only up to $\mathcal{O}(\omega)$ within the hydrodynamic limit, the second term in Eq. \ref{eq:pertu_edo_2} shall not be considered, and the perturbation equation becomes simply

\begin{equation}
\left ( \frac{\sqrt{na}}{u^2}\phi^{\prime}\right )^{\prime} = \left ( \frac{\sqrt{na}}{u^2}\Phi_0^{\prime}\right )^{\prime} + \omega \left ( \frac{\sqrt{na}}{u^2}\Phi_1^{\prime}\right )^{\prime} = 0 \ ,
\end{equation}

\noindent so that we must have, separately for $i=0,1$,

\begin{equation}
 \left ( \frac{\sqrt{na}}{u^2}\Phi_i^{\prime}\right )^{\prime} = 0 \ ,
\end{equation}

\noindent whose solution is

\begin{equation}
 \Phi_i = c_i + k_i \int \frac{u^2}{\sqrt{na}}\, \mathrm{d} u \ ,
\end{equation}

\noindent where $c_i$ and $k_i$ are integration constants. Consequently, the general solution up to leading order in $\omega$ is, according to Eq. \ref{eq:sol_omega_power_2},

\begin{equation}
 \phi = \left (c_0 + \omega c_1 \right ) + \left ( k_0 + \omega k_1 \right ) \int \frac{u^2}{\sqrt{na}}\, \mathrm{d} u \ .
 \label{eq:sol_omega_power_general_2}
\end{equation}

Once again, since we are interested in the $u \rightarrow 0$ and $u \rightarrow 1$ limits of the solution to impose the boundary conditions, we can expand the integral in $u$ around these extreme values, which yields, up to leading order in the respective expansions,

\begin{equation}
\int \frac{u^2}{\sqrt{na}}\, \mathrm{d} u  \approx \frac{u^3}{3} \ , \ \left (u \rightarrow 0 \right ) \ ;
\end{equation}

\begin{equation}
\int \frac{u^2}{\sqrt{na}}\, \mathrm{d} u  \approx \frac{3}{Q^2-3} \sqrt{\frac{2 (u-1)}{\beta -1}} \ , \ \left (u \rightarrow 1 \right ) \ .
\end{equation}

The boundary condition at the AdS boundary then fixes the first pair of integration constants,

\begin{equation}
\begin{gathered}
  \phi (u\rightarrow0)  \approx \left (c_0 + \omega c_1 \right ) + \left ( k_0 + \omega k_1 \right ) \lim_{u\rightarrow0} \frac{u^3}{3} =  \phi^{(0)} 
  \\
  \Rightarrow \left (c_0 + \omega c_1 \right ) = \phi^{(0)} \ .
\end{gathered}
\end{equation}

Near the horizon, one has

\begin{equation}
   \phi \approx \phi^{(0)} + \left ( k_0 + \omega k_1 \right ) \frac{3}{Q^2-3} \sqrt{\frac{2 (u-1)}{\beta -1}} \ , \ (u\rightarrow1) \ .
   \label{eq:sol_general_nh_2}
\end{equation}

Now, if we expand the near-horizon solution of Eq. \ref{eq:sol_nh_2} in $\omega$, one gets, up to $\mathcal{O}(\omega)$,

\begin{equation}
   \phi \propto 1 -i \frac{3\omega}{r_0|Q^2-3|} \sqrt{\frac{2(u-1)}{\beta-1}} \ , \ (u\rightarrow1) \ ,
\end{equation}

\noindent so that Eq. \ref{eq:sol_general_nh_2} fixes the proportionality according to

\begin{equation}
\phi \approx \phi^{(0)} - i \phi^{(0)} \frac{3\omega}{r_0|Q^2-3|} \sqrt{\frac{2(u-1)}{\beta-1}} \ , \ (u\rightarrow1) \ ,
\label{eq:sol_nh_power_omega_2}
\end{equation}

\noindent and we can then determine the second pair of integration constants, by comparing Eqs. \ref{eq:sol_general_nh_2} and \ref{eq:sol_nh_power_omega_2},

\begin{equation}
\begin{gathered}
 \left ( k_0 + \omega k_1 \right ) \frac{1}{Q^2-3} = -i \phi^{(0)}\frac{\omega}{r_o}\frac{1}{|Q^2-3|} 
 \\
 \Rightarrow  \left ( k_0 + \omega k_1 \right ) = -i \phi^{(0)}\frac{\omega}{r_o} \frac{Q^2-3}{|Q^2-3|} \ ,
\end{gathered}
\end{equation}

\noindent so that the full solution is

\begin{equation}
 \phi = \phi^{(0)} \left ( 1 - i \frac{\omega}{r_0} \frac{Q^2-3}{|Q^2-3|} \int \frac{u^2}{\sqrt{na}}\, \mathrm{d} u \right ) \ , \ 0 \leq u \leq 1 \ .
\end{equation}

Accordingly, the full time-dependent perturbation is given by

\begin{equation}
\varphi = g^{xx}h_{xy} = \phi^{(0)} e^{-i\omega t} \left ( 1 - i \frac{\omega}{r_0} \frac{Q^2-3}{|Q^2-3|} \int \frac{u^2}{\sqrt{na}}\, \mathrm{d} u \right )  \ , \ 0 \leq u \leq 1 \ ,
\end{equation}

\noindent which is asymptotically given by

\begin{equation}
g^{xx}h_{xy} \approx \phi^{(0)} e^{-i\omega t} \left ( 1 - i \frac{\omega}{r_0} \frac{Q^2-3}{|Q^2-3|}\frac{u^3}{3} \right )  \ , \  (u \rightarrow 0) \ .
\label{eq:perturb_asym_solution_2}
\end{equation}

Comparing now Eq. \ref{eq:perturb_asym_solution_2} to Eq. \ref{eq:perturb_asym_def_2}, and identifying $h_{xy}^{(0)} =  \phi^{(0)} e^{-i \omega t}$, we promptly obtain

\begin{equation}
h_{xy}^{(1)} = \frac{-i\omega}{3r_0}  \frac{Q^2-3}{|Q^2-3|} \ .
\end{equation}

Thus, substituting the resulting $h_{xy}^{(1)}$ above in Eq. \ref{eq:eta_s_geral_2} yields

\begin{equation} \label{eta-sRatio}
\frac{\eta}{s}=\color{black}\frac{9}{\left(1-\frac{2^{4/3}}{\chi_{Q}}+\frac{\chi_{Q}}{2^{1/3}}\right)^{2}} \left(\frac{3-Q^{2}}{\left|3-Q^{2}\right|}\right)\ ,
\end{equation}

\noindent for $\color{black}\chi_Q=\left(-7-27Q^{2}+3\sqrt{9+42Q^{2}+81Q^{4}}\right)^{1/3}$, where the $\beta = \beta(Q)$ parameter was substituted.

Now, for $\eta/s$ to be a positive quantity, which it must be according to the second law of thermodynamics, $Q$ must be in the range

\begin{equation}
0 < Q < \sqrt{3} \ , 
\end{equation}

\noindent in which case the $\eta/s$ ratio is in accordance with the KSS bound. Notice that $Q$ must be a positive quantity, so that the $-\sqrt{3} < Q < 0$ interval, which also satisfies the $3-Q^2 > 0$, bound was not considered in the result above.

As $\beta = \beta(Q)$, the range above also implies a range for $\beta$, constraining the allowed deformations of the AdS$_4$-RN black branes. This is a very interesting result, and worthy further investigation, namely in the AdS/CMT (Condensed Matter Theory) correspondence context, as discussed in \cite{ads4}.

\chapter{Generalized actions}
\label{sec:gen_actions}

As we discussed before, the bound is saturated for standard GR gravity, so that sufficiently generic corrections to the Einstein--Hilbert action may lead to a \emph{violation} of the conjectured KSS bound (Eq. \ref{eq:eta_s_bound}). Now, since corrections to GR gravity are expected in a quantum theory of gravity, it may be the case that the KSS bound will be violated, which would violate the conjectured universality. However, if the bound is proven to be correct, it would present itself as a constraint on possible corrections to standard GR gravity, since modifications of standard GR which violate the bound would not be possible. In either case, it is thus clear why the study of generalized actions which lead to corrections to standard GR are relevant.

In this section, we shall consider a generalization of the Einstein--Hilbert gravitational action, and discuss the consequences to the KSS bound. Following \cites{viol_1, viol_2}, we shall consider a gravitational theory described by the classical action with Gauss--Bonnet terms,

\begin{equation}
S = \frac{1}{16 \pi G_5} \int \mathrm{d}^5x \sqrt{-g} \left ( R - 2 \Lambda + \frac{\lambda_{GB}}{2} L^2 \left ( R^2 - 4 R_{\mu \nu} R^{\mu \nu} + R_{\mu \nu \rho \sigma } R^{\mu \nu \rho \sigma} \right )\right ) \ ,
\label{eq:g_b}
\end{equation}

\noindent where here $\Lambda = -\frac{6}{L^2}$, according to Eq. \ref{eq:lambda_def}, with $D+2 = 5$; and $\lambda_{GB}$ is a constant parameter associated to the Gauss--Bonnet gravity.

By calculating the $\eta/s$ ratio in the gravitational background which is a solution to the generalized action of Eq. \ref{eq:g_b}, in a manner similar to what we did in Secs. \ref{sec:eta_s_ads_mgd_5} and \ref{sec:eta_s_ads_mgd_4}, it was found in \cite{viol_1},

\begin{equation}
 \frac{\eta}{s} = \frac{1}{4\pi} \left ( 1 - 4 \lambda_{GB} \right ) \ ,
 \label{eq:viol_gb}
\end{equation}

\noindent which is a non-perturbative result in $\lambda_{GB}$, instead of representing a linear correction. In fact, it is easy to see from Eq. \ref{eq:viol_gb} that the KSS bound is violated for $\lambda_{GB} > 0$, going yet to zero as $\lambda_{GB} \rightarrow 1/4$, in which case the viscosity vanishes. On the other hand, $\lambda_{GB}$ must be bounded above by $1/4$ to a boundary CFT to exist \cite{viol_1}, so that it is guaranteed that $\eta/s \geq 0$. 

Now, in \cite{viol_2}, it was argued that for $\lambda_{GB} > 9/100$, the gravitational theory is inconsistent, as it leads to microcausality violation. Therefore, consistent 4-dimensional boundary theories with 5-dimensional Gauss--Bonnet gravitational duals must be such that $\lambda_{GB} \leq 9/100$, so that one has

\begin{equation}
 \frac{\eta}{s} \geq \frac{1}{4\pi} \left ( \frac{16}{25} \right ) \ .
 \label{eq:eta_s_bound_gb}
\end{equation}

Notice that the bound of Eq. \ref{eq:eta_s_bound_gb} is 36\% smaller than the original KSS bound of Eq. \ref{eq:eta_s_bound}, thus still representing a violation to the conjecture universal bound. In \cite{viol_2}, such difference is discussed within two possibilities: first, Gauss--Bonnet gravity with  $\lambda_{GB} \leq 9/100$ is consistent as a classical limit of a quantum theory of gravity within the string landscape \cite{landscape}, so that the violation in the KSS bound is an inherent feature of nature; or, second, an even subtler inconsistency in the theory may be present in the range $0 < \lambda_{GB} \leq 9/100$, in which case further investigation is necessary.

We now turn the discussion to the actions leading to the deformed AdS black branes studied in the last two sections. Both the deformed AdS$_5$--Schwarzschild and the deformed AdS$_4$--RN black brane solutions were constructed via a deformation of the original AdS$_5$--Schwarzschild and AdS$_4$--Reissner-Nordström black branes, respectively. As detailed in \cite{ads5} and \cite{ads4}, this was implemented by using the ADM formalism \cite{adm} as well as the geometric deformation methods \cites{c_o_r_2014, ads4}, provided the bulk Weyl fluid is identified with the stress-energy tensor on the boundary theory \cite{ads_memb}.

Once in this last approach, although fundamental actions were not explicitly constructed, it is reasonable to pose the question on whether or not the KSS bound is supposed to hold, or if its violated. As we saw, the answer depends on the particular dimension, and we were able to find a violation to the KSS bound for reasonable values of the deformation parameter $\beta$ in the 5D case, which is a very interesting result. Further investigation is due, including the implications of the aforementioned results to particular holographic correspondences. This will be done in future works.

As of now, we shall start to consider yet another class of holographic correspondences: that between gravity and fluid dynamics.
 
\chapter{Fluid/gravity correspondence}
\label{sec:fgcorr}

The fluid/gravity correspondence will be the topic of the remainder of the thesis. In a few words, it is a holographic correspondence which establishes a very precise relationship between fluid dynamics on a fixed $(3+1)$-dimensional background, and gravity in $4+1$ dimensions, which is why we can refer to the correspondence as holographic, as we discussed before. More precisely, the gravitational theory the correspondence considers is standard GR with negative cosmological constant, so that, mathematically, it asserts that the EFE with negative cosmological constant in $d+1$ dimensions capture the (generalized) Navier--Stokes (NS) equations in $d$ dimensions. The main goal of this section is to provide an overview of this correspondence, following the works \cites{hubeny_izy, minwalla, hubeny_hard}. 

As we shall see, the correspondence is established in such a way that given an arbitrary fluid dynamical solution, it is possible to systematically construct a corresponding asymptotically AdS black hole spacetime with a regular horizon whose dynamics is closely related to that of the fluid flow. Now, since a given fluid solution leads to a correspondent (and generic!) black hole solution, there are no longer requirements of symmetry limiting the gravitational dynamics. On the other hand, like we did before, we can also use tools from the gravitational theory to study aspects of the fluids, which is of course of great relevance, both theoretical and practical, as the correspondence then provides a new perspective on a wide class of unsolved problems on both its sides.

Fluid/gravity correspondence is built upon AdS/CFT duality, which was discussed and practically employed in the previous sections. Here we remind one of its main aspects: different asymptotically AdS bulk geometries correspond to different states in the boundary gauge theory, with the pure AdS bulk geometry corresponding to the vacuum state of the boundary theory. A deformation in the bulk geometry (keeping the AdS asymptotic behavior) then corresponds to excited states. More importantly in what follows, remind that a large (i.e., with a radius $r_0$ much greater than the AdS radius $L$, $r_0 \gg L$) SAdS black hole corresponds to a thermal state in the gauge theory. As we discussed in Sec. \ref{sec:gkp_w}, perturbations are related to the stress tensor expectation value in the field theory. 

The physical macroscopic properties of the boundary gauge theory (such as pressure, entropy, energy density, temperature, etc.) are captured by its stress tensor, which in turn is induced by the bulk geometry and can be extracted by Brown--York \cites{natsuume, brown-york} procedures. Notice, however. the important difference between the stress tensors we shall consider: the \emph{bulk} stress tensor of the EFE is null, so that the bulk solutions $g_{A B}$ will generally be vacuum black holes spacetimes with no matter content (except for the negative cosmological constant accounting to the asymptotically AdS behavior). Distinctly, the \emph{boundary} stress tensor $T^{\mu\nu}$ is non-zero and contains the matter content of the boundary theory. Most importantly, the boundary stress tensor conservation \emph{does not} curve the boundary spacetime, which is non-dynamical and fixed to be flat.

In sum, the boundary fluid is characterized by the boundary stress tensor $T^{\mu\nu}(x^\mu)$, whilst the bulk geometry is given by the bulk metric $g_{AB}(r,x^\mu)$, where $x^\mu = (t, x^i)$ represents the boundary spacetime directions. Naturally, the bulk dynamics is determined by the bulk EFE, whilst the boundary dynamics is given by stress tensor conservation, $\nabla_\mu T^{\mu\nu} = 0$, realized through the NS equation after the consideration of the proper constitutive equation, as discussed in Sec. \ref{sec:fluids}.

Let us first analyze the global thermal equilibrium state described by the SAdS$_5$ black hole, whose metric is, according to Eq. \ref{eq:sads_5_r},

\begin{equation}
\begin{gathered}
ds^2_{SAdS_5} = -r^2 f(r) \mathrm{d}t^2 + \frac{1}{r^2 f(r)} \mathrm{d}r^2 + r^2 \left (\mathrm{d}x^2 + \mathrm{d}y^2 + \mathrm{d}z^2 \right ) \ ,
\\
f(r) = 1 - \frac{r_0^4}{r^4} \ ,
\end{gathered}
\end{equation}

\noindent where the horizon is located at $r=r_0$. The black hole temperature, according to Eq. \ref{eq:sadstemp}, is given by $T = r_0/\pi$, from which it is easy to see that the temperature is linearly proportional to the black hole size, which confirms that such a black hole is thermodynamically stable, and therefore can corresponded to the state of thermal equilibrium. Now, this metric is translationally invariant in the boundary spatial directions $x^i$, so that, by scaling $r$ and boosting in $\mathbb{R}^{3,1}$ with normalized and constant $4$-velocity $u^\mu$, one is able to generate a family of solutions parameterized under $u^\mu$ and $T$,

\begin{equation}
ds^2 =-2\, u_{\mu}\, dx^{\mu} \,  dr 
+ r^2\, \left( \eta_{\mu\nu} + \frac{\pi^4 \, T^4}{r^4} \, u_{\mu}\, u_{\nu} \right) \, dx^{\mu} \,  dx^
{\nu}  \ .
\label{SAdSzero}
\end{equation}

In a few sections we shall perform a similar procedure, in which case the explicit calculations will be shown.

Now, on the boundary, the stress tensor induced by the bulk metric, according to the Brown--York procedure \cite{brown-york} (which we will also discuss in more detail later on) is,

\begin{equation}
T^{\mu \nu} =  \pi^4 \, T^4 \, \left(\eta^{\mu \nu} + 4 \, u^\mu \, u^\nu \right) \ ,
\label{TzeroO}
\end{equation}

\noindent which corresponds to a perfect fluid at temperature $T$, moving with velocity $u^{\mu}$ on the flat boundary background $\eta_{\mu\nu}$. In fact, as a perfect fluid, there is no dissipation in the system. Now, to capture more general systems, we need to go away from equilibrium and start considering dissipation, which is very important as it allows the state to stabilize at late times.

Naturally, to contain dissipative terms, the stress tensor must contain derivatives of $T$ and $u^{\mu}$, according to the gradient expansion previously discussed. Motivated by hydrodynamics, we shall consider those variations to be within the \emph{long wavelength regime}, i.e., the scale of variation $L$ of $T$ and $u^{\mu}$ must be large compared to the microscopic scale $1/T$. From this, we immediately have a natural small parameter, $\epsilon \equiv \frac{1}{L \, T} \ll 1$, so that boundary derivatives scale accordingly $(\partial_\mu u_\nu)^n , \ldots , \partial_\mu^n  \, u_\nu \sim \epsilon^n$. \cite{minwalla}. More on this, known as the \emph{hydrodynamic limit}, will be discussed later on.

As before, we write the stress tensor in a gradient expansion,

\begin{equation}
T^{\mu \nu} =  \pi^4 \, T^4 \, \left(\eta^{\mu \nu} + 4 \, u^\mu \, u^\nu \right) + \Pi_{(1)}^{\mu \nu}+ \Pi_{(2)}^{\mu \nu} + \ldots \ ,
\label{TtwoO}
\end{equation}

\noindent where $\Pi_{(1)}^{\mu \nu}$ contains dissipative terms built from first-order derivatives, $\partial^\mu  u^\nu$; $\Pi_{(2)}^{\mu \nu}$ contains the second order dissipative terms, and so forth. With the inclusion of dissipation in the constitutive equation as in Eq. \ref{TtwoO}, the conservation of the stress tensor determines its dynamics through the generalized NS equations. Now, since one has a conformal fluid, the stress tensor has to be Weyl covariant, as well as generally covariant in the boundary directions, which allows the employment of the Weyl covariant formalism \cite{weylcov}, which yields the $d$-dimensional dissipative stress tensor for a fluid on a fixed background $\gamma_{\mu\nu}$, to second order,

\begin{equation}
\begin{split}
T^{\mu\nu} =&\ P \, \left(\gamma^{\mu\nu}+d  \, u^\mu  \, u^\nu \right) 
-2 \, \eta \, \sigma^{\mu\nu}\\
&+2 \, \eta  \,  \left[\tau_1 \,  u^{\lambda} \, \mathcal{D}_{\lambda}\sigma^{\mu \nu} -\tau_\epsilon \, (\omega^{\mu}{}_{\lambda} \, \sigma^{\lambda \nu}+\omega^\nu{}_\lambda \,  \sigma^{\lambda\mu}) \right]
+ \xi_C \, C^{\mu\alpha\nu\beta} \, u_\alpha  \, u_\beta \\ 
&+ \xi_\sigma \, [ \sigma^{\mu}_{\ \lambda} \,\sigma^{\lambda \nu}
- \frac{ P^{\mu \nu}}{d-1} \sigma_{\alpha \beta} \, \sigma^{\alpha \beta}]
+\xi_\omega \, [ \omega^{\mu}_{\ \lambda} \,\omega^{\lambda \nu}
+ \frac{ P^{\mu \nu}}{d-1} \omega_{\alpha \beta} \, \omega^{\alpha \beta}] \ ,
\label{ddim2T}
\end{split}
\end{equation}

\noindent where we introduced now the pressure $P$ and several quantities built from the velocity $u^\mu$ and the background metric $\gamma_{\mu\nu}$:  $\sigma_{\mu\nu}$ and $\omega_{\mu\nu}$ are the shear and the vorticity of the fluid;  $P^{\mu\nu} = \gamma^{\mu\nu} + u^\mu  \, u^\nu $ is the spatial projector; $\mathcal{D}_{\lambda}$ is the Weyl-covariant derivative and $C_{\mu\nu\alpha\beta}$ is the Weyl tensor for $\gamma_{\mu\nu}$. $\eta$ is the fluid shear viscosity, and the five second-order transport coefficients were labeled as $\tau_1$, $\tau_\epsilon$, $\xi_C$, $\xi_\sigma$, and $\xi_\omega$. As we mentioned before, these transport coefficients depend on the microscopic structure of the fluid, thus cannot be determined by hydrodynamics. Nonetheless, as we will shortly see, the bulk dual determines these transport coefficients uniquely, which could have been anticipated, since we already calculated $\eta$ employing linear response theory and the methods of AdS/CFT in previous sections. The fact of this naturally arising also in the fluid/gravity context only further confirms its relevance.

Now our task becomes to find a bulk solution to the full bulk EFE with arbitrarily large deviations from the stationary SAdS black hole in the long-wavelength regime. To construct such solutions, let us now suppose that the parameters $T$ and $u^\mu$ describing the black hole in Eq. \ref{SAdSzero} are spacetime-dependent fields which vary slowly --- that is, with respect to $\epsilon$ --- in $x^\mu$. In such a setup, at every point $x^\mu_P$, the geometry is approximately that of a black hole with constant temperature $T(x_P)$ and velocity $u^\mu(x_P)$. The spacetime region in the neighborhood of a fixed point $x^\mu_P$ but extended over all $r$ is referred to as a \emph{tube}, so that in the long-wavelength regime the bulk geometry \emph{tubewise approximates} a planar black hole with specific (constant) temperature and velocity. Thus, if we are able to patch the tubes together, we will then have constructed a non-uniform and time-evolving black hole! 

Now, naturally, if we simply replace $u_\mu$ and $T$ in Eq. \ref{SAdSzero} by arbitrarily-spacetime dependent fields $T(x^\mu)$, $u^\mu(x^\nu)$, the metric will no longer solve Einstein's equations, given the arbitrary dependence included. However, such a metric \emph{approaches} a solution in the limit of infinitely slow variations, so that we can use it as a starting point for an iterative solution. If we call this first metric $g^{(0)}_{ab}$, and realize ``slow variations'' as such that obey

\begin{equation}
\frac{\partial_\mu  \log T}{T} \sim \mathcal{O}(\epsilon) \ , \qquad \frac{\partial_\mu u}{T} \sim \mathcal{O}(\epsilon) 
\label{} \ ,
\end{equation}	

\noindent we can use $\epsilon$ as an expansion parameter and expand the metric and the fields $u_\mu(x^\nu)$ and $T(x^\mu)$ as

\begin{equation}
 g_{ab} = \sum_{k=0}^\infty \, \epsilon^k \, g^{(k)}_{ab} \ , \qquad
 T =\sum_{k=0}^\infty \, \epsilon^k \, T^{(k)}  \ , \qquad
u_{\mu} =\sum_{k=0}^\infty \, \epsilon^k \, u_{\mu}^{(k)} \ . 
\label{expand}
\end{equation}

Now, substituting these series into the EFE we can find a solution order by order in $\epsilon$. Thus, the term $g^{(k)}_{ab}$ will be such that Einstein's equations will be satisfied to ${\mathcal O}(\epsilon^k)$, as long as $T(x^\mu)$ and $u^\mu(x^\nu)$ obey a certain \emph{set of equations of motion}, which are \emph{precisely the stress tensor conservation equations} of boundary fluid dynamics at ${\mathcal O}(\epsilon^{k-1})$! With this we can therefore construct a metric up to arbitrary order in $\epsilon$. Details on the procedure can be found on \cite{minwalla}.

In fact, it was shown \cites{minwalla, hubeny_hard} that the dynamical set of equations splitting from the bulk EFE (namely $G_{\mu\nu} = 0$ and $G_{rr} =0$) are all of the form

\begin{equation}
{\mathbb H}\left[g^{(0)}(u^{(0)}_\mu, T^{(0)})\right] g^{(k)}= s_k  \ ,
\label{schemEeq}
\end{equation}

\noindent where ${\mathbb H}$ is a second-order, ultra-local in the boundary directions, linear differential operator in $r$ alone, and $s_k$ are regular source terms built from $g^{(n)}$ with $n \le k-1$. At a given point $x^{\mu}_P$, the precise form of ${\mathbb H}$ depends only on the \emph{local} values of $T$ and $u^\mu$ and \emph{not} on their derivatives at $x^{\mu}_P$. Further, the same homogeneous operator ${\mathbb H}$ appears at every  order in perturbation theory, which then allows us to find an explicit solution to Eq. \ref{schemEeq} systematically at any order. However, as reflect of the nonlinear nature of the theory, the source term $s_k$ gets more complicated with each order. It is important to notice that the solution is subjected to regularity in the interior and asymptotically AdS behavior as boundary conditions. 

Applying this procedure, which can be fully found in \cite{minwalla}, one obtain black hole spacetimes which actually correspond to not just a single solution, but to a \emph{continuously-infinite} set of approximate solutions, specified by $T(x^\mu)$ and $u^{\mu}(x^\nu)$. Notice, still, that the metric is not fully explicit, since we need to use a given solution to fluid dynamics relating $T(x^\mu)$ and $u^{\mu}(x^\nu)$ \emph{as input}. Still, given such a solution, the construction guarantees that the bulk geometry indeed describes a regular black hole.

To illustrate the procedure, here we present the bulk metric up to first order in $\epsilon$,

\begin{eqnarray}
ds^2 &=&-2\, u_{\mu}\, dx^{\mu} dr 
+ r^2\, \left( \eta_{\mu\nu} + \left[1-f\left(\frac{r}{\pi T}\right)\right] \, u_{\mu}\, u_{\nu} \right) \, dx^{\mu}dx^{\nu} \nonumber \\
&+& 2r \left[ \frac{r}{\pi T} \, F\left(\frac{r}{\pi T}\right)\, \sigma_{\mu\nu} +\frac{1}{3} \, u_{\mu}u_{\nu} \,\partial_{\lambda} u^{\lambda}  -  \frac{1}{2}\, u^{\lambda}\partial_{\lambda}\left(u_\nu u_{\mu}\right)\right] \, dx^{\mu} dx^{\nu} ,
\label{metfirstO}
\end{eqnarray}

\noindent where $f(r) = 1 - r_0/r$, $F(r)$ is  

\begin{equation}
F(r) \equiv \int_r^{\infty}\, dx \,\frac{x^2+x+1}{x (x+1) \left(x^2+1\right)} =\frac{1}{4}\, \left[\pi + \ln\left(\frac{(1+r)^2(1+r^2)}{r^4}\right) - 2\,\arctan(r)\right] \ ,
\end{equation}

\noindent and $\sigma^{\mu\nu}= P^{\mu \alpha} P^{\nu \beta} \, 
\, \partial_{(\alpha} u_{\beta)}
-\frac{1}{3} \, P^{\mu \nu} \, \partial_\alpha u^\alpha $ is the shear.

$T(x^\mu)$ and $u_\mu(x^\nu)$ are arbitrary \emph{slowly-varying} functions satisfying the conservation equation $\nabla_\mu T^{\mu\nu} = 0$ for the perfect fluid stress tensor of Eq. \ref{TzeroO}. As mentioned above, this bulk solution is tubewise approximated by a planar black hole, which means that in each tube (a small neighborhood of a point $x^{\mu}_P$ extended in the radial direction), the radial dependence of the metric is approximately that of a boosted planar black hole at some constant temperature $T$ and horizon velocity $u^\mu$, with corrections suppressed by the rate of variation $\epsilon$.  And, most importantly: these parameters vary from one point $x^{\mu}$ to another according to fluid dynamics. Physically, the solution in Eq. \ref{metfirstO} describes a dynamically evolving black hole with a planar and non-uniform event horizon. Also, dissipation will cause the black hole to approach a stationary SAdS$_5$ solution at late times.

On the boundary, the stress tensor up to first order can be easily obtained from  the bulk metric of Eq. \ref{metfirstO}, taking the form

\begin{equation}
T^{\mu \nu} =\pi^4 \, T^4
\left( 4\,  u^\mu u^\nu +\eta^{\mu \nu}
\right)  - 2 \,\pi^3 \, T^3 \, \sigma^{\mu \nu} \ .
\label{TfirstO}
\end{equation}

Like before, the first two terms describe a perfect fluid with pressure $P =\pi^4 \, T^4$, which, from thermodynamics, imply an entropy density $s=4 \, \pi^4 \, T^3$.  On the other hand, the shear viscosity $\eta$ is obtained from the coefficient of $\sigma^{\mu\nu}$, which yields $ \eta = \pi^3 \, T^3$. With these values, it is easy to see that this system saturates the KSS limit discussed above $\eta/s = 1/(4 \pi)$.

Finally, as argued in \cite{hubeny_izy}, one of the fortunate outcomes of the fluid/gravity correspondence is that it provides a new perspective on the black hole membrane paradigm, beyond the discussion of Sec. \ref{sec:brane_memb_ads_cft} since, within the fluid/gravity correspondence framework, the full spacetime evolution is mapped to the dynamics of a conformal fluid living \emph{on a membrane} located precisely \emph{on the boundary} of the spacetime, which is by itself unambiguously defined. This so-called \emph{membrane at the end of the universe} picture, can therefore be seen as a natural consequence of the holographic nature of the fluid/gravity correspondence, as well as a solution to long-standing questions.

With this, we finish this discussion of the fluid/gravity correspondence, as originally established. In the next chapter, we shall explore an alternative view on the correspondence, which emerged from a different context. There we will discuss in further details some of the ideas used in this section and explicitly present calculations analogous to the ones we briefly mentioned here.

\chapter{Another view on fluid/gravity}
\label{sec:further}

In the following sections, we shall explore an alternative view on the fluid/gravity correspondence, which is essentially different from what was discussed in Sec. \ref{sec:fgcorr}, as it will become clear. More specifically, we will realize the establishment of a holographic correspondence between the Rindler spacetime and an incompressible fluid living \emph{on the horizon}. 

First we will discuss in which sense it is possible to reduce the vacuum Einstein Field Equations (EFE) to the incompressible Navier--Stokes (NS) equation on a cutoff surface. As it will be shown, there exists a vacuum EFE solution in $(D+2)$ dimensions for every solution of the incompressible NS equation in $(D+1)$ dimensions, which establishes a holographic correspondence between the theories. Further, as it was shown in \cite{fg2}, the near-horizon expansion in gravity is mathematically identical to the hydrodynamic expansion in fluid dynamics, which then allows us to state in a mathematical precise way that \emph{horizons} are incompressible fluids. Our preliminary presentation will follow that of references \cites{fg2, fg3}, in which the chosen boundary conditions fix a flat induced metric on the cutoff surface.

Afterward, we will generalize the boundary conditions, following those introduced in references \cites{daniel1, daniel2, daniel3}. The major difference in these boundary conditions is the allowance of the induced metric on the cutoff surface to fluctuate, in a sense which will be detailed later on. In fact, these boundary conditions led to the discovery of soft hair excitations in the black hole horizons, which was proved to hold in arbitrary dimensions in \cite{daniel3}. Following the same procedure as in the preliminary presentation, but with these new boundary conditions, we arrived at a generalization of the incompressible NS equation describing the dual fluid at a \emph{soft-hairy horizon}. These are original results.

In what follows, we shall use lowercase Greek indices $\mu = \{0, ..., D+1 \}$, to label full (D+2)-dimensional spacetime coordinates, whilst lowercase Latin indices $i = \{1, ..., D\}$, will denote purely spatial D-dimensional coordinates, excluding the radial direction.

\section{The hydrodynamic limit of fluid dynamics}
\label{sec:hydro2}

Before we begin, it is important to complement the exposition of Sec. \ref{sec:hydro} and discuss a little further the hydrodynamic limit of fluid dynamics. In component notation, the incompressible NS equation is given by \cite{landau_fluids}

\begin{equation}
\begin{gathered}
\partial^i v_i = 0 \ ;
\\
\partial_t v_i - \eta \partial^2 v_i + \partial_i P + v^j \partial_j v_i = 0 \ ,
\end{gathered}
\end{equation}

\noindent where $i = 1, ..., D$ and $\eta$ now denotes the kinematic viscosity. The equation is solved by an incompressible fluid described by the pair $(v_i, P)$, where $v_i = v_i(t, x_i)$ is the velocity field; and $P = P(t, x^i)$ is the pressure field. Let us now consider a non-relativistic rescaling of the coordinates, as well as a rescaling of the amplitude of these fields under the parameter $\epsilon$, given by

\begin{equation}
\begin{gathered}
v_i^{(\epsilon)}(t, x^i) = \epsilon v_i(\epsilon^2 t, \epsilon x^i) \ , 
\\
P^{(\epsilon)}(t, x^i) = \epsilon^2 P(\epsilon^2 t, \epsilon x^i) \ .
\end{gathered}
\end{equation}

In fact, this transformation explicitly amounts to making the maps

\begin{equation}
\begin{gathered}
v_i \mapsto v_i^{(\epsilon)} = \epsilon v_i \ ; \ 
P \mapsto P^{(\epsilon)} = \epsilon^2 P \ ;
\\
\left\{\begin{matrix}
x^i \mapsto x^i = \epsilon x^i\\ 
\partial_i \mapsto \partial_i = \epsilon \partial_i
\end{matrix}\right. \ ; \ 
\left\{\begin{matrix}
t \mapsto t = \epsilon^2 t \\ 
\partial_t \mapsto \partial_t = \epsilon^2 \partial_t
\end{matrix}\right. \ ,
\label{eq:scale}
\end{gathered}
\end{equation}

\noindent under which the original incompressible NS equation is mapped to,

\begin{equation}
\begin{gathered}
\partial^i v_i = 0 \mapsto \partial^i v_i^{(\epsilon)} = (\epsilon \partial^i)(\epsilon v_i) = \epsilon^2 (\partial^i v_i) = 0 \ .
\\
\therefore \ \partial^i v_i^{(\epsilon)} = 0 \ ;
\end{gathered}
\end{equation}

\noindent and

\begin{equation}
\begin{gathered}
\partial_t v_i - \eta \partial^2 v_i + \partial_i P + v^j v_j v_i = 0 \mapsto \partial_t v_i^{(\epsilon)} - \eta \partial^2 v_i^{(\epsilon)} + \partial_i P^{(\epsilon)} + (v^j)^{\epsilon} v_j ^{\epsilon} v_i^{(\epsilon)}  
\\
= (\epsilon^2 \partial_t) (\epsilon v_i) - \eta (\epsilon^2 \partial^2) (\epsilon v_i) + (\epsilon \partial_i) (\epsilon^2 P) + (\epsilon v^j)(\epsilon v_j)(\epsilon v_i)  
\\
= \epsilon^3 (\partial_t v_i - \eta \partial^2 v_i + \partial_i P + v^j v_j v_i) = 0 \ .
\\
\therefore \ \partial_t v_i^{(\epsilon)} - \eta \partial^2 v_i^{(\epsilon)} + \partial_i P^{(\epsilon)} + (v^j)^{\epsilon} v_j ^{\epsilon} v_i^{(\epsilon)} = 0 \ .
\end{gathered}
\end{equation}

Therefore, the pair $(v_i^{(\epsilon)}, P^{(\epsilon)})$ is also a solution to the incompressible NS equation, which can be seen as a family of solutions built from the original solution pair $(v_i, P)$ under a parametrization by $\epsilon$. 

The hydrodynamic limit is achieved under the hydrodynamic $\epsilon$-scaling with the limit $\epsilon \rightarrow 0$, in which case the higher-derivative corrections to the NS equation which may appear for real fluids become irrelevant, since they scale in order $\mathcal{O}(\epsilon^2)$ or higher \cite{landau_fluids}. For this reason, the incompressible NS equation is adequate to describe any fluid in the hydrodynamic limit.

Finally, notice that the hydrodynamic scaling summarized in Eq. \ref{eq:scale} is such that

\begin{equation}
 v_i^{(\epsilon)} \sim \mathcal{O}(\epsilon) \ ; \ \   P^{(\epsilon)} \sim \mathcal{O}(\epsilon^2) \ ; \ \  \partial_i \sim \mathcal{O}(\epsilon) \ ; \ \  \partial_t \sim \mathcal{O}(\epsilon^2) \ .
 \label{eq:hydro_scale}
\end{equation}

These are important relations which will be used when we employ the hydrodynamic expansion to the solution of the EFE.

\section{The general setup}
\label{sec:setup}

From now on, we will consider geometries in $(D+2)$ dimensions, covered by the chart $\{t, r, x^i\}$, possessing an outer boundary, the cutoff co-dimension one hypersurface $\Sigma_c$, which is defined by

\begin{equation}
 \Sigma_c : r-r_c = 0 \ .
\end{equation}

We will denote by $x^a = (t, x^i)$ the coordinates on $\Sigma_c$. The unit normal vector on $\Sigma_c$ is defined according to

\begin{equation}
 n_{\mu} = \frac{\partial_{\mu}(r-r_c)}{\sqrt{g^{\nu \sigma}\partial_{\nu}(r-r_c)\partial_{\sigma}(r-r_c)}} \ ,
\end{equation}

\noindent where $x^{\mu} = (t, r, x^i)$ are the coordinates on the full $(D+2)$-dimensional spacetime of metric $g_{\mu \nu}$. Notice that

\begin{equation}
\partial_{\mu} (r - r_c) =
\left\{\begin{matrix}
1, \ \ \mu = r
\\
0, \ \ \mu = a
\end{matrix}\right. \ ,
\end{equation}

\noindent so that

\begin{equation}
g^{\nu \sigma}\partial_{\nu}(r-r_c)\partial_{\sigma}(r-r_c) = g^{r r} \ .
\end{equation}

Therefore,

\begin{equation}
n_{\mu}\mathrm{d}x^{\mu} = \frac{1}{\sqrt{g^{rr}}} \mathrm{d}r \ .
\label{eq:norm_vect_down}
\end{equation}

On the other hand, one has

\begin{equation}
n^{\mu} = g^{\mu \nu}n_{\nu} = \frac{g^{\mu r}}{\sqrt{g^{rr}}} \ , 
\end{equation}

\noindent so that

\begin{equation}
n^{\mu}\partial_{\mu} = \frac{1}{\sqrt{g^{rr}}} \left ( g^{rt} \partial_t + g^{rr}\partial_r + g^{ri} \partial_i \right ) \ .
\label{eq:norm_vect_up}
\end{equation}

The induced metric on $\Sigma_c$ is defined as,

\begin{equation}
 \gamma_{\mu \nu} = g_{\mu \nu} - n_{\mu}n_{\nu} \ ,
 \label{eq:induc_met}
\end{equation}

\noindent in terms of which it is possible to obtain the extrinsic curvature

\begin{equation}
 K_{\mu \nu} = \frac{1}{2} \mathcal{L}_n \gamma_{\mu \nu} = \frac{1}{2} \left ( n^{\sigma} \partial_{\sigma} \gamma_{\mu \nu} + \gamma_{\sigma \nu}\partial_{\mu}n^{\sigma} + \gamma_{\mu \sigma} \partial_{\nu} n^{\sigma} \right ) \ .
 \label{eq:ext_curv}
\end{equation}

Naturally, we will be ultimately interested in $\gamma_{a b}$ and $K_{a b}$, that is, the induced metric and the extrinsic curvature \emph{on $\Sigma_c$}, which are easily obtained by evaluating the tensors defined above at the cutoff, that is, by fixing $r = r_c$. Finally, we will also need to calculate the Brown--York stress tensor \cite{brown-york} on $\Sigma_c$, defined in terms of the extrinsic curvature as

\begin{equation}
 T_{a b}^{BY} = 2 \left ( \gamma_{a b} K - K_{a b} \right ) \ ,
 \label{eq:brown_york}
\end{equation}

\noindent where $K = \gamma^{a b} K_{a b}$, and we are using units in which $G = 1/16\pi$. 

In what follows, we will consider two different boundary conditions for the induced metric on $\Sigma_c$: first keeping it fixed, and then allowing it to fluctuate. Either choice, which corresponds to a boundary condition imposed to the $(D+2)$-dimensional geometry, will lead to interesting yet different results in the fluid-gravity correspondence context, as we will start to construct in the next section.

\section{A fixed induced metric}
\label{sec:fixed}

In this section, we will follow the construction presented in references \cites{fg2, fg3}, and present their main results, whilst introducing the conceptual and operational basis of what we shall do in Sec. \ref{sec:varying}. Namely, we consider as a boundary condition for the $(D+2)$-dimensional spacetime, a fixed induced metric on $\Sigma_c$, given by

\begin{equation}
\gamma_{a b} \mathrm{d}x^a \mathrm{d}x^b = -r_c \mathrm{d} t^2 + \delta_{ij}\mathrm{d}x^i \mathrm{d}x^j \ , 
\label{eq:bc1_down}
\end{equation}

\noindent where $\delta_{ij}$ is the $D$-dimensional boundary metric of $\Sigma_c$ -- which is just the $D$-dimensional Kronecker delta. Notice that this corresponds to \emph{fixing a flat induced metric on $\Sigma_c$}, since $\gamma_{ab} = \eta_{ab}$, if we identify $\sqrt{r_c}$ as the speed of light.

A $(D+2)$-dimensional geometry which is straightforwardly compatible with this boundary condition is the \emph{Rindler spacetime}

\begin{equation}
 g_{\mu \nu} \mathrm{d}x^\mu \mathrm{d}x^\nu = -r \mathrm{d}t^2 + 2 \mathrm{d}t \mathrm{d}r + \delta_{ij}\mathrm{d}x^i \mathrm{d}x^j \ , 
 \label{eq:rindler_st}
\end{equation}

\noindent where we are using ingoing Rindler coordinates $\{ r, t, x^i \}$. Notice that by setting $t = 2 \ln (X + T)$ and $4r = X^2 - T^2$, one gets $g_{\mu \nu} \mathrm{d}x^\mu \mathrm{d}x^\nu  = -\mathrm{d}T^2 + \mathrm{d}X^2 + \delta_{ij}\mathrm{d}x^i \mathrm{d}x^j$, so one has only flat spacetime in different coordinates. It is then easy to see that the boundary condition is satisfied at $r=r_c$. 

From the definition $g^{\mu \sigma}g_{\sigma \nu} = \delta^{\mu}_{\nu}$, it is easy to see that

\begin{equation}
 g^{\mu \nu} \partial_{\mu} \partial_{\nu} = 2 \partial_t \partial_r + r \partial_r^2 + \delta^{i j} \partial_i \partial_j \ . 
\end{equation}

Now, substituting the components of the inverse metric above in Eqs. \ref{eq:norm_vect_down} and \ref{eq:norm_vect_up}, respectively, yields

\begin{equation}
n_{\mu}\mathrm{d}x^{\mu} = \frac{1}{\sqrt{r}} \mathrm{d}r \ ; \ \ 
n^{\mu}\partial_{\mu} = \frac{1}{\sqrt{r}} \partial_t + \sqrt{r} \partial_r \ .
\end{equation}

From Eq. \ref{eq:induc_met}, then, one has

\begin{equation}
 \gamma_{\mu \nu} \mathrm{d}x^{\mu} \mathrm{d}x^{\nu} = -r \mathrm{d}t^2 - \frac{1}{r} \mathrm{d}r^2 + 2 \mathrm{d}t \mathrm{d}r + \delta_{ij}\mathrm{d}x^i \mathrm{d}x^j \ , 
\end{equation}

\noindent which clearly reduces to Eq. \ref{eq:bc1_down} at $r = r_c$. The same happens for the inverse induced metric $\gamma^{\mu \nu} = g^{\mu \nu} - n^{\mu}n^{\nu}$,

\begin{equation}
 \gamma^{\mu \nu} \partial_{\mu} \partial_{\nu} = 2 \partial_t \partial_r + r \partial_r^2 + \delta^{i j} \partial_i \partial_j - \left( r\partial_r^2 + \frac{1}{r} \partial_t^2 + 2 \partial_t \partial_r \right ) =  -\frac{1}{r} \partial_t^2 + \delta^{i j} \partial_i \partial_j \ ,
\end{equation}

\noindent which at $\Sigma_c$ becomes simply

\begin{equation}
\gamma^{ab}\partial_a \partial_b = -\frac{1}{r_c} \partial_t^2 + \delta^{ij}\partial_i\partial_j \ .
\label{eq:bc1_up}
\end{equation}

Now, we shall calculate the extrinsic curvature, according to Eq. \ref{eq:ext_curv}. The three terms (noticing that the induced metric components do not depend on $t$) are,

\begin{equation}
\frac{1}{2} n^{\sigma} \partial_{\sigma} \gamma_{\mu \nu} = \frac{1}{2} \left ( \frac{1}{\sqrt{r}} \cancelto{^0}{\partial_t \gamma_{\mu \nu}} + \sqrt{r} \partial_r \gamma_{\mu \nu} \right ) = \frac{\sqrt{r}}{2} \partial_r \gamma_{\mu \nu} \ ;
\end{equation}

\begin{equation}
\frac{1}{2} \gamma_{\sigma \nu}\partial_{\mu}n^{\sigma} = \frac{1}{2} \left ( \gamma_{t \nu} \partial_{\mu} \left ( \frac{1}{\sqrt{r}} \right ) + \gamma_{r \nu} \partial_\mu \left ( \sqrt{r} \right ) \right ) \ ; 
\end{equation}

\begin{equation}
\frac{1}{2} \gamma_{\mu \sigma}\partial_{\nu}n^{\sigma} = \frac{1}{2} \left ( \gamma_{\mu t} \partial_{\nu} \left ( \frac{1}{\sqrt{r}} \right ) + \gamma_{\mu r} \partial_\nu \left ( \sqrt{r} \right ) \right ) \ ,
\end{equation}

\noindent so that one has

\begin{equation} 
\begin{aligned}
K_{\mu \nu}\mathrm{d}x^\mu \mathrm{d}x^\nu &= \frac{\sqrt{r}}{2} \partial_r \gamma_{\mu \nu} \mathrm{d}x^\mu \mathrm{d}x^\nu + \frac{1}{2} \left ( \gamma_{t \nu} \partial_{r} \left ( \frac{1}{\sqrt{r}} \right ) + \gamma_{r \nu} \partial_r \left ( \sqrt{r} \right ) \right ) \mathrm{d}x^\nu \mathrm{d}r 
\\
&+ \frac{1}{2} \left ( \gamma_{\mu t} \partial_{r} \left ( \frac{1}{\sqrt{r}} \right ) + \gamma_{\mu r} \partial_r \left ( \sqrt{r} \right ) \right ) \mathrm{d}x^\mu \mathrm{d}r 
\\
&= \frac{\sqrt{r}}{2} \left ( \partial_r \left (-r \right) \mathrm{d}t^2 + \partial_r \left ( \frac{1}{r} \right ) \mathrm{d}r^2 \right ) + \frac{1}{2} \left (  \frac{-\gamma_{t \nu}}{2r^{3/2}} + \frac{\gamma_{r \nu}}{2\sqrt{r}} \right ) \mathrm{d}x^\nu \mathrm{d}r 
\\
&+ \frac{1}{2} \left (  \frac{-\gamma_{\mu t}}{2r^{3/2}} + \frac{\gamma_{\mu r}}{2\sqrt{r}} \right ) \mathrm{d}x^\mu \mathrm{d}r  
\\
\Rightarrow K_{\mu \nu}\mathrm{d}x^\mu \mathrm{d}x^\nu &= -\frac{\sqrt{r}}{2} \left ( \mathrm{d}t^2 + \frac{1}{r^2} \mathrm{d}r^2 \right ) + \frac{1}{4\sqrt{r}} \left [ \left (\gamma_{r \nu} - \frac{\gamma_{t\nu}}{r}\right ) \mathrm{d}x^\nu + \left (\gamma_{\mu r} - \frac{\gamma_{\mu t}}{r}\right ) \mathrm{d}x^\mu  \right ] \mathrm{d}r \ .
\\
\end{aligned}
\end{equation}

Now, remind that we are interested in the extrinsic curvature on $\Sigma_c$, that is, $K_{ab}\mathrm{d}x^a\mathrm{d}x^b$, in which the terms $K_{\mu r}\mathrm{d}x^\mu\mathrm{d}r$ do not appear (since $r = r_c = \text{cte} \Rightarrow \mathrm{d}r = 0$). Therefore, one has, simply,

\begin{equation}
 K_{ab}\mathrm{d}x^a\mathrm{d}x^b = -\frac{\sqrt{r_c}}{2} \mathrm{d}t^2 \ . 
\end{equation}

Using Eq. \ref{eq:bc1_up}, one obtains the trace of the extrinsic curvature,

\begin{equation}
\begin{gathered}
 K = \gamma^{ab}K_{ab} = \gamma^{tt} K_{tt} = \left ( -\frac{1}{r_c} \right ) \left (  -\frac{\sqrt{r_c}}{2} \right ) 
 \\
 \Rightarrow K = \frac{1}{2\sqrt{r_c}} \ .
\end{gathered}
\end{equation}

Now, from Eq. \ref{eq:brown_york}, we can easily calculate the Brown--York stress tensor,

\begin{equation}
\begin{aligned}
T_{a b}^{BY}\mathrm{d}x^a \mathrm{d}x^b &= 2 \left ( \gamma_{a b} K - K_{a b} \right ) \mathrm{d}x^a \mathrm{d}x^b 
\\
&= \left ( \frac{1}{\sqrt{r_c}} \gamma_{ab} - 2 K_{ab} \right )\mathrm{d}x^a \mathrm{d}x^b  
\\
&= \frac{1}{\sqrt{r_c}} \left ( -r_c \mathrm{d}t^2 + \delta_{ij} \mathrm{d}x^i \mathrm{d}x^j \right ) - 2 \left ( -\frac{\sqrt{r_c}}{2} \mathrm{d}t^2 \right )  
\\
\Rightarrow T_{a b}^{BY} \mathrm{d}x^a \mathrm{d}x^b &= \frac{1}{\sqrt{r_c}}  \delta_{ij} \mathrm{d}x^i \mathrm{d}x^j \ .
\end{aligned}
\end{equation}

On the other hand, the stress tensor of a perfect fluid is given by

\begin{equation}
 T_{ab}^{PF} = \rho u_a u_b + p \left (\gamma_{ab} + u_a u_b \right ) \ , 
 \label{eq:perfect_fluid}
\end{equation}

\noindent where $\rho$ and $p$ are respectively the energy density and the pressure of the fluid in the local rest frame; and $u_a$ is the normalized fluid velocity. Notice that with the choice $\rho = 0; \  p = r_c^{-1/2}; \  u_t = r_c^{1/2}; \  u_i = 0$, one has

\begin{equation}
\begin{gathered}
T_{a b}^{PF} \mathrm{d}x^a \mathrm{d}x^b = \left ( \frac{1}{\sqrt{r_c}} \right ) \left ( \left ( -r_c \mathrm{d}t^2 + \delta_{ij} \mathrm{d}x^i \mathrm{d}x^j \right ) + \left ( r_c \mathrm{d}t^2 \right ) \right ) 
\\
\Rightarrow T_{a b}^{PF} \mathrm{d}x^a \mathrm{d}x^b = \frac{1}{\sqrt{r_c}}  \delta_{ij} \mathrm{d}x^i \mathrm{d}x^j \ .
\end{gathered}
\end{equation}

Therefore, 

\begin{equation}
T_{a b}^{BY} = T_{ab}^{PF} \ , \ \ \text{with:} \ \ \rho = 0; \  p = \frac{1}{\sqrt{r_c}}; \  u_t = \sqrt{r_c}; \  u_i = 0 \ ,
\label{eq:by=pf_1}
\end{equation}

\noindent that is, the Brown--York stress tensor has the form of a perfect fluid on $\Sigma_c$, with the fluid properties as specified.

\subsection{Boosting the metric}
\label{sec:boost1}

Now we shall perform two diffeomorphisms on Rindler spacetime of Eq. \ref{eq:rindler_st}. The first is a constant shift of the radial coordinate (which replaces the Rindler horizon from $r=0$ to $r=r_h$), as well as a constant scaling of the temporal coordinate,

\begin{equation}
\begin{gathered}
\left\{\begin{matrix}
r \mapsto r - r_h
\\
t \mapsto \alpha t
\\
\end{matrix}\right. \ ; \ \ 
\alpha = \left (1 - \dfrac{r_h}{r_c} \right )^{-1/2} \ .
\end{gathered}
\end{equation}

The Rindler spacetime metric is accordingly transformed as

\begin{equation}
g_{\mu \nu} \mathrm{d}x^\mu \mathrm{d}x^\nu  \mapsto g_{\mu \nu} \mathrm{d}x^\mu \mathrm{d}x^\nu  = -\alpha^2 (r-r_h) \mathrm{d}t^2 + 2\alpha \mathrm{d}t\mathrm{d}r + \delta_{ij}\mathrm{d}x^i \mathrm{d}x^j \ .
\end{equation}

The second diffeomorphism is a constant boost, which accounts to the coordinate transformation $x^\mu = (t, r, x^i) \mapsto x^{\tilde{\mu}} = (\tilde{t}, \tilde{r}, x^{\tilde{i}})$, according to

\begin{equation}
\left\{\begin{matrix}
t \mapsto \tilde{t} = \gamma \left (t - \dfrac{v_i x^i}{r_c} \right )
\\
x^i \mapsto x^{\tilde{i}} = x^i - \gamma t v_i + \left ( \gamma - 1 \right ) \dfrac{v_i v_j}{v^2} x^j 
\\
r \mapsto \tilde{r} = r
\end{matrix}\right. \ ; \ \ 
\gamma = \left (1 - \dfrac{v^2}{r_c} \right )^{-1/2} \ .
\end{equation}

Remind (Sec. \ref{sec:tensors_transf}) that the metric components will change under the transformation $x^\mu \mapsto x^{\tilde{\mu}}$ according to

\begin{equation}
\begin{gathered}
g_{\tilde{\mu}\tilde{\nu}} = \frac{\partial x^\mu}{\partial x^{\tilde{\mu}}} \frac{\partial x^\nu}{\partial x^{\tilde{\nu}}} g_{\mu \nu} 
\\
\Rightarrow g_{\tilde{\mu}\tilde{\nu}} = \frac{\partial t}{\partial x^{\tilde{\mu}}} \frac{\partial t}{\partial x^{\tilde{\nu}}} g_{tt} + \left ( \frac{\partial t}{\partial x^{\tilde{\mu}}} \frac{\partial r}{\partial x^{\tilde{\nu}}} +  \frac{\partial r}{\partial x^{\tilde{\mu}}} \frac{\partial t}{\partial x^{\tilde{\nu}}} \right ) g_{tr} + \frac{\partial x^i}{\partial x^{\tilde{\mu}}} \frac{\partial x^j}{\partial x^{\tilde{\nu}}} g_{ij} 
\\
\Rightarrow g_{\tilde{\mu}\tilde{\nu}} = -\alpha^2(r-r_h)\frac{\partial t}{\partial x^{\tilde{\mu}}} \frac{\partial t}{\partial x^{\tilde{\nu}}} + \alpha \left ( \frac{\partial t}{\partial x^{\tilde{\mu}}} \frac{\partial r}{\partial x^{\tilde{\nu}}} +  \frac{\partial r}{\partial x^{\tilde{\mu}}} \frac{\partial t}{\partial x^{\tilde{\nu}}} \right ) + \delta_{ij} \frac{\partial x^i}{\partial x^{\tilde{\mu}}} \frac{\partial x^j}{\partial x^{\tilde{\nu}}}  \ ,
\label{eq:transfs}
\end{gathered}
\end{equation}

\noindent since only the components $g_{tt}$, $g_{tr} = g_{rt}$ and $g_{ij}$ are non-zero. To calculate the transformation above, it is necessary to consider the inverse transformations, which in this case are only the inverse boosts, achieved by simply changing the sign of the velocity components,

\begin{equation}
\left\{\begin{matrix}
t = \gamma \left (\tilde{t} + \dfrac{v_i x^i}{r_c} \right )
\\
x^i = x^{\tilde{i}} + \gamma \tilde{t} v_i + \left ( \gamma - 1 \right ) \dfrac{v_i v_j}{v^2} x^{\tilde{j}} 
\\
r = \tilde{r}
\end{matrix}\right. \ ; \ \ 
\end{equation}

So that one gets

\begin{equation}
\begin{gathered}
\frac{\partial t}{\partial \tilde{t}} = \gamma \ ; \ \frac{\partial t}{\partial \tilde{r}} = 0 \ ; \  \frac{\partial t}{\partial x^{\tilde{i}}} = \frac{\gamma}{r_c}v_{\tilde{i}} \ ;
\\
\frac{\partial x^i}{\partial \tilde{t}} = \gamma v^i \ ; \ \frac{\partial x^i}{\partial \tilde{r}} = 0 \ ; \  \frac{\partial x^i}{\partial x^{\tilde{j}}} = \delta^i_{\tilde{j}} + \left ( \gamma - 1 \right ) \frac{v^i v_{\tilde{j}}}{v^2} \ ;
\\
\frac{\partial r}{\partial \tilde{t}} = 0 \ ; \ \frac{\partial r}{\partial \tilde{r}} = 1 \ ; \ \frac{\partial r}{\partial x^{\tilde{i}}} = 0 \ .
\end{gathered}
\end{equation}

Therefore, each component of the transformed metric is

\begin{equation}
\begin{aligned}
g_{\tilde{t}\tilde{t}} &= -\alpha^2(r-r_h) \frac{\partial t}{\partial \tilde{t}} \frac{\partial t}{\partial \tilde{t}} + \alpha \left ( \frac{\partial t}{\partial \tilde{t}} \cancelto{^0}{\frac{\partial r}{\partial \tilde{t}}} +  \cancelto{^0}{\frac{\partial r}{\partial \tilde{t}}} \frac{\partial t}{\partial \tilde{t}} \right ) + \delta_{ij} \frac{\partial x^i}{\partial \tilde{t}} \frac{\partial x^j}{\partial \tilde{t}} 
\\
&= -\alpha^2(r-r_h) \left (\gamma \right )^2 + \delta_{ij} \left (\gamma v^i \right ) \left ( \gamma v^j \right ) 
\\
&= -\gamma^2\alpha^2(r-r_h) + \gamma^2v^2 
\\
\Rightarrow g_{\tilde{t}\tilde{t}} &= \gamma^2 \left ( v^2 - \alpha^2 \left ( r - r_h \right ) \right ) \ ;
\end{aligned}
\end{equation}

\begin{equation}
\begin{aligned}
g_{\tilde{t}\tilde{r}} &= -\alpha^2(r-r_h) \frac{\partial t}{\partial \tilde{t}} \cancelto{^0}{\frac{\partial t}{\partial \tilde{r}}} + \alpha \left ( \frac{\partial t}{\partial \tilde{t}} \frac{\partial r}{\partial \tilde{r}} +  \cancelto{^0}{\frac{\partial r}{\partial \tilde{t}}} \cancelto{^0}{\frac{\partial t}{\partial \tilde{r}}} \ \right ) + \delta_{ij} \frac{\partial x^i}{\partial \tilde{t}} \cancelto{^0}{\frac{\partial x^j}{\partial \tilde{r}}} 
\\
&= \alpha \left ( \gamma \right ) \left ( 1 \right ) 
\\
\Rightarrow g_{\tilde{t}\tilde{r}} &= \gamma \alpha  \ ;
\end{aligned}
\end{equation}

\begin{equation}
\begin{aligned}
g_{\tilde{t}\tilde{i}} &= -\alpha^2(r-r_h) \frac{\partial t}{\partial \tilde{t}} \frac{\partial t}{\partial x^{\tilde{i}}} + \alpha \left ( \frac{\partial t}{\partial \tilde{t}} \cancelto{^0}{\frac{\partial r}{\partial x^{\tilde{i}}}} + \cancelto{^0}{\frac{\partial r}{\partial \tilde{t}}} \frac{\partial t}{\partial x^{\tilde{i}}} \right ) + \delta_{kj} \frac{\partial x^k}{\partial \tilde{t}} \frac{\partial x^j}{\partial x^{\tilde{i}}} 
\\
&= -\alpha^2(r-r_h) \left ( \frac{\gamma}{r_c}v_{\tilde{i}} \right )\left (\gamma \right ) + \delta_{kj} \left (\gamma v^k \right ) \left ( \delta^j_{\tilde{i}} + \left ( \gamma - 1 \right ) \frac{v^j v_{\tilde{i}}}{v^2} \right ) 
\\
&= -\gamma^2\alpha^2(r-r_h) \frac{v_{\tilde{i}}}{r_c} + \gamma v_j \delta^j_{\tilde{i}} + \gamma \left ( \gamma - 1 \right ) \frac{v_j v^j}{v^2} v_{\tilde{i}}  
\\
&= -\gamma^2\alpha^2(r-r_h) \frac{v_{\tilde{i}}}{r_c} + \gamma v_{\tilde{i}} + \gamma^2 v_{\tilde{i}} -\gamma v_{\tilde{i}}  
\\
&= \gamma^2 \left (1 - \frac{\alpha^2}{r_c} \left (r - r_h \right ) \right ) v_{\tilde{i}} = \gamma^2 \left (1 - \frac{r - r_h}{r_c-r_h}  \right ) v_{\tilde{i}}  
\\
&= \gamma^2 \left (\frac{r_c - r}{r_c-r_h}  \right ) v_{\tilde{i}} =  \gamma^2 \left (\frac{r_c - r}{r_c \left (1-\frac{r_c}{r_h} \right )}  \right ) v_{\tilde{i}} 
\\
\Rightarrow g_{\tilde{t}\tilde{i}} &= \frac{\gamma^2 \alpha^2}{r_c} \left ( r_c - r \right ) v_{\tilde{i}} \ ;
\end{aligned}
\end{equation}

\begin{equation}
\begin{aligned}
g_{\tilde{r}\tilde{r}} &= -\alpha^2(r-r_h) \cancelto{^0}{\frac{\partial t}{\partial \tilde{r}}} \cancelto{^0}{\frac{\partial t}{\partial \tilde{r}}} + \alpha \left ( \cancelto{^0}{\frac{\partial t}{\partial \tilde{r}}} \frac{\partial r}{\partial \tilde{r}} + \frac{\partial r}{\partial \tilde{r}} \cancelto{^0}{\frac{\partial t}{\partial \tilde{r}}} \  \right ) + \delta_{ij} \cancelto{^0}{\frac{\partial x^i}{\partial \tilde{r}}} \cancelto{^0}{\frac{\partial x^j}{\partial \tilde{r}}} 
\\
\Rightarrow g_{\tilde{r}\tilde{r}} &=  0 \ ;
\end{aligned}
\end{equation}

\begin{equation}
\begin{aligned}
g_{\tilde{r}\tilde{i}} &= -\alpha^2(r-r_h) \cancelto{^0}{\frac{\partial t}{\partial \tilde{r}}} \frac{\partial t}{\partial x^{\tilde{i}}} + \alpha \left ( \cancelto{^0}{\frac{\partial t}{\partial \tilde{r}}} \cancelto{^0}{\frac{\partial r}{\partial x^{\tilde{i}}}} +  \frac{\partial r}{\partial \tilde{r}} \frac{\partial t}{\partial x^{\tilde{i}}} \ \right ) + \delta_{kj} \cancelto{^0}{\frac{\partial x^k}{\partial \tilde{r}}} \frac{\partial x^j}{\partial x^{\tilde{i}}}  
\\
&= \alpha \left ( 1 \right ) \left ( \frac{\gamma}{r_c} v_{\tilde{i}}\right ) \Rightarrow
\\ 
\Rightarrow g_{\tilde{r}\tilde{i}} &=  \frac{\gamma \alpha}{r_c}v_{\tilde{i}} \ ;
\end{aligned}
\end{equation}

\begin{equation}
\begin{aligned}
g_{\tilde{i}\tilde{j}} &= -\alpha^2(r-r_h) \frac{\partial t}{\partial x^{\tilde{i}}} \frac{\partial t}{\partial x^{\tilde{j}}} + \alpha \left ( \frac{\partial t}{\partial x^{\tilde{i}}} \cancelto{^0}{\frac{\partial r}{\partial x^{\tilde{j}}}} +  \cancelto{^0}{\frac{\partial r}{\partial x^{\tilde{i}}}} \frac{\partial t}{\partial x^{\tilde{j}}} \right ) + \delta_{kl} \frac{\partial x^k}{\partial x^{\tilde{i}}} \frac{\partial x^l}{\partial x^{\tilde{j}}} 
\\
&= -\alpha^2(r-r_h) \left ( \frac{\gamma}{r_c} v_{\tilde{i}} \right )
\left ( \frac{\gamma}{r_c} v_{\tilde{j}} \right ) + \delta_{kl} \left ( \delta^k_{\tilde{i}} + \left ( \gamma - 1 \right ) \frac{v^k v_{\tilde{i}}}{v^2} \right ) \left ( \delta^l_{\tilde{j}} + \left ( \gamma - 1 \right ) \frac{v^l v_{\tilde{j}}}{v^2} \right )  
\\
&= - \frac{\gamma^2 \alpha^2}{r_c^2} \left ( r - r_h \right ) v_{\tilde{i}} v_{\tilde{j}} + \delta_{kl} \left ( \delta^k_{\tilde{i}} \delta^l_{\tilde{j}} + \left ( \gamma -1 \right )\delta^k_{\tilde{i}} \frac{v_{\tilde{j}} v^l}{v^2} + \left ( \gamma -1 \right )\delta^l_{\tilde{j}} \frac{v_{\tilde{i}} v^k}{v^2} + \left ( \gamma -1 \right )^2 \frac{v_{\tilde{i}}v_{\tilde{j}}}{v^2} \frac{v^k v^l}{v^2} \right ) 
\\
&= - \frac{\gamma^2 \alpha^2}{r_c^2} \left ( r - r_h \right ) v_{\tilde{i}} v_{\tilde{j}} + \delta_{\tilde{i}\tilde{j}} + \left ( \gamma -1 \right ) \frac{v_{\tilde{j}} \delta_{\tilde{i}l} v^l}{v^2} + \left ( \gamma -1 \right ) \frac{v_{\tilde{i}} \delta_{\tilde{j}k} v^k}{v^2} + \left ( \gamma -1 \right )^2 \frac{v_{\tilde{i}}v_{\tilde{j}}}{v^2} \frac{v_l v^l}{v^2} 
\\
&= - \frac{\gamma^2 \alpha^2}{r_c^2} \left ( r - r_h \right ) v_{\tilde{i}} v_{\tilde{j}} + \delta_{\tilde{i}\tilde{j}} + \left ( \gamma -1 \right ) \frac{v_{\tilde{j}} v_{\tilde{i}}}{v^2} + \left ( \gamma -1 \right ) \frac{v_{\tilde{i}} v_{\tilde{j}}}{v^2} + \left ( \gamma -1 \right )^2 \frac{v_{\tilde{i}}v_{\tilde{j}}}{v^2} 
\\
&= - \frac{\gamma^2 \alpha^2}{r_c^2} \left ( r - r_h \right ) v_{\tilde{i}} v_{\tilde{j}} + \delta_{\tilde{i}\tilde{j}} + \left [ 2\left(\gamma-1\right) + \left(\gamma -1\right)^2 \right ] \frac{v_{\tilde{i}}v_{\tilde{j}}}{v^2} 
\\
&= - \frac{\gamma^2 \alpha^2}{r_c^2} \left ( r - r_h \right ) v_{\tilde{i}} v_{\tilde{j}} + \delta_{\tilde{i}\tilde{j}} + \left ( 2\gamma-2 + \gamma^2 -2\gamma + 1 \right ) \frac{v_{\tilde{i}}v_{\tilde{j}}}{v^2} 
\\
&= - \frac{\gamma^2 \alpha^2}{r_c^2} \left ( r - r_h \right ) v_{\tilde{i}} v_{\tilde{j}} + \delta_{\tilde{i}\tilde{j}} + \left ( \gamma^2 - 1 \right ) \frac{v_{\tilde{i}}v_{\tilde{j}}}{v^2} 
\\
&= \delta_{\tilde{i}\tilde{j}} - \frac{\gamma^2 \alpha^2}{r_c^2} \left ( r - r_h - \frac{\gamma^2 -1}{v^2}\frac{r_c^2}{\gamma^2 \alpha^2} \right ) v_{\tilde{i}} v_{\tilde{j}} 
\\
&= \delta_{\tilde{i}\tilde{j}} - \frac{\gamma^2 \alpha^2}{r_c^2} \left ( r - r_h - \frac{\frac{\gamma^2 v^2}{r_c}}{v^2}\frac{r_c^2}{\gamma^2 \alpha^2} \right ) v_{\tilde{i}} v_{\tilde{j}} 
\\
&= \delta_{\tilde{i}\tilde{j}} - \frac{\gamma^2 \alpha^2}{r_c^2} \left ( r - r_h - \frac{r_c}{\alpha^2} \right ) v_{\tilde{i}} v_{\tilde{j}} = \delta_{\tilde{i}\tilde{j}} - \frac{\gamma^2 \alpha^2}{r_c^2} \left ( r - r_h - \left(r_c-r_h\right) \right ) v_{\tilde{i}} v_{\tilde{j}} 
\\
\Rightarrow g_{\tilde{i}\tilde{j}} &= \delta_{\tilde{i}\tilde{j}} - \frac{\gamma^2 \alpha^2}{r_c^2}  \left (r -r_c \right ) v_{\tilde{i}} v_{\tilde{j}}  \ ,
\end{aligned}
\end{equation}

\noindent where we used, at the 10th line, the useful identity,

\begin{equation}
\begin{gathered}
\gamma^2 - 1  = \frac{1}{1 - \frac{v^2}{r_c}} - 1 = \frac{1 - 1 + \frac{v^2}{r_c}}{1 - \frac{v^2}{r_c}} = \frac{\frac{v^2}{r_c}}{1 - \frac{v^2}{r_c}} = \gamma^2 \frac{v^2}{r_c} 
\\
\Rightarrow \gamma^2 - 1 = \frac{\gamma^2 v^2}{r_c} \ .
\label{eq:id_gamma2-1}
\end{gathered}
\end{equation}

To keep the same sign convention in the components $g_{\tilde{r}\tilde{i}}$ and $g_{\tilde{t}\tilde{i}}$ as that of references \cites{fg2, fg3}, we will also consider the transformation $x^{\tilde{i}} \mapsto -x^{\tilde{i}}$, which simply changes the sign of these two components. After dropping the tildes for simplicity, through the coordinates relabel $(\tilde{t}, \tilde{r}, x^{\tilde{i}}) \mapsto (t, r, x^i) = (\tilde{t}, \tilde{r}, x^{\tilde{i}})$, one has the metric components transformed under the diffeomorphisms,

\begin{equation}
\begin{gathered}
g_{tt} = \gamma^2 \left ( v^2 - \alpha^2 \left ( r - r_h \right ) \right ) \ ; \ 
g_{tr} = \gamma \alpha  \ ; \ 
g_{ti} = \frac{\gamma^2 \alpha^2}{r_c} \left ( r - r_c \right ) v_i \ ; \ 
\\
g_{rr} = 0; \ 
g_{ri} = -\frac{\gamma \alpha}{r_c}v_i \ ; \ 
g_{ij} = \delta_{ij} - \frac{\gamma^2 \alpha^2}{r_c^2}  \left (r -r_c \right ) v_i v_j \ ,
\label{eq:boosted_metric_comps}
\end{gathered}
\end{equation}

\noindent or, more explicitly, the transformed metric (from now on referred to simply as ``the boosted metric'', although we remind that the boost was not the only transformation performed) is,

\begin{equation}
\begin{aligned}
g_{\mu \nu} \mathrm{d}x^\mu \mathrm{d}x^\nu &= \gamma^2 \left ( v^2 - \alpha^2 \left ( r - r_h \right ) \right ) \mathrm{d}t^2 + 2\gamma \alpha \mathrm{d}t\mathrm{d}r + 2\frac{\gamma^2 \alpha^2}{r_c} \left ( r - r_c \right ) v_i \mathrm{d}x^i \mathrm{d}t \ +  
\\
&-2\frac{\gamma \alpha}{r_c}v_i \mathrm{d}x^i \mathrm{d}r + \left ( \delta_{ij} - \frac{\gamma^2 \alpha^2}{r_c^2}  \left (r -r_c \right ) v_i v_j \right ) \mathrm{d}x^i \mathrm{d}x^j \ .
\label{eq:boosted_metric}
\end{aligned}
\end{equation}

Notice that the boosted metric of Eq. \ref{eq:boosted_metric} still describes the same Rindler spacetime of Eq. \ref{eq:rindler_st}, but described in a complicated, shifted, rescaled and boosted coordinate system. Nonetheless, it is, namely, still an \emph{exact} solution to the vacuum EFE. 

Our goal now is to calculate the Brown--York stress tensor for the boosted metric, as we did for the metric in simple ingoing Rindler coordinates. Naturally, we do not expect many expressive changes, but this result will be important for what comes next.

The first step is to calculate the inverse metric components $g^{\mu \nu}$, which are necessary, for example, to normalize the normal vector on $\Sigma_c$. To do so, we will use the definition $g^{\mu \sigma}g_{\sigma \nu} = \delta^{\mu}_{\nu}$, as usual. Notice, however, that the boosted metric of Eq. \ref{eq:boosted_metric} has only one single null component ($g_{rr}$), so that the computation of its inverse will not be so simple. For this reason, we left the explicit calculation to Appendix \ref{ap:inverse}, and will present here only the final result,

\begin{equation}
\begin{gathered}
g^{tt} = \frac{\gamma^2 v^2}{r_c^2} \ ; \ 
g^{tr} = \frac{\gamma}{\alpha}  \ ; \ 
g^{ti} = \frac{\gamma^2}{r_c} v^i \ ; \ 
\\
g^{rr} = r-r_h; \ 
g^{ri} = \frac{\gamma}{\alpha}v^i \ ; \ 
g^{ij} = \delta^{ij} + \frac{\gamma^2}{r_c} v^i v^j \ ,
\label{eq:boosted_metric_inv_comps}
\end{gathered}
\end{equation}

\noindent that is, 

\begin{equation}
\begin{aligned}
g^{\mu \nu} \partial_\mu \partial_\nu &= \frac{\gamma^2 v^2}{r_c^2} \partial_t^2 + 2\frac{\gamma}{\alpha}\partial_t\partial_r + 2\frac{\gamma^2}{r_c} v^i \partial_i \partial_t + \left (r - r_h \right ) \partial_r^2 \  +
\\
&+2\frac{\gamma}{\alpha}v^i \partial_i \partial_r + \left ( \delta^{ij} + \frac{\gamma^2}{r_c} v^i v^j \right ) \partial_i \partial_j \ .
\label{eq:boosted_metric_inv}
\end{aligned}
\end{equation}

Now we can proceed to the calculation of the Brown--York stress tensor, following the procedure adopted in the last section. Substituting the components of the inverse metric above in Eqs. \ref{eq:norm_vect_down} and \ref{eq:norm_vect_up}, respectively, yields

\begin{equation}
n_{\mu}\mathrm{d}x^{\mu} = \frac{1}{\sqrt{r - r_h}} \mathrm{d}r \ ; \ \ 
n^{\mu}\partial_{\mu} = \frac{1}{\sqrt{r-r_h}} \left ( \frac{\gamma}{\alpha} \partial_t + \left (r - r_h \right ) \partial_r + \frac{\gamma}{\alpha}v^i \partial_i \right ) \ .
\end{equation}

From Eq. \ref{eq:induc_met}, then, one has

\begin{equation}
\begin{aligned}
\gamma_{\mu \nu} \mathrm{d}x^{\mu} \mathrm{d}x^{\nu} &= \gamma^2 \left ( v^2 - \alpha^2 \left ( r - r_h \right ) \right ) \mathrm{d}t^2 + 2\gamma \alpha \mathrm{d}t\mathrm{d}r + 2\frac{\gamma^2 \alpha^2}{r_c} \left ( r - r_c \right ) v_i \mathrm{d}x^i \mathrm{d}t   
\\
&-\frac{1}{r-r_h} \mathrm{d}r^2 -2\frac{\gamma \alpha}{r_c}v_i \mathrm{d}x^i \mathrm{d}r + \left ( \delta_{ij} - \frac{\gamma^2 \alpha^2}{r_c^2}  \left (r -r_c \right ) v_i v_j \right ) \mathrm{d}x^i \mathrm{d}x^j \ , 
\end{aligned}
\end{equation}

\noindent whilst for the inverse induced metric $\gamma^{\mu \nu} = g^{\mu \nu} - n^{\mu}n^{\nu}$, one has

\begin{equation}
\begin{aligned}
\gamma^{\mu \nu} \partial_{\mu} \partial_{\nu} &= \frac{\gamma^2 v^2}{r_c^2} \partial_t^2 + 2\frac{\gamma}{\alpha}\partial_t\partial_r + 2\frac{\gamma^2}{r_c} v^i \partial_i \partial_t + \left (r - r_h \right ) \partial_r^2 \ 
\\
&+ 2\frac{\gamma}{\alpha}v^i \partial_i \partial_r + \left ( \delta^{ij} + \frac{\gamma^2}{r_c} v^i v^j \right ) \partial_i \partial_j \  
\\
&- \frac{1}{r-r_h}\left( \frac{\gamma^2}{\alpha^2} \partial_t^2 + \left (r-r_h \right )^2 \partial_r^2 + \frac{\gamma^2}{\alpha^2} v^i v^j \partial_i \partial_j  \right ) \ 
\\
&- \frac{1}{r- r_h} \left ( 2 \frac{\gamma}{\alpha} \left (r - r_h \right ) \partial_t \partial_r + 2 \frac{\gamma^2}{\alpha^2} v^i \partial_i \partial_t + 2 \frac{\gamma}{\alpha} \left ( r - r_h \right ) v^i \partial_i\partial_r \right ) 
\\
\Rightarrow \gamma^{\mu \nu} \partial_{\mu} \partial_{\nu} &= \gamma^2 \left ( \frac{v^2}{r_c^2} - \frac{1}{\alpha^2 \left (r-r_h \right )} \right ) \partial_t^2 + 2\gamma^2 \left ( \frac{1}{r_c} - \frac{1}{\alpha^2 \left (r - r_h \right )} \right ) v^i \partial_i \partial_t \  
\\
&+ \left ( \delta^{ij} + \gamma^2 \left ( \frac{1}{r_c} - \frac{1}{\alpha^2 \left (r - r_h \right )} \right ) v^i v^j \right ) \partial_i \partial_j \ .
\label{eq:inducedup}
\end{aligned}
\end{equation}

Notice that on $\Sigma_c$ one has $\alpha^2 \left (r - r_h \right ) = \alpha^2 \left (r_c - r_h \right )= r_c$, thus both the induced metric and its inverse reduce to Eqs. \ref{eq:bc1_down} and \ref{eq:bc1_up}, as it should be.

The first term of the extrinsic curvature, Eq. \ref{eq:ext_curv}, is

\begin{equation}
\frac{1}{2} n^{\sigma} \partial_{\sigma} \gamma_{\mu \nu} = \frac{1}{2\sqrt{r-r_h}} \left ( \frac{\gamma}{\alpha} \cancelto{^0}{\partial_t \gamma_{\mu \nu}} + \left (r - r_h \right ) \partial_r \gamma_{\mu \nu} + \frac{\gamma}{\alpha} v^i \cancelto{^ 0}{\partial_i \gamma_{\mu \nu}} \ \ \ \right ) = \frac{\sqrt{r-r_h}}{2} \partial_r \gamma_{\mu \nu} \ ,
\end{equation}

\noindent where we used the fact that the $\gamma_{\mu \nu}$ components only have radial dependence. Thus, 

\begin{equation}
\begin{gathered}
\frac{1}{2} n^{\sigma} \partial_{\sigma} \gamma_{\mu \nu} \mathrm{d}x^\mu \mathrm{d}x^\nu = \frac{\sqrt{r-r_h}}{2} \partial_r \gamma_{\mu \nu}  
\\
= \frac{\sqrt{r-r_h}}{2} \left [ \partial_r \left ( \gamma^2 \left ( v^2 - \alpha^2 \left ( r - r_h \right ) \right ) \right ) \mathrm{d}t^2 + \partial_r \left ( 2\frac{\gamma^2 \alpha^2}{r_c} \left ( r - r_c \right ) v_i\right ) \mathrm{d}x^i \mathrm{d}t \right ] 
\\
+ \frac{\sqrt{r-r_h}}{2} \left [ \partial_r \left (-\frac{1}{r-r_h} \right ) \mathrm{d}r^2 + \partial_r \left ( \delta_{ij} - \frac{\gamma^2 \alpha^2}{r_c^2}  \left (r -r_c \right ) v_i v_j \right ) \mathrm{d}x^i \mathrm{d}x^j \right ]  
\\
= \frac{\sqrt{r-r_h}}{2} \left [ -\gamma^2 \alpha^2 \mathrm{d}t^2 + 2 \frac{\gamma^2 \alpha^2}{r_c}v_i \mathrm{d}x^i \mathrm{d}t + \frac{1}{\left (r-r_h \right )^2} \mathrm{d}r^2 - \frac{\gamma^2 \alpha^2}{r_c^2} v_i v_j \mathrm{d}x^i \mathrm{d}x^j \right ] \ . 
\label{eq:ec1}
\end{gathered}
\end{equation}

The second and third terms on Eq. \ref{eq:ext_curv} are such that
 
\begin{equation}
\begin{gathered}
\frac{1}{2} \left ( \gamma_{\sigma \nu}\partial_{\mu}n^{\sigma} + \gamma_{\mu \sigma} \partial_{\nu}n^{\sigma} \right ) \mathrm{d}x^\mu \mathrm{d}x^\nu 
= \frac{1}{2} \left (  \gamma_{\sigma \nu}\partial_{r}n^{\sigma} \mathrm{d}r \mathrm{d}x^\nu + \gamma_{\mu \sigma} \partial_{r}n^{\sigma} \mathrm{d}x^\mu \mathrm{d}r \right ) \ ,
\end{gathered}
\end{equation}

\noindent since the components $n^\mu$ have only radial dependence, and therefore $\partial_\mu n^\sigma \mathrm{d}x^\mu = \partial_r n^\sigma \mathrm{d}r$. Now, although these terms are present in $K_{\mu \nu}\mathrm{d}x^\mu \mathrm{d}x^\nu$, they vanish once we consider $K_{ab}\mathrm{d}x^a \mathrm{d}x^b$, since at $\Sigma_c$ one has $r = r_c = \text{cte}$. Naturally, the same happens with the third term of Eq. \ref{eq:ec1}. Therefore,

\begin{equation}
\begin{aligned}
K_{ab}\mathrm{d}x^a \mathrm{d}x^b &= \frac{\sqrt{r_c-r_h}}{2} \left ( -\gamma^2 \alpha^2 \mathrm{d}t^2 + 2 \frac{\gamma^2 \alpha^2}{r_c}v_i \mathrm{d}x^i \mathrm{d}t - \frac{\gamma^2 \alpha^2}{r_c^2} v_i v_j \mathrm{d}x^i \mathrm{d}x^j \right )  
\\
&= \frac{\sqrt{r_c-r_h}}{2} \gamma^2 \alpha^2 \left ( -\mathrm{d}t^2 + \frac{2}{r_c}v_i \mathrm{d}x^i \mathrm{d}t - \frac{1}{r_c^2} v_i v_j \mathrm{d}x^i \mathrm{d}x^j \right )  
\\
&= \frac{\gamma^2 \alpha}{2} \sqrt{\frac{r_c \left ( r_c-r_h\right )}{r_c-r_h}} \left ( -\mathrm{d}t^2 + \frac{2}{r_c}v_i \mathrm{d}x^i \mathrm{d}t - \frac{1}{r_c^2} v_i v_j \mathrm{d}x^i \mathrm{d}x^j \right ) 
\\
\Rightarrow K_{ab}\mathrm{d}x^a \mathrm{d}x^b &= \gamma^2 \alpha \frac{\sqrt{r_c}}{2} \left ( - \mathrm{d}t^2 + \frac{2}{r_c}v_i \mathrm{d}x^i \mathrm{d}t - \frac{1}{r_c^2} v_i v_j \mathrm{d}x^i \mathrm{d}x^j \right ) \ . 
\end{aligned}
\end{equation}

The trace of the extrinsic curvature is

\begin{equation}
\begin{gathered}
K = \gamma^{ab}K_{ab} = \gamma^{tt} K_{tt} + \gamma^{ij} K_{ij} 
\\
= \left ( -\frac{1}{r_c} \right ) \left ( -\gamma^2 \alpha \frac{\sqrt{r_c}}{2} \right ) + \left ( \delta^{ij} \right ) \left ( -\gamma^2 \alpha \frac{\sqrt{r_c}}{2} \frac{v_iv_j}{r_c^2} \right )   
\\
= \gamma^2 \alpha \frac{\sqrt{r_c}}{2 r_c} \left (1 - \frac{v^2}{r_c} \right ) = \frac{\alpha}{2 \sqrt{r_c}} \gamma^2 \left ( \frac{1}{\gamma^2} \right ) 
\\
\Rightarrow K = \frac{\alpha}{2\sqrt{r_c}} \ ,
\end{gathered}
\end{equation}

\noindent with which it is possible to calculate the Brown--York stress tensor, according to Eq. \ref{eq:brown_york},

\begin{equation}
\begin{aligned}
T_{a b}^{BY} \mathrm{d}x^a \mathrm{d}x^b &= 2 \left ( \gamma_{a b} K - K_{a b} \right ) \mathrm{d}x^a \mathrm{d}x^b  
\\
&= \left (\frac{\alpha}{\sqrt{r_c}} \gamma_{ab} - 2 K_{ab} \right ) \mathrm{d}x^a \mathrm{d}x^b 
\\
&= \frac{\alpha}{\sqrt{r_c}} \left ( -r_c \mathrm{d}t^2 + \delta_{ij} \mathrm{d}x^i \mathrm{d}x^j \right )  
\\
&- 2 \left [ \gamma^2 \alpha \frac{\sqrt{r_c}}{2} \left ( - \mathrm{d}t^2 + \frac{2}{r_c}v_i \mathrm{d}x^i \mathrm{d}t - \frac{1}{r_c^2} v_i v_j \mathrm{d}x^i \mathrm{d}x^j \right ) \right ] 
\\
&= \alpha \sqrt{r_c} \left ( \gamma^2 - 1 \right ) \mathrm{d}t^2 -2\gamma^2\frac{\alpha}{\sqrt{r_c}} v_i \mathrm{d}x^i \mathrm{d}t + \frac{\alpha}{\sqrt{r_c}} \left ( \delta_{ij} + \frac{\gamma^2}{r_c}v_i v_j \right )  \mathrm{d}x^i \mathrm{d}x^j 
\\
&= \alpha \sqrt{r_c} \left ( \frac{\gamma^2 v^2}{r_c} \right) \mathrm{d}t^2 -2\gamma^2\frac{\alpha}{\sqrt{r_c}} v_i \mathrm{d}x^i \mathrm{d}t + \frac{\alpha}{\sqrt{r_c}} \left ( \delta_{ij} + \frac{\gamma^2}{r_c}v_i v_j \right )  \mathrm{d}x^i \mathrm{d}x^j 
\\
\Rightarrow T_{a b}^{BY} \mathrm{d}x^a \mathrm{d}x^b &= \frac{\alpha}{\sqrt{r_c}} \left [ \gamma^2 v^2 \mathrm{d}t^2 -2\gamma^2 v_i \mathrm{d}x^i \mathrm{d}t + \left ( \delta_{ij} + \frac{\gamma^2}{r_c}v_i v_j \right )  \mathrm{d}x^i \mathrm{d}x^j\right ] \ .
\end{aligned}
\end{equation}

We shall now consider a perfect fluid characterized by $\rho = 0; \  p = \alpha r_c^{-1/2}; \  u_a = \gamma \left ( -r_c^{1/2} , r_c^{-1/2} v_i \right )$. The stress tensor for this fluid is, according to Eq. \ref{eq:perfect_fluid},

\begin{equation}
\begin{aligned}
T_{a b}^{PF} \mathrm{d}x^a \mathrm{d}x^b &= \frac{\alpha}{\sqrt{r_c}} \left ( -r_c \mathrm{d}t^2 + \delta_{ij} \mathrm{d}x^i \mathrm{d}x^j \right ) \ 
\\
&+ \frac{\alpha}{\sqrt{r_c}} \gamma^2 \left ( -\sqrt{r_c} \mathrm{d}t + \frac{1}{\sqrt{r_c}} v_i \mathrm{d}x^i \right ) \left ( -\sqrt{r_c} \mathrm{d}t + \frac{1}{\sqrt{r_c}} v_j \mathrm{d}x^j \right )  
\\
&= \frac{\alpha}{\sqrt{r_c}} \left ( -r_c \mathrm{d}t^2 + \delta_{ij} \mathrm{d}x^i \mathrm{d}x^j \right ) + \frac{\alpha}{\sqrt{r_c}} \gamma^2 \left ( r_c \mathrm{d}t^2 - 2v_i\mathrm{d}x^i\mathrm{d}t + \frac{1}{r_c}v_iv_j \mathrm{d}x^i \mathrm{d}x^j \right )  
\\
&= \alpha \sqrt{r_c} \left ( \gamma^2 - 1 \right ) \mathrm{d}t^2 -2\gamma^2\frac{\alpha}{\sqrt{r_c}} v_i \mathrm{d}x^i \mathrm{d}t + \frac{\alpha}{\sqrt{r_c}} \left ( \delta_{ij} + \frac{\gamma^2}{r_c}v_i v_j \right )  \mathrm{d}x^i \mathrm{d}x^j 
\\
\Rightarrow T_{a b}^{PF} \mathrm{d}x^a \mathrm{d}x^b &= \frac{\alpha}{\sqrt{r_c}} \left [ \gamma^2 v^2 \mathrm{d}t^2 -2\gamma^2 v_i \mathrm{d}x^i \mathrm{d}t + \left ( \delta_{ij} + \frac{\gamma^2}{r_c}v_i v_j \right )  \mathrm{d}x^i \mathrm{d}x^j\right ] \ .
\end{aligned}
\end{equation}

Therefore, one has

\begin{equation}
T_{a b}^{BY} = T_{ab}^{PF} \ , \ \ \text{with:} \ \ \rho = 0 ; \  p = \frac{\alpha}{\sqrt{r_c}}; \  u_t = -\gamma \sqrt{r_c} \ ; \  u_i = \frac{\gamma}{\sqrt{r_c}}v_i \ .
\label{eq:by=pf_2}
\end{equation}

Notice that one has the same correspondence between the Brown--York stress tensor and that of a perfect fluid on $\Sigma_c$ as in Eq. \ref{eq:by=pf_1}, although the transformations performed (boost, rescaling and shift) changed the properties of the fluid, namely: a change on its pressure (which was rescaled by $\alpha$), and the appearance of spatial components on its velocity, which can be seen as a direct consequence of the boost. It is interesting to notice, on the other hand, that nothing changed on the energy density: it remained zero, which makes sense considering that the changes in the fluid had origin on the diffeomorphisms performed.

It is possible to explicitly introduce the fluid pressure as a parameter of the metric components, by substituting $\alpha = p \sqrt{r_c}$ in Eq. \ref{eq:boosted_metric_comps}: 

\begin{equation}
\begin{gathered}
g_{tt} = \gamma^2 \left ( v^2 - p^2 r_c \left ( r - r_h \right ) \right ) \ ; \ 
g_{tr} = \gamma p \sqrt{r_c}  \ ; \ 
g_{ti} = \gamma^2 p^2 \left ( r - r_c \right ) v_i \ ; \ 
\\
g_{rr} = 0; \ 
g_{ri} = -\frac{\gamma p}{\sqrt{r_c}}v_i \ ; \ 
g_{ij} = \delta_{ij} - \frac{\gamma^2 p^2}{r_c}  \left (r -r_c \right ) v_i v_j \ ,
\label{eq:boosted_metric_presure_comps}
\end{gathered}
\end{equation}

\noindent or, more explicitly,

\begin{equation}
\begin{aligned}
g_{\mu \nu} \mathrm{d}x^\mu \mathrm{d}x^\nu &= \gamma^2 \left ( v^2 - p^2 r_c \left ( r - r_h \right ) \right ) \mathrm{d}t^2 + 2 \gamma p \sqrt{r_c} \mathrm{d}t\mathrm{d}r + 2\gamma^2 p^2 \left ( r - r_c \right ) v_i \mathrm{d}x^i \mathrm{d}t \  
\\
&-2\frac{\gamma p}{\sqrt{r_c}}v_i \mathrm{d}x^i \mathrm{d}r + \left ( \delta_{ij} - \frac{\gamma^2 p^2}{r_c}  \left (r -r_c \right ) v_i v_j  \right) \mathrm{d}x^i \mathrm{d}x^j \ .
\label{eq:boosted_metric_presure}
\end{aligned}
\end{equation}

We can do the same for the inverse metric components

\begin{equation}
\begin{gathered}
g^{tt} = \frac{\gamma^2 v^2}{r_c^2} \ ; \ 
g^{tr} = \frac{\gamma}{p\sqrt{r_c}}  \ ; \ 
g^{ti} = \frac{\gamma^2}{r_c} v^i \ ; \ 
\\
g^{rr} = r-r_h \ ; \ 
g^{ri} = \frac{\gamma}{p\sqrt{r_c}}v^i \ ; \ 
g^{ij} = \delta^{ij} + \frac{\gamma^2}{r_c} v^i v^j \ ,
\label{eq:boosted_metric_inv_presure_comps}
\end{gathered}
\end{equation}

\noindent that is,

\begin{equation}
\begin{aligned}
g^{\mu \nu} \partial_\mu \partial_\nu &= \frac{\gamma^2 v^2}{r_c^2} \partial_t^2 + 2\frac{\gamma}{p\sqrt{r_c}}\partial_t\partial_r + 2\frac{\gamma^2}{r_c} v^i \partial_i \partial_t + \left (r - r_h \right ) \partial_r^2 \
\\
&+2\frac{\gamma}{p\sqrt{r_c}}v^i \partial_i \partial_r + \left ( \delta^{ij} + \frac{\gamma^2}{r_c} v^i v^j \right ) \partial_i \partial_j \ .
\label{eq:boosted_metric_inv_presure}
\end{aligned}
\end{equation}

It is important to stress that Eq. \ref{eq:boosted_metric_presure} is still only flat spacetime in a coordinate system which allowed its parametrization under the (constant) pressure $p$ and velocity $v_i$. It is, therefore, still an exact solution to the vacuum EFE. Nevertheless, this particular parametrization is the most appropriate to perform the hydrodynamic expansion to the metric, as will be done in what follows. 

\subsection{The hydrodynamic expansion}
\label{sec:he1}

We shall now promote the up to now constant velocity $v_i$ and pressure $p$ to spacetime ($x^a$) dependent fields, using the hydrodynamic expansion to yield a near-equilibrium configuration. The procedure adopted here --- that is, the performance of a boost with constant velocity which is afterward promoted to a velocity field --- is similar to what is done in reference \cite{minwalla}, although we will perform a different kind of expansion.

By promoting $v_i$ and $p$ to arbitrarily $x^a$-dependent fields and simply considering $v_i = v_i(x^a), \ p = p(x^a)$ in Eq. \ref{eq:boosted_metric_presure}, will make the resulting metric to no longer solve the vacuum EFE, since the metric components will carry an arbitrary spacetime dependence over which one has no control. On the other hand, if we promote $v_i$ and $p$ to slowly varying fields of $x^a$ and with an amplitude small enough as to be seen as \emph{perturbations around a background in which the vacuum EFE are exactly solved}, we can perturbatively solve the vacuum EFE in the perturbation parameter we choose. To achieve this, we will employ the hydrodynamic expansion, by making,

\begin{equation}
 v_i \mapsto v_i(x^a) = v_i^{(\epsilon)}(x^a) \ ; \ \ \ p \mapsto p(x^a) = \frac{1}{\sqrt{r_c}} \left ( 1 + \frac{P^{(\epsilon)}(x^a)}{r_c} \right ) \ ,
 \label{eq:promotion}
\end{equation}

\noindent where $v_i^{(\epsilon)}(x^a) = \epsilon v_i(\epsilon^2 t, \epsilon x^i)$ and $P^{(\epsilon)}(x^a) = \epsilon^2P(\epsilon^2 t, \epsilon x^i)$ express the hydrodynamic limit of $v^i$ and $P$, as discussed in Sec. \ref{sec:hydro2}. Notice that Eq. \ref{eq:promotion} can be seen as a promotion of $v_i$ and $p$ to spacetime fields composed of small (in the hydrodynamic expansion sense) fluctuations around the equilibrium background $v_i = 0, \  p = r_c^{-1/2}$, which, according to Eq. \ref{eq:by=pf_1}, is the configuration reproducing Rindler spacetime, an exact solution to the EFE.

Therefore, since we are considering perturbations scaled under the hydrodynamic parameter $\epsilon$, we can perform the hydrodynamic expansion to the metric of Eq. \ref{eq:boosted_metric_presure} after the performance of Eq. \ref{eq:promotion}, and the resulting metric under this expansion will be guaranteed to solve the vacuum EFE  up to the desired order in $\epsilon$. In reference \cite{fg3}, the construction is extended to arbitrary order in $\epsilon$, whilst in this work we will perform the expansion up to $\mathcal{O}(\epsilon^2)$.

In the hydrodynamic expansion (in which the quantities scale according to Eq. \ref{eq:hydro_scale}), notice that

\begin{equation}
\begin{gathered}
p = \frac{\alpha}{r_c} = \frac{1}{\sqrt{r_c - r_h}} = \frac{1}{\sqrt{r_c}} \left ( 1 + \frac{P^{(\epsilon)}}{r_c}\right ) 
\\
\Rightarrow \sqrt{\frac{r_c-r_h}{r_c}} = \left ( 1 + \frac{P^{(\epsilon)}}{r_c} \right )^{-1} 
\\
\Rightarrow 1 - \frac{r_h}{r_c} = \left ( 1 + \frac{P^{(\epsilon)}}{r_c} \right )^{-2} = 1 - \frac{2 P^{(\epsilon)}}{r_c} + \mathcal{O}(\epsilon^4) 
\\
\Rightarrow r_h = r_c \left ( 1 - 1 + \frac{2 P^{(\epsilon)}}{r_c} \right ) + \mathcal{O}(\epsilon^4) 
\\
\Rightarrow r_h = 2P^{(\epsilon)} + \mathcal{O}(\epsilon^4) \ .
\end{gathered} 
\end{equation}

And, since $\delta^{ij}v_i^{(\epsilon)}v_j^{(\epsilon)} \equiv (v^2)^{(\epsilon)} \sim \mathcal{O}(\epsilon^2)$, we also have

\begin{equation}
\begin{gathered}
\gamma = \left (1-\frac{(v^2)^{(\epsilon)}}{r_c}\right )^{-1/2} = 1 + \frac{(v^2)^{(\epsilon)}}{2r_c} + \mathcal{O}(\epsilon^4) \ ;
\end{gathered} 
\end{equation}

\begin{equation}
\begin{gathered}
\gamma^2 = \left (1-\frac{(v^2)^{(\epsilon)}}{r_c}\right )^{-1} = 1 + \frac{(v^2)^{(\epsilon)}}{r_c} + \mathcal{O}(\epsilon^4) \ .
\end{gathered} 
\end{equation}

For the remaining of this section, we will drop the superscript $(\epsilon)$ from $v_i^{(\epsilon)}$, $P^{(\epsilon)}$ and $(v^2)^{(\epsilon)}$ for simplicity, but bear in mind that all the time we are working in the hydrodynamic limit, so that our quantities scale according to Eq. \ref{eq:hydro_scale}. Now all one has to do is to perform the hydrodynamic expansion on the metric, which we will do for each component of Eq. \ref{eq:boosted_metric_presure_comps}, considering the performance of Eq. \ref{eq:promotion} and the expansion of the important quantities above. This yields,

\begin{equation}
\begin{aligned}
g_{tt} &= \gamma^2 \left ( v^2 - p^2 r_c \left ( r - r_h \right ) \right ) 
\\
&= \left ( 1 + \frac{v^2}{r_c} + \mathcal{O}(\epsilon^4) \right ) \left ( v^2 - \frac{1}{r_c}\left (1 + \frac{P}{r_c} \right )^2 r_c \left [r - \left (2P + \mathcal{O}(\epsilon^4) \right ) \right ]\right )  
\\
&= \left ( 1 + \frac{v^2}{r_c}\right ) \left ( v^2 - \left (1 + \frac{2P}{r_c} \right )  \left (r - 2P \right )\right ) + \mathcal{O}(\epsilon^4)   
\\
&= \left ( 1 + \frac{v^2}{r_c}\right ) \left ( v^2 -  r + 2P - 2P\frac{r}{r_c} \right ) + \mathcal{O}(\epsilon^4)   
\\
&= v^2 -  r + 2P - 2P\frac{r}{r_c} - v^2 \frac{r}{r_c} + \mathcal{O}(\epsilon^4)  
\\
&=-r +  v^2 \left (1 -\frac{r}{r_c} \right ) + 2P \left (1 -\frac{r}{r_c} \right ) + \mathcal{O}(\epsilon^4) 
\\
\Rightarrow g_{tt} &= -r + \left (1 -\frac{r}{r_c} \right )\left (v^2 + 2P\right ) + \mathcal{O}(\epsilon^4) \ ;
\end{aligned}
\end{equation}

\begin{equation}
\begin{aligned}
g_{tr} &= \gamma p \sqrt{r_c}  
\\
&= \left ( 1 + \frac{v^2}{2r_c} + \mathcal{O}(\epsilon^4) \right )\frac{1}{\sqrt{r_c}}\left (1 + \frac{P}{r_c} \right )\sqrt{r_c}  
\\
&= \left ( 1 + \frac{v^2}{2r_c} \right )\left (1 + \frac{P}{r_c} \right ) + \mathcal{O}(\epsilon^4) =
\\ 
&= 1 + \frac{P}{r_c} + \frac{v^2}{2r_c} + \mathcal{O}(\epsilon^4) 
\\
\Rightarrow g_{tr} &= 1 + \frac{v^2 + 2P}{2r_c}  + \mathcal{O}(\epsilon^4) \ ;
\end{aligned}
\end{equation}

\begin{equation}
\begin{aligned}
g_{ti} &= \gamma^2 p^2 \left ( r - r_c \right ) v_i  
\\
&= \left ( 1 + \frac{v^2}{r_c} + \mathcal{O}(\epsilon^4) \right ) \left ( r - r_c \right ) \frac{1}{r_c} \left (1 + \frac{P}{r_c} \right )^2 v_i  
\\
&= \left ( 1 + \frac{v^2}{r_c} \right ) \left (\frac{r - r_c}{r_c}\right ) \left (1 + \frac{2P}{r_c} \right )v_i + \mathcal{O}(\epsilon^4)  
\\
&= \left ( 1 + \frac{v^2}{r_c} \right ) \left (\frac{r}{r_c}-1\right ) v_i + \mathcal{O}(\epsilon^3) 
\\
\Rightarrow g_{ti} &= -v_i \left ( 1 - \frac{r}{r_c} \right ) + \mathcal{O}(\epsilon^3) \ ; 
\end{aligned}
\end{equation}

\begin{equation}
\begin{aligned}
g_{rr} = 0;
\end{aligned}
\end{equation}

\begin{equation}
\begin{aligned}
g_{ri} &= -\frac{\gamma p}{\sqrt{r_c}}v_i  
\\
&= - \left( 1 + \frac{v^2}{2r_c} + \mathcal{O}(\epsilon^4) \right ) \frac{1}{r_c} \left ( 1 + \frac{P}{r_c} \right ) v_i 
\\
&= - \left( 1 + \frac{v^2}{2r_c} \right ) \frac{v_i}{r_c} + \mathcal{O}(\epsilon^3) 
\\
\Rightarrow g_{ri} &= -\frac{v_i}{r_c} +  \mathcal{O}(\epsilon^3) \ ; 
\end{aligned}
\end{equation}

\begin{equation}
\begin{aligned}
g_{ij} &= \delta_{ij} - \frac{\gamma^2 p^2}{r_c}  \left (r -r_c \right ) v_i v_j  
\\
&= \delta_{ij} - \left( 1 + \frac{v^2}{2r_c} + \mathcal{O}(\epsilon^4) \right )\frac{1}{r_c} \left ( 1 + \frac{P}{r_c} \right ) \left (\frac{r - r_c}{r_c} \right ) v_i v_j  
\\
&= \delta_{ij} - \left( 1 + \frac{v^2}{2r_c} \right ) \left (\frac{r}{r_c} - 1 \right ) \frac{v_i v_j}{r_c} + \mathcal{O}(\epsilon^4)  
\\
&= \delta_{ij} - \left (\frac{r}{r_c} - 1 \right ) \frac{v_i v_j}{r_c} + \mathcal{O}(\epsilon^4) 
\\
\Rightarrow g_{ij} &= \delta_{ij} + \frac{1}{r_c}\left (1-\frac{r}{r_c} \right )v_iv_j + \mathcal{O}(\epsilon^4)  \ .
\end{aligned}
\end{equation}

Therefore, the full metric --- written in a way that one has $\mathcal{O}(\epsilon^0)$ terms in the first line and terms an order higher in $\epsilon$ in each of the following lines --- is

\begin{equation}
\begin{aligned}
g_{\mu \nu} \mathrm{d}x^\mu \mathrm{d}x^\nu = &-r \mathrm{d}t^2 + 2 \mathrm{d}t \mathrm{d}r + \delta_{ij}\mathrm{d}x^i\mathrm{d}x^j \ 
\\
&-2 \left ( 1 -  \frac{r}{r_c} \right ) v_i \mathrm{d}x^i \mathrm{d}t - \frac{2}{r_c}v_i \mathrm{d}x^i \mathrm{d}r \ 
\\
&+ \left ( 1 - \frac{r}{r_c}\right ) \left [ \left (v^2 + 2P \right ) \mathrm{d}t^2 + \frac{1}{r_c} v_i v_j \mathrm{d}x^i \mathrm{d}x^j \right ] + \frac{\left (v^2 + 2P \right )}{r_c}\mathrm{d}t \mathrm{d}r \ 
\\
&+ \mathcal{O}(\epsilon^3) \ .
\label{eq:metric_expansion}
\end{aligned}
\end{equation}

It is easy to see that the boundary condition of Eq. \ref{eq:bc1_down} is met on $\Sigma_c$. 

Just like before, we are now interested in calculating the Brown--York stress tensor associated to the metric of Eq. \ref{eq:metric_expansion}. To do this, we first have to calculate the inverse metric components, which is done by expanding the components of Eq. \ref{eq:boosted_metric_inv_presure_comps} in the hydrodynamic expansion, as we did above. one has

\begin{equation}
\begin{aligned}
g^{tt} &= \frac{\gamma^2 v^2}{r_c^2}  
\\
&= \left( 1 + \frac{v^2}{r_c} + \mathcal{O}(\epsilon^4) \right ) \frac{v^2}{r_c^2} 
\\
&= \left( 1 + \frac{v^2}{r_c} \right ) \frac{v^2}{r_c^2}  + \mathcal{O}(\epsilon^4) 
\\
\Rightarrow g^{tt} &= \frac{v^2}{r_c^2} + \mathcal{O}(\epsilon^4) \ ;
\end{aligned}
\end{equation}

\begin{equation}
\begin{aligned}
g^{tr} &= \frac{\gamma}{p\sqrt{r_c}}  
\\
&= \left( 1 + \frac{v^2}{2r_c} + \mathcal{O}(\epsilon^4) \right ) \left ( 1 + \frac{P}{r_c} \right)^{-1} 
\\
&= \left( 1 + \frac{v^2}{2r_c} \right ) \left ( 1 - \frac{P}{r_c} \right) + \mathcal{O}(\epsilon^4) 
\\
&= 1 - \frac{P}{r_c} + \frac{v^2}{2r_c} + \mathcal{O}(\epsilon^4) 
\\
\Rightarrow g^{tr} &= 1 + \frac{v^2 - 2P}{2r_c} + \mathcal{O}(\epsilon^4) \ ;
\end{aligned}
\end{equation}

\begin{equation}
\begin{aligned}
g^{ti} &= \frac{\gamma^2}{r_c} v^i  
\\
&= \frac{1}{r_c} \left( 1 + \frac{v^2}{r_c} + \mathcal{O}(\epsilon^4) \right ) v^i 
\\
&= \frac{1}{r_c} \left( 1 + \frac{v^2}{r_c} \right ) v^i  + \mathcal{O}(\epsilon^4) 
\\
\Rightarrow g^{ti} &= \frac{1}{r_c}v^i + \mathcal{O}(\epsilon^4) \ ;
\end{aligned}
\end{equation}

\begin{equation}
\begin{aligned}
g^{rr} &= r-r_h 
\\
&= r - \left ( 2P + \mathcal{O}(\epsilon^4) \right ) 
\\
\Rightarrow g^{rr} &= r - 2P + \mathcal{O}(\epsilon^4) \ ;
\end{aligned}
\end{equation}

\begin{equation}
\begin{aligned}
g^{ri} &= \frac{\gamma}{p\sqrt{r_c}}v^i 
\\
&= \left( 1 + \frac{v^2}{2r_c} + \mathcal{O}(\epsilon^4) \right ) \left ( 1 + \frac{P}{r_c} \right)^{-1} v^i 
\\
&= \left( 1 + \frac{v^2}{2r_c} \right ) \left ( 1 - \frac{P}{r_c} \right) v^i + \mathcal{O}(\epsilon^4) 
\\
&= \left( 1 + \frac{v^2}{2r_c} \right ) v^i + \mathcal{O}(\epsilon^3) 
\\
\Rightarrow g^{ri} &= v^i + \mathcal{O}(\epsilon^3)  \ ;
\end{aligned}
\end{equation}

\begin{equation}
\begin{aligned}
g^{ij} &= \delta^{ij} + \frac{\gamma^2}{r_c} v^i v^j  
\\
&= \delta^{ij} + \frac{1}{r_c} \left( 1 + \frac{v^2}{r_c} + \mathcal{O}(\epsilon^4) \right ) v^i v^j 
\\
\Rightarrow g^{ij} &= \delta^{ij} + \frac{1}{r_c} v^i v^j + \mathcal{O}(\epsilon^4) \ .
\end{aligned}
\end{equation}

We shall not explicitly write $g^{\mu \nu}\partial_\mu \partial_\nu$ because the hydrodynamic scaling of the partial derivatives (here the basis vectors of the tangent space), according to Eq. \ref{eq:hydro_scale}, may lead us to erroneously not account some of the components. Therefore, in the hydrodynamic expansion it is better to work only with the components of the upper indices quantities.

Eq. \ref{eq:norm_vect_down} then yields

\begin{equation}
\begin{aligned}
n_{\mu}\mathrm{d}x^{\mu} &= \frac{1}{\sqrt{g^{rr}}} \mathrm{d}r 
\\
&= \frac{1}{\sqrt{r-2P + \mathcal{O}(\epsilon^4)}} \mathrm{d}r 
\\
&= \frac{1}{\sqrt{r}} \left ( 1 - \frac{2P}{r} + \mathcal{O}(\epsilon^4) \right )^{-1/2} \mathrm{d}r  
\\
&= \frac{1}{\sqrt{r}} \left ( 1 + \frac{P}{r} + \mathcal{O}(\epsilon^4) \right ) \mathrm{d}r 
\\
\Rightarrow n_{\mu}\mathrm{d}x^{\mu} &=  \frac{1}{\sqrt{r}} \left ( 1 + \frac{P}{r} \right ) \mathrm{d}r + \mathcal{O}(\epsilon^4) \ .
\end{aligned}
\end{equation}

Similarly, from Eq. \ref{eq:norm_vect_up} one has the $n^\mu$ components

\begin{equation}
\begin{aligned}
n^t &= \frac{1}{\sqrt{r-2P + \mathcal{O}(\epsilon^4)}} g^{rt} 
\\
&= \frac{1}{\sqrt{r}} \left ( 1 + \frac{P}{r} \right ) \left ( 1 + \frac{v^2 - 2P}{2r_c} + \mathcal{O}(\epsilon^4)\right ) + \mathcal{O}(\epsilon^4)  
\\
&= \frac{1}{\sqrt{r}} \left ( 1 + \frac{P}{r} + \frac{v^2}{2r_c} - \frac{P}{r_c} \right ) + \mathcal{O}(\epsilon^4) 
\\
\Rightarrow n^t &= \frac{1}{\sqrt{r}} \left ( 1 + \frac{v^2}{2r_c} \right ) + \mathcal{O}(\epsilon^4) \ ;
\end{aligned}
\end{equation}

\begin{equation}
\begin{aligned}
n^r &= \frac{1}{\sqrt{r-2P + \mathcal{O}(\epsilon^4)}} g^{rr} 
\\
&= \frac{1}{\sqrt{r}} \left ( 1 + \frac{P}{r} \right ) \left ( r - 2P + \mathcal{O}(\epsilon^4) \right ) + \mathcal{O}(\epsilon^4)  
\\
&= \frac{1}{\sqrt{r}} \left ( r - 2P + P \right ) + \mathcal{O}(\epsilon^4) 
\\
\Rightarrow n^r &= \frac{1}{\sqrt{r}} \left (r - P \right ) + \mathcal{O}(\epsilon^4) \ ;
\end{aligned}
\end{equation} 

\begin{equation}
\begin{aligned}
n^i &= \frac{1}{\sqrt{r-2P + \mathcal{O}(\epsilon^4)}} g^{ri} 
\\
&= \frac{1}{\sqrt{r}} \left ( 1 + \frac{P}{r} \right ) \left ( v^i + \mathcal{O}(\epsilon^3)\right )+ \mathcal{O}(\epsilon^4) 
\\
\Rightarrow n^i &=  \frac{1}{\sqrt{r}} v^i + \mathcal{O}(\epsilon^3) \ .
\end{aligned}
\end{equation}

The induced metric, according to Eq. \ref{eq:induc_met}, is

\begin{equation}
\begin{aligned}
\gamma_{\mu \nu} \mathrm{d}x^{\mu} \mathrm{d}x^{\nu} &= -r \mathrm{d}t^2 + 2 \mathrm{d}t \mathrm{d}r - \frac{1}{r}\mathrm{d}r^2 + \delta_{ij}\mathrm{d}x^i\mathrm{d}x^j \ 
\\
&-2 \left ( 1 -  \frac{r}{r_c} \right ) v_i \mathrm{d}x^i \mathrm{d}t - \frac{2}{r_c}v_i \mathrm{d}x^i \mathrm{d}r \ 
\\
&+ \left ( 1 - \frac{r}{r_c}\right ) \left [ \left (v^2 + 2P \right ) \mathrm{d}t^2 + \frac{1}{r_c} v_i v_j \mathrm{d}x^i \mathrm{d}x^j \right ] + \frac{\left (v^2 + 2P \right )}{r_c}\mathrm{d}t \mathrm{d}r - \frac{2P}{r^2} \mathrm{d}r^2 \ 
\\
&+ \mathcal{O}(\epsilon^3) \ , 
\end{aligned}
\end{equation}

\noindent which clearly reduces to Eq. \ref{eq:bc1_down} on $\Sigma_c$. 

For the components of the inverse induced metric $\gamma^{\mu \nu} = g^{\mu \nu} - n^{\mu}n^{\nu}$, one has

\begin{equation}
\begin{aligned}
\gamma^{tt} &= \frac{v^2}{r_c^2} - \frac{1}{r} \left (1 + \frac{v^2}{2 r_c}\right )^2 + \mathcal{O}(\epsilon^4) 
\\
&= \frac{v^2}{r_c^2} - \frac{1}{r} \left (1 + \frac{v^2}{r_c}\right ) + \mathcal{O}(\epsilon^4) 
\\
\Rightarrow \gamma^{tt} &= \frac{v^2}{r_c} \left (\frac{1}{r_c} - \frac{1}{r} \right ) - \frac{1}{r} + \mathcal{O}(\epsilon^4) \ ;
\label{eq:iim_first}
\end{aligned}
\end{equation}

\begin{equation}
\begin{aligned}
\gamma^{tr} &= 1 + \frac{v^2 - 2P}{2r_c} - \frac{1}{r} \left ( 1 + \frac{v^2}{2r_c}\right )\left (r - P \right ) + \mathcal{O}(\epsilon^4) 
\\
&= 1 + \frac{v^2 - 2P}{2r_c} - \frac{1}{r} \left ( r - P + \frac{v^2r}{2r_c}\right ) + \mathcal{O}(\epsilon^4) 
\\
\Rightarrow \gamma^{tr} &= P \left ( \frac{1}{r} - \frac{1}{r_c} \right ) + \mathcal{O}(\epsilon^4) \ ;
\end{aligned}
\end{equation}

\begin{equation}
\begin{aligned}
\gamma^{ti} &= \frac{1}{r_c}v^i - \frac{1}{r} \left ( 1 + \frac{v^2}{2r_c}\right ) v^i +  \mathcal{O}(\epsilon^3) 
\\
\Rightarrow \gamma^{ti} &= \left (\frac{1}{r_c} - \frac{1}{r} \right ) v^i  +  \mathcal{O}(\epsilon^3) \ ;
\end{aligned}
\end{equation}

\begin{equation}
\begin{aligned}
\gamma^{rr} &= r - 2P - \frac{1}{r} \left (r - P \right )^2 + \mathcal{O}(\epsilon^4)  
\\
&= r - 2P - \frac{1}{r} \left (r^2 - 2Pr \right ) + \mathcal{O}(\epsilon^4) 
\\
\Rightarrow \gamma^{rr} &= \mathcal{O}(\epsilon^4) \ ;
\end{aligned}
\end{equation}

\begin{equation}
\begin{aligned}
\gamma^{ri} &= v^i - \frac{1}{r} \left (r - P \right )v^i + \mathcal{O}(\epsilon^3) 
\\
\Rightarrow \gamma^{ri} &= \mathcal{O}(\epsilon^3) \ ;
\end{aligned}
\end{equation}

\begin{equation}
\begin{aligned}
\gamma^{ij} &= \delta^{ij} + \frac{1}{r_c} v^i v^j - \frac{1}{r}v^i v^j + \mathcal{O}(\epsilon^4) 
\\
\Rightarrow \gamma^{ij} &= \delta^{ij} + \left (\frac{1}{r_c} - \frac{1}{r} \right ) v^i v^j + \mathcal{O}(\epsilon^4) \ .
\label{eq:iim_last}
\end{aligned}
\end{equation}

On $\Sigma_c$, it is easy to see that

\begin{equation}
\gamma^{tt} = -\frac{1}{r_c} + \mathcal{O}(\epsilon^4) \ ; \ \ \gamma^{ti} = \mathcal{O}(\epsilon^3) \ ; \ \ \gamma^{ij} = \delta^{ij} + \mathcal{O}(\epsilon^4) \ ,
\end{equation}

\noindent as expected. 

The first term of the extrinsic curvature, Eq. \ref{eq:ext_curv}, is

\begin{equation}
\frac{1}{2} n^{\sigma} \partial_{\sigma} \gamma_{\mu \nu} = \frac{1}{2\sqrt{r}} \left (  \left ( 1 + \frac{v^2}{2r_c} \right ) \partial_t \gamma_{\mu \nu} + \left ( r - P \right ) \partial_r \gamma_{\mu \nu} + v^i \partial_i \gamma_{\mu \nu} \right ) + \mathcal{O}(\epsilon^4) \ .
\end{equation}

Notice that now the $\gamma_{\mu \nu}$ components have not only radial dependence, but also spacetime dependence due to their parametrization with the velocity and pressure fields. In fact, one has

\begin{equation}
\begin{aligned}
\partial_t \gamma_{tt} &= \partial_t \left ( -r + \left ( 1-\frac{r}{r_c}\right )\left ( v^2 + 2P\right ) + \mathcal{O}(\epsilon^4) \right )  
\\
&= \left ( 1-\frac{r}{r_c}\right ) \partial_t \left (v^2 + 2P \right ) + \mathcal{O}(\epsilon^4) 
\\
\Rightarrow \partial_t \gamma_{tt} &= \mathcal{O}(\epsilon^4) \ ;
\end{aligned}
\end{equation}

\begin{equation}
\begin{aligned}
\partial_t \gamma_{ti} &= \partial_t \left ( -v_i \left (1-\frac{r}{r_c}\right ) + \mathcal{O}(\epsilon^3) \right )  
\\
&= \left (\frac{r}{r_c}-1\right ) \partial_t v_i + \mathcal{O}(\epsilon^3) 
\\
\Rightarrow \partial_t \gamma_{ti} &= \mathcal{O}(\epsilon^3) \ ;
\end{aligned}
\end{equation}

\begin{equation}
\begin{aligned}
\partial_t \gamma_{ij} &= \partial_t \left ( \delta_{ij} + \frac{1}{r_c} \left ( 1-\frac{r}{r_c}\right ) v_i v_j + \mathcal{O}(\epsilon^4) \right )  
\\
&= \frac{1}{r_c} \left ( 1-\frac{r}{r_c}\right ) \partial_t \left ( v_i v_j \right ) + \mathcal{O}(\epsilon^4)  
\\
\Rightarrow \partial_t \gamma_{ij} &= \mathcal{O}(\epsilon^4) \ ;
\end{aligned}
\end{equation}

\begin{equation}
\begin{aligned}
\partial_r \gamma_{tt} &= \partial_r \left ( -r + \left ( 1-\frac{r}{r_c}\right )\left ( v^2 + 2P\right ) + \mathcal{O}(\epsilon^4) \right ) 
\\
\Rightarrow \partial_r \gamma_{tt} &= -1 - \frac{v^2 + 2P}{r_c} + \mathcal{O}(\epsilon^4)  \ ;
\end{aligned}
\end{equation}

\begin{equation}
\begin{aligned}
\partial_r \gamma_{ti} &= \partial_r \left ( -v_i \left (1-\frac{r}{r_c}\right ) + \mathcal{O}(\epsilon^3) \right ) 
\\
\Rightarrow \partial_r \gamma_{ti} &= \frac{1}{r_c}v_i + \mathcal{O}(\epsilon^3) \ ;
\end{aligned}
\end{equation}

\begin{equation}
\begin{aligned}
\partial_r \gamma_{ij} &= \partial_r \left ( \delta_{ij} + \frac{1}{r_c} \left ( 1-\frac{r}{r_c}\right ) v_i v_j + \mathcal{O}(\epsilon^4) \right )  
\\
\Rightarrow \partial_r \gamma_{ij} &= - \frac{1}{r_c^2} v_i v_j + \mathcal{O}(\epsilon^4) \ ;
\end{aligned}
\end{equation}

\begin{equation}
\begin{aligned}
\partial_i \gamma_{tt} &= \partial_i \left ( -r + \left ( 1-\frac{r}{r_c}\right )\left ( v^2 + 2P\right ) + \mathcal{O}(\epsilon^4) \right ) 
\\
&= \left ( 1 - \frac{r}{r_c} \right ) \partial_i \left ( v^2 + 2P\right ) + \mathcal{O}(\epsilon^4) 
\\
\Rightarrow \partial_i \gamma_{tt} &= \mathcal{O}(\epsilon^3)  \ ;
\end{aligned}
\end{equation}

\begin{equation}
\begin{aligned}
\partial_i \gamma_{tj} &= \partial_i \left ( -v_j \left (1-\frac{r}{r_c}\right ) + \mathcal{O}(\epsilon^3) \right ) 
\\
\Rightarrow \partial_i \gamma_{tj} &=  -\left (1-\frac{r}{r_c}\right ) \partial_i v_j + \mathcal{O}(\epsilon^3) \ ;
\end{aligned}
\end{equation}

\begin{equation}
\begin{aligned}
\partial_i \gamma_{jk} &= \partial_i \left ( \delta_{jk} + \frac{1}{r_c} \left ( 1-\frac{r}{r_c}\right ) v_j v_k + \mathcal{O}(\epsilon^4) \right )  
\\
\Rightarrow \partial_i \gamma_{jk} &= \mathcal{O}(\epsilon^3) \ ,
\end{aligned}
\end{equation}

\noindent where we used the hydrodynamic scaling for the partial derivatives, according to Eq. \ref{eq:hydro_scale}. We did not calculate the derivatives of the form $\partial_\sigma \gamma_{\mu r}$, because, like before, we are ultimately only interested in $K_{ab}\mathrm{d}x^a \mathrm{d}x^b$, therefore the terms $\frac{1}{2} n^{\sigma} \partial_{\sigma} \gamma_{\mu r}\mathrm{d}x^\mu \mathrm{d}r $, which are present in $K_{\mu \nu}\mathrm{d}x^\mu \mathrm{d}x^\nu$ but not in $K_{ab}\mathrm{d}x^a \mathrm{d}x^b$, shall not be considered. Therefore, one has, on $\Sigma_c$,

\begin{equation}
\begin{aligned}
\left. \frac{1}{2} n^{\sigma} \partial_{\sigma} \gamma_{\mu \nu}\mathrm{d}x^\mu \mathrm{d}x^\nu \right\rvert_{r = r_c} &= \left. \frac{1}{2} \left ( n^r \left ( -1 - \frac{v^2 + 2P}{r_c}\right )\mathrm{d}t^2 + n^r \left ( \frac{2}{r_c}v_i\right )\mathrm{d}x^i \mathrm{d}t \right )\right\rvert_{r = r_c} \ 
\\
&+ \left. \frac{1}{2} \left ( n^r \left (- \frac{1}{r_c^2} v_i v_j \right )\mathrm{d}x^i \mathrm{d}x^j \right )\right\rvert_{r = r_c} \ 
\\
&+\left.  \frac{1}{2} \left ( n^i \left ( -2\left (1-\frac{r}{r_c}\right ) \partial_i v_j \right )\mathrm{d}x^j \mathrm{d}t  \right )\right\rvert_{r = r_c} + \mathcal{O}(\epsilon^3)  
\\
&=\frac{\left (r_c - P \right )}{2\sqrt{r_c}} \left ( \left ( -1 - \frac{v^2 + 2P}{r_c}\right )\mathrm{d}t^2 + \frac{2}{r_c}v_i \mathrm{d}x^i \mathrm{d}t \right ) \ 
\\
&+ \frac{\left (r_c - P \right )}{2\sqrt{r_c}} \left (- \frac{1}{r_c^2} v_i v_j \mathrm{d}x^i \mathrm{d}x^j \right ) \ 
\\
&- \frac{1}{\sqrt{r_c}} v^i \left (1-\frac{r}{r_c}\right ) \partial_i v_j \mathrm{d}x^j \mathrm{d}t + \mathcal{O}(\epsilon^3)  
\\
&=\frac{1}{2\sqrt{r_c}} \left ( \left ( -r_c - v^2 - 2P + P \right )\mathrm{d}t^2 + 2v_i\mathrm{d}x^i \mathrm{d}t \right ) \
\\
&- \frac{1}{2\sqrt{r_c}} \left (  \frac{1}{r_c} v_i v_j \mathrm{d}x^i \mathrm{d}x^j \right ) + \mathcal{O}(\epsilon^3) 
\\
\Rightarrow \left. \frac{1}{2} n^{\sigma} \partial_{\sigma} \gamma_{\mu \nu}\mathrm{d}x^\mu \mathrm{d}x^\nu \right\rvert_{r = r_c} &= \frac{1}{2\sqrt{r_c}} \left ( -\left (r_c+ v^2+ P \right )\mathrm{d}t^2 + 2v_i\mathrm{d}x^i \mathrm{d}t - \frac{1}{r_c} v_i v_j \mathrm{d}x^i \mathrm{d}x^j \right ) \ 
\\ 
&+ \mathcal{O}(\epsilon^3) \ .
\label{eq:ec1_hydro}
\end{aligned}
\end{equation}

The second and third terms on Eq. \ref{eq:ext_curv} are such that
 
\begin{equation}
\begin{gathered}
\frac{1}{2} \left ( \gamma_{\sigma \nu}\partial_{\mu}n^{\sigma} + \gamma_{\mu \sigma} \partial_{\nu}n^{\sigma} \right ) \mathrm{d}x^\mu \mathrm{d}x^\nu 
\\
= \frac{1}{2} \gamma_{\sigma \nu} \left (\partial_{t}n^{\sigma} \mathrm{d}t \mathrm{d}x^\nu + \partial_{r}n^{\sigma} \mathrm{d}r \mathrm{d}x^\nu + \partial_{i}n^{\sigma} \mathrm{d}x^i \mathrm{d}x^\nu \right ) \ 
\\
+ \frac{1}{2} \gamma_{\mu \sigma} \left (\partial_{t}n^{\sigma} \mathrm{d}t \mathrm{d}x^\mu + \partial_{r}n^{\sigma} \mathrm{d}r \mathrm{d}x^\mu + \partial_{i}n^{\sigma} \mathrm{d}x^i \mathrm{d}x^\mu \right ) \ , 
\end{gathered}
\end{equation}

\noindent since we now have radial as well as spacetime dependence on the components $n^\mu$. Now, as we did before, we will not consider the terms of the form $\partial_r n^\sigma \mathrm{d}r$, since they are not present in $K_{ab}\mathrm{d}x^a \mathrm{d}x^b$. As for what remains, we must calculate the derivatives

\begin{equation}
\begin{aligned}
\partial_t n^t &= \partial_t \left (\frac{1}{\sqrt{r}} \left ( 1 + \frac{v^2}{2 r_c} \right ) + \mathcal{O}(\epsilon^4) \right )  
\\
&= \frac{1}{2r_c\sqrt{r}} \partial_t v^2 + \mathcal{O}(\epsilon^4) 
\\
\Rightarrow \partial_t n^t &= \mathcal{O}(\epsilon^4) \ ;
\end{aligned}
\end{equation}

\begin{equation}
\begin{aligned}
\partial_t n^r &= \partial_t \left (\frac{1}{\sqrt{r}} \left ( r-P \right ) + \mathcal{O}(\epsilon^4) \right )  
\\
&= -\frac{1}{\sqrt{r}} \partial_t P + \mathcal{O}(\epsilon^4) 
\\
\Rightarrow \partial_t n^r &= \mathcal{O}(\epsilon^4) \ ;
\end{aligned}
\end{equation}

\begin{equation}
\begin{aligned}
\partial_t n^i &= \partial_t \left (\frac{1}{\sqrt{r}} v^i  + \mathcal{O}(\epsilon^3) \right )  
\\
&= \frac{1}{\sqrt{r}} \partial_t v^i + \mathcal{O}(\epsilon^3) 
\\
\Rightarrow \partial_t n^i &= \mathcal{O}(\epsilon^3) \ ;
\end{aligned}
\end{equation}

\begin{equation}
\begin{aligned}
\partial_i n^t &= \partial_i \left (\frac{1}{\sqrt{r}} \left ( 1 + \frac{v^2}{2 r_c} \right ) + \mathcal{O}(\epsilon^4) \right )  
\\
&= \frac{1}{2r_c\sqrt{r}} \partial_i v^2 + \mathcal{O}(\epsilon^4) 
\\
\Rightarrow \partial_i n^t &= \mathcal{O}(\epsilon^3) \ ;
\end{aligned}
\end{equation}

\begin{equation}
\begin{aligned}
\partial_i n^r &= \partial_i \left (\frac{1}{\sqrt{r}} \left ( r-P \right ) + \mathcal{O}(\epsilon^4) \right )  
\\
&= -\frac{1}{\sqrt{r}} \partial_i P + \mathcal{O}(\epsilon^4) 
\\
\Rightarrow \partial_i n^r &= \mathcal{O}(\epsilon^3) \ ;
\end{aligned}
\end{equation}

\begin{equation}
\begin{aligned}
\partial_i n^j &= \partial_i \left (\frac{1}{\sqrt{r}} v^j  + \mathcal{O}(\epsilon^3) \right )  
\\
&= \frac{1}{\sqrt{r}} \partial_i v^j + \mathcal{O}(\epsilon^3) 
\\
\Rightarrow \partial_i n^j &= \frac{1}{\sqrt{r}} \partial_i v^j + \mathcal{O}(\epsilon^3) \ .
\end{aligned}
\end{equation}

Therefore,

\begin{equation}
\begin{aligned}
\left. \frac{1}{2} \left ( \gamma_{\sigma \nu}\partial_{\mu}n^{\sigma} + \gamma_{\mu \sigma} \partial_{\nu}n^{\sigma} \right ) \mathrm{d}x^\mu \mathrm{d}x^\nu \right\rvert_{r = r_c} &= \frac{1}{2} \left ( \gamma_{\sigma \nu} \partial_{i}n^\sigma + \gamma_{i\sigma} \partial_\nu n^\sigma \right ) \mathrm{d}x^i \mathrm{d}x^\nu + \mathcal{O}(\epsilon^3) 
\\
&= \frac{1}{2} \left ( \gamma_{j\nu} \partial_{i}n^j + \gamma_{ij} \partial_\nu n^j \right ) \mathrm{d}x^i \mathrm{d}x^\nu + \mathcal{O}(\epsilon^3) 
\\
&= \frac{1}{2} \left ( \gamma_{jk} \partial_{i}n^j + \gamma_{ij} \partial_k n^j \right ) \mathrm{d}x^i \mathrm{d}x^k + \mathcal{O}(\epsilon^3) 
\\
&= \frac{1}{\sqrt{2r_c}} \left (\partial_i \left (v^j \delta_{jk}\right ) + \partial_k \left ( v^j \delta_{ji} \right ) \right )  \mathrm{d}x^i \mathrm{d}x^k +  \mathcal{O}(\epsilon^3) 
\\
\Rightarrow \left. \frac{1}{2} \left ( \gamma_{\sigma \nu}\partial_{\mu}n^{\sigma} + \gamma_{\mu \sigma} \partial_{\nu}n^{\sigma} \right ) \mathrm{d}x^\mu \mathrm{d}x^\nu \right\rvert_{r = r_c} &= \frac{1}{\sqrt{2r_c}} \left ( \partial_i v_j + \partial_j v_i \right )\mathrm{d}x^i \mathrm{d}x^j + \mathcal{O}(\epsilon^3) \ .
\label{eq:ec23_hydro}
\end{aligned}
\end{equation}

Eqs. \ref{eq:ec1_hydro} and \ref{eq:ec23_hydro} then yield the extrinsic curvature on $\Sigma_c$,

\begin{equation}
 K_{ab}\mathrm{d}x^a \mathrm{d}x^b = -\frac{\left ( P + v^2 + r_c\right )}{2\sqrt{r_c}} \mathrm{d}t^2 + \frac{1}{\sqrt{r_c}}v_i \mathrm{d}x^i \mathrm{d}t + \frac{1}{2\sqrt{r_c}} \left ( \partial_i v_j + \partial_j v_i - \frac{1}{r_c} v_i v_j \right ) \mathrm{d}x^i \mathrm{d}x^j + \mathcal{O}(\epsilon^3) \ ,
\end{equation}

\noindent whose trace is

\begin{equation}
\begin{aligned}
 K &= \gamma^{ab}K_{ab} = \gamma^{tt} K_{tt} + \gamma^{it} K_{it} + \gamma^{ij} K_{ij} 
 \\
 &= \left ( -\frac{1}{r_c} \right ) \left ( -\frac{\left ( P + v^2 + r_c\right )}{2\sqrt{r_c}} \right ) + \left ( \delta^{ij} \right ) \left [ \frac{1}{2\sqrt{r_c}} \left ( \partial_i v_j + \partial_j v_i - \frac{1}{r_c} v_i v_j \right )\right ] + \mathcal{O}(\epsilon^4)  
 \\
 &= \frac{P + r_c}{2r_c^{3/2}} + \frac{v^2}{2r_c^{3/2}} + \frac{1}{\sqrt{r_c}} \partial^i v_i - \frac{v^2}{2r_c^{3/2}} + \mathcal{O}(\epsilon^4) 
 \\
 \Rightarrow K &= \frac{1}{\sqrt{r_c}} \left (\frac{P + r_c}{2r_c} + \partial^i v_i \right ) + \mathcal{O}(\epsilon^4) \ ,
\end{aligned}
\end{equation}

\noindent with which is possible to calculate the Brown--York stress tensor, according to Eq. \ref{eq:brown_york},

\begin{equation}
\begin{aligned}
T_{a b}^{BY} \mathrm{d}x^a \mathrm{d}x^b &= 2 \left ( \gamma_{a b} K - K_{a b} \right ) \mathrm{d}x^a \mathrm{d}x^b  
\\
&= \left [ \left ( \frac{2}{\sqrt{r_c}} \left (\frac{P + r_c}{2r_c} + \partial^i v_i \right ) + \mathcal{O}(\epsilon^4) \right ) \gamma_{ab} - 2 K_{ab} \right ] \mathrm{d}x^a \mathrm{d}x^b 
\\
&= \left ( \frac{P + r_c}{r_c^{3/2}} \right ) \left ( -r_c \mathrm{d}t^2 + \delta_{ij} \mathrm{d}x^i \mathrm{d}x^j \right ) + \frac{2}{\sqrt{r_c}}\partial^i v_i \gamma_{ab}\mathrm{d}x^a \mathrm{d}x^b \ 
\\
&+ \frac{\left ( P + v^2 + r_c\right )}{\sqrt{r_c}} \mathrm{d}t^2 - \frac{2}{\sqrt{r_c}}v_i \mathrm{d}x^i \mathrm{d}t - \frac{1}{\sqrt{r_c}} \left ( \partial_i v_j + \partial_j v_i - \frac{1}{r_c} v_i v_j \right ) \mathrm{d}x^i \mathrm{d}x^j + \mathcal{O}(\epsilon^3)  
\\
&= \frac{v^2}{\sqrt{r_c}} \mathrm{d}t^2 +  \frac{1}{r_c^{3/2}} P \delta_{ij} \mathrm{d}x^i \mathrm{d}x^j + \frac{1}{\sqrt{r_c}}\delta_{ij} \mathrm{d}x^i \mathrm{d}x^j + \frac{2}{\sqrt{r_c}}\partial^i v_i \gamma_{ab}\mathrm{d}x^a \mathrm{d}x^b \ 
\\
&- \frac{2}{\sqrt{r_c}}v_i \mathrm{d}x^i \mathrm{d}t + \frac{1}{r_c^{3/2}} \left (v_i v_j - r_c \left (\partial_iv_j + \partial_j v_i \right ) \right ) \mathrm{d}x^i \mathrm{d}x^j  + \mathcal{O}(\epsilon^3) 
\\
\Rightarrow T_{a b}^{BY} \mathrm{d}x^a \mathrm{d}x^b &= \frac{1}{\sqrt{r_c}} \delta_{ij} \mathrm{d}x^i \mathrm{d}x^j - \frac{2}{\sqrt{r_c}} v_i \mathrm{d}x^i \mathrm{d}t + \frac{v^2}{\sqrt{r_c}} \mathrm{d}t^2 \ 
\\
&+ \frac{\left ( v_i v_j + P \delta_{ij} - r_c \left (\partial_iv_j + \partial_j v_i \right ) \right )}{r_c^{3/2}} \mathrm{d}x^i \mathrm{d}x^j + \frac{2}{\sqrt{r_c}} \partial^i v_i \gamma_{ab} \mathrm{d}x^a \mathrm{d}x^b + \mathcal{O}(\epsilon^3) \ .
\end{aligned}
\end{equation}

\subsection{The dual fluid}
\label{sec:df1}

We shall now follow reference \cite{fg2}, and impose the conservation of the Brow--York stress tensor on $\Sigma_c$, that is,

\begin{equation}
 \nabla^a T_{ab} = \partial^a T_{ab} = 0 \ ,
 \label{eq:by_cons1}
\end{equation}

\noindent where one has $\nabla_a = \partial_a$ since the affine connection coefficients associated to $\gamma_{ab}$ --- whose components are constant --- are all null.

Notice that Eq. \ref{eq:by_cons1} represents a constraint on $\Sigma_c$ which necessarily must be satisfied to guarantee that the EFE are solved perturbatively in $\epsilon$. In that sense, the conservation of the Brow--York stress tensor may be seen as an integrability condition of the EFE, as discussed in \cite{fg3}. Moreover, once the constraints are satisfied on $\Sigma_c$, it is possible to evolve the solution in the radial direction, provided the absence of singularities, which was shown to be the case in \cites{fg1, fg2}. Therefore, in this sense, we can reduce the EFE to Eq. \ref{eq:by_cons1}, which then yields, at $\mathcal{O}(\epsilon^2)$,

\begin{equation}
\begin{gathered}
\partial^a T_{ab} = \gamma^{tt}\partial_t T_{tt} + \gamma^{ij} \partial_j T_{it} + \mathcal{O}(\epsilon^4) = 0 
\\
\Rightarrow  \left ( -\frac{1}{r_c} \right ) \partial_t \left (\frac{v^2}{\sqrt{r_c}} - \frac{2 r_c}{\sqrt{r_c}} \partial^i v_i \right ) + \left (\delta^{ij} \right ) \partial_j \left ( - \frac{1}{\sqrt{r_c}} v_i \right )  + \mathcal{O}(\epsilon^4) = 0 
\\
\Rightarrow - \frac{1}{\sqrt{r_c}}\partial^iv_i + \mathcal{O}(\epsilon^4) =  0 
\\
\Rightarrow \partial^iv_i  = \mathcal{O}(\epsilon^4) \ ,
\label{eq:incomp_1}
\end{gathered}
\end{equation}

\noindent which is satisfied by an incompressible fluid, up to $\mathcal{O}(\epsilon^4)$! On the other hand, at $\mathcal{O}(\epsilon^3)$ one has

\begin{equation}
\begin{gathered}
\partial^a T_{aj} = \gamma^{tt}\partial_t T_{tj} + \gamma^{ik} \partial_k T_{ij} + \mathcal{O}(\epsilon^4) = 0 
\\
\Rightarrow  \left ( -\frac{1}{r_c} \right ) \partial_t \left ( - \frac{1}{\sqrt{r_c}} v_j \right )  \ 
\\
+ \left (\delta^{ik} \right ) \partial_k \left ( \frac{1}{\sqrt{r_c}}\delta_{ij} + \frac{\left ( v_i v_j + P \delta_{ij} - r_c \left (\partial_iv_j + \partial_j v_i \right )\right )}{r_c^{3/2}} + \frac{2}{\sqrt{r_c}} \delta_{ij} \partial^i v_i \right ) + \mathcal{O}(\epsilon^4) = 0 
\\
\Rightarrow \frac{1}{r_c^{3/2}} \left ( \partial_t v_j + \partial^i \left (v_i v_j \right ) + \partial^i \left (P \delta_{ij} \right ) - r_c \left (\partial^i \partial_i v_j + \partial^i \partial_j v_i \right )\right ) + \mathcal{O}(\epsilon^4) = 0 
\\
\Rightarrow \partial_t v_j + v_i \partial^i v_j + v_j \partial^i v_i + \partial_j P - r_c \partial^2 v_j -r_c \partial_j \partial^iv_i = \mathcal{O}(\epsilon^4) 
\\
\Rightarrow \partial_t v_j - \eta \partial^2 v_j + \partial_jP + v^i \partial_i v_j = \mathcal{O}(\epsilon^4) \ ,
\label{eq:ns_1}
\end{gathered}
\end{equation}

\noindent where we identified the kinematic viscosity $\eta = r_c$ and used Eq. \ref{eq:incomp_1} at the third and fifth lines to obtain the incompressible Navier--Stokes equation for a fluid of velocity field $v_i$ and pressure field $P$.

This is precisely the sense in which the vacuum Einstein Field Equations are reduced to the incompressible Navier--Stokes equation, within the setup and boundary conditions considered up to now. In this sense, one can say that there is an incompressible fluid on $\Sigma_c$, whose velocity and pressure fields parameterize the full bulk spacetime, according to Eq. \ref{eq:metric_expansion}.

In \cite{fg2}, the authors perform a near-horizon expansion of the Rindler spacetime initially considered. In this context, the near-horizon limit is achieved by taking the acceleration of $\Sigma_c$ to infinity, as this results in pushing the hypersurface to its future horizon. It is then shown that \emph{the near-horizon expansion is mathematically equivalent to the hydrodynamic expansion}. Therefore, in the near-horizon limit, we can see $\Sigma_c$ as being located at the horizon, which then establishes the precise sense in which \emph{the horizon} --- and not only an arbitrary hypersurface at constant radius --- can be seen as an incompressible fluid. 

\section{A varying induced metric: soft hairy fluid/gravity}
\label{sec:varying}

In this section (whose results are original), we shall consider a generalization of the boundary conditions employed in the last section. These new boundary conditions were introduced in the references \cites{daniel1, daniel2} for the 3-dimensional case, and in \cite{daniel3} for arbitrary dimensions, which is the case we are specifically interested here, so that we will closely follow the conventions of this work in our approach.

In \cite{daniel3}, the authors consider a metric in $(D+2) \geq 3$ dimensions with a non-extremal horizon, within the assumption that the horizon is free from singularities, which then allows a Taylor expansion in the near-horizon region under the radius $\rho$ as the expansion parameter, such that the horizon is at $\rho = 0$. The spacetime under consideration is covered by the chart $\left \{\tau, \rho, x^i\right \}$ --- where $x^i$ are the spatial coordinates transverse to the horizon ---, and is constrained by the following near-horizon boundary conditions,

\begin{equation}
g_{\tau \tau} = - \kappa^2 \rho^2 \ ; \ \ g_{\rho \rho} = 1 + \mathcal{O}(\rho^2) \ ; \ \ g_{\rho i} = \mathcal{O}(\rho) \ ; \ \ g_{ij} = \Omega_{ij} + \mathcal{O}(\rho^2) \ ,
\label{eq:bc_nh}
\end{equation}

\noindent where $\kappa$ is the (constant) Rindler acceleration, taken to be greater than zero to guarantee the non-extremality; and $\Omega_{ij} = \Omega_{ij}(t, x^i)$ is the metric transverse to the horizon, such that $\det \Omega_{ij} \neq 0$ to guarantee its non-singularity. An important special case is that in which the boundary metric $\Omega_{ij}$ does not depend explicitly on time, that is, $\Omega_{ij} = \Omega_{ij}(x^i)$. This is guaranteed if we consider constant surface gravity, which is the case under interest here. Therefore, we shall consider a time-independent boundary metric.

As shown in \cite{daniel3}, these near-horizon boundary conditions lead to the discovery of an infinite set of near horizon symmetries and associated soft hair excitations, in the sense of \cite{sh}, featuring in any non-extremal horizon, which holds valid in any spacetime dimension greater than two. The main goal of this section is then to investigate the relationship between fluids and the soft-hairy horizon generated by this choice of boundary conditions. The developments and results in this section are original, 
and were published in \cite{sh_fg_mine}.

Before we begin, it is important to notice that the boundary conditions of Eq. \ref{eq:bc_nh} were specifically introduced to constraint the $(D+2)$-dimensional metric \emph{in the near-horizon expansion}. However, as we discussed in Sec. \ref{sec:df1}, it was shown in \cites{fg1, fg2} that the near-horizon and hydrodynamic expansions are mathematically identical. For this reason, we can reproduce the procedure adopted in the last section for the fixed induced metric --- namely, performing the hydrodynamic expansion --- also for these new boundary conditions. In practice, we will perform a hydrodynamic expansion considering the boundary condition $g_{ij} = \Omega_{ij}$ (which is the relevant condition here) to hold in a generic hypersurface $\Sigma_c$, which will in the near-horizon limit be identified with the horizon.

Therefore, for our computation purposes, we shall consider the same general setup described in Sec. \ref{sec:setup}, and choose as a boundary condition for the $(D+2)$-dimensional spacetime the following induced metric on $\Sigma_c$,

\begin{equation}
\gamma_{a b} \mathrm{d}x^a \mathrm{d}x^b = -r_c \mathrm{d} t^2 + \Omega_{ij}\mathrm{d}x^i \mathrm{d}x^j \ , 
\label{eq:bc2_down}
\end{equation}

\noindent where $\Omega_{ij} = \Omega_{ij}(x^i)$ is the $D$-dimensional boundary metric of $\Sigma_c$, which, we remark, shall be considered as time-independent, although it does depend on $x^i$, being, therefore, allowed to fluctuate. Although the physical interpretation here in principle differs from the one in \cite{daniel3}, the near-horizon--hydrodynamic equivalence of \cites{fg1, fg2}  mathematically guarantees the same interpretation.

We shall now redo the same procedure as in the last section, but with this important change in the boundary conditions, which, as we shall see, will lead to interesting generalizations.

Like before, we start by considering a $(D+2)$-dimensional spacetime covered by the ingoing Rindler coordinates $\{ r, t, x^i \}$, which nonetheless cannot be seen as Rindler spacetime itself, since we are now considering $\Omega_{ij}$ as the spacelike part of the metric,

\begin{equation}
 g_{\mu \nu} \mathrm{d}x^\mu \mathrm{d}x^\nu = -r \mathrm{d}t^2 + 2 \mathrm{d}t \mathrm{d}r + \Omega_{ij}\mathrm{d}x^i \mathrm{d}x^j \ .
 \label{eq:rindler_st_new}
\end{equation}

We will assume that the metric of Eq. \ref{eq:rindler_st_new} exactly solves the vacuum EFE, which is not true for an arbitrary spacetime dependence of $\Omega_{ij}$, but we will consider to be the case here. In fact, the relevant specific forms of $\Omega_{ij}$ which lead to the appearance of soft hair excitations are such that this assumption holds \cite{daniel3}, so that one has no loss of generality within the context considered here.

From the definition $g^{\mu \sigma}g_{\sigma \nu} = \delta^{\mu}_{\nu}$, it is easy to see that the inverse metric is

\begin{equation}
 g^{\mu \nu} \partial_{\mu} \partial_{\nu} = 2 \partial_t \partial_r + r \partial_r^2 + \Omega^{i j} \partial_i \partial_j \ ,
\end{equation}

\noindent where, naturally, $\Omega^{ij} \Omega_{jk} = \delta^i_k$. 

Substituting the components of the inverse metric above in Eqs. \ref{eq:norm_vect_down} and \ref{eq:norm_vect_up}, respectively, yields

\begin{equation}
n_{\mu}\mathrm{d}x^{\mu} = \frac{1}{\sqrt{r}} \mathrm{d}r \ ; \ \ 
n^{\mu}\partial_{\mu} = \frac{1}{\sqrt{r}} \partial_t + \sqrt{r} \partial_r \ ,
\end{equation}

\noindent so that from Eq. \ref{eq:induc_met} one has

\begin{equation}
 \gamma_{\mu \nu} \mathrm{d}x^{\mu} \mathrm{d}x^{\nu} = -r \mathrm{d}t^2 - \frac{1}{r} \mathrm{d}r^2 + 2 \mathrm{d}t \mathrm{d}r + \Omega_{ij}\mathrm{d}x^i \mathrm{d}x^j \ , 
\end{equation}

\noindent which reduces to Eq. \ref{eq:bc2_down} at $r = r_c$. This is also the case of the inverse induced metric $\gamma^{\mu \nu} = g^{\mu \nu} - n^{\mu}n^{\nu}$,

\begin{equation}
 \gamma^{\mu \nu} \partial_{\mu} \partial_{\nu} = 2 \partial_t \partial_r + r \partial_r^2 + \Omega^{i j} \partial_i \partial_j - \left( r\partial_r^2 + \frac{1}{r} \partial_t^2 + 2 \partial_t \partial_r \right ) =  -\frac{1}{r} \partial_t^2 + \Omega^{i j} \partial_i \partial_j \ ,
\end{equation}

\noindent which becomes on $\Sigma_c$ simply

\begin{equation}
\gamma^{ab}\partial_a \partial_b = -\frac{1}{r_c} \partial_t^2 + \Omega^{ij}\partial_i\partial_j \ .
\label{eq:bc2_up}
\end{equation}

Since the components $n^\mu$ have only radial dependence, one has $\partial_\mu n^\sigma \mathrm{d}x^\mu = \partial_r n^\sigma \mathrm{d}r$, so that the second and third terms of the extrinsic curvature (Eq. \ref{eq:ext_curv}) will be present in $K_{\mu \nu}\mathrm{d}x^\mu \mathrm{d}x^\nu$ but not in $K_{ab}\mathrm{d}x^a \mathrm{d}x^b$, as we discussed several times in Sec. \ref{sec:fixed}. Therefore, the only remaining term is the first one,

\begin{equation}
\begin{aligned}
\frac{1}{2} n^{\sigma} \partial_{\sigma} \gamma_{\mu \nu} \mathrm{d}x^\mu \mathrm{d}x^\nu &= \frac{1}{2} \left ( \sqrt{r} \partial_r \gamma_{\mu \nu} \right ) \mathrm{d}x^\mu \mathrm{d}x^\nu  
\\
&= \frac{1}{2} \left [ + \sqrt{r_c} \left ( \partial_r (-r) \mathrm{d}t^2 + \partial_r \left ( -\frac{1}{r} \right ) \mathrm{d}r^2 \right ) \right ] 
\\
\Rightarrow \frac{1}{2} n^{\sigma} \partial_{\sigma} \gamma_{\mu \nu} \mathrm{d}x^\mu \mathrm{d}x^\nu &= \frac{1}{2} \left [ \sqrt{r_c} \left ( - \mathrm{d}t^2 + \left ( \frac{1}{r^2} \right ) \mathrm{d}r^2 \right ) \right ] \ ,
\end{aligned}
\end{equation}

\noindent so that the intrinsic curvature on $\Sigma_c$ is

\begin{equation}
K_{a b}\mathrm{d}x^a \mathrm{d}x^b = -\frac{\sqrt{r_c}}{2} \mathrm{d}t^2 \ , 
\end{equation}

\noindent whose trace is given by

\begin{equation}
\begin{gathered}
K = \gamma^{tt} K_{tt} = \left ( -\frac{1}{r_c}\right ) \left (-\frac{\sqrt{r_c}}{2} \right ) 
\\
\Rightarrow K = \frac{1}{2 \sqrt{r_c}} \ .
\end{gathered}
\end{equation}

We can now calculate the Brown--York stress tensor, according to Eq. \ref{eq:brown_york},

\begin{equation}
\begin{aligned}
T_{a b}^{BY}\mathrm{d}x^a \mathrm{d}x^b &= 2 \left ( \gamma_{a b} K - K_{a b} \right ) \mathrm{d}x^a \mathrm{d}x^b 
\\
&= \left ( \frac{1}{ \sqrt{r_c}} \gamma_{ab} - 2 K_{ab} \right )\mathrm{d}x^a \mathrm{d}x^b  
\\
&= \frac{1}{\sqrt{r_c}} \left ( -r_c \mathrm{d}t^2 + \Omega_{ij} \mathrm{d}x^i \mathrm{d}x^j \right ) + \sqrt{r_c} \mathrm{d}t^2  
\\
\Rightarrow T_{a b}^{BY} \mathrm{d}x^a \mathrm{d}x^b &= \frac{1}{\sqrt{r_c}}\Omega_{ij} \mathrm{d}x^i \mathrm{d}x^j \ .
\end{aligned}
\end{equation}

If we now consider a perfect fluid such that $p = r_c^{-1/2} \ ; \  u_t = r_c^{1/2}$ and $u_i = 0$, one has, according to Eq. \ref{eq:perfect_fluid}, the following stress tensor,

\begin{equation}
\begin{aligned}
T_{a b}^{PF} \mathrm{d}x^a \mathrm{d}x^b &= \frac{1}{\sqrt{r_c}} \left [ \left ( -r_c \mathrm{d}t^2 + \Omega_{ij} \mathrm{d}x^i \mathrm{d}x^j \right ) + \left ( r_c \mathrm{d}t^2 \right ) \right ] 
\\
\Rightarrow T_{a b}^{PF} \mathrm{d}x^a \mathrm{d}x^b &= \frac{1}{\sqrt{r_c}}\Omega_{ij} \mathrm{d}x^i \mathrm{d}x^j \ .
\end{aligned}
\end{equation}

Therefore, the Brown--York stress tensor has the form of a perfect fluid on $\Sigma_c$,

\begin{equation}
T_{a b}^{BY} = T_{ab}^{PF} \ ,
\label{eq:by=pf_quase_1}
\end{equation}

\noindent where the perfect fluid is such that

\begin{equation}
p = \frac{1}{\sqrt{r_c}} \ ; \ \ u_t = \sqrt{r_c} \ ; \ \ u_i = 0 \ .
\label{eq:fluid_bc_new1}
\end{equation}

Notice that one has precisely the same background as before, in which the Brown--York stress tensor is precisely equal to that of perfect fluid with vanishing energy density, just like in Eq. \ref{eq:by=pf_1}. That is of major importance, as it allows the performance of the hydrodynamic expansion which will be done below.

\subsection{Boosting the metric}

We shall now consider a family of $(D+2)$-dimensional metrics parameterized under constant velocity $v_i$, which is achieved through the performance on Eq. \ref{eq:rindler_st_new} of the same diffeomorphisms presented in Sec. \ref{sec:boost1}, to which we collectively refer as ``boost'', although they consist of rescales and linear shifts as well. Now, since these diffeomorphisms do not involve any derivatives, it is clear that the resulting boosted metric components will have the same form as in Eq. \ref{eq:boosted_metric_comps}, but with the change $\delta_{ij} \mapsto \Omega_{ij}(x^i)$,

\begin{equation}
\begin{gathered}
g_{tt} = \gamma^2 \left ( v^2 - \alpha^2 \left ( r - r_h \right ) \right ) \ ; \ 
g_{tr} = \gamma \alpha  \ ; \ 
g_{ti} = \frac{\gamma^2 \alpha^2}{r_c} \left ( r - r_c \right ) v_i \ ; \ 
\\
g_{rr} = 0; \ 
g_{ri} = -\frac{\gamma \alpha}{r_c}v_i \ ; \ 
g_{ij} = \Omega_{ij} - \frac{\gamma^2 \alpha^2}{r_c^2}  \left (r -r_c \right ) v_i v_j \ ,
\label{eq:boosted_metric_comps_2}
\end{gathered}
\end{equation}

\noindent or, more explicitly,

\begin{equation}
\begin{aligned}
g_{\mu \nu} \mathrm{d}x^\mu \mathrm{d}x^\nu &= \gamma^2 \left ( v^2 - \alpha^2 \left ( r - r_h \right ) \right ) \mathrm{d}t^2 + 2\gamma \alpha \mathrm{d}t\mathrm{d}r + 2\frac{\gamma^2 \alpha^2}{r_c} \left ( r - r_c \right ) v_i \mathrm{d}x^i \mathrm{d}t \  
\\
&-2\frac{\gamma \alpha}{r_c}v_i \mathrm{d}x^i \mathrm{d}r + \left ( \Omega_{ij} - \frac{\gamma^2 \alpha^2}{r_c^2}  \left (r -r_c \right ) v_i v_j \right ) \mathrm{d}x^i \mathrm{d}x^j \ .
\label{eq:boosted_metric_2}
\end{aligned}
\end{equation}

Just like before, the performance of the diffeomorphisms only changes the parametrization of the metric components, so that the boosted metric of Eq. \ref{eq:boosted_metric_2} is still an exact solution to the vacuum EFE, within our initial assumption. 

As discussed in Appendix \ref{ap:inverse}, the inverse boosted metric components are

\begin{equation}
\begin{gathered}
g^{tt} = \frac{\gamma^2 v^2}{r_c^2} \ ; \ 
g^{tr} = \frac{\gamma}{\alpha}  \ ; \ 
g^{ti} = \frac{\gamma^2}{r_c} v^i \ ; \ 
\\
g^{rr} = r-r_h; \ 
g^{ri} = \frac{\gamma}{\alpha}v^i \ ; \ 
g^{ij} = \Omega^{ij} + \frac{\gamma^2}{r_c} v^i v^j \ ,
\label{eq:boosted_metric_inv_comps_2}
\end{gathered}
\end{equation}

\noindent that is, 

\begin{equation}
\begin{aligned}
g^{\mu \nu} \partial_\mu \partial_\nu &= \frac{\gamma^2 v^2}{r_c^2} \partial_t^2 + 2\frac{\gamma}{\alpha}\partial_t\partial_r + 2\frac{\gamma^2}{r_c} v^i \partial_i \partial_t + \left (r - r_h \right ) \partial_r^2 \  
\\
&+2\frac{\gamma}{\alpha}v^i \partial_i \partial_r + \left ( \Omega^{ij} + \frac{\gamma^2}{r_c} v^i v^j \right ) \partial_i \partial_j \ .
\label{eq:boosted_metric_inv_2}
\end{aligned}
\end{equation}

Now we can proceed to the calculation of the Brown--York stress tensor, as done in the last section. Substituting the components of the inverse metric in Eqs. \ref{eq:norm_vect_down} and \ref{eq:norm_vect_up}, respectively, yields

\begin{equation}
n_{\mu}\mathrm{d}x^{\mu} = \frac{1}{\sqrt{r - r_h}} \mathrm{d}r \ ; \ \ 
n^{\mu}\partial_{\mu} = \frac{1}{\sqrt{r-r_h}} \left ( \frac{\gamma}{\alpha} \partial_t + \left (r - r_h \right ) \partial_r + \frac{\gamma}{\alpha}v^i \partial_i \right ) \ .
\end{equation}

From Eq. \ref{eq:induc_met}, then, one has

\begin{equation}
\begin{aligned}
\gamma_{\mu \nu} \mathrm{d}x^{\mu} \mathrm{d}x^{\nu} &= \gamma^2 \left ( v^2 - \alpha^2 \left ( r - r_h \right ) \right ) \mathrm{d}t^2 + 2\gamma \alpha \mathrm{d}t\mathrm{d}r + 2\frac{\gamma^2 \alpha^2}{r_c} \left ( r - r_c \right ) v_i \mathrm{d}x^i \mathrm{d}t 
\\
&-\frac{1}{r-r_h} \mathrm{d}r^2 -2\frac{\gamma \alpha}{r_c}v_i \mathrm{d}x^i \mathrm{d}r + \left ( \Omega_{ij} - \frac{\gamma^2 \alpha^2}{r_c^2}  \left (r -r_c \right ) v_i v_j \right ) \mathrm{d}x^i \mathrm{d}x^j \ , 
\end{aligned}
\end{equation}

\noindent whilst for the inverse induced metric $\gamma^{\mu \nu} = g^{\mu \nu} - n^{\mu}n^{\nu}$, one has, similarly to Eq. \ref{eq:inducedup},

\begin{equation}
\begin{aligned}
\gamma^{\mu \nu} \partial_{\mu} \partial_{\nu} &= \gamma^2 \left ( \frac{v^2}{r_c^2} - \frac{1}{\alpha^2 \left (r-r_h \right )} \right ) \partial_t^2 + 2\gamma^2 \left ( \frac{1}{r_c} - \frac{1}{\alpha^2 \left (r - r_h \right )} \right ) v^i \partial_i \partial_t \ 
\\
&+ \left ( \Omega^{ij} + \gamma^2 \left ( \frac{1}{r_c} - \frac{1}{\alpha^2 \left (r - r_h \right )} \right ) v^i v^j \right ) \partial_i \partial_j \ .
\end{aligned}
\end{equation}

Notice that on $\Sigma_c$ one has $\alpha^2 \left (r - r_h \right ) = \alpha^2 \left (r_c - r_h \right )= r_c$, thus both the induced metric and its inverse reduce to Eqs. \ref{eq:bc2_down} and \ref{eq:bc2_up}.

As we are ultimately interested in the extrinsic curvature on $\Sigma_c$, we shall compute the first term of Eq. \ref{eq:ext_curv} already evaluated at $r = r_c$,

\begin{equation}
\begin{aligned}
\left. \frac{1}{2} n^{\sigma} \partial_{\sigma} \gamma_{\mu \nu}\mathrm{d}x^\mu \mathrm{d}x^\nu \right\rvert_{r = r_c} &= \left. \frac{1}{2\sqrt{r-r_h}} \left ( n^r \partial_r \gamma_{\mu \nu} + n^i \partial_i \gamma_{\mu \nu} \right ) \mathrm{d}x^\mu \mathrm{d}x^\nu \right\rvert_{r = r_c}  
\\
&= \left. \frac{1}{2\sqrt{r-r_h}} \left ( \left (r - r_h \right ) \partial_r \gamma_{\mu \nu} + \frac{\gamma}{\alpha} v^i \partial_i \gamma_{\mu \nu} \right ) \mathrm{d}x^\mu \mathrm{d}x^\nu \right\rvert_{r = r_c}  
\\
&= \left. \frac{1}{2\sqrt{r-r_h}} \frac{\gamma}{\alpha} \left ( v^k \partial_k \Omega_{ij} \right ) \mathrm{d}x^i \mathrm{d}x^j \right\rvert_{r = r_c} \ 
\\
&+ \left . \frac{\sqrt{r-r_h}}{2} \left ( -\gamma^2 \alpha^2 \mathrm{d}t^2 + 2 \frac{\gamma^2 \alpha^2}{r_c}v_i \mathrm{d}x^i \mathrm{d}t \right ) \right\rvert_{r = r_c} 
\\
&+ \left . \frac{\sqrt{r-r_h}}{2} \left ( \frac{1}{\left (r-r_h \right )^2} \mathrm{d}r^2 - \frac{\gamma^2 \alpha^2}{r_c^2} v_i v_j \mathrm{d}x^i \mathrm{d}x^j \right ) \right\rvert_{r = r_c}  
\\
&= \frac{1}{2\sqrt{r_c-r_h}} \frac{\gamma}{\alpha} \left ( v^k \partial_k \Omega_{ij} \right ) \mathrm{d}x^i \mathrm{d}x^j +
\\
&+ \frac{\gamma^2 \alpha^2}{2}\sqrt{r_c-r_h} \left ( - \mathrm{d}t^2 + \frac{2}{r_c}v_i \mathrm{d}x^i \mathrm{d}t - \frac{1}{r_c^2} v_i v_j \mathrm{d}x^i \mathrm{d}x^j \right ) 
\\
\Rightarrow \left. \frac{1}{2} n^{\sigma} \partial_{\sigma} \gamma_{\mu \nu}\mathrm{d}x^\mu \mathrm{d}x^\nu \right\rvert_{r = r_c} &=  \frac{\gamma}{2\sqrt{r_c}} v^k \partial_k  \Omega_{ij} \mathrm{d}x^i \mathrm{d}x^j \ 
\\
&+ \gamma^2 \alpha \frac{\sqrt{r_c}}{2} \left ( - \mathrm{d}t^2 + \frac{2}{r_c}v_i \mathrm{d}x^i \mathrm{d}t - \frac{1}{r_c^2} v_i v_j \mathrm{d}x^i \mathrm{d}x^j \right ) \ .
\end{aligned}
\end{equation}

The second and third terms on Eq. \ref{eq:ext_curv} are such that
 
\begin{equation}
\begin{gathered}
\frac{1}{2} \left ( \gamma_{\sigma \nu}\partial_{\mu}n^{\sigma} + \gamma_{\mu \sigma} \partial_{\nu}n^{\sigma} \right ) \mathrm{d}x^\mu \mathrm{d}x^\nu 
= \frac{1}{2} \left (  \gamma_{\sigma \nu}\partial_{r}n^{\sigma} \mathrm{d}r \mathrm{d}x^\nu + \gamma_{\mu \sigma} \partial_{r}n^{\sigma} \mathrm{d}x^\mu \mathrm{d}r \right ) \ ,
\end{gathered}
\end{equation}

\noindent therefore, since the components $n^\mu$ have only radial dependence, one has $\partial_\mu n^\sigma \mathrm{d}x^\mu = \partial_r n^\sigma \mathrm{d}r$, so that these terms will not be present in $K_{ab}\mathrm{d}x^a \mathrm{d}x^b$. Consequently, the extrinsic curvature on $\Sigma_c$ is simply

\begin{equation}
K_{ab}\mathrm{d}x^a \mathrm{d}x^b = \frac{\gamma}{2\sqrt{r_c}} v^k \partial_k  \Omega_{ij}  \mathrm{d}x^i \mathrm{d}x^j +  \gamma^2 \alpha \frac{\sqrt{r_c}}{2} \left ( - \mathrm{d}t^2 + \frac{2}{r_c}v_i \mathrm{d}x^i \mathrm{d}t - \frac{1}{r_c^2} v_i v_j \mathrm{d}x^i \mathrm{d}x^j \right ) \ .
\end{equation}

The trace of the extrinsic curvature is

\begin{equation}
\begin{aligned}
K &= \gamma^{ab}K_{ab} = \gamma^{tt} K_{tt} + \gamma^{ij} K_{ij} 
\\
&= \left ( -\frac{1}{r_c} \right ) \left ( -\gamma^2 \alpha \frac{\sqrt{r_c}}{2} \right ) + \left ( \Omega^{ij} \right ) \left ( -\gamma^2 \alpha \frac{\sqrt{r_c}}{2} \frac{v_iv_j}{r_c^2} + \frac{\gamma}{2\sqrt{r_c}}v^k \partial_k  \Omega_{ij} \right )   
\\
&=\gamma^2 \alpha \frac{\sqrt{r_c}}{2 r_c} \left (1 - \frac{v^2}{r_c} \right ) + \frac{\gamma}{2\sqrt{r_c}}\Omega^{ij} v^k \partial_k  \Omega_{ij}   
\\
\Rightarrow K &= \frac{1}{2\sqrt{r_c}} \left ( \alpha + \gamma \Omega^{ij} v^k \partial_k  \Omega_{ij}  \right ) \ ,
\end{aligned}
\end{equation}

\noindent with which is possible to calculate the Brown--York stress tensor, according to Eq. \ref{eq:brown_york},

\begin{equation}
\begin{aligned}
T_{a b}^{BY} \mathrm{d}x^a \mathrm{d}x^b &= 2 \left ( \gamma_{a b} K - K_{a b} \right ) \mathrm{d}x^a \mathrm{d}x^b  
\\
&= \left [ \frac{1}{\sqrt{r_c}} \left ( \alpha + \gamma \Omega^{kl} v^m \partial_m \Omega_{kl}  \right ) \gamma_{ab} - 2 K_{ab} \right ] \mathrm{d}x^a \mathrm{d}x^b 
\\
&= \frac{\alpha}{\sqrt{r_c}} \left ( -r_c \mathrm{d}t^2 + \Omega_{ij} \mathrm{d}x^i \mathrm{d}x^j \right ) + \frac{\gamma}{\sqrt{r_c}} \Omega^{ij} v^k \partial_k  \Omega_{ij} \left ( -r_c \mathrm{d}t^2 + \Omega_{ij} \mathrm{d}x^i \mathrm{d}x^j \right ) \ 
\\
& - \frac{\gamma}{\sqrt{r_c}} v^k \partial_k  \Omega_{ij} \mathrm{d}x^i \mathrm{d}x^j -  \gamma^2 \alpha \sqrt{r_c} \left ( - \mathrm{d}t^2 + \frac{2}{r_c}v_i \mathrm{d}x^i \mathrm{d}t - \frac{1}{r_c^2} v_i v_j \mathrm{d}x^i \mathrm{d}x^j \right )  
\\
&= \alpha \sqrt{r_c} \left ( \gamma^2 - 1 \right ) \mathrm{d}t^2 -2\gamma^2\frac{\alpha}{\sqrt{r_c}} v_i \mathrm{d}x^i \mathrm{d}t + \frac{\alpha}{\sqrt{r_c}} \left ( \Omega_{ij} + \frac{\gamma^2}{r_c}v_i v_j \right )  \mathrm{d}x^i \mathrm{d}x^j \
\\
&- \gamma \sqrt{r_c} \Omega^{kl} v^m \partial_m \Omega_{kl} \mathrm{d}t^2 + \frac{\gamma}{\sqrt{r_c}} \left ( \Omega^{kl} v^m \partial_m \Omega_{kl} \Omega_{ij} - v^k \partial_k  \Omega_{ij}  \right ) \mathrm{d}x^i \mathrm{d}x^j 
\\
&= \alpha \sqrt{r_c} \left ( \frac{\gamma^2 v^2}{r_c} \right) \mathrm{d}t^2 -2\gamma^2\frac{\alpha}{\sqrt{r_c}} v_i \mathrm{d}x^i \mathrm{d}t + \frac{\alpha}{\sqrt{r_c}} \left ( \Omega_{ij} + \frac{\gamma^2}{r_c}v_i v_j \right )  \mathrm{d}x^i \mathrm{d}x^j \ 
\\
&- \gamma \sqrt{r_c} \Omega^{kl} v^m \partial_m \Omega_{kl} \mathrm{d}t^2 + \frac{\gamma}{\sqrt{r_c}} \left ( \Omega^{kl} v^m \partial_m \Omega_{kl} \Omega_{ij} - v^k \partial_k  \Omega_{ij}  \right ) \mathrm{d}x^i \mathrm{d}x^j 
\\
\Rightarrow T_{a b}^{BY} \mathrm{d}x^a \mathrm{d}x^b &= \frac{\alpha}{\sqrt{r_c}} \left [ \gamma^2 v^2 \mathrm{d}t^2 -2\gamma^2 v_i \mathrm{d}x^i \mathrm{d}t + \left ( \Omega_{ij} + \frac{\gamma^2}{r_c}v_i v_j \right )  \mathrm{d}x^i\mathrm{d}x^j\right ] 
\\
&- \gamma \sqrt{r_c} \Omega^{ij} v^k \partial_k  \Omega_{ij} \mathrm{d}t^2 + \frac{\gamma}{\sqrt{r_c}} \left ( \Omega^{kl} v^m \partial_m \Omega_{kl} \Omega_{ij} - v^k \partial_k  \Omega_{ij}  \right ) \mathrm{d}x^i \mathrm{d}x^j \ .
\end{aligned}
\end{equation}

If we now consider a perfect fluid with $\rho = -\gamma \Omega^{ij}v^k \partial_k  \Omega_{ij} r_c^{-1/2}$; $p = \left( \alpha + \gamma \Omega^{ij}v^k \partial_k  \Omega_{ij} \right ) r_c^{-1/2}$ and $u_a = \gamma \left( -r_c^{1/2} , r_c^{-1/2} v_i\right)$, its stress tensor is, according to Eq. \ref{eq:perfect_fluid}, given by

\begin{equation}
\begin{aligned}
T_{a b}^{PF} \mathrm{d}x^a \mathrm{d}x^b &= \frac{1}{\sqrt{r_c}} \left ( -\gamma \Omega^{kl}v^m \partial_m \Omega_{kl} \right ) \gamma^2 \left ( -\sqrt{r_c} \mathrm{d}t^2 + \frac{1}{\sqrt{r_c}} v_i \mathrm{d}x^i \right ) \left ( -\sqrt{r_c} \mathrm{d}t^2 + \frac{1}{\sqrt{r_c}} v_j \mathrm{d}x^j \right ) \ 
\\
&+ \frac{1}{\sqrt{r_c}} \left ( \alpha + \gamma \Omega^{kl}v^m \partial_m \Omega_{kl} \right ) \left ( -r_c \mathrm{d}t^2 + \Omega_{ij} \mathrm{d}x^i \mathrm{d}x^j \right ) \ 
\\
&+ \frac{1}{\sqrt{r_c}} \left ( \alpha + \gamma \Omega^{kl}v^m \partial_m \Omega_{kl} \right ) \gamma^2 \left ( -\sqrt{r_c} \mathrm{d}t^2 + \frac{1}{\sqrt{r_c}} v_i \mathrm{d}x^i \right ) \left ( -\sqrt{r_c} \mathrm{d}t^2 + \frac{1}{\sqrt{r_c}} v_j \mathrm{d}x^j \right )  
\\
&= \frac{1}{\sqrt{r_c}} \left ( \alpha + \gamma \Omega^{kl}v^m \partial_m \Omega_{kl} \right ) \left ( -r_c \mathrm{d}t^2 + \Omega_{ij} \mathrm{d}x^i \mathrm{d}x^j \right ) \ 
\\
&+ \frac{1}{\sqrt{r_c}} \alpha \gamma^2 \left ( r_c \mathrm{d}t^2 - 2v_i\mathrm{d}x^i\mathrm{d}t + \frac{1}{r_c}v_iv_j \mathrm{d}x^i \mathrm{d}x^j \right )  
\\
&= \alpha \sqrt{r_c} \left ( \gamma^2 - 1 \right ) \mathrm{d}t^2 -2\gamma^2\frac{\alpha}{\sqrt{r_c}} v_i \mathrm{d}x^i \mathrm{d}t + \frac{\alpha}{\sqrt{r_c}} \left ( \Omega_{ij} + \frac{\gamma^2}{r_c}v_i v_j \right )  \mathrm{d}x^i \mathrm{d}x^j \ 
\\
&-\gamma \sqrt{r_c} \Omega^{kl}v^m \partial_m \Omega_{kl} \mathrm{d}t^2 + \frac{\gamma}{\sqrt{r_c}} \Omega^{kl}v^m \partial_m \Omega_{kl}\Omega_{ij} \mathrm{d}x^i \mathrm{d}x^j 
\\
\Rightarrow T_{a b}^{PF} \mathrm{d}x^a \mathrm{d}x^b &= \frac{\alpha}{\sqrt{r_c}} \left [ \gamma^2 v^2 \mathrm{d}t^2 -2\gamma^2 v_i \mathrm{d}x^i \mathrm{d}t + \left ( \Omega_{ij} + \frac{\gamma^2}{r_c}v_i v_j \right )  \mathrm{d}x^i\mathrm{d}x^j\right ] 
\\
&- \gamma \sqrt{r_c} \Omega^{kl} v^m \partial_m \Omega_{kl} \mathrm{d}t^2 + \frac{\gamma}{\sqrt{r_c}} \Omega^{kl} v^m \partial_m \Omega_{kl} \Omega_{ij} \mathrm{d}x^i \mathrm{d}x^j \ .
\end{aligned}
\end{equation}

Therefore, the Brown--York stress tensor has the form of a perfect fluid on $\Sigma_c$, plus a correction,

\begin{equation}
T_{a b}^{BY} = T_{ab}^{PF} + \tilde{T}_{ab} \ ,
\label{eq:by=pf_quase_2}
\end{equation}

\noindent where the perfect fluid is such that,

\begin{equation}
\rho = -\frac{\gamma}{\sqrt{r_c}} \Omega^{ij}v^k \partial_k  \Omega_{ij} \ ; \ \ p = \frac{1}{\sqrt{r_c}} \left ( \alpha + \gamma \Omega^{ij}v^k \partial_k  \Omega_{ij} \right ) \ ; \ \ u_t = -\gamma \sqrt{r_c} \ ; \ \ u_i = \frac{\gamma}{\sqrt{r_c}} v_i \ ,
\label{eq:fluid_bc_new2}
\end{equation}

\noindent and

\begin{equation}
 \tilde{T}_{ab}\mathrm{d}x^a \mathrm{d}x^b = -\frac{\gamma}{\sqrt{r_c}}v^k \partial_k  \Omega_{ij} \mathrm{d}x^i \mathrm{d}x^j \ .
\end{equation}

By considering the new boundary conditions, here one has a generalization, as we got a fluid with a non-zero energy density $\rho$. This seems odd, but will soon be fixed by the imposition of the boundary condition. The changes in the other parameters of the fluid with respect to the results we got for the non-boosted metric, once again are a direct consequence of the diffeomorphisms performed.

Motivated by the hydrodynamic expansion that we shall perform in what follows, we will explicitly introduce the fluid pressure as a parameter of the metric components, by substituting $\alpha = \sqrt{r_c} \left ( p + \rho \right) $ in Eq. \ref{eq:boosted_metric_comps_2}: 

\begin{equation}
\begin{gathered}
g_{tt} = \gamma^2 \left ( v^2 - r_c\left(p + \rho \right)^2\left ( r - r_h \right ) \right ) \ ; \ 
g_{tr} = \gamma \sqrt{r_c} \left(p + \rho \right)  \ ; \ 
g_{ti} = \gamma^2 \left(p + \rho \right)^2 \left ( r - r_c \right ) v_i \ ; \ 
\\
g_{rr} = 0; \ 
g_{ri} = -\frac{\gamma}{\sqrt{r_c}}\left(p + \rho \right)v_i \ ; \ 
g_{ij} = \Omega_{ij} - \frac{\gamma^2}{r_c} \left(p + \rho \right)^2  \left (r -r_c \right ) v_i v_j \ ,
\label{eq:boosted_metric_presure_comps_2}
\end{gathered}
\end{equation}

\noindent or, more explicitly,

\begin{equation}
\begin{aligned}
g_{\mu \nu} \mathrm{d}x^\mu \mathrm{d}x^\nu &= \gamma^2 \left ( v^2 - r_c\left(p + \rho \right)^2\left ( r - r_h \right ) \right ) \mathrm{d}t^2 + 2 \gamma \sqrt{r_c} \left(p + \rho \right) \mathrm{d}t\mathrm{d}r \ 
\\
&+ 2\gamma^2 \left(p + \rho \right)^2 \left ( r - r_c \right ) v_i \mathrm{d}x^i \mathrm{d}t -2\frac{\gamma}{\sqrt{r_c}}\left(p + \rho \right)v_i  \mathrm{d}x^i \mathrm{d}r \ 
\\
&+ \left (\Omega_{ij} - \frac{\gamma^2}{r_c} \left(p + \rho \right)^2  \left (r -r_c \right ) v_i v_j \right) \mathrm{d}x^i \mathrm{d}x^j \ .
\label{eq:boosted_metric_presure_2}
\end{aligned}
\end{equation}

We can do the same for the inverse metric components,

\begin{equation}
\begin{gathered}
g^{tt} = \frac{\gamma^2 v^2}{r_c^2} \ ; \ 
g^{tr} = \frac{\gamma}{\sqrt{r_c}(p + \rho)}  \ ; \ 
g^{ti} = \frac{\gamma^2}{r_c} v^i \ ; \ 
\\
g^{rr} = r-r_h \ ; \ 
g^{ri} = \frac{\gamma}{\sqrt{r_c}(p + \rho)}v^i \ ; \ 
g^{ij} = \Omega^{ij} + \frac{\gamma^2}{r_c} v^i v^j \ ,
\label{eq:boosted_metric_inv_presure_comps_2}
\end{gathered}
\end{equation}

\noindent that is,

\begin{equation}
\begin{aligned}
g^{\mu \nu} \partial_\mu \partial_\nu &= \frac{\gamma^2 v^2}{r_c^2} \partial_t^2 + 2\frac{\gamma}{\sqrt{r_c}(p + \rho)}\partial_t\partial_r + 2\frac{\gamma^2}{r_c} v^i \partial_i \partial_t + \left (r - r_h \right ) \partial_r^2 \  
\\
&+2\frac{\gamma}{\sqrt{r_c}(p + \rho)}v^i \partial_i \partial_r + \left ( \Omega^{ij} + \frac{\gamma^2}{r_c} v^i v^j \right ) \partial_i \partial_j \ .
\label{eq:boosted_metric_inv_presure_2}
\end{aligned}
\end{equation}

\subsection{The hydrodynamic expansion}
\label{sec:he2}

Like we did in Sec. \ref{sec:fixed}, we shall now promote the constant velocity $v_i$ and pressure $p$ to slowly varying fields of $x^a$ with an amplitude small enough as to be seen as perturbations around a background in which the vacuum EFE are exactly solved, so that we can perturbatively solve the vacuum EFE in the hydrodynamic expansion,

\begin{equation}
 v_i \mapsto v_i(x^a) = v_i^{(\epsilon)}(x^a) \ ; \ \ \ p \mapsto p(x^a) = \frac{1}{\sqrt{r_c}} \left ( 1 + \frac{P^{(\epsilon)}(x^a)}{r_c} \right ) \ ,
\end{equation}

\noindent so that, like we did before, we perform a promotion of $v_i$ and $p$ to spacetime fields composed of small (in the hydrodynamic expansion sense) fluctuations around the equilibrium background $v_i = 0, \  p = r_c^{-1/2}$, which, according to Eq. \ref{eq:by=pf_quase_1}, is the configuration reproducing Eq. \ref{eq:rindler_st_new}, assumed to be an exact solution to the EFE. 

In the hydrodynamic expansion (in which the quantities scale according to Eq. \ref{eq:hydro_scale}), one has, just like before,

\begin{equation}
r_h = 2P^{(\epsilon)} + \mathcal{O}(\epsilon^4) \ ;
\end{equation}

\begin{equation}
\gamma = \left (1-\frac{(v^2)^{(\epsilon)}}{r_c}\right )^{-1/2} = 1 + \frac{(v^2)^{(\epsilon)}}{2r_c} + \mathcal{O}(\epsilon^4) \ ;
\end{equation}

\begin{equation}
\begin{gathered}
\gamma^2 = \left (1-\frac{(v^2)^{(\epsilon)}}{r_c}\right )^{-1} = 1 + \frac{(v^2)^{(\epsilon)}}{r_c} + \mathcal{O}(\epsilon^4) \ ,
\end{gathered} 
\end{equation}

\noindent where $\Omega^{ij}v_i^{(\epsilon)}v_j^{(\epsilon)} \equiv (v^2)^{(\epsilon)} \sim \mathcal{O}(\epsilon^2)$. Consequently, 

\begin{equation}
\begin{aligned}
\rho &= -\frac{\gamma}{\sqrt{r_c}} \Omega^{ij}v^k \partial_k  \Omega_{ij} 
\\
&= -\frac{1}{\sqrt{r_c}} \left (1 + \frac{(v^2)^{(\epsilon)}}{2r_c} + \mathcal{O}(\epsilon^4) \right ) \Omega^{ij}v^k \partial_k  \Omega_{ij} 
\\
\Rightarrow \rho &= -\frac{1}{\sqrt{r_c}} \Omega^{ij}v^k \partial_k  \Omega_{ij} + \mathcal{O}(\epsilon^4) \ ,
\end{aligned}
\end{equation}

\noindent so that

\begin{equation}
\begin{aligned}
 p + \rho &= \frac{1}{\sqrt{r_c}} \left ( 1 + \frac{P}{r_c} - \Omega^{ij}v^k \partial_k\Omega_{ij} \right ) + \mathcal{O}(\epsilon^4) \ ;
\label{eq:p+rho}
\end{aligned}
\end{equation}

\begin{equation}
(p + \rho)^2 = \frac{1}{r_c} \left ( 1 + \frac{2P}{r_c} - 2\Omega^{ij}v^k \partial_k\Omega_{ij} \right ) + \mathcal{O}(\epsilon^4) \ .
\label{eq:p+rho^2}
\end{equation}

For the remaining of this section, we shall also drop the superscript $(\epsilon)$ from $v_i^{(\epsilon)}$, $P^{(\epsilon)}$ and $(v^2)^{(\epsilon)}$ for simplicity, although we will keep working in the hydrodynamic limit. The performance of the hydrodynamic expansion on the metric components of Eq. \ref{eq:boosted_metric_presure_comps_2} yields

\begin{equation}
\begin{aligned}
g_{tt} &= \gamma^2 \left ( v^2 - r_c\left(p + \rho \right)^2\left ( r - r_h \right ) \right )  
\\
&= \left ( 1 + \frac{v^2}{r_c} + \mathcal{O}(\epsilon^4) \right ) \left ( v^2 - \left ( \left ( 1 + \frac{2P}{r_c} - 2\Omega^{ij}v^k \partial_k\Omega_{ij} \right ) + \mathcal{O}(\epsilon^4) \right )\left [r - \left (2P + \mathcal{O}(\epsilon^4) \right ) \right ]\right )  
\\
&= \left ( 1 + \frac{v^2}{r_c}\right ) \left ( v^2 - \left ( 1 + \frac{2P}{r_c} - 2\Omega^{ij}v^k \partial_k\Omega_{ij} \right )  \left (r - 2P \right )\right ) + \mathcal{O}(\epsilon^4)   
\\
&= \left ( 1 + \frac{v^2}{r_c}\right ) \left ( v^2 -  r + 2P - 2P\frac{r}{r_c} + 2r \Omega^{ij}v^k \partial_k\Omega_{ij} \right ) + \mathcal{O}(\epsilon^4)   
\\
&= v^2 -  r + 2P - 2P\frac{r}{r_c} + 2r \Omega^{ij}v^k \partial_k\Omega_{ij} - v^2 \frac{r}{r_c} + \mathcal{O}(\epsilon^4)  
\\
&=-r +  v^2 \left (1 -\frac{r}{r_c} \right ) + 2P \left (1 -\frac{r}{r_c} \right ) + 2r \Omega^{ij}v^k \partial_k\Omega_{ij} + \mathcal{O}(\epsilon^4) 
\\
\Rightarrow g_{tt} &= -r + \left (1 -\frac{r}{r_c} \right )\left (v^2 + 2P\right ) + 2r \Omega^{ij}v^k \partial_k\Omega_{ij} + \mathcal{O}(\epsilon^4) \ ;
\end{aligned}
\end{equation}

\begin{equation}
\begin{aligned}
g_{tr} &= \gamma \sqrt{r_c} \left(p + \rho \right)  
\\
&= \left ( 1 + \frac{v^2}{2r_c} + \mathcal{O}(\epsilon^4) \right ) \left ( \left ( 1 + \frac{P}{r_c} - \Omega^{ij}v^k \partial_k\Omega_{ij} \right ) + \mathcal{O}(\epsilon^4) \right )   
\\
&= \left ( 1 + \frac{v^2}{2r_c} \right )\left (1 + \frac{P}{r_c} - \Omega^{ij}v^k \partial_k\Omega_{ij} \right ) + \mathcal{O}(\epsilon^4) =
\\ 
&= 1 + \frac{P}{r_c} - \Omega^{ij}v^k \partial_k\Omega_{ij} + \frac{v^2}{2r_c} + \mathcal{O}(\epsilon^4) 
\\
\Rightarrow g_{tr} &= 1 + \frac{v^2 + 2P}{2r_c} - \Omega^{ij}v^k \partial_k\Omega_{ij} + \mathcal{O}(\epsilon^4) \ ;
\end{aligned}
\end{equation}

\begin{equation}
\begin{aligned}
g_{ti} &= \gamma^2 \left(p + \rho \right)^2 \left ( r - r_c \right ) v_i  
\\
&= \left ( 1 + \frac{v^2}{r_c} + \mathcal{O}(\epsilon^4) \right ) \left (\frac{1}{r_c} \left ( 1 + \frac{2P}{r_c} - 2\Omega^{ij}v^k \partial_k\Omega_{ij} \right ) + \mathcal{O}(\epsilon^4) \right )\left ( r - r_c \right )v_i  
\\
&= \left ( 1 + \frac{v^2}{r_c} \right ) \left (\frac{r - r_c}{r_c}\right ) \left (1 + \frac{2P}{r_c} - 2\Omega^{ij}v^k \partial_k\Omega_{ij} \right )v_i + \mathcal{O}(\epsilon^4)  
\\
&= \left ( 1 + \frac{v^2}{r_c} \right ) \left (\frac{r}{r_c}-1\right ) v_i + \mathcal{O}(\epsilon^3) 
\\
\Rightarrow g_{ti} &= -v_i \left ( 1 - \frac{r}{r_c} \right ) + \mathcal{O}(\epsilon^3) \ ; 
\end{aligned}
\end{equation}

\begin{equation}
\begin{aligned}
g_{rr} = 0;
\end{aligned}
\end{equation}

\begin{equation}
\begin{aligned}
g_{ri} &= -\frac{\gamma}{\sqrt{r_c}}\left(p + \rho \right)v_i  
\\
&= - \left( 1 + \frac{v^2}{2r_c} + \mathcal{O}(\epsilon^4) \right ) \frac{1}{\sqrt{r_c}} \left ( \frac{1}{\sqrt{r_c}} \left ( 1 + \frac{P}{r_c} - \Omega^{ij}v^k \partial_k\Omega_{ij} \right ) + \mathcal{O}(\epsilon^4) \right ) v_i 
\\
&=  - \left( 1 + \frac{v^2}{2r_c} + \mathcal{O}(\epsilon^4) \right ) \frac{1}{r_c}\left (1 + \frac{P}{r_c} - \Omega^{ij}v^k \partial_k\Omega_{ij} \right ) v_i + \mathcal{O}(\epsilon^4) 
\\
&= - \left( 1 + \frac{v^2}{2r_c} \right ) \frac{v_i}{r_c} + \mathcal{O}(\epsilon^3) 
\\
\Rightarrow g_{ri} &= -\frac{v_i}{r_c} +  \mathcal{O}(\epsilon^3) \ ; 
\end{aligned}
\end{equation}

\begin{equation}
\begin{aligned}
g_{ij} &=  \Omega_{ij} - \frac{\gamma^2}{r_c} \left(p + \rho \right)^2  \left (r -r_c \right ) v_i v_j  
\\
&= \Omega_{ij} - \left( 1 + \frac{v^2}{2r_c} + \mathcal{O}(\epsilon^4) \right ) \left (\frac{1}{r_c} \left ( 1 + \frac{2P}{r_c} - 2\Omega^{ij}v^k \partial_k\Omega_{ij} \right ) + \mathcal{O}(\epsilon^4) \right ) \left (\frac{r - r_c}{r_c} \right ) v_i v_j  
\\
&= \Omega_{ij} - \left( 1 + \frac{v^2}{2r_c} \right )\left (\frac{r}{r_c} - 1 \right ) \left (1 + \frac{2P}{r_c} - 2\Omega^{ij}v^k \partial_k\Omega_{ij}\right ) \frac{v_i v_j}{r_c} + \mathcal{O}(\epsilon^4)  
\\
&= \Omega_{ij} - \left( 1 + \frac{v^2}{2r_c} \right ) \left (\frac{r}{r_c} - 1 \right ) \frac{v_i v_j}{r_c} + \mathcal{O}(\epsilon^4)  
\\
&= \Omega_{ij} - \left (\frac{r}{r_c} - 1 \right ) \frac{v_i v_j}{r_c} + \mathcal{O}(\epsilon^4) 
\\
\Rightarrow g_{ij} &= \Omega_{ij} + \frac{1}{r_c}\left (1-\frac{r}{r_c} \right )v_iv_j + \mathcal{O}(\epsilon^4)  \ .
\end{aligned}
\end{equation}

Therefore, the full metric is

\begin{equation}
\begin{aligned}
g_{\mu \nu} \mathrm{d}x^\mu \mathrm{d}x^\nu = &-r \mathrm{d}t^2 + 2 \mathrm{d}t \mathrm{d}r + \Omega_{ij}\mathrm{d}x^i\mathrm{d}x^j \ 
\\
&-2 \left ( 1 -  \frac{r}{r_c} \right ) v_i \mathrm{d}x^i \mathrm{d}t - \frac{2}{r_c}v_i \mathrm{d}x^i \mathrm{d}r \ 
\\
&+ \left ( 1 - \frac{r}{r_c}\right ) \left [ \left (v^2 + 2P \right ) \mathrm{d}t^2 + \frac{1}{r_c} v_i v_j \mathrm{d}x^i \mathrm{d}x^j \right ] + \frac{\left (v^2 + 2P \right )}{r_c}\mathrm{d}t \mathrm{d}r \ 
\\
&+ 2r \Omega^{ij}v^k \partial_k\Omega_{ij} \mathrm{d}t^2 - 2\Omega^{ij}v^k \partial_k\Omega_{ij} \mathrm{d}t \mathrm{d}r \ 
\\
&+ \mathcal{O}(\epsilon^3) \ ,
\label{eq:metric_expansion_2}
\end{aligned}
\end{equation}

\noindent where the third and fourth lines have terms of $\mathcal{O}(\epsilon^2)$.

Notice that, by setting $r=r_c$ in Eq. \ref{eq:metric_expansion_2}, one has the following induced metric on $\Sigma_c$,

\begin{equation}
\gamma_{a b} \mathrm{d}x^a \mathrm{d}x^b = -r_c \left( 1 - 2\Omega^{ij}v^k \partial_k\Omega_{ij} \right )\mathrm{d} t^2 + \Omega_{ij}\mathrm{d}x^i \mathrm{d}x^j + \mathcal{O}(\epsilon^3) \ .
\label{eq:bc2_down_gen}
\end{equation}

However, according to the boundary condition as expressed in Eq. \ref{eq:bc2_down} --- namely, that we must have a constant $\kappa$ ---, we must impose

\begin{equation}
\Omega^{ij}v^k \partial_k\Omega_{ij} = 0 \ ,
\label{eq:salvador}
\end{equation}

\noindent so that the induced metric is in fact,

\begin{equation}
\gamma_{a b} \mathrm{d}x^a \mathrm{d}x^b = -r_c \mathrm{d} t^2 + \Omega_{ij}\mathrm{d}x^i \mathrm{d}x^j + \mathcal{O}(\epsilon^3) \ , 
\end{equation}

\noindent exactly as imposed by Eq. \ref{eq:bc2_down}, up to $\mathcal{O}(\epsilon^3)$.

Notice that the constraint of Eq. \ref{eq:salvador} appeared after the imposition of the chosen boundary conditions, being, therefore, of major importance to guarantee consistency. Also, it could be relaxed to $\Omega^{ij}v^k \partial_k\Omega_{ij} = \text{constant}$, and $\kappa$ would still be constant. But, to satisfy the specific boundary condition in Eq. \ref{eq:bc2_down}, and to fix the unjustified energy density acquired by the perfect fluid after the boost (Eq. \ref{eq:fluid_bc_new2}), we must set the constant to zero.

Lastly, notice that Eq.\ref{eq:salvador} may be seen as a tensor density equation, which is then easily promoted to a covariant expression after the multiplication by the weight at both sides. However, the physical meaning of such a constraint is not entirely clear. At first, it seems to imply a constraint on the allowed directions of the velocity field $v^k$. We shall keep this constraint and carry on with the calculation, but bearing in mind that further investigation on the meaning of this condition and its physical content is necessary and shall be conducted in different contexts in future works.

By applying the constraint  of Eq. \ref{eq:salvador}, some of the results calculated above are simplified. Namely, Eqs. \ref{eq:p+rho} and \ref{eq:p+rho^2} reduce respectively to

\begin{equation}
p + \rho = \frac{1}{\sqrt{r_c}} \left ( 1 + \frac{P}{r_c} \right ) + \mathcal{O}(\epsilon^4) \ ;
\end{equation}

\begin{equation}
(p + \rho)^2 = \frac{1}{r_c} \left ( 1 + \frac{2P}{r_c} \right ) + \mathcal{O}(\epsilon^4) \ .
\end{equation}

Prior to calculating the Brown--York stress tensor associated to the metric of Eq. \ref{eq:metric_expansion_2}, one has to calculate the inverse metric components, which is done by expanding the components of Eq. \ref{eq:boosted_metric_inv_presure_comps_2} in the hydrodynamic expansion, which yields

\begin{equation}
\begin{aligned}
g^{tt} &= \frac{\gamma^2 v^2}{r_c^2} 
\\
\Rightarrow g^{tt} &= \frac{v^2}{r_c^2} + \mathcal{O}(\epsilon^4) \ ;
\end{aligned}
\end{equation}

\begin{equation}
\begin{aligned}
g^{tr} &= \frac{\gamma}{\sqrt{r_c}(p + \rho)}  
\\
&= \left( 1 + \frac{v^2}{2r_c} + \mathcal{O}(\epsilon^4) \right ) \frac{1}{\sqrt{r_c}} \left ( \frac{1}{\sqrt{r_c}} \left ( 1 + \frac{P}{r_c} \right ) + \mathcal{O}(\epsilon^4)\right)^{-1} 
\\
&= \left( 1 + \frac{v^2}{2r_c} \right ) \left ( 1 + \frac{P}{r_c} \right)^{-1} + \mathcal{O}(\epsilon^4) 
\\
&= \left( 1 + \frac{v^2}{2r_c} \right ) \left ( 1 - \frac{P}{r_c}  \right) + \mathcal{O}(\epsilon^4) 
\\
&= 1 - \frac{P}{r_c} + \frac{v^2}{2r_c} + \mathcal{O}(\epsilon^4) 
\\
\Rightarrow g^{tr} &= 1 + \frac{v^2 - 2P}{2r_c} + \mathcal{O}(\epsilon^4) \ ;
\end{aligned}
\end{equation}

\begin{equation}
\begin{aligned}
g^{ti} &= \frac{\gamma^2}{r_c} v^i 
\\
\Rightarrow g^{ti} &= \frac{1}{r_c}v^i + \mathcal{O}(\epsilon^4) \ ;
\end{aligned}
\end{equation}

\begin{equation}
\begin{aligned}
g^{rr} &= r-r_h 
\\
\Rightarrow g^{rr} &= r - 2P + \mathcal{O}(\epsilon^4) \ ;
\end{aligned}
\end{equation}

\begin{equation}
\begin{aligned}
g^{ri} &= \frac{\gamma}{\sqrt{r_c}(p + \rho)}v^i 
\\
&= \left( 1 + \frac{v^2}{2r_c} + \mathcal{O}(\epsilon^4) \right ) \frac{1}{\sqrt{r_c}} \left ( \frac{1}{\sqrt{r_c}} \left ( 1 + \frac{P}{r_c} \right ) + \mathcal{O}(\epsilon^4)\right)^{-1}v^i 
\\
&= \left( 1 + \frac{v^2}{2r_c} \right ) \left ( 1 + \frac{P}{r_c} \right)^{-1}v^i  + \mathcal{O}(\epsilon^4) 
\\
&= \left( 1 + \frac{v^2}{2r_c} \right ) \left ( 1 - \frac{P}{r_c} \right)v^i + \mathcal{O}(\epsilon^4) 
\\
&= \left( 1 + \frac{v^2}{2r_c} \right ) v^i + \mathcal{O}(\epsilon^4) 
\\
\Rightarrow g^{ri} &= v^i + \mathcal{O}(\epsilon^3) \ ;
\end{aligned}
\end{equation}

\begin{equation}
\begin{aligned}
g^{ij} &= \Omega^{ij} + \frac{\gamma^2}{r_c} v^i v^j 
\\
\Rightarrow g^{ij} &= \Omega^{ij} + \frac{1}{r_c} v^i v^j + \mathcal{O}(\epsilon^4) \ .
\end{aligned}
\end{equation}

Eq. \ref{eq:norm_vect_down} then yields, exactly as in Sec. \ref{sec:he1},

\begin{equation}
\begin{aligned}
n_{\mu}\mathrm{d}x^{\mu} &= \frac{1}{\sqrt{g^{rr}}} \mathrm{d}r = \frac{1}{\sqrt{r-2P + \mathcal{O}(\epsilon^4)}} \mathrm{d}r 
\\
\Rightarrow n_{\mu}\mathrm{d}x^{\mu} &=  \frac{1}{\sqrt{r}} \left ( 1 + \frac{P}{r} \right ) \mathrm{d}r + \mathcal{O}(\epsilon^4) \ .
\end{aligned}
\end{equation}

Similarly, from Eq. \ref{eq:norm_vect_up} one has the $n^\mu$ components

\begin{equation}
\begin{aligned}
n^t &= \frac{1}{\sqrt{r-2P + \mathcal{O}(\epsilon^4)}} g^{rt} 
\\
&= \frac{1}{\sqrt{r}} \left ( 1 + \frac{P}{r} \right ) \left ( 1 + \frac{v^2 - 2P}{2r_c} + \mathcal{O}(\epsilon^4) \right ) + \mathcal{O}(\epsilon^4)  
\\
&= \frac{1}{\sqrt{r}} \left ( 1 -\frac{P}{r} + \frac{v^2}{2r_c} +\frac{P}{r_c} \right ) + \mathcal{O}(\epsilon^4) 
\\
\Rightarrow n^t &= \frac{1}{\sqrt{r}} \left ( 1 + \frac{v^2}{2r_c} \right ) + \mathcal{O}(\epsilon^4) \ ;
\end{aligned}
\end{equation} 

\begin{equation}
\begin{aligned}
n^r &= \frac{1}{\sqrt{r-2P + \mathcal{O}(\epsilon^4)}} g^{rr} 
\\
&= \frac{1}{\sqrt{r}} \left ( 1 + \frac{P}{r} \right ) \left ( r - 2P + \mathcal{O}(\epsilon^4) \right ) + \mathcal{O}(\epsilon^4)  
\\
&= \frac{1}{\sqrt{r}} \left ( r - 2P + P \right ) + \mathcal{O}(\epsilon^4) 
\\
\Rightarrow n^r &= \frac{1}{\sqrt{r}} \left (r - P \right ) + \mathcal{O}(\epsilon^4) \ ;
\end{aligned}
\end{equation} 

\begin{equation}
\begin{aligned}
n^i &= \frac{1}{\sqrt{r-2P + \mathcal{O}(\epsilon^4)}} g^{ri} 
\\
&= \frac{1}{\sqrt{r}} \left ( 1 + \frac{P}{r} \right ) \left ( v^i + \mathcal{O}(\epsilon^3)\right )+ \mathcal{O}(\epsilon^4) 
\\
\Rightarrow n^i &=  \frac{1}{\sqrt{r}} v^i + \mathcal{O}(\epsilon^3) \ .
\end{aligned}
\end{equation}

The induced metric, according to Eq. \ref{eq:induc_met}, is

\begin{equation}
\begin{aligned}
\gamma_{\mu \nu} \mathrm{d}x^{\mu} \mathrm{d}x^{\nu} &= -r \mathrm{d}t^2 + 2 \mathrm{d}t \mathrm{d}r - \frac{1}{r}\mathrm{d}r^2 + \Omega_{ij}\mathrm{d}x^i\mathrm{d}x^j \ 
\\
&-2 \left ( 1 -  \frac{r}{r_c} \right ) v_i \mathrm{d}x^i \mathrm{d}t - \frac{2}{r_c}v_i \mathrm{d}x^i \mathrm{d}r \ 
\\
&+ \left ( 1 - \frac{r}{r_c}\right ) \left [ \left (v^2 + 2P \right ) \mathrm{d}t^2 + \frac{1}{r_c} v_i v_j \mathrm{d}x^i \mathrm{d}x^j \right ] + \frac{\left (v^2 + 2P \right )}{r_c}\mathrm{d}t \mathrm{d}r - \frac{2P}{r^2} \mathrm{d}r^2 \ 
\\
&+ \mathcal{O}(\epsilon^3) \ , 
\label{eq:induced_metric_2}
\end{aligned}
\end{equation}

\noindent which clearly reduces to Eq. \ref{eq:bc2_down} on $\Sigma_c$. 

For the components of the inverse induced metric $\gamma^{\mu \nu} = g^{\mu \nu} - n^{\mu}n^{\nu}$, one has, similarly to Eqs. \ref{eq:iim_first} to \ref{eq:iim_last},

\begin{equation}
\begin{aligned}
\gamma^{tt} &= \frac{v^2}{r_c^2} - \frac{1}{r} \left (1 + \frac{v^2}{2 r_c}\right )^2 + \mathcal{O}(\epsilon^4) 
\\
&= \frac{v^2}{r_c^2} - \frac{1}{r} \left (1 + \frac{v^2}{r_c} \right ) + \mathcal{O}(\epsilon^4) 
\\
\Rightarrow \gamma^{tt} &= \frac{v^2}{r_c} \left (\frac{1}{r_c} - \frac{1}{r} \right ) - \frac{1}{r} + \mathcal{O}(\epsilon^4) \ ;
\label{eq:iim_first_2}
\end{aligned}
\end{equation}

\begin{equation}
\begin{aligned}
\gamma^{tr} &= 1 + \frac{v^2 - 2P}{2r_c} - \frac{1}{r} \left ( 1 + \frac{v^2}{2r_c} \right )\left (r - P \right ) + \mathcal{O}(\epsilon^4) 
\\
&= 1 + \frac{v^2 - 2P}{2r_c} - \frac{1}{r} \left ( r - P + \frac{v^2r}{2r_c}\right ) + \mathcal{O}(\epsilon^4) 
\\
\Rightarrow \gamma^{tr} &= P \left ( \frac{1}{r} - \frac{1}{r_c} \right ) + \mathcal{O}(\epsilon^4) \ ;
\end{aligned}
\end{equation}

\begin{equation}
\begin{aligned}
\gamma^{ti} &= \frac{1}{r_c}v^i - \frac{1}{r} \left ( 1 + \frac{v^2}{2r_c}\right ) v^i +  \mathcal{O}(\epsilon^3) 
\\
\Rightarrow \gamma^{ti} &= \left (\frac{1}{r_c} - \frac{1}{r} \right ) v^i  +  \mathcal{O}(\epsilon^3) \ ;
\end{aligned}
\end{equation}

\begin{equation}
\begin{aligned}
\gamma^{rr} &= r - 2P - \frac{1}{r} \left (r - P \right )^2 + \mathcal{O}(\epsilon^4)  
\\
&= r - 2P - \frac{1}{r} \left (r^2 - 2Pr \right ) + \mathcal{O}(\epsilon^4) 
\\
\Rightarrow \gamma^{rr} &= \mathcal{O}(\epsilon^4) \ ;
\end{aligned}
\end{equation}

\begin{equation}
\begin{aligned}
\gamma^{ri} &= v^i - \frac{1}{r} \left (r - P \right )v^i + \mathcal{O}(\epsilon^3) 
\\
\Rightarrow \gamma^{ri} &= \mathcal{O}(\epsilon^3) \ ;
\end{aligned}
\end{equation}

\begin{equation}
\begin{aligned}
\gamma^{ij} &= \Omega^{ij} + \frac{1}{r_c} v^i v^j - \frac{1}{r}v^i v^j + \mathcal{O}(\epsilon^4) 
\\
\Rightarrow \gamma^{ij} &= \Omega^{ij} + \left (\frac{1}{r_c} - \frac{1}{r} \right ) v^i v^j + \mathcal{O}(\epsilon^4) \ .
\label{eq:iim_last_2}
\end{aligned}
\end{equation}

Before we proceed, let us make an important remark. Since we are now working in a background with a metric whose components are no longer constant, we must be very careful to guarantee that the quantities we are considering are manifestly covariant. Namely, the connection coefficients associated to the induced metric $\gamma_{\mu \nu}$ are no longer null, so that they must be taken into account together with partial derivatives to construct covariant derivatives. 

In this context, it may be enlightening to consider a slightly different way to calculate the extrinsic curvature of that presented in Eq. \ref{eq:ext_curv}. Since the Lie derivative is connection independent, we may simply substitute the partial derivatives by covariant derivatives, which will yield precisely the same result. Notice, however, that due to the metric compatibility condition, $\nabla_\mu \gamma_{\nu \sigma} = 0$, so that the first term in the extrinsic curvature vanishes, and we are left with

\begin{equation}
K_{\mu \nu} = \frac{1}{2} \left (\gamma_{\sigma \nu} \nabla_\mu n^\sigma + \gamma_{\mu \sigma} \nabla_\nu n^\sigma \right ) \ ,
\label{eq:ext_curv_2}
\end{equation}

\noindent which will be our working definition of the extrinsic curvature in this section. We stress that this is entirely equivalent to our previous definition of the extrinsic curvature in terms of the Lie derivative, given its connection independence.

Just like before, we will be only interested in the extrinsic curvature on $\Sigma_c$, that is, we are interested in the components

\begin{equation}
\begin{gathered}
K_{tt} = \left .\gamma_{t \sigma} \nabla_t n^\sigma \right\rvert_{r = r_c} \ ;
\\
K_{ti} = \left . \frac{1}{2}\left ( \gamma_{t \sigma} \nabla_i n^\sigma + \gamma_{i \sigma} \nabla_t n^\sigma \right ) \right\rvert_{r = r_c} \ ;
\\
K_{ij} = \left . \frac{1}{2}\left ( \gamma_{j\sigma} \nabla_i n^\sigma + \gamma_{i \sigma} \nabla_j n^\sigma \right ) \right\rvert_{r = r_c} \ .
\label{eq:k_comps}
\end{gathered}
\end{equation}

To calculate the components above, we must first calculate the connection coefficients (Christoffel symbols) associated to $\gamma_{\mu \nu}$, given by

\begin{equation}
\Gamma^\sigma_{\mu \nu} = \frac{1}{2} \gamma^{\sigma \lambda} \left ( \partial_\mu \gamma_{\nu \lambda} + \partial_\nu \gamma_{\lambda \mu} - \partial_\lambda \gamma_{\mu \nu} \right ) \ ,
\end{equation}

\noindent which are, according to the components of Eqs. \ref{eq:iim_first_2} to \ref{eq:iim_last_2} and Eq. \ref{eq:induced_metric_2},

\begin{equation}
\begin{aligned}
\Gamma^t_{tt} &= \frac{1}{2} \gamma^{t\mu} \left ( \partial_t \gamma_{t\mu} + \partial_t \gamma_{t \mu} - \partial_\mu \gamma_{tt} \right )  
\\
&= \frac{1}{2} \gamma^{tt} \left ( \partial_t \gamma_{tt} \right ) + \frac{1}{2} \gamma^{tr} \left ( 2 \partial_t \gamma_{tr} - \partial_r \gamma_{tt} \right ) + \frac{1}{2} \gamma^{ti} \left (2\partial_t \gamma_{ti} - \partial_i \gamma_{tt} \right ) 
\\
\Rightarrow \Gamma^t_{tt} &= \frac{1}{2} \left ( \frac{1}{r} - \frac{1}{r_c} \right ) P + \mathcal{O}(\epsilon^4) \ ;
\end{aligned}
\end{equation}

\begin{equation}
\begin{aligned}
\Gamma^t_{tr} &= \frac{1}{2} \gamma^{tt} \left ( \partial_r \gamma_{tt} \right ) + \frac{1}{2} \gamma^{tr} \left ( \partial_t \gamma_{rr}\right ) + \frac{1}{2} \gamma^{ti} \left (\partial_t \gamma_{ri} + \partial_r \gamma_{ti} - \partial_i \gamma_{tr} \right ) 
\\
\Rightarrow \Gamma^t_{tr} &= \frac{1}{2r} \left ( 1 + \frac{v^2 + 2P}{r_c} \right ) + \mathcal{O}(\epsilon^4) \ ;
\end{aligned}
\end{equation}

\begin{equation}
\begin{aligned}
\Gamma^t_{ti} &= \frac{1}{2} \gamma^{tt} \left ( \partial_i \gamma_{tt} \right ) + \frac{1}{2} \gamma^{tr} \left ( \partial_t \gamma_{ir} + \partial_i \gamma_{tr} - \partial_r \gamma_{ti} \right ) + \frac{1}{2} \gamma^{tj} \left (\partial_t \gamma_{ij} + \partial_i \gamma_{tj} - \partial_j \gamma_{ti} \right ) 
\\
\Rightarrow \Gamma^t_{ti} &= \mathcal{O}(\epsilon^3) \ ;
\end{aligned}
\end{equation}

\begin{equation}
\begin{aligned}
\Gamma^t_{rr} &= \frac{1}{2} \gamma^{tt} \left ( 2 \partial_r \gamma_{rt} - \partial_t \gamma_{rr} \right ) + \frac{1}{2} \gamma^{tr} \left ( \partial_r \gamma_{rr}\right ) + \frac{1}{2} \gamma^{ti} \left ( 2 \partial_r \gamma_{ri} - \partial_i \gamma_{rr} \right ) 
\\
\Rightarrow \Gamma^t_{rr} &= \frac{1}{2r^2} \left ( \frac{1}{r} - \frac{1}{r_c} \right ) P + \mathcal{O}(\epsilon^4) \ ;
\end{aligned}
\end{equation}

\begin{equation}
\begin{aligned}
\Gamma^t_{ri} &= \frac{1}{2} \gamma^{tt} \left ( \partial_r \gamma_{it} + \partial_i \gamma_{rt} - \partial_t \gamma_{ri} \right ) + \frac{1}{2} \gamma^{tr} \left ( \partial_i \gamma_{rr}\right ) + \frac{1}{2} \gamma^{tj} \left ( \partial_r \gamma_{ij} + \partial_i \gamma_{rj} - \partial_j \gamma_{ri} \right ) 
\\
\Rightarrow \Gamma^t_{ri} &= -\frac{1}{2r r_c}v_i + \mathcal{O}(\epsilon^3) \ ;
\end{aligned}
\end{equation}

\begin{equation}
\begin{aligned}
\Gamma^t_{ij} &= \frac{1}{2} \gamma^{tt} \left (\partial_i \gamma_{jt} + \partial_j \gamma_{it} - \partial_t \gamma_{ij} \right ) + \frac{1}{2} \gamma^{tr} \left ( \partial_i \gamma_{jr} + \partial_j \gamma_{ir} - \partial_r \gamma_{ij}\right ) \ 
\\
&+ \frac{1}{2} \gamma^{tk} \left ( \partial_i \gamma_{jk} + \partial_j \gamma_{ik} - \partial_k \gamma_{ij} \right ) 
\\
\Rightarrow \Gamma^t_{ij} &= \frac{1}{2} \left [ \left ( 1 - \frac{r}{r_c} \right ) \frac{1}{r} \left ( \partial_i v_j + \partial_j v_i \right )
\right ] \ 
\\
&+ \frac{1}{2} \left [ \left (\frac{1}{r_c} - \frac{1}{r} \right ) v^k \left ( \partial_i \Omega_{jk} + \partial_j \Omega_{ik} - \partial_k \Omega_{ij} \right )\right ] + \mathcal{O}(\epsilon^4) \ ;
\end{aligned}
\end{equation}

\begin{equation}
\begin{aligned}
\Gamma^r_{tt} &= \frac{1}{2} \gamma^{r\mu} \left ( \partial_t \gamma_{t\mu} + \partial_t \gamma_{t\mu} - \partial_\mu \gamma_{tt} \right )  
\\
&= \frac{1}{2} \gamma^{rt} \left ( 2 \partial_t \gamma_{tt} - \partial_t \gamma_{tt} \right ) + \mathcal{O}(\epsilon^4) 
\\
\Rightarrow \Gamma^r_{tt} &= \mathcal{O}(\epsilon^4) \ ;
\end{aligned}
\end{equation}

\begin{equation}
\begin{aligned}
\Gamma^r_{tr} &= \frac{1}{2} \gamma^{rt} \left ( \partial_t \gamma_{rt} + \partial_r \gamma_{tt} - \partial_t \gamma_{tr} \right ) + \mathcal{O}(\epsilon^4) 
\\
\Rightarrow \Gamma^r_{tr} &= \frac{1}{2} \left (\frac{1}{r_c} - \frac{1}{r} \right )P + \mathcal{O}(\epsilon^4) \ ;
\end{aligned}
\end{equation}

\begin{equation}
\begin{aligned}
\Gamma^r_{ti} &= \frac{1}{2} \gamma^{rt} \left ( \partial_t \gamma_{it} + \partial_i \gamma_{tt} - \partial_t\gamma_{ti} \right ) + \mathcal{O}(\epsilon^4) 
\\
\Rightarrow \Gamma^r_{ti} &= \mathcal{O}(\epsilon^4) \ ;
\end{aligned}
\end{equation}

\begin{equation}
\begin{aligned}
\Gamma^r_{rr} &= \frac{1}{2} \gamma^{rt} \left ( 2\partial_r \gamma_{rt} - \partial_t \gamma_{rr} \right ) + \mathcal{O}(\epsilon^4) 
\\
\Rightarrow \Gamma^r_{rr} &= \mathcal{O}(\epsilon^4) \ ;
\end{aligned}
\end{equation}

\begin{equation}
\begin{aligned}
\Gamma^r_{ri} &= \frac{1}{2} \gamma^{rt} \left ( \partial_r \gamma_{it} + \partial_i \gamma_{rt} - \partial_t \gamma_{ri} \right ) + \mathcal{O}(\epsilon^4) 
\\
\Rightarrow \Gamma^r_{ri} &= \mathcal{O}(\epsilon^3) \ ;
\end{aligned}
\end{equation}

\begin{equation}
\begin{aligned}
\Gamma^r_{ij} &= \frac{1}{2} \gamma^{rt} \left ( \partial_i \gamma_{jt} + \partial_j \gamma_{it} - \partial_t \gamma_{ij} \right ) + \mathcal{O}(\epsilon^4) 
\\
\Rightarrow \Gamma^r_{ij} &= \mathcal{O}(\epsilon^4) \ ;
\end{aligned}
\end{equation}

\begin{equation}
\begin{aligned}
\Gamma^i_{tt} &= \frac{1}{2} \gamma^{i\mu} \left ( \partial_t \gamma_{t\mu} + \partial_t \gamma_{t\mu} - \partial_\mu \gamma_{tt} \right ) 
\\
&= \frac{1}{2} \gamma^{it} \left ( \partial_t \gamma_{tt} \right ) + \frac{1}{2} \gamma^{ij} \left ( 2 \partial_t \gamma_{tj} - \partial_j \gamma_{tt} \right ) + \mathcal{O}(\epsilon^4) 
\\
\Rightarrow \Gamma^i_{tt} &= \mathcal{O}(\epsilon^4) \ ;
\end{aligned}
\end{equation}

\begin{equation}
\begin{aligned}
\Gamma^i_{tr} &= \frac{1}{2} \gamma^{it} \left ( \partial_t \gamma_{rt} + \partial_r \gamma_{tt} - \partial_t \gamma_{tr} \right ) + \frac{1}{2} \gamma^{ij} \left ( \partial_t \gamma_{rj} + \partial_r \gamma_{tj} - \partial_j \gamma_{tr} \right ) + \mathcal{O}(\epsilon^4) 
\\
\Rightarrow \Gamma^i_{tr} &= \frac{1}{2r} v^i +  \mathcal{O}(\epsilon^4) \ ;
\end{aligned}
\end{equation}

\begin{equation}
\begin{aligned}
\Gamma^i_{tj} &= \frac{1}{2} \gamma^{it} \left ( \partial_t \gamma_{jt} + \partial_j \gamma_{tt} - \partial_t \gamma_{tj} \right ) + \frac{1}{2} \gamma^{ik} \left ( \partial_t \gamma_{jk} + \partial_j \gamma_{tk} - \partial_k \gamma_{tj} \right ) + \mathcal{O}(\epsilon^4) 
\\
\Rightarrow \Gamma^i_{tj} &= \frac{1}{2} \left ( \frac{r}{r_c} - 1 \right ) \Omega^{ik} \left ( \partial_j v_k - \partial_k v_j \right ) + \mathcal{O}(\epsilon^4) \ ;
\end{aligned}
\end{equation}

\begin{equation}
\begin{aligned}
\Gamma^i_{rr} &= \frac{1}{2} \gamma^{it} \left ( 2\partial_r \gamma_{rt} - \partial_t \gamma_{rr} \right ) + \frac{1}{2} \gamma^{ij} \left (2 \partial_r \gamma_{rj} - \partial_j \gamma_{rr} \right ) + \mathcal{O}(\epsilon^4) 
\\
\Rightarrow \Gamma^i_{rr} &= \mathcal{O}(\epsilon^4) \ ;
\end{aligned}
\end{equation}

\begin{equation}
\begin{aligned}
\Gamma^i_{rj} &= \frac{1}{2} \gamma^{it} \left ( \partial_r \gamma_{jt} + \partial_j \gamma_{rt} - \partial_t \gamma_{rj} \right ) + \frac{1}{2} \gamma^{ik} \left ( \partial_r \gamma_{jk} + \partial_j \gamma_{rk} - \partial_k \gamma_{rj} \right ) + \mathcal{O}(\epsilon^4) 
\\
\Rightarrow \Gamma^i_{rj} &= - \frac{1}{2r_c} \left (  \frac{1}{r}v^i v_j + \Omega^{ik} \left ( \partial_j v_k - \partial_k v_j \right ) \right) + \mathcal{O}(\epsilon^4) \ ;
\end{aligned}
\end{equation}

\begin{equation}
\begin{aligned}
\Gamma^i_{jk} &= \frac{1}{2} \gamma^{it} \left ( \partial_j \gamma_{kt} + \partial_k \gamma_{jt} - \partial_t \gamma_{jk}\right ) + \frac{1}{2} \gamma^{il} \left ( \partial_j \gamma_{kl} + \partial_k\gamma_{jl} - \partial_l \gamma_{jk} \right ) + \mathcal{O}(\epsilon^4) 
\\
\Rightarrow \Gamma^i_{jk} &= \frac{1}{2} \Omega^{il} \left ( \partial_j \Omega_{kl} + \partial_k \Omega_{jl} - \partial_l \Omega_{jk} \right ) + \mathcal{O}(\epsilon^4) \ .
\end{aligned}
\end{equation}

Apart from the connection coefficients above, to calculate the covariant derivative of the $n^\mu$ components we also need their partial derivatives

\begin{equation}
\begin{aligned}
\partial_t n^t &= \partial_t \left ( \frac{1}{\sqrt{r}} \left ( 1 + \frac{v^2}{2r_c} \right ) + \mathcal{O}(\epsilon^4) \right ) 
\\
\Rightarrow \partial_t n^t &= \mathcal{O}(\epsilon^4) \ ;
\end{aligned}
\end{equation}

\begin{equation}
\begin{aligned}
\partial_t n^r &= \partial_t \left (\frac{1}{\sqrt{r}} \left ( r-P \right ) + \mathcal{O}(\epsilon^4) \right ) 
\\
\Rightarrow \partial_t n^r &= \mathcal{O}(\epsilon^4) \ ;
\end{aligned}
\end{equation}

\begin{equation}
\begin{aligned}
\partial_t n^i &= \partial_t \left (\frac{1}{\sqrt{r}} v^i  + \mathcal{O}(\epsilon^3) \right ) 
\\
\Rightarrow \partial_t n^i &= \mathcal{O}(\epsilon^3) \ ;
\end{aligned}
\end{equation}

\begin{equation}
\begin{aligned}
\partial_r n^t &= \partial_r \left ( \frac{1}{\sqrt{r}} \left ( 1 + \frac{v^2}{2r_c} \right ) + \mathcal{O}(\epsilon^4) \right ) 
\\
\Rightarrow \partial_r n^t &= -\frac{1}{2r^{3/2}} \left (1 + \frac{v^2}{2r_c} \right ) + \mathcal{O}(\epsilon^4) \ ;
\end{aligned}
\end{equation}

\begin{equation}
\begin{aligned}
\partial_r n^r &= \partial_r \left (\frac{1}{\sqrt{r}} \left ( r-P \right ) + \mathcal{O}(\epsilon^4) \right ) 
\\
\Rightarrow \partial_r n^r &= \frac{1}{2\sqrt{r}} \left (1 + \frac{P}{r} \right ) + \mathcal{O}(\epsilon^4) \ ;
\end{aligned}
\end{equation}

\begin{equation}
\begin{aligned}
\partial_r n^i &= \partial_r\left (\frac{1}{\sqrt{r}} v^i  + \mathcal{O}(\epsilon^3) \right ) 
\\
\Rightarrow \partial_r n^i &= -\frac{1}{2r^{3/2}} v^i + \mathcal{O}(\epsilon^3) \ ;
\end{aligned}
\end{equation}

\begin{equation}
\begin{aligned}
\partial_i n^t &= \partial_i \left ( \frac{1}{\sqrt{r}} \left ( 1 + \frac{v^2}{2r_c} \right ) + \mathcal{O}(\epsilon^4) \right ) 
\\
\Rightarrow \partial_i n^t &= \mathcal{O}(\epsilon^3) \ ;
\end{aligned}
\end{equation}

\begin{equation}
\begin{aligned}
\partial_i n^r &= \partial_i \left (\frac{1}{\sqrt{r}} \left ( r-P \right ) + \mathcal{O}(\epsilon^4) \right ) 
\\
\Rightarrow \partial_i n^r &= \mathcal{O}(\epsilon^3) \ ;
\end{aligned}
\end{equation}

\begin{equation}
\begin{aligned}
\partial_i n^j &= \partial_i \left (\frac{1}{\sqrt{r}} v^j  + \mathcal{O}(\epsilon^3) \right ) 
\\
\Rightarrow \partial_i n^j &= \frac{1}{\sqrt{r}} \partial_i v^j + \mathcal{O}(\epsilon^3) \ .
\end{aligned}
\end{equation}

Thus, one has the covariant derivative of the $n^\mu$ components,

\begin{equation}
\begin{aligned}
\nabla_t n^t &= \partial_t n^t + \Gamma^t_{tt}n^t + \Gamma^t_{tr}n^r + \Gamma^t_{ti}n^i 
\\
\Rightarrow \nabla_t n^t &= \frac{1}{2\sqrt{r}} \left ( \frac{P + v^2 + r_c}{r_c}\right ) + \mathcal{O}(\epsilon^4) \ ;
\end{aligned}
\end{equation}

\begin{equation}
\begin{aligned}
\nabla_r n^t &= \partial_r n^t + \Gamma^t_{rt}n^t + \Gamma^t_{rr}n^r + \Gamma^t_{ri}n^i 
\\
\Rightarrow \nabla_r n^t &= \frac{1}{2r^{3/2}} \left ( \frac{1}{r} + \frac{1}{r_c}\right ) P + \mathcal{O}(\epsilon^4) \ ;
\end{aligned}
\end{equation}

\begin{equation}
\begin{aligned}
\nabla_i n^t &= \partial_i n^t + \Gamma^t_{it}n^t + \Gamma^t_{ir}n^r + \Gamma^t_{ij}n^j 
\\
\Rightarrow \nabla_i n^t &= -\frac{1}{2\sqrt{r}} \frac{1}{r_c} v_i + \mathcal{O}(\epsilon^3) \ ;
\end{aligned}
\end{equation}

\begin{equation}
\begin{aligned}
\nabla_t n^r &= \partial_t n^r + \Gamma^r_{tt}n^t + \Gamma^r_{tr}n^r + \Gamma^r_{ti}n^i 
\\
\Rightarrow \nabla_t n^r &= \frac{\sqrt{r}}{2} \left ( \frac{1}{r_c} - \frac{1}{r}\right ) P + \mathcal{O}(\epsilon^4) \ ;
\end{aligned}
\end{equation}

\begin{equation}
\begin{aligned}
\nabla_r n^r &= \partial_r n^r + \Gamma^r_{rt}n^t + \Gamma^r_{rr}n^r + \Gamma^r_{ri}n^i 
\\
\Rightarrow \nabla_r n^r &= \frac{1}{2\sqrt{r}} \left ( 1 + \frac{P}{r_c} \right ) + \mathcal{O}(\epsilon^4) \ ;
\end{aligned}
\end{equation}

\begin{equation}
\begin{aligned}
\nabla_i n^r &= \partial_i n^r + \Gamma^r_{it}n^t + \Gamma^r_{ir}n^r + \Gamma^r_{ij}n^j 
\\
\Rightarrow \nabla_i n^r &= \mathcal{O}(\epsilon^3) \ ;
\end{aligned}
\end{equation}

\begin{equation}
\begin{aligned}
\nabla_t n^i &= \partial_t n^i + \Gamma^i_{tt}n^t + \Gamma^i_{tr}n^r + \Gamma^i_{tj}n^j 
\\
\Rightarrow \nabla_t n^i &= \frac{1}{2\sqrt{r}} v^i+ \mathcal{O}(\epsilon^3) \ ;
\end{aligned}
\end{equation}

\begin{equation}
\begin{aligned}
\nabla_r n^i &= \partial_r n^i + \Gamma^i_{rt}n^t + \Gamma^i_{rr}n^r + \Gamma^i_{rj}n^j 
\\
\Rightarrow \nabla_r n^i &= \mathcal{O}(\epsilon^3) \ ;
\end{aligned}
\end{equation}

\begin{equation}
\begin{aligned}
\nabla_j n^i &= \partial_j n^i + \Gamma^i_{jt}n^t + \Gamma^i_{jr}n^r + \Gamma^i_{jk}n^k 
\\
\Rightarrow \nabla_j n^i &= \frac{1}{\sqrt{r}} \left [ \nabla_jv^i - \frac{1}{2r_c}v_j v^i + \frac{1}{2} \Omega^{ik} \left (\partial_kv_j - \partial_jv_k \right )\right ] + \mathcal{O}(\epsilon^4) \ .
\end{aligned}
\end{equation}

Now we can finally calculate the components of Eq. \ref{eq:k_comps},

\begin{equation}
\begin{aligned}
K_{tt} &= \left .\gamma_{\sigma t} \nabla_t n^\sigma \right\rvert_{r = r_c}  
\\
&= \left. \left ( \gamma_{tt} \nabla_t n^t + \gamma_{tr} \nabla_t n^r+ \gamma_{ti} \nabla_t n^i \right )\right\rvert_{r = r_c}  
\\
&= \left .\left ( -r + \left ( 1 - \frac{r}{r_c} \right ) \left (v^2 + 2P \right ) \right )\frac{1}{2\sqrt{r}}\left (\frac{P + v^2 + r_c}{r_c} \right )\right\rvert_{r = r_c} \ 
\\
&+ \left .\left ( 1 + \frac{v^2 + 2P}{2 r_c} \right ) \frac{\sqrt{r}}{2} \left ( \frac{1}{r_c} - \frac{1}{r}\right )P\right\rvert_{r = r_c} \ 
\\
&+ \left .\left (\frac{r}{r_c} - 1 \right )\frac{1}{2\sqrt{r}}v^i v_i \right\rvert_{r = r_c} + \mathcal{O}(\epsilon^4) 
\\
\Rightarrow K_{tt} &= -\frac{\left ( P + v^2 + r_c \right )}{2\sqrt{r_c}} + \mathcal{O}(\epsilon^4) \ ;
\end{aligned}
\end{equation}

\begin{equation}
\begin{aligned}
K_{ti} &= \left .\frac{1}{2} \left ( \gamma_{t \sigma} \nabla_i n^\sigma + \gamma_{i\sigma} \nabla_t n^\sigma \right ) \right\rvert_{r = r_c}  
\\
&= \left.\frac{1}{2} \left ( \gamma_{tt} \nabla_i n^t + \gamma_{tr} \nabla_i n^r+ \gamma_{tj} \nabla_i n^j \right )\right\rvert_{r = r_c} \ 
\\
&+ \left.\frac{1}{2} \left ( \gamma_{it} \nabla_t n^t + \gamma_{ir} \nabla_t n^r+ \gamma_{ij} \nabla_t n^j \right )\right\rvert_{r = r_c}  
\\
&= \left.\frac{1}{2} (-r)\left( -\frac{1}{2 \sqrt{r}} \frac{1}{r_c} v_i\right ) \right\rvert_{r = r_c} + \mathcal{O}(\epsilon^4) \ +
\\
&+ \left.\frac{1}{2} \left [ \left (\frac{r}{r_c} - 1 \right ) \frac{1}{2\sqrt{r}} v_i + \left (\Omega_{ij} + \left (1 - \frac{r}{r_c} \right )\frac{1}{r_c}v_i v_j \right )\frac{1}{2\sqrt{r}} v^j \right ]\right\rvert_{r = r_c} + \mathcal{O}(\epsilon^3) 
\\
\Rightarrow K_{it} &= \frac{1}{2\sqrt{r_c}}v_i + \mathcal{O}(\epsilon^3)
\end{aligned}
\end{equation}

\begin{equation}
\begin{aligned}
K_{ij} &= \left . \frac{1}{2}\left ( \gamma_{j\sigma} \nabla_i n^\sigma + \gamma_{i \sigma} \nabla_j n^\sigma \right ) \right\rvert_{r = r_c}  
\\
&= \left. \frac{1}{2}\left ( \gamma_{jt} \nabla_i n^t + \gamma_{jr} \nabla_i n^r+ \gamma_{jk} \nabla_i n^k \right )\right\rvert_{r = r_c} \ 
\\
&+ \left. \frac{1}{2}\left ( \gamma_{it} \nabla_j n^t + \gamma_{ir} \nabla_j n^r+ \gamma_{ik} \nabla_j n^k \right )\right\rvert_{r = r_c} 
\\
&= \left. \frac{1}{2}\left [\left (\frac{r}{r_c}-1 \right )v_j \left (-\frac{1}{2\sqrt{r}} \frac{1}{r_c} v_i \right ) \right ] \right\rvert_{r = r_c}  \ 
\\
&+ \left . \frac{1}{2}\left (\Omega_{jk} + \left (1 - \frac{r}{r_c} \right )\frac{1}{r_c}v_j v_k \right ) \frac{1}{\sqrt{r}} \left (\nabla_iv^k - \frac{1}{2r_c}v_i v^k + \frac{1}{2} \Omega^{kl} \left (\partial_lv_i - \partial_iv_l \right ) \right ) \right\rvert_{r = r_c} \ 
\\
&+ \left. \frac{1}{2}\left [\left (\frac{r}{r_c}-1 \right )v_i \left (-\frac{1}{2\sqrt{r}} \frac{1}{r_c} v_j \right ) \right ] \right\rvert_{r = r_c}  \ 
\\
&+ \left . \frac{1}{2}\left (\Omega_{ik} + \left (1 - \frac{r}{r_c} \right )\frac{1}{r_c}v_i v_k \right ) \frac{1}{\sqrt{r}} \left (\nabla_jv^k - \frac{1}{2r_c}v_j v^k + \frac{1}{2} \Omega^{kl} \left ( \partial_lv_j - \partial_jv_l \right ) \right ) \right\rvert_{r = r_c} \ 
\\
&+ \mathcal{O}(\epsilon^4) 
\\
&= \frac{1}{\sqrt{r_c}} \left ( \frac{1}{2}\left (\nabla_iv_j + \nabla_jv_i \right ) - \frac{1}{2r_c} v_i v_j + \frac{1}{4} \left (-\partial_i v_j + \partial_j v_i - \partial_j v_i + \partial_i v_j \right ) \right ) \ 
\\
&+ \mathcal{O}(\epsilon^4) 
\\
\Rightarrow K_{ij} &= \frac{1}{\sqrt{r_c}} \left ( \frac{1}{2}\left (\nabla_iv_j + \nabla_jv_i \right ) - \frac{1}{2r_c} v_i v_j \right ) + \mathcal{O}(\epsilon^4) \ .
\end{aligned}
\end{equation}

Therefore, the extrinsic curvature on $\Sigma_c$ is

\begin{equation}
\begin{aligned}
K_{ab}\mathrm{d}x^a \mathrm{d}x^b &= -\frac{\left ( P + v^2 + r_c\right )}{2\sqrt{r_c}} \mathrm{d}t^2 + \frac{1}{\sqrt{r_c}}v_i \mathrm{d}x^i \mathrm{d}t \ 
\\
&+ \frac{1}{2\sqrt{r_c}} \left ( \nabla_i v_j + \nabla_j v_i - \frac{1}{r_c} v_i v_j \right ) \mathrm{d}x^i \mathrm{d}x^j + \mathcal{O}(\epsilon^3) \ ,
\end{aligned}
\end{equation}

\noindent whose trace is

\begin{equation}
\begin{aligned}
 K &= \gamma^{ab}K_{ab} = \gamma^{tt} K_{tt} + \gamma^{it} K_{it} +  \gamma^{ij} K_{ij} =
 \\
 &= \left ( -\frac{1}{r_c}  \right ) \left ( -\frac{\left ( P + v^2 + r_c\right )}{2\sqrt{r_c}} \right )  \ 
 \\
 &+ \left ( \Omega^{ij} \right ) \left [ \frac{1}{2\sqrt{r_c}} \left ( \nabla_i v_j + \nabla_j v_i - \frac{1}{r_c} v_i v_j \right ) \right ] + \mathcal{O}(\epsilon^4)  
 \\
 &= \frac{(P + r_c + v^2)}{2r_c^{3/2}} + \frac{1}{\sqrt{r_c}} \nabla^iv_i - \frac{v^2}{2r_c^{3/2}} +  \mathcal{O}(\epsilon^4) 
 \\
 \Rightarrow K &= \frac{1}{2\sqrt{r_c}} \left (\frac{P + r_c}{r_c} + 2\nabla^i v_i \right ) + \mathcal{O}(\epsilon^4) \ .
\end{aligned}
\end{equation}

The first term of the Brown--York stress tensor, according to Eq. \ref{eq:brown_york}, is such that

\begin{equation}
\begin{aligned}
2 K \gamma_{ab} \mathrm{d}x^a \mathrm{d}x^b &= \frac{1}{\sqrt{r_c}} \left (\frac{P + r_c}{r_c} + 2\nabla^i v_i \right ) \left ( -r_c \mathrm{d}t^2 \right ) \ 
\\
&+ \frac{1}{\sqrt{r_c}} \left (\frac{P + r_c}{r_c} + 2\nabla^i v_i \right ) \left (\Omega_{ij}\mathrm{d}x^i \mathrm{d}x^j \right ) + \mathcal{O}(\epsilon^4)  
\\
&= \left [ -\frac{r_c}{\sqrt{r_c}} \left ( \frac{P + r_c}{r_c} + 2\nabla^i v_i \right )\right ]\mathrm{d}t^2 \ 
\\
&+ \frac{1}{\sqrt{r_c}} \left [ \frac{P + r_c}{r_c} + 2\nabla^k v_k\right ] \Omega_{ij}\mathrm{d}x^i \mathrm{d}x^j + \mathcal{O}(\epsilon^4) 
\\
\Rightarrow 2 K \gamma_{ab} \mathrm{d}x^a \mathrm{d}x^b &= \left [ -\frac{\left (P + r_c \right )}{\sqrt{r_c}} - 2\sqrt{r_c}\nabla^i v_i\right ]\mathrm{d}t^2 \ 
\\
&+ \frac{1}{\sqrt{r_c}} \left [ \frac{P + r_c}{r_c} + 2\nabla^k v_k \right ] \Omega_{ij}\mathrm{d}x^i \mathrm{d}x^j + \mathcal{O}(\epsilon^4) \ , 
\end{aligned}
\end{equation}

\noindent whilst the second term is simply

\begin{equation}
\begin{aligned}
-2K_{ab}\mathrm{d}x^a \mathrm{d}x^b &= \left [\frac{\left ( P+ r_c\right )}{\sqrt{r_c}} + \frac{ v^2 }{\sqrt{r_c}} \right ]\mathrm{d}t^2 - \frac{2}{\sqrt{r_c}}v_i \mathrm{d}x^i \mathrm{d}t \ 
\\
&- \frac{1}{\sqrt{r_c}} \left ( \nabla_i v_j + \nabla_j v_i - \frac{1}{r_c} v_i v_j \right ) \mathrm{d}x^i \mathrm{d}x^j + \mathcal{O}(\epsilon^3) \ ,
\end{aligned}
\end{equation}

\noindent so that the Brown--York stress tensor on $\Sigma_c$ is

\begin{equation}
\begin{aligned}
T_{a b}^{BY} \mathrm{d}x^a \mathrm{d}x^b &= \frac{1}{\sqrt{r_c}} \Omega_{ij} \mathrm{d}x^i \mathrm{d}x^j - \frac{2}{\sqrt{r_c}} v_i \mathrm{d}x^i \mathrm{d}t + \frac{v^2}{\sqrt{r_c}} \mathrm{d}t^2 \ 
\\
&+ \frac{1}{r_c^{3/2}} \left ( v_i v_j + P \Omega_{ij} - r_c \left (\nabla_i v_j + \nabla_j v_i \right ) \right ) \mathrm{d}x^i \mathrm{d}x^j \ 
\\
&+ \frac{2}{\sqrt{r_c}} \nabla^kv_k \Omega_{ij}\mathrm{d}x^i \mathrm{d}x^j - 2\sqrt{r_c} \nabla^iv_i \mathrm{d}t^2 + \mathcal{O}(\epsilon^3) \ .
\end{aligned}
\end{equation}

\subsection{The dual fluid}
\label{sec:df2}

To obtain the dual fluid, we shall impose the conservation of the Brow--York stress tensor on $\Sigma_c$

\begin{equation}
 \nabla^a T_{ab} = 0 \ .
 \label{eq:by_cons2}
\end{equation}

Like we had before, this may be seen as an integrability condition of the EFE, and also represents a constraint on $\Sigma_c$ which necessarily must be satisfied to guarantee that the EFE are solved perturbatively in $\epsilon$.

The metric and its inverse components on $\Sigma_c$ are no longer constant,

\begin{equation}
\begin{gathered}
\gamma_{tt} = -r_c + 2r_c \Omega^{ij}v^k \partial_k\Omega_{ij} + \mathcal{O}(\epsilon^4)\ ; \ \ \gamma_{ij} = \Omega_{ij} + \mathcal{O}(\epsilon^4) \ ;
\\
\gamma^{tt} = -\frac{1}{r_c} -\frac{2}{r_c} \Omega^{ij}v^k \partial_k\Omega_{ij} + \mathcal{O}(\epsilon^4) \ ; \ \ \gamma^{ij} = \Omega^{ij} + \mathcal{O}(\epsilon^4) \ ,
\end{gathered}
\end{equation}

\noindent so that the associated connection coefficients, given by

\begin{equation}
\Gamma^c_{ab} = \frac{1}{2} \gamma^{cd} \left ( \partial_a \gamma_{bd} + \partial_b \gamma_{da} - \partial_d \gamma_{ab} \right ) \ ,
\end{equation}

\noindent are no longer trivially null,

\begin{equation}
\begin{aligned}
\Gamma^t_{tt} &= \frac{1}{2} \gamma^{td} \left ( \partial_t \gamma_{td} + \partial_t \gamma_{dt} - \partial_d \gamma_{tt} \right ) 
\\
&= \frac{1}{2} \gamma^{tt} \left ( \partial_t \gamma_{tt} + \partial_t \gamma_{tt} - \partial_t \gamma_{tt} \right ) = \frac{1}{2} \gamma^{tt} \partial_t \gamma_{tt}  
\\
&= \frac{1}{2} \gamma^{tt} \partial_t \left ( -r_c + 2r_c \Omega^{ij}v^k \partial_k\Omega_{ij} + \mathcal{O}(\epsilon^4)\right ) 
\\
\Rightarrow \Gamma^t_{tt} &= \mathcal{O}(\epsilon^4) \ ;
\end{aligned}
\end{equation}

\begin{equation}
\begin{aligned}
\Gamma^t_{ti} &= \frac{1}{2} \gamma^{td} \left ( \partial_t \gamma_{id} + \partial_i \gamma_{dt} - \partial_d \cancelto{^0}{\gamma_{ti}} \ \ \ \right )  
\\
&= \frac{1}{2} \gamma^{tt} \left ( \partial_t \cancelto{^0}{\gamma_{it}} \  + \partial_i \gamma_{tt} \right ) = \frac{1}{2} \gamma^{tt} \partial_i \gamma_{tt} 
\\
&= \frac{1}{2} \gamma^{tt} \partial_i \left ( -r_c + \mathcal{O}(\epsilon^4)\right ) 
\\
&= \frac{1}{2} \left ( -\frac{1}{r_c} \right ) \partial_i \left ( -r_c \right ) + \mathcal{O}(\epsilon^4) 
\\
\Rightarrow \Gamma^t_{ti} &= \Gamma^t_{it} = \mathcal{O}(\epsilon^4) \ ;
\end{aligned}
\end{equation}

\begin{equation}
\begin{aligned}
\Gamma^t_{ij} &= \frac{1}{2} \gamma^{td} \left ( \partial_i \gamma_{jd} + \partial_j \gamma_{di} - \partial_d \gamma_{ij} \right )  
\\
&= \frac{1}{2} \gamma^{tt} \left ( \partial_i \cancelto{^0}{\gamma_{jt}} \ + \partial_j \cancelto{^0}{\gamma_{ti}} \ - \cancelto{^0}{\partial_t \gamma_{ij}} \ \ \  \right ) = - \frac{1}{2} \gamma^{tt}  \partial_t \gamma_{ij} 
\\
\Rightarrow \Gamma^t_{ij} &= \Gamma^t_{ji} = \mathcal{O}(\epsilon^4) \ ;
\end{aligned}
\end{equation}

\begin{equation}
\begin{aligned}
\Gamma^i_{tt} &= \frac{1}{2} \gamma^{ij} \left ( \partial_t \cancelto{^0}{\gamma_{tj}} \ + \partial_t \cancelto{^0}{\gamma_{jt}} \ - \partial_j \gamma_{tt} \right ) = - \frac{1}{2} \gamma^{ij} \partial_j \gamma_{tt} 
\\
&= - \frac{1}{2} \Omega^{ij} \partial_j \left (-r_c \right ) + \mathcal{O}(\epsilon^4) 
\\
\Rightarrow \Gamma^i_{tt} &=  \mathcal{O}(\epsilon^4) \ ;
\end{aligned}
\end{equation}

\begin{equation}
\begin{aligned}
\Gamma^i_{tj} &= \frac{1}{2} \gamma^{ik} \left ( \cancelto{^0}{\partial_t \gamma_{jk}} + \partial_j \cancelto{^0}{\gamma_{kt}} - \partial_k \cancelto{^0}{\gamma_{tj}} \ \ \ \right )  
\\
\Rightarrow \Gamma^i_{tj} &= \Gamma^i_{jt} = \mathcal{O}(\epsilon^4) \ ;
\end{aligned}
\end{equation}

\begin{equation}
\begin{aligned}
\Gamma^i_{jk} &= \frac{1}{2} \gamma^{il} \left ( \partial_j \gamma_{kl} + \partial_k \gamma_{lj} - \partial_l \gamma_{jk} \right ) 
\\
\Rightarrow \Gamma^i_{jk} &=  \Gamma^i_{kj} =  \frac{1}{2} \Omega^{il} \left ( \partial_j \Omega_{kl} + \partial_k \Omega_{lj} \right ) - \frac{1}{2} \partial^i \Omega_{jk} + \mathcal{O}(\epsilon^4) \ .
\end{aligned}
\end{equation}

Therefore, we no longer have the equivalence between the covariant and partial derivatives as in Eq. \ref{eq:by_cons1}, so that we must consider the covariant conservation of the Brown--York stress tensor according to Eq. \ref{eq:by_cons2}.

Notice that we explicitly wrote the terms of $\mathcal{O}(\epsilon^3)$ constructed with a partial derivative since, as seen in Sec. \ref{sec:df1}, we will be interested in the equation arising from the Brown--York stress tensor conservation at $\mathcal{O}(\epsilon^3)$, in which case it is important to explicitly account the connection coefficients of this order. 

As discussed in Sec. \ref{sec:df1}, once the constraint of Eq. \ref{eq:by_cons2} is satisfied on $\Sigma_c$, it is possible to evolve the solution in the radial direction, which precisely establishes the sense in which we can reduce the EFE to Eq. \ref{eq:by_cons2}. Therefore, at $\mathcal{O}(\epsilon^2)$ one has

\begin{equation}
\begin{gathered}
\nabla^a T_{at} = \gamma^{tt}\nabla_t T_{tt} + \gamma^{ij} \nabla_j T_{it} = 0 
\\
\Rightarrow \left ( -\frac{1}{r_c} + \mathcal{O}(\epsilon^4) \right ) \overbrace{\left ( \partial_t T_{tt} - \Gamma^t_{tt} T_{tt} - \Gamma^k_{tt} T_{tk}  - \Gamma^t_{tt} T_{tt}  - \Gamma^k_{tt}T_{tk} \right )}^{\mathcal{O}(\epsilon^4)} \ 
\\
+ \Omega^{ij} \Big( \partial_j T_{it} - \overbrace{\Gamma^t_{ji} T_{tt}}^{\mathcal{O}(\epsilon^4)} - \Gamma^k_{ji} T_{tk} - \overbrace{\Gamma^t_{jt} T_{it}}^{\mathcal{O}(\epsilon^4)} - \overbrace{\Gamma^k_{jt}T_{ik}}^{\mathcal{O}(\epsilon^4)}  \Big) = 0 
\\
\Rightarrow \Omega^{ij} \left ( \partial_j \left (-\frac{1}{\sqrt{r_c}} v_i \right ) - \Gamma^k_{ji} \left (-\frac{1}{\sqrt{r_c}} v_k \right ) \right ) + \mathcal{O}(\epsilon^4) = 0 
\\
\Rightarrow \partial^i v_i - \Omega^{ij} \Gamma^k_{ji} v_k = \partial^i v_i - \Omega^{ij} \left ( \partial_j v_i - \nabla_j v_i \right ) = \mathcal{O}(\epsilon^4) 
\\
\Rightarrow \nabla^i v_i = \mathcal{O}(\epsilon^4) \ .
\label{eq:incomp_2}
\end{gathered}
\end{equation}

Eq. \ref{eq:incomp_2} above states the incompressibility of the fluid up to $\mathcal{O}(\epsilon^4)$ in a covariant manner, thus being valid in the more general background we are considering! At $\mathcal{O}(\epsilon^3)$, one has

\begin{equation}
\begin{gathered}
\nabla^a T_{aj} = \gamma^{tt}\nabla_t T_{tj} + \gamma^{ik} \nabla_k T_{ij} + \mathcal{O}(\epsilon^4) =
\\
\Rightarrow \left( -\frac{1}{r_c} + \mathcal{O}(\epsilon^4) \right ) \Big ( \partial_t T_{tj} - \overbrace{\Gamma^t_{tt} T_{tj}}^{\mathcal{O}(\epsilon^5)} - \overbrace{\Gamma^k_{tt}}^{\mathcal{O}(\epsilon^4)} T_{jk}  - \overbrace{\Gamma^t_{tj} T_{tt}}^{\mathcal{O}(\epsilon^5)} - \overbrace{\Gamma^k_{tj}}^{\mathcal{O}(\epsilon^4)}T_{tk} \Big ) \ 
\\
+ \Omega^{ik} \Big( \partial_k T_{ij} -\overbrace{\Gamma^t_{ki}}^{\mathcal{O}(\epsilon^4)} T_{tj} - \Gamma^l_{ki} T_{jl} - \overbrace{\Gamma^t_{kj}}^{\mathcal{O}(\epsilon^4)} T_{ti}  - \Gamma^l_{kj}T_{il} \Big) = 0 
\\
\Rightarrow \left ( -\frac{1}{r_c} \right ) \left ( \partial_t \left (-\frac{1}{\sqrt{r_c}} v_j \right )  \right ) 
\\
+ \Omega^{ik} \left [ \partial_k \left ( \frac{1}{\sqrt{r_c}} \Omega_{ij} + \frac{1}{r_c^{3/2}} \left ( v_i v_j + P \Omega_{ij} - r_c \left (\nabla_i v_j + \nabla_j v_i \right )\right ) \right ) \right ] 
\\
- \Omega^{ik} \Gamma^l_{ki}  \left (\frac{1}{\sqrt{r_c}} \Omega_{jl} + \frac{1}{r_c^{3/2}} \left ( v_j v_l + P \Omega_{jl} - r_c \left (\nabla_j v_l + \nabla_l v_j \right )\right ) \right ) 
\\
- \Omega^{ik} \Gamma^l_{kj}  \left ( \frac{1}{\sqrt{r_c}} \Omega_{il} + \frac{1}{r_c^{3/2}} \left ( v_i v_l + P \Omega_{il} - r_c \left (\nabla_i v_l + \nabla_i v_j \right ) \right ) \right ) 
\\
+ \frac{2}{\sqrt{r_c}}\Omega^{ik} \Big [ \partial_k \Big ( \overbrace{\nabla^jv_j}^{\mathcal{O}(\epsilon^4)} \Omega_{ij} \Big ) \Big ] - \frac{2}{\sqrt{r_c}}\Omega^{ik} \Big [ \Gamma^l_{ki} \Big(\overbrace{\nabla^mv_m}^{\mathcal{O}(\epsilon^4)}\Omega_{jl} \Big ) - \Gamma^l_{kj} \Big (\overbrace{\nabla^mv_m}^{\mathcal{O}(\epsilon^4)} \Big ) \Omega_{il} \Big ) \Big ] \ 
\\
+ \mathcal{O}(\epsilon^4) = 0 
\\
\Rightarrow \partial_t v_j + r_c \Omega^{ik} \overbrace{\left( \partial_k \Omega_{ij} - \Gamma^l_{ki} \Omega_{lj} - \Gamma^l_{kj} \Omega_{il} \right )}^{= \nabla_k \Omega_{ij}} + \Omega^{ik} \overbrace{\left( \partial_k (v_iv_j) - \Gamma^l_{ki} (v_lv_j) - \Gamma^l_{kj} ( v_iv_l ) \right )}^{= \nabla_k (v_i v_j )} 
\\
+ \Omega^{ik} \overbrace{\left( \partial_k \left ( P\Omega_{ij} \right ) - \Gamma^l_{ki} \left (P \Omega_{lj} \right ) - \Gamma^l_{kj} \left ( P\Omega_{il} \right ) \right )}^{= \nabla_k \left (P \Omega_{ij} \right)} 
\\
-r_c \Omega^{ik} \overbrace{\left( \partial_k \left ( \nabla_i v_j + \nabla_j v_i \right ) - \Gamma^l_{ki} \left ( \nabla_l v_j + \nabla_j v_l \right ) - \Gamma^l_{kj} \left ( \nabla_i v_l + \nabla_l v_i \right ) \right )}^{= \nabla_k \left (\nabla_i v_j + \nabla_j v_i \right )} 
\\
- r_c \Omega^{ik} \overbrace{\left( \partial_k \left ( \partial_t \Omega_{ij} \right ) - \Gamma^l_{ki} \left ( \partial_t \Omega_{lj} \right ) - \Gamma^l_{kj} \left ( \partial_t \Omega_{li} \right ) \right )}^{=\nabla_k \left ( \partial_t\Omega_{ij} \right )} = \mathcal{O}(\epsilon^4) 
\\
\Rightarrow \nabla_t v_j + v_j \overbrace{\nabla^i v_i}^{\mathcal{O}(\epsilon^4)} + v_i \nabla^i v_j+ \nabla_j P -r_c \nabla^2 v_j -r_c \nabla^i \left (\nabla_j v_i \right ) = \mathcal{O}(\epsilon^4) 
\\
\Rightarrow \nabla_t v_j - r_c \nabla^2 v_j + \nabla_j P + v^i \nabla_i v_j-r_c \nabla^i \nabla_j v_i = \mathcal{O}(\epsilon^4) \ .
\end{gathered}
\end{equation}

The commutator of the covariant derivatives is such that

\begin{equation}
\begin{gathered}
\nabla_i \nabla_j v^i - \nabla_j \nabla_i v^i  = R_{ij}v^i 
\\
\Rightarrow \nabla^i \nabla_j v_i  - \nabla_j \nabla^i v_i = R\indices{^{i}_{j}} v_i 
\\
\Rightarrow \nabla^i \nabla_j v_i = R\indices{^{i}_{j}} v_i + \nabla_j \overbrace{\nabla^i v_i}^{\mathcal{O}(\epsilon^4)} = R\indices{^{i}_{j}} v_i + \mathcal{O}(\epsilon^4) \ ,
\end{gathered}
\end{equation}

\noindent where $R\indices{^{i}_{j}}$ denote the Ricci tensor components. Therefore, one has

\begin{equation}
\begin{gathered}
\nabla_t v_j - \eta \left (\nabla^2 v_j + R\indices{^{i}_{j}} v_i \right ) + \nabla_j P + v^i \nabla_i v_j = \mathcal{O}(\epsilon^4) \ ,
\label{eq:ns_2}
\end{gathered}
\end{equation}

\noindent where we identified $\eta = r_c$ as the kinematic viscosity of the fluid whose velocity and pressure fields satisfy Eq. \ref{eq:ns_2}, which can be seen as a manifestly covariant generalization of the incompressible Navier--Stokes equation, up to $\mathcal{O}(\epsilon^4)$, thus generalizing the description of the fluid located in the hypersurface featuring the boundary conditions we imposed --- which, we stress again, is directly related to soft hair excitations! The additional term, proportional to the Ricci tensor, is a direct consequence of the non-commutativity of the covariant derivatives.

We then have a precise sense in which the vacuum Einstein Field Equations are reduced to the generalized incompressible Navier--Stokes equation describing a fluid on $\Sigma_c$. Further, considering the near-horizon--hydrodynamic expansion equivalence of \cite{fg2}, as we already discussed, we can again realize $\Sigma_c$ as being located at the horizon, which then establishes the precise sense in which \emph{the soft-hairy horizon is a generalized incompressible fluid}. These results are original, 
and were published in \cite{sh_fg_mine}.

Holography is currently a very hot topic in theoretical physics, which makes the holographic landscape quite rich and diverse. Apart from the ones we discussed, there are many other holographic constructions in the literature, which show how beautiful, useful and broadly applicable this principle is. For some examples of holographic constructions, see \cites{24, 57, 62_f/g_cmt, 10_cft3, 102_hol, 109_hol, 6_nastase, 20_nastase, 33_nastase, 57_nastase, 4_boschi, 5_boschi, 6_boschi, 7_boschi}.

\newpage

\vspace*{\fill}
%
%

With this, we finish Part II of the thesis. At this point, we are familiar with the second wonder of gravity: holography, which tell us that gravity is more intimately related to different theories than we could initially think, what is not only surprising and useful, but conceptually very beautiful. Now we know the two particular wonders of gravity addressed in the thesis. We hope the journey was informative and fun, and that hopefully it is now clear how rich, intriguing, useful, surprising and beautiful gravity can be. In the next section, we finish the thesis with concluding remarks.

\

\expandafter\pgfornament\expandafter{89}
\vspace*{\fill}

%
%

\chapter{Conclusion}
\label{sec:conclusions}

Before we conclude the thesis, let us quickly sum up everything that was presented and discussed.

The fundamental aspects of General Relativity (GR), especially its mathematical aspects, of major importance for the whole thesis, are presented for completeness and self-containment in Appendix \ref{ap:gr}, since the discussion therein is way too long and displaced from the main text, which was nonetheless started with a brief and physically-guided discussion of GR, providing the basis for the following derivation of the Einstein Field Equations (EFE). The Schwarzschild solution was then derived and analyzed, which opened the way to the introduction of the concept of a black hole and its astrophysical possibility. 

We then provided a brief discussion on extra dimensions and the bulk EFE, which served to motivate braneworld scenarios and more specifically the Randall--Sundrum model, the first example of braneworld, in which the fundamental framework was established. We then derived the effective EFE on the brane, presenting the explicit calculations. In what followed, we presented the Minimal Geometric Deformation (MGD) method and its extension (EMGD), which provided generalized metrics whose parameters were constrained using the classical tests of GR, presented and developed in a way that makes them fully applicable to arbitrary spherically symmetric metrics.
Next, we reconsidered braneworld scenarios with an important change: a variable tension, whose implications were discussed and illustrated through the construction of a MGD black string, whose singularities at the bulk were calculated. We closed the discussion by presenting two applications of braneworld models, namely to the study of glueball condensates and to the quantum portrait of black holes.

Afterward, we provided a short discussion on black hole thermodynamics, the AdS spacetime and black branes, which play a major role in AdS/CFT duality. After presenting the conceptual and operational basis of linear response theory, hydrodynamics formalism, AdS/CFT and its methods, we discussed the calculation of the $\eta/s$ ratio in the AdS$_5$--Schwarzschild gravitational background. The calculation is only schematically discussed, as the explicit calculation was carried out later on for generalized black branes. We then proceeded to the discussion of the important relationship between braneworld scenarios, the membrane paradigm and the AdS/CFT correspondence, which allowed us to transliterate the braneworld scenario (in which the deformed AdS black branes are constructed) into the AdS/CFT language. After this discussion, we presented the explicit calculation of the $\eta/s$ ratio in both the 5-dimensional and 4-dimensional deformed AdS black branes, whose results were used to constrain the parameters of the black brane. Afterward, we discussed a generalized gravitational action and its consequence: the violation of the KSS bound.

Next, after discussing the fluid/gravity correspondence as initially proposed, we focused on an alternative view of it, which established a holographic correspondence between the Rindler spacetime and an incompressible fluid living on the horizon. We then extended the analysis to the generalized scenario of soft hairy horizons, which yielded novel generalized results. 

At the beginning of each chapter/section, it was stated whether its content is original, or a presentation of existing literature. The sections which contain original results are:

\begin{itemize}
 \item Sec. \ref{sec:emgd_ct}: constraints on the values of the EMGD deformation parameter according to the classical tests of GR. Results in \cite{glueballs};
 \item Sec. \ref{sec:mbg_bs_s}: calculation of the MGD black string bulk singularities. Results to be further explored in future works, and eventually published;
 \item Sec. \ref{sec:apps_branes}: discussion of published results \cite{glueballs} and \cite{quantum_portrait} (the original contribution was the (E)MGD modeling of the gravitational setup);
 \item Sec. \ref{sec:eta_s_ads_mgd_5}: calculation of $\eta/s$ in the deformed AdS$_5$--Schwarzschild black brane gravitational background and constraint of its deformation parameter. Results in \cite{ads5};
 \item Sec. \ref{sec:eta_s_ads_mgd_4}: calculation of $\eta/s$ in the deformed AdS$_4$--Reissner--Nordstr\"om black brane gravitational background and constraint of its deformation parameter. Results in \cite{ads4};
 \item Sec. \ref{sec:varying}: generalization of the fluid-gravity correspondence between Rindler spacetime and incompressible fluid dynamics to encompass soft hairy horizons. Results in \cite{sh_fg_mine}.
\end{itemize}

In conclusion, we hope the reader enjoyed the journey through the dark chest of wonders of gravity, and could appreciate how beautiful the ideas discussed are. Many open questions regarding gravity remain, but we hope to have shed light on some of them --- or, even better, contributed to the formulation of new questions! 

Once again: gravity is wonderful --- and now that the chest is open, you know some of the reasons why!

\cleardoublepage
\phantomsection
\addcontentsline{toc}{chapter}{Bibliography}
\bibliography{bibliografia}


%
%

\appendix

\chapter{The mathematical basis of General Relativity}
\label{ap:gr} 

In this appendix we will present the main mathematical formalism behind the general theory of relativity, which provides the most fundamental basis for this research project and is therefore of enormous importance. Still, since the presentation here is quite long, formal and detailed, it was suppressed from the main text and placed in an appendix, which by no means make it less important --- quite the opposite, actually. Our presentation will closely follow, in style and content, the texts of Carroll \cite{carroll}, Wald \cite{wald} and the lectures of Dr. Frederic Schuller \cites{gravlight, gravlightvd}. 

Before we delve into the mathematical details, let us quickly remind the core concepts and ideas behind General Relativity (GR).

In GR, gravity becomes a geometrical manifestation of spacetime curvature, which is produced by mass/energy and momentum. Such relationship is governed by the Einstein Field Equations (EFE), a set of partial differential equations relating symmetric tensors,

\begin{equation}
 R_{\mu \nu} - \frac{1}{2} R g_{\mu \nu} + \Lambda g_{\mu \nu} = \frac{8 \pi G}{c^4} T_{\mu \nu} \ ,
 \label{eq:efe_si}
\end{equation}

\noindent where $ R_{\mu \nu}$ is the Ricci curvature tensor; $R$ is the Ricci (curvature) scalar; $g_{\mu \nu}$ is the metric tensor; $\Lambda$ is the cosmological constant; $G$ is the gravitational constant; $c$ is the speed of light in vacuum and $T_{\mu \nu}$ is the energy-momentum tensor. The derivation of the EFE was presented in Sec. \ref{sec:EFEqs}, after the introduction of the energy-momentum tensor in Sec. \ref{sec:em_tensor}, followed by a discussion on the cosmological constant in Sec. \ref{sec:cc_efe}.The purpose of this Appendix is to present the precise mathematical definition and meaning of the remaining terms in Eq. \ref{eq:efe_si}.

GR also establishes how matter responds to spacetime curvature: free particles move along \emph{geodesics}, which are the paths of shortest distance between points in spacetime. Geodesics are the generalization of straight lines, which are the geodesics in flat space --- though in curved spacetime a straight line in general is not the shortest path between two points. The paths of such particles obey the \emph{geodesic equation},

\begin{equation}
 \frac{d^2 x^\mu}{d \lambda^2} + \Gamma^\mu_{\rho \sigma}  \frac{d x^\rho}{d \lambda}  \frac{d x^\sigma}{d \lambda} = 0 \ ,
 \label{eq:geo}
\end{equation}

\noindent where $x^{\mu}(\lambda)$ are the coordinates of the particle path, as a curve parameterized with $\lambda$, and $\Gamma^\mu_{\rho \sigma}$ are the \emph{connection coefficients} known as as \emph{Christoffel symbols}. In Sec. \ref{sec:Geodesics}, Eq. \ref{eq:geo} will be fully derived and the consequences of geodesics as the path followed by free particles in spacetime will be detailed. It will then be clear that, in GR, free particles subjected to a gravitational field (i.e., to the curvature of spacetime) feel no acceleration --- they move through geodesics in a free fall. In that sense, in such a feature of the theory lies the claim that gravity is not a force --- exerted on a particle therefore deflecting it from its straight-line motion ---, but something entirely different: a geometrical manifestation of spacetime. 

The construction of GR by Albert Einstein may be thought of as being guided and motivated by two fundamental sets of ideas. The first and more precise one is the \emph{principle of equivalence}, which may be stated as the fact that gravity affects every body in the same manner, and all bodies therefore fall in the same way in a gravitational field. Such principle, most precisely characterized as the \emph{weak equivalence principle}, was already present in Newtonian physics, through the statement that gravitational mass (the $m$ in Eq. \ref{eq:isl}) and inertial mass ($m$ in \ref{eq:n2l}) are equivalent. 

Realizing that there is no way for any observer in a closed box to distinguish between uniform acceleration and an external gravitational field, Einstein generalized the principle of equivalence in what is known as the \emph{Einstein principle of equivalence}, which states that in small regions of spacetime, the laws of physics reduce to that of SR, that is, the laws of a flat spacetime, even if it is globally curved. This idea suggests that gravity is in fact not a force field having spacetime as its background, but a property \emph{intrinsic to spacetime itself}, which is non-locally manifested as the deviation from flat spacetime geometry of SR.

The second motivating principle of GR encodes a set of rather philosophical and less precise ideas proposed by Ernst Mach: the \emph{Mach principle}. It may be stated as the principle that the large scale matter content of the universe directly determines locally inertial frames. A way to exemplify the principle is through the following experimental fact: if one is standing still looking at the stars, which are seen as not moving, one's arms freely stand still at the side of the body. Nevertheless, if one starts spinning, the stars will also be seen as spinning, whilst the arms are pushed away from their initial position. Therefore, there must be some relationship between the first state (in which both the stars and the arms are in rest) and the second state (in which the stars are seen to move, as well as the arms are pushed away from rest), such that the movement of the far-away distribution of mass as locally seen, somehow influenced the local inertia.

Einstein accepted such a view and interpreted it a way that the structure of spacetime itself would have to be affected by mass distributions. The formal establishment of such ideas culminated in the EFE, the equation satisfied by the spacetime metric, in which it is clear the motivation provided by Mach's principle, through the precise formulation of how the matter content of the universe influences spacetime geometry, in a way that the metric will be a dynamical variable responding to energy and momentum, and no longer a static background. The final structure of GR is in agreement with these Mach's ideas, even though some aspects of the theory soundly violates the principle, partly due to the variety of ways in which this vague principle may be stated. 
 
As it may have been made clear by the brief introduction above, within the framework of GR there is an underlying notion around which the whole theory is constructed: the \emph{spacetime}. Intuitively, one may have some grasp of the concept of spacetime as a structure in which space and time are sewn together in a $4$-dimensional single space. Of course, though, such a notion lacks of the precision necessary to build a theory, so that we can define spacetime somewhat more precisely as it is presented in Lecture 1 of \cite{gravlightvd},

\begin{mydef}
Spacetime is a four-dimensional differential manifold with a smooth atlas carrying a torsion-free connection compatible with a Lorentzian metric and a time orientation satisfying the Einstein equations.
\label{def:st}
\end{mydef}

The main goal of this appendix will be that of clarifying the notion of spacetime, by detailing the definition presented above with mathematical precision. This will provide a better understanding of curvature and its measurement, which will then lead us to the establishment of the precise relationship between matter, curvature of spacetime and its manifestation as gravity. For now, an important remark must be done on the dimensionality defined above: although we will present the basis of GR as constructed around a 4-dimensional manifold as a model to spacetime, in the main thesis text we generalized the notions to higher dimensional spacetimes. Whenever the dimension of the manifold is important, it will be remarked. Most of time, though, the developments are independent of dimensionality, so that the discussion may be carried out naturally and independently of the manifold dimension.

\section{Manifolds}

Mathematical analysis provides the full precise framework in which functions and their operations --- like differentiation and integration --- are defined in $\mathbb{R}^n$, the n-dimensional Euclidean space of positive-definite metric $\delta_{i j}$. But, as already made clear, in GR the interest lies on curved spaces, on which it is necessary to establish analogous notions which allow the performance in this spaces of the same operations on functions defined in $\mathbb{R}^n$. 

Such task is achieved through the notion of \emph{manifold}, which, simply put, is a space that although may be topologically complicated (curved, for example) in a global scale, it locally resembles an Euclidean space. In fact, a manifold is constructed through a smooth sew of locally Euclidean (flat) regions to compose an $n$-dimensional space, that may be globally curved. Such construction provides the establishment of a \emph{local differential structure}, which allows the analysis of functions defined on the Manifold as locally seen as defined in an Euclidean space, to which all the tools of $\mathbb{R}^n$ analysis can then be applied.

To precisely define what is meant by ``a space consisting of pieces locally resembling $\mathbb{R}^n$ and smoothly sewn together'', we must first precisely define some basic notions, which will be done in a crude mathematical way, as follows,

\begin{mydef}
 Let $A$ and $B$ be two sets. A \emph{map} $f : A \rightarrow B$ is an object such that $\forall a \in A$, $\exists! \  f(a) \in B$.
 The sets $A$ and $B$ are respectively called the \emph{domain} and \emph{codomain} of the map $f$, and the subset $I \subset B$  consisting of all possible $f(a)$ is its \emph{image}.
\end{mydef}

Therefore, a map can be easily seen as a generalization of the common idea of a \emph{function} (which here denotes a particular map from $\mathbb{R}^n$ to $\mathbb{R}$).

\begin{mydef}
A map $f : A \rightarrow B$ satisfying $\forall a, b \in A, a \neq b \Rightarrow f(a) \neq f(b)$ is called \emph{injective}.
\end{mydef}

\begin{mydef}
A map $f : A \rightarrow B$ such that $\forall b \in B, \exists a \in A \mid f(a) = b$ is called \emph{surjective}.
\end{mydef}

\begin{mydef}
A map is called \emph{bijective} or \emph{invertible} if it is both injective and surjective.
\end{mydef}

Therefore, a bijective map is such that every element of its codomain is mapped from one and only one element of its domain.

\begin{mydef}
Given two maps $f : A \rightarrow B$ and $g : B \rightarrow C$, the map $g \circ f : A \rightarrow C$, defined by $(g \circ f)(a) = g(f(a)), \forall a \in A$, is called the \emph{composition} of the maps $g$ and $f$.
\end{mydef}

\begin{mydef}
Given a bijective map $f : A \rightarrow B$, the map $f^{-1} : B \rightarrow A$, such that $(f^{-1}\circ f)(a) = a$ is called the \emph{inverse map} of $f$.
\end{mydef}

Notice that inverse maps are only defined for bijective maps, hence their denomination as ``invertible''.

A map $f: \mathbb{R}^m \rightarrow \mathbb{R}^n$ may be conceived as a set of $n$ functions of $m$ variables, $\varphi^i: \mathbb{R}^m \rightarrow \mathbb{R}$, $i = {1, 2, \cdots, n}$.

\begin{mydef}
A function $\varphi: \mathbb{R}^m \rightarrow \mathbb{R}$ is said of \emph{class $C^N$} if its $N$-th derivative exists and is continuous.
\end{mydef}

Accordingly, a map $f: \mathbb{R}^m \rightarrow \mathbb{R}^n$ consisting of $n$ such functions $\varphi^i$ is also called of class $C^N$, if each function $\varphi^i$ is of class $C^N$.

\begin{mydef}
A map of \emph{class $C^{\infty}$} is called \emph{smooth}.
\end{mydef}

Now, to finally establish the precise way in which locally Euclidean spaces are sewn together to form a manifold, some more definitions are necessary,

\begin{mydef}
Let $y \in \mathbb{R}^n$ be a fixed point and $r \in \mathbb{R_+}$. The set of $n$-tuples $B_r(y) = \left \{ (x_1, x_2, \cdots , x_n) \mid \sum_{i=1}^{n} (x_i - y_i)^2 < r^2 \right \}$ is called an \emph{open ball}.
\end{mydef}

Notice that an open ball may be thought of as the interior region of a $n$-sphere of radius $r$ centered in $y$.

\begin{mydef}
A set $V \subset \mathbb{R}^n$ such that $\forall y \in V, \exists r \in \mathbb{R_+} \mid B_r(y) \subseteq V$ is called an \emph{open set}.
\end{mydef}

Such definition states that a set is said to be \emph{open in $\mathbb{R}^n$} if and only if for \emph{every point} of the set, there exists an open ball of radius $r > 0$ which lies entirely in the set. Intuitively it makes sense why such a set is open --- it cannot contain points of the boundary of a given region in $\mathbb{R}^n$. Alternatively, an open set can be thought of as being built from an union of open balls. More formally, an open set is the element of the so-called \emph{standard topology}, built on  $\mathbb{R}^n$.

\begin{mydef}
Let $M$ be a set; $D \subset M$ a subset and $f : D\rightarrow I \subset \mathbb{R}^n$ a bijective map, such that $I$, its image, is an open set in $\mathbb{R}^n$. The pair $(D, f)$ is called a \emph{chart} or \emph{coordinate system}.
\end{mydef}

Notice that, as defined above, $D$ is an open set in $M$.

\begin{mydef}
Let $A$ be an index set. The collection of charts $\{ (D_\alpha, f_\alpha) \}$, $\alpha \in A$ satisfying,
\begin{itemize}
 \item $\bigcup_{\alpha \in A} D_\alpha = M$; 
 \item $D_\alpha \cap D_\beta \neq \varnothing \Rightarrow (f_\alpha \circ f_\beta^{-1}) : I_\beta(D_\alpha \cap D_\beta) \subset \mathbb{R}^n \rightarrow I_\alpha(D_\alpha \cap D_\beta) \subset \mathbb{R}^n$ is a surjective map, where $I_\alpha(D_\alpha \cap D_\beta)$ denotes the subset of map $f_\alpha$ image into which the elements of its domain intersection $D_\alpha \cap D_\beta$ are mapped and $f_\alpha, f_\beta$ are of class $C^{\infty}$. 
\end{itemize}
 Is called a \emph{class $C^\infty$ (smooth) atlas}.
 \label{def:atlas}
\end{mydef}

This complicated definition states something very simple: a smooth atlas is a collection of maps which must both cover the whole set $M$ in which they are defined, and be such that the regions of overlapping charts must be related by smooth maps. This is how charts are guaranteed to be smoothly sewn together to cover the whole set $M$. In many important uses, it takes more than one chart to cover the whole set, indexed by the elements of the set $A$.

Now everything is set to finally present the definition which formalizes the concept of manifold and its property of local resemblance to $\mathbb{R}^n$,

\begin{mydef} 
Let $M$ be a set and $ \{ (D_\alpha, f_\alpha) \}$, $\alpha \in A$ a smooth atlas, where $A$ is an index set. If the atlas is maximal, i.e., contains all compatible charts, the tuple $(M, \{ (D_\alpha, f_\alpha) \} )$ is called a \emph{$n$-dimensional manifold}.
\label{def:mfld}
\end{mydef}

And this is a manifold: a set with a well defined smooth atlas, which guarantees the local resemblance to $\mathbb{R}^n$. Now, it is important to remark that the definition of spacetime, as presented in Def. \ref{def:st}, makes clear reference to a \emph{differentiable manifold}. Although it was not mentioned, the definition of manifold presented above is not absolutely general, so that it refers specifically to differentiable manifolds. As this is precisely what is interesting for GR, any further generality is for now expendable.

Specifically, it is the second condition established on Def. \ref{def:atlas} for an atlas which characterizes a manifold as differentiable. In fact, one can locally apply methods from calculus to manifolds, because each point of an $n$-dimensional manifold is homeomorphic to  $\mathbb{R}^n$ \cite{gamelin}, which provides a local differential structure on the local coordinate system. Now, to guarantee that such a structure is globally induced, it is necessary to provide a smooth composition of maps on chart intersections. That is, in the chart intersections, the defined coordinates must be differentiable with relation to the coordinates defined by each overlapping charts. The smooth maps which relate these different coordinates are called \emph{transition maps}, and, as defined in Def. \ref{def:atlas}, they guarantee a global differential structure on differentiable manifolds.

As a final remark, it may be natural the question of why using a manifold to describe spacetime. In fact, in classical physics, the notion of continuity of curves is of major importance, since classical paths are expected to be everywhere continuous. Now, the manifold, as defined above, serves as the minimal mathematical structure in which continuity may be promptly (and locally) analyzed, although the space may be topologically complicated.

\subsection{Vectors}

In curved spacetime, vectors are not objects which may be freely transported and stretched. Instead, each vector is placed in a specified point of spacetime, and there only. In fact, to each point $p$ of the manifold $M$ modeling spacetime, a set of all possible vectors located precisely at such a point is defined as the \emph{tangent space at $p$}, denoted $T_p$. To construct the tangent space using notions intrinsic to the manifold, let us consider the space of all class $C^{\infty}$ functions on the manifold, $f : M \rightarrow \mathbb{R}$, and the set of all curves passing through $p$, $\gamma: \mathbb{R} \rightarrow M$, with $p \in I_\gamma \subset M$, the image of the curve $\gamma$. Now, each curve may be used to define an operator acting on the functions space: the \emph{directional derivative} at $p$, through the action $f \mapsto \dfrac{d f}{d \lambda}$, where $\lambda$ is a parameter for the curve. We then define,

\begin{mydef}
 The space of directional derivatives operators along curves through $p$ is the \emph{tangent space $T_p$}.
\end{mydef}

In fact, as a differential operator, the directional derivatives form a vector space. Also, within a chart of coordinates $x^{\mu}$, it is possible to write any directional derivative as a linear combination of partial derivatives $\partial_\mu \equiv \dfrac{\partial}{\partial x^\mu}$ at point $p$ (which is itself a directional derivative along curves with all other coordinates $x^\nu, \nu \neq \mu$ kept constant),

\begin{equation}
\frac{d}{d \lambda} = \frac{d x^{\mu}}{d \lambda} \partial_{\mu} \ .
 \label{eq:dirder}
\end{equation}

In flat space, the components of a vector $\bm{V(} \lambda \bm{)}$ tangent to a parameterized curve of coordinates $x^\mu(\lambda)$ are given by

\begin{equation}
 V^\mu = \frac{d x^\mu}{d \lambda} \ .
 \label{eq:tanvec}
\end{equation}

Comparing Eqs. \ref{eq:tanvec} and \ref{eq:dirder}, and realizing that $\bm{V} = V^\mu \bm{\hat{e}}_\mu$, it is clear that in an arbitrary manifold, one has the basis $\{ \bm{\hat{e}}_\mu = \partial_\mu \}$ spanning the vector space of vectors tangent to the curve of parameter $\lambda$, $\bm{V} = \dfrac{d}{d \lambda}$. The basis $\{ \partial_\mu \}$ is called the \emph{coordinate basis} of $T_p$ as, by definition, it generalizes the idea of the basis vectors pointing in the direction of the coordinate axes.

By imposing that a vector $\bm{V} = V^\mu \partial_{\mu}$ must be invariant under a change of coordinates $x^\mu \rightarrow x^{\hat{\mu}}$ --- and since the basis vectors, according to the chain rule, transform as $\partial_{\tilde{\mu}} = \dfrac{\partial x^\mu}{\partial x^{\tilde{\mu}}} \partial_\mu $ ---, one has the \emph{vector transformation law},

\begin{equation}
 V^\mu \rightarrow V^{\tilde{\mu}} = \frac{\partial x^{\tilde{\mu}}}{\partial x^\mu}  V^\mu \ .
 \label{eq:transvec}
\end{equation}

\subsection{Dual vectors}

From the tangent vector space $T_p$ constructed as defined above, it is possible to define the \emph{cotangent space}, denoted $T_p^*$, which is the dual space associated to $T_p$. The cotangent space $T_p^*$ consists in the set of linear maps from $T_p$ to $\mathbb{R}$. That is, $v \in T_p^*$ acts in $\bm{U}, \bm{V} \in T_p$ \emph{preserving linear combinations}, i.e., according to $v(\alpha \bm{U} + \beta \bm{V}) = \alpha v(\bm{U}) + \beta v(\bm{V}) \in \mathbb{R}$, for $\alpha, \beta \in \mathbb{R}$.

The most natural example of a dual vector is the gradient of a function $f$, denoted by $\mathrm{d}f$. In fact, the action of the gradient --- whose components are $\partial_\mu f$ --- on the vector tangent to a curve precisely gives the directional derivative of the function,

\begin{equation}
 \mathrm{d}f \left ( \frac{d}{d \lambda}\right ) = \partial_\mu f \left ( \frac{\partial x^\mu}{\partial \lambda} \right ) = \frac{d f}{d \lambda} \ .
 \label{eq:gradfun}
\end{equation}

Setting $\{ \bm{\hat{e}}^\mu \}$ as a basis for $T_p^*$, the proper action of dual vectors in vectors is guaranteed by imposing that $\bm{\hat{e}}^\mu ( \bm{\hat{e}}_\nu ) = \delta^\mu_\nu$. Now, following the procedure with which we constructed the basis for the tangent space as the partial derivatives of the coordinate maps $x^\mu$, the natural choice of basis for the cotangent space is then the gradient of such functions, i.e., $\{ \bm{\hat{e}}^\mu = \mathrm{d}x^\mu \}$. In fact, Eq. \ref{eq:gradfun} confirms that such a choice is consistent with our demands,

\begin{equation}
\mathrm{d}x^\mu (\partial_\nu) = \frac{\partial x^\mu}{\partial x^\nu} = \delta^\mu_\nu \ .
\end{equation}

The imposition of invariance of a dual vector $v = v_\mu \mathrm{d}x^\mu$ under $x^\mu \rightarrow x^{\hat{\mu}}$, as well as the transformation of the basis vectors as $\mathrm{d}x^{\tilde{\mu}} = \dfrac{\partial x^{\tilde{\mu}}}{\partial x^\mu} \mathrm{d}x^\mu $, directly lead to the \emph{dual vector transformation law},

\begin{equation}
 v_\mu \rightarrow v_{\tilde{\mu}} = \frac{\partial x^\mu}{\partial x^{\tilde{\mu}}}  v_\mu \ .
 \label{eq:transdualvec}
\end{equation}

\subsection{Tensors}
\label{sec:tensors_transf}

The concepts of vectors and dual vectors meet an immediate generalization in the concept of a \emph{tensor} $T$ of rank $(k, l)$, defined as a multilinear map from $k$ dual vectors and $l$ vectors to $\mathbb{R}$,

\begin{equation}
T: \underbrace{T_p^* \times \cdots \times T_p^*}_\text{k times} \times \underbrace{T_p \times \cdots \times T_p}_\text{l times} \rightarrow \mathbb{R} \ .
\end{equation}

Tensors form a vector space, whose coordinate basis consist of the tensor product between tangent and cotangent basis vectors $\{ \partial_{\mu_1} \otimes \cdots \otimes  \partial_{\mu_k} \otimes \mathrm{d}x^{\nu_1} \otimes \cdots \otimes \mathrm{d}x^{\nu_l} \}$. The components of a tensor $T$ in a given coordinate basis can be obtained by its action on the respective basis vectors $T\indices{^{\mu_1 \cdots \mu_k}_{\nu_1 \cdots \nu_l}} = T(\mathrm{d}x^{\mu_1}, \cdots, \mathrm{d}x^{\mu_k}, \partial_{\nu_1}, \cdots, \partial_{\nu_l})$. Therefore, an arbitrary tensor is then written as $T = T\indices{^{\mu_1 \cdots \mu_k}_{\nu_1 \cdots \nu_l}} \partial_{\mu_1} \otimes \cdots \otimes  \partial_{\mu_k} \otimes \mathrm{d}x^{\nu_1} \otimes \cdots \otimes \mathrm{d}x^{\nu_l}$.

Tensors transform according to a very well-defined transformation law, 

\begin{equation}
 T\indices{^{\mu_1 \cdots \mu_k}_{\nu_1 \cdots \nu_l}} \rightarrow T\indices{^{\tilde{\mu_1} \cdots \tilde{\mu_k}}_{\tilde{\nu_1} \cdots \tilde{\nu_l}}} = \frac{\partial x^{\tilde{\mu_1}}}{\partial x^{\mu_1}} \cdots \frac{\partial x^{\tilde{\mu_k}}}{\partial x^{\mu_k}} \frac{\partial x^{\nu_1}}{\partial x^{\tilde{\nu_1}}} \cdots \frac{\partial x^{\nu_l}}{\partial x^{\tilde{\nu_l}}} T\indices{^{\mu_1 \cdots \mu_k}_{\nu_1 \cdots \nu_l}} \ .
 \label{eq:transtens}
\end{equation}

In fact, since vectors are rank-$(1, 0)$ tensors and dual vectors are rank-$(0, 1)$ tensors, it is easy to realize that their respective transformation laws given by Eqs. \ref{eq:transvec} and \ref{eq:transdualvec} are only particular cases of Eq. \ref{eq:transtens}. Also, such a transformation makes clear why scalars (which are rank-$(0, 0)$ tensors) are invariant under changes of basis.

In a curved spacetime it is not possible to use Cartesian coordinates, so that it is very important to guarantee the \emph{coordinate invariance} of all equations, that is, the equations must hold unchanged independently of the coordinate system used. The construction of the manifold presented above makes it even clearer the importance of this invariance, since, as seen, different charts are established in different regions of a manifold, so that the coordinate invariance guarantees the consistency of the equations throughout the manifold.

And this is why tensors are so important and widely used in GR: because \emph{an equation between tensors will hold in every coordinate system}, thanks to the special way tensors transform, according to Eq. \ref{eq:transtens}. This way, it is possible to express relationships between physical quantities without establishing any coordinate system at all, since, provided such quantities transform as tensors, a change of coordinates will induce the same transformation to all quantities, in both sides of the tensor equation, so that their relationship will always hold unchanged. Such coordinate-invariant relationships are often called  \emph{covariant}, and it will always be our aim to construct covariant equations and objects in spacetime.

\subsection{The metric tensor}
\label{sec:metric}

Perhaps the most important tensor in GR is the metric tensor, a rank-$(0, 2)$ tensor whose components are $g_{\mu \nu}$. It is the solution to the EFE, and all information regarding the spacetime geometry is encoded in it, as will be explored in the next section. In a way, the metric tensor shows how Pythagoras's theorem --- which provides the distance between two points in flat space --- changes in a curved space. In that sense, the metric tensor may be seen as the provider of the proper way to measure distances in curved manifolds. In fact, the metric carries the information necessary to determine the manifold curvature, and the alteration in the way to measure distances is encoded in this notion.

The metric is related to the line element $ds^2 = g_{\mu \nu} dx^\mu dx^\nu$, which in the context of GR is an infinitesimal invariant interval between spacetime points. If $ds^2> 0$, we say that the points are \emph{spacelike separated}; if $ds^2 = 0$, the points are \emph{null separated} and for $ds^2 < 0$, \emph{timelike separated}. The same nomenclature is used to classify vectors and paths through spacetime. Within the context of spacetime diagrams and light cones, as it is lively discussed at \cite{gr_atob}, it is clear that light follow null trajectories, while massive particles go only through timelike paths. 

On the other hand, the metric, as a proper rank-$(0, 2)$ tensor, is defined as 

\begin{equation}
 ds^2 = g_{\mu \nu} \mathrm{d}x^\mu \otimes \mathrm{d}x^\nu \ .
 \label{eq:ds2}
\end{equation}

Within such a precise definition, it is clear that $ds^2$ is the metric tensor, expanded in the coordinate cotangent space basis, whose components are $g_{\mu \nu}$. It would therefore be more appropriate to replace ``$ds^2$'' by ``$g$''. But, as ``$g$'' is already used to denote the metric determinant, we keep the notation from the geometrical notion of the line element, by keeping clear that in fact the metric is a rank-$(0, 2)$ tensor, a bilinear map from two vectors $\bm{U}, \bm{V} \in T_p$ to the real numbers, by the definition of the inner product as

\begin{equation}
\bm{U \cdot V} = g_{\mu \nu} U^\mu V^\nu \ .
 \label{eq:innerprod}
\end{equation}

Of course, though, the relationship between the metric and the line element is not to be forgotten. Taking, for instance, the metric of $\mathbb{R}^3$ ($ds^2 =  \delta_{i j} \mathrm{d}x^i \otimes \mathrm{d}x^j  = \mathrm{d}x^2  + \mathrm{d}y^2 + \mathrm{d}z^2$, where we denote $\mathrm{d}x^2 \equiv \mathrm{d}x \otimes \mathrm{d}x$), in spherical coordinates $\{ r, \theta, \varphi \}$, one obtains $ds^2 = \mathrm{d}r^2 + r^2 \mathrm{d}\theta^2 + r^2 \sin^2\theta \mathrm{d}\varphi^2$. Now, we can directly obtain the metric of the non-euclidean manifold $S^2$, the $2$-sphere, by setting $r=1$ and $\mathrm{d}r = 0$ in $\mathbb{R}^3$ metric in spherical coordinates, to get

\begin{equation}
 ds^2 = \mathrm{d}\theta^2 + \sin^2\theta \mathrm{d}\varphi^2 \ , 
 \label{eq:s2metric}
\end{equation}

\noindent which is exactly the line element in a sphere, within the geometrical intuition.

The metric of flat spacetime, the \emph{Minkowski space} in which SR is set up, has components written in matrix form as

\begin{equation}
\eta_{\mu \nu} =
\begin{pmatrix}
-1 & 0 & 0 & 0 \\ 
0 & 1 & 0 & 0 \\ 
0 & 0 & 1 & 0 \\ 
0 & 0 & 0 & 1
\end{pmatrix}
\ .
 \label{eq:minkmatr}
\end{equation}

In its explicit form as a line element, the flat spacetime metric in cartesian coordinates is then $ds^2 = -\mathrm{d}t^2 + \mathrm{d}x^2  + \mathrm{d}y^2 + \mathrm{d}z^2$.

The \emph{canonical form} of a metric is such that its components are those of a diagonal matrix with eigenvalues $-1, 0$ or $1$. When in canonical form, we can then characterize the metric with its \emph{signature}, which is the number of positive and negative eigenvalues. The metric for Minkowski space of Eq. \ref{eq:minkmatr}, for example, is naturally in the canonical form, and has a signature $(-, +, +, +)$. Metrics with such a signature, of a single minus, are called \emph{Lorentzian} or \emph{pseudo-Riemannian}. In GR, spacetime manifolds always carry a Lorentzian metric, and should therefore more precisely be called \emph{pseudo-Riemannian manifolds}.

A very important feature of the metric is that it is always possible to put it into its canonical form \emph{at some point $p \in M$}. This is done by choosing a coordinate system $x^{\hat{\mu}}$ such that the metric $g_{\hat{\mu}\hat{\nu}}$ takes its canonical form. This is possible $\forall p \in M$, and the chart functions $x^{\hat{\mu}}$ are called the \emph{locally inertial coordinates}, while the associated basis vectors form the \emph{local Lorentz frame}. In fact, in such coordinates the components of the metric are locally at $p$ precisely those of the Minkowski space in first order. Such feature formalizes the idea that a manifold may be locally seen as a flat space. On the other hand, the curvature of the manifold is such that, in general, its second derivatives are non-vanishing. In the next section this idea will become clear.

The metric is generally constructed in a way that its determinant $| g_{\mu \nu} | = g$ is nonzero. This allows us to define the \emph{inverse metric} $g^{\mu \nu}$, a rank-$(2, 0)$ tensor, by definition satisfying $g^{\mu \nu} g_{\nu \sigma} = \delta^\mu_\sigma$. This allows the \emph{lowering} and \emph{raising} indices operations, which are respectively defined as $U_\nu = g_{\mu \nu} U^\mu$ and $U^\nu = g^{\mu \nu} U_\mu$. Notice how a vector is transformed into a dual vector and vice versa, with generally different components. Although such operations are defined for any tensor, they are of special interest for vectors and dual vectors, as they provide an operational way to compute the inner product of two vectors as $\bm{U \cdot V} = U_\nu V^\nu$.

\section{Curvature}

At this point, almost all notions presented on Def. \ref{def:st} of spacetime have been precisely defined. Now it is time to formalize the notion of curvature, and how to measure its occurrence on manifolds. Of course, as we already motivated, GR is a geometrical theory in which gravity is seen as the manifestation of spacetime curvature, so that the precise mathematical definition of curvature is of greatest importance in the context of GR. Therefore, this establishment will be the aim of this section.

\subsection{Connection and covariant derivatives}

Curvature directly depends on an object called \emph{connection}, which, as will soon be shown, provides a way to relate vectors of different tangent vector spaces. On the other hand, the connection is introduced to make possible the construction of a coordinate-independent derivative operator. In fact, the partial derivative $\partial_\mu$ when acting on tensor fields, for instance a vector $V^\mu \in T_p$, does not transform like a tensor, since the change of coordinates $x^\mu \rightarrow x^{\tilde{\mu}}$ leads to the transformation

\begin{equation}
 \partial_\mu V^\nu \rightarrow \partial_{\tilde{\mu}}V^{\tilde{\nu}} = \left ( \frac{\partial x^\mu}{\partial x^{\tilde{\mu}}} \partial_\mu \right ) \left ( \frac{\partial  x^{\tilde{\nu}}}{\partial x^\nu} V^\nu  \right ) = \frac{\partial x^\mu}{\partial x^{\tilde{\mu}}} \frac{\partial  x^{\tilde{\nu}}}{\partial x^\nu} \partial_\mu V^\nu + \frac{\partial x^\mu}{\partial x^{\tilde{\mu}}} V^\mu \frac{\partial^2  x^{\tilde{\nu}}}{\partial x^\nu \partial x^\mu} \ .
 \label{eq:notatens}
\end{equation}

The last term of Eq. \ref{eq:notatens} is what makes the transformation law non-tensorial. To construct a \emph{covariant derivative} $\nabla_\mu$, a non-tensor correction must be added to the partial derivative, linearly to the original tensor, which will then provide a proper tensor object. $\nabla_\mu$ is then defined by

\begin{equation}
 \nabla_\mu V^\nu = \partial_\mu V^\nu + \Gamma^\nu_{\mu \lambda} V^\lambda \ .
 \label{eq:covder}
\end{equation}

In Eq. \ref{eq:covder}, $\Gamma^\nu_{\mu \lambda}$ are the \emph{connection coefficients}, non-tensorial objects which transform in the exact way to make the covariant derivative transformation law perfectly tensorial, in opposition to that of Eq. \ref{eq:notatens}. In fact, the non-tensorial transformation of the connection coefficients is given by

\begin{equation}
\Gamma^\nu_{\mu \lambda} \rightarrow \Gamma^{\tilde{\nu}}_{\tilde{\mu} \tilde{\lambda}} = \frac{\partial x^\mu}{\partial x^{\tilde{\mu}}} \frac{\partial x^\lambda}{\partial x^{\tilde{\lambda}}} \frac{\partial  x^{\tilde{\nu}}}{\partial x^\nu} \Gamma^\nu_{\mu \lambda} + \frac{\partial x^\mu}{\partial x^{\tilde{\mu}}} \frac{\partial x^\lambda}{\partial x^{\tilde{\lambda}}} \frac{\partial^2  x^{\tilde{\nu}}}{\partial x^\mu \partial x^\lambda} \ ,
\end{equation}

\noindent which then guarantees the transformation of the covariant derivative to be tensorial,

\begin{equation}
 \nabla_\mu V^\nu \rightarrow \nabla_{\tilde{\mu}} V^{\tilde{\nu}} =  \frac{\partial x^\mu}{\partial x^{\tilde{\mu}}} \frac{\partial  x^{\tilde{\nu}}}{\partial x^\nu} \nabla_\mu V^\nu \ .
 \label{eq:transfcovder}
\end{equation}

Notice that the covariant derivative, as above defined, does indeed generalize the action of a partial derivative: it maps a rank-$(k, l)$ tensor into a rank-$(k, l+1)$ tensor, in a linear way and respecting the product rule. And by construction it is guaranteed to transform like a tensor, which is of great usefulness, as already discussed.

A similar method is used to define the covariant derivative of a dual vector $v_\mu \in T_p^*$, 

\begin{equation}
 \nabla_\mu v_\nu = \partial_\mu v_\nu - \Gamma^\lambda_{\mu \nu} v_\lambda \ .
 \label{eq:covderdual}
\end{equation}

For higher-rank tensors, the covariant derivative is quite simple: for each upper index a positive connection term is added, whilst for each lower index we add a negative connection term, as follows,

\begin{multline}
 \nabla_\rho T\indices{^{\mu_1 \cdots \mu_k}_{\nu_1 \cdots \nu_l}} =
 \partial_\rho T\indices{^{\mu_1 \cdots \mu_k}_{\nu_1 \cdots \nu_l}} 
 \\
 + \Gamma^{\mu_1}_{\rho \lambda} T\indices{^{\lambda \cdots \mu_k}_{\nu_1 \cdots \nu_l}} + \cdots + \Gamma^{\mu_k}_{\rho \lambda} T\indices{^{\mu_1 \cdots \lambda}_{\nu_1 \cdots \nu_l}} 
 \\
 - \Gamma^{\lambda}_{\rho \nu_1} T\indices{^{\mu_1 \cdots \mu_k}_{\lambda \cdots \nu_l}} - \cdots -  \Gamma^{\lambda}_{\rho \nu_l} T\indices{^{\mu_1 \cdots \mu_k}_{\nu_1 \cdots \lambda}} \ .
 \end{multline}

In $4$ dimensions, the connection $\Gamma^\nu_{\mu \lambda}$ consists of a set of $64$ independent coefficients, and it may be defined in various ways. But there is a natural way to construct a connection from the metric, and this connection is the one used in GR. Its construction follows two properties, known as \emph{torsion-free} and \emph{metric-compatibility}.

A torsion-free connection is such that it is symmetric in its lower indices, that is $\Gamma^\mu_{\nu \lambda} = \Gamma^\mu_{(\nu \lambda)}$, in which case the \emph{torsion tensor}, defined as $T^\mu_{\nu \lambda} = \Gamma^\mu_{\nu \lambda} - \Gamma^\mu_{\lambda \nu}$ is identically null. In Def. \ref{def:st} it is clearly stated that spacetime is a manifold which carries such a torsion-free connection, and now it is clear what it means. On the other hand, metric compatibility is the feature of a connection with respect to which the covariant derivative of the metric is globally null $\nabla_\sigma g_{\mu \nu} = 0$. Now, by using these properties, after some algebraic manipulations it is easy to get an explicit expression for the connection as a function of the metric, according to,

\begin{equation}
 \Gamma^\rho_{\mu \nu} = \frac{1}{2} g^{\rho \sigma} \left ( \partial_\mu g_{\nu \sigma} + \partial_\nu g_{\sigma \mu} - \partial_\sigma g_{\mu \nu} \right ) \ .
 \label{eq:christsymb}
\end{equation}

The connection coefficients $\Gamma^\rho_{\mu \nu}$ are called \emph{Christoffel symbols}, and we refer to the connection constructed from the metric as in Eq. \ref{eq:christsymb} as \emph{Christoffel connection}. In flat spacetime, the Christoffel connection constructed from $\eta_{\mu \nu}$ is implicitly chosen, which give, in cartesian coordinates, null Christoffel symbols. On the other hand, the symbols are non-vanishing for curvilinear charts. In curved manifolds it is also possible to make the Christoffel symbols vanish at a point, through the choice of locally inertial coordinates as aforementioned, such that $ g_{\hat{\mu} \hat{\nu}} \rightarrow \eta_{\hat{\mu} \hat{\nu}}$. 

\subsection{Geodesics}
\label{sec:Geodesics}

The covariant derivative, as it is expected by an object which generalizes partial derivatives, provides a way to quantify the rate of change of a given tensor from its configuration in case it was \emph{parallel transported}, i.e., transported along a path while it is kept constant. In flat spaces, it is possible to freely move vectors at will, whilst in curved spaces vectors are only defined in their local $T_p$, so that parallel transport in such spaces require the connection to relate vectors of different tangent vector spaces, hence the term ``connection''. As it turns out, a major difference between parallel transporting a vector between two points in flat and curved spaces, is that its result depends on the path taken between the points in curved space, so that there is no natural way to move vectors. The consequence of such a fact is that there is no natural way to operate with vectors of different $T_p$.

More precisely, the parallel transport is defined as the way of keeping a tensor constant as it is moved through a given path. Thus, the constancy of a tensor $T\indices{^{\mu_1 \cdots \mu_k}_{\nu_1 \cdots \nu_l}}$ along a parameterized path $x^\mu(\lambda)$ is expressed by the \emph{parallel transport equation},

\begin{equation}
 \frac{d x^\rho}{d \lambda} \nabla_\rho T\indices{^{\mu_1 \cdots \mu_k}_{\nu_1 \cdots \nu_l}} = 0 \ .
\end{equation}

For a vector $V^\mu$, the parallel transport equation gets the form

\begin{equation}
  \frac{d x^\rho}{d \lambda} \nabla_\rho V^\mu = 0 \Rightarrow \frac{d}{d \lambda} V^\mu + \Gamma^\mu_{\rho \sigma} \frac{d x^\rho}{d \lambda} V^\sigma = 0 \ .
 \label{eq:dircovder}
\end{equation}

The concept of parallel transport is used to define the \emph{geodesic}, which is the generalization to curved manifolds of the notion of a straight line in Euclidean spaces, which, although generally defined as ``the curve of shortest length between two points'', may also be equivalently defined as ``the path along which its tangent vector is parallel transported (i.e., kept unchanged)''. This last definition is particularly useful when applied to the generalized concept of a geodesic, as follows: since the tangent vector to a parameterized curve $x^\mu(\lambda)$ is $\dfrac{d x^\mu}{d \lambda}$, it will be parallel transported if $\dfrac{d x^\rho}{d \lambda} \nabla_\rho \dfrac{d x^\mu}{d \lambda} = 0 $, which then, according to Eq. \ref{eq:dircovder}, leads to

\begin{equation}
   \frac{d^2 x^\mu}{d \lambda^2} + \Gamma^\mu_{\rho \sigma} \frac{d x^\rho}{d \lambda} \frac{d x^\sigma}{d \lambda} = 0 \ .
 \label{eq:geodeq}
\end{equation}

Eq. \ref{eq:geodeq} is the \emph{geodesic equation}, whose solution provides the geodesic curves $x^\mu(\lambda)$. In fact, for an Euclidean space in which we choose Cartesian coordinates, one has that the Christoffel symbols are all null, so that the geodesic equation becomes $\dfrac{d^2 x^\mu}{d \lambda^2} = 0$, the equation of a straight line!

It is important to mention that the parameter $\lambda$ for timelike geodesics satisfying Eq. \ref{eq:geodeq} must be what is called an \emph{affine parameter}, i.e., of the form $\lambda = \alpha \tau + \beta$, where $\alpha, \beta \in \mathbb{R}$ and $\tau$ is the proper time, defined as the spacetime interval elapsed for a particle moving through a timelike path with fixed spacial coordinates. Naturally, this quantity is negative, so that we define $d\tau^2 = -ds^2$, and therefore $\tau = \int \sqrt{-ds^2}$. The proper time along a timelike path in spacetime will then correspond to the actual time elapsed as measured by an observer moving along this path.

As aforementioned, the major importance of geodesics lies in the fact that these are the curves which free test particles follow, that is, unaccelerated bodies which do not cause themselves any noticeable gravitational field --- i.e., that do not influence the spacetime geometry through which they move. In fact, the geodesic equation is the curved-space generalization of Newton's second law for null forces (i.e., for free particles).

\subsection{The Riemann tensor}

As discussed above, although it is known that geometrical information of a manifold is encoded in the metric tensor, there is no useful way to extract any information regarding a manifold curvature directly from the connection $\Gamma^\mu_{\nu \rho}$, as it can be zero or not even for flat spaces, depending of the chosen charts. Thus, it is necessary to define an object which effectively measures the curvature of a manifold. This is achieved though the \emph{Riemann (curvature) tensor}, defined in terms of the connection according to
\begin{equation}
R\indices{^{\rho}_{\sigma \mu \nu}} = \partial_\mu \Gamma^\rho_{\nu \sigma} - \partial_\nu \Gamma^\rho_{\mu \sigma} + \Gamma^\rho_{\mu \lambda}\Gamma^\lambda_{\nu \sigma} - \Gamma^\rho_{\nu \lambda}\Gamma^\lambda_{\mu \sigma} \ .
\label{eq:riemtens}
\end{equation}

It is clear from Eq. \ref{eq:riemtens} that the Riemann tensor is antisymmetric in its last two indices, i.e., $R\indices{^{\rho}_{\sigma \mu \nu}} = - R\indices{^{\rho}_{\sigma \nu \mu}}$. As it is constructed from the Christoffel connection, which is a function of the metric, it is clear that the curvature retrieved from the Riemann tensor will be that which the metric carries itself. Such retrieval then allows us to make sense of curvature through the notion that Euclidean or Minkowski spaces are flat by merely analyzing their metric, in the following way: \emph{the Riemann tensor will be identically zero if and only if one can construct a coordinate system whose metric components are globally constant}. Then we say that the space with such a structure is flat.

Notice that the following statement is twofold. The converse of the statement is straightforward: if in a given coordinate system the metric tensor components are globally constant, the definitions of the Christoffel symbols (Eq. \ref{eq:christsymb}) and of the Riemann tensor (Eq. \ref{eq:riemtens}) directly lead to the implication $\partial_\rho g_{\mu \nu} = 0 \Rightarrow \Gamma^\rho_{\mu \nu} = 0 \Rightarrow R\indices{^{\rho}_{\sigma \mu \nu}} = 0$. The statement that $R\indices{^{\rho}_{\sigma \mu \nu}} = 0$ implies the existence of a coordinate system with a metric globally constant is subtler. But notice that $R\indices{^{\rho}_{\sigma \mu \nu}} = 0$ is a tensor equation, so that it is independent of the chosen coordinates. Therefore, a vanishing Riemann tensor is a necessary condition for the existence of coordinates such that the associated metric is globally constant --- and if it is true for a given coordinate system, it will be true for every possible chart, and is therefore a feature of the manifold itself. 

Therefore, this is the ultimate check of curvature:  if, for a given metric, the Riemann tensor is identically null, $R\indices{^{\rho}_{\sigma \mu \nu}} = 0$, the space is definitely flat. If not, it is certainly curved.

There are several symmetries which allow us to reduce the $256$ components of the $4$-dimensional Riemann tensor to a smaller set of independent components. Such symmetries are easily seen with the Riemann tensor with all lower indices, $R_{\rho \sigma \mu \nu} = g_{\rho \lambda} R\indices{^{\lambda}_{\sigma \mu \nu}}$. The symmetries follow,

\vspace{-2mm}

\begin{equation}
\begin{gathered}
R_{\rho \sigma \mu \nu} = - R_{\sigma \rho \mu \nu} ~;
\\
R_{\rho \sigma \mu \nu} = - R_{\rho \sigma \nu \mu}~;
\\
R_{\rho \sigma \mu \nu} =  R_{\mu \nu \rho \sigma}~;
\\
R_{\rho [ \sigma \mu \nu ]} = 0~.
\end{gathered}
\end{equation}

Altogether, the symmetries leave only $20$ independent components for the Riemann tensor. It also satisfies the \emph{Bianchi identity} $\nabla_{[\lambda}R_{\rho \sigma] \mu \nu} = 0$.

From the Riemann tensor, we define the \emph{Ricci tensor}, constructed from the following contraction $R_{\mu \nu} = R\indices{^{\lambda}_{\mu \lambda \nu}}$, which can be explicitly written as

\begin{equation}
 R_{\mu \nu} = \partial_\sigma \Gamma^\sigma_{\mu \nu} - \partial_\nu \Gamma^\sigma_{\mu \sigma} + \Gamma^\sigma_{\sigma \lambda}\Gamma^\lambda_{\mu \nu} - \Gamma^\sigma_{\nu \lambda}\Gamma^\lambda_{\mu \sigma} \ .
 \label{eq:ricci_tens}
\end{equation}

The Ricci tensor is symmetric $R_{\mu \nu} = R_{\nu \mu}$. Its trace is called the \emph{Ricci (curvature) scalar}, defined as $R = R\indices{^\mu_\mu} = g^{\mu \nu} R_{ \mu \nu}$. $R_{\mu \nu}$ and $R$ encode all the independent non-vanishing traces of the Riemann tensor, so that the remaining trace-free parts compose the \emph{Weyl tensor} $ C_{\rho \sigma \mu \nu} $, which is of great importance in some parts of the main text. The Weyl tensor carries all the symmetries of the Riemann tensor above presented, and is invariant under conformal transformations. It is defined in $n$ dimensions as

\begin{equation}
 C_{\rho \sigma \mu \nu} = R_{\rho \sigma \mu \nu} - \left ( \frac{2}{n-2} \right ) \left ( g_{\rho[\mu}R_{\nu]\sigma} - g_{\sigma[\mu}R_{\nu]\rho} \right ) + \left ( \frac{2}{(n-1)(n-2)} \right ) g_{\rho [\mu} g_{\nu ] \sigma} R \ .
 \label{eq:weyl}
\end{equation}

Or, alternatively, depending on the convention for antisymmetrization, we may have
\begin{equation}
 C_{\rho \sigma \mu \nu} = R_{\rho \sigma \mu \nu} - \left ( \frac{1}{n-2} \right ) \left ( g_{\rho[\mu}R_{\nu]\sigma} - g_{\sigma[\mu}R_{\nu]\rho} \right ) + \left ( \frac{1}{(n-1)(n-2)} \right ) g_{\rho [\mu} g_{\nu ] \sigma} R \ .
 \label{eq:weyl_2}
\end{equation}

Now, given the symmetries of the Riemann tensor, notice that it is possible to write the Bianchi identity as $\nabla_\lambda R_{\rho \sigma \mu \nu} + \nabla_\rho R_{ \sigma \lambda \mu \nu} + \nabla_\sigma R_{\lambda \rho \mu \nu} = 0$. By contracting twice with the inverse metric, $ g^{\nu \sigma} g^{\mu \lambda} (\nabla_\lambda R_{\rho \sigma \mu \nu} + \nabla_\rho R_{ \sigma \lambda \mu \nu} + \nabla_\sigma R_{\lambda \rho \mu \nu}) = 0 $, we arrive at $\nabla^\mu R_{\rho \mu} - \nabla_\rho R + \nabla^\nu R_{\rho \nu} = 0$, which can be written as $\nabla^\mu R_{\rho \mu} = \dfrac{1}{2} \nabla_\rho R = \dfrac{1}{2} \nabla^\mu g_{\rho \mu} R \Rightarrow \nabla^\mu (R_{\rho \mu} - \dfrac{1}{2}R g_{\rho \mu}) = 0$. We then define the \emph{Einstein tensor},

\begin{equation}
 G_{\mu \nu} = R_{\mu \nu} - \frac{1}{2} R g_{\mu \nu}  \ ,
 \label{eq:einstens}
\end{equation}

\noindent which allows us to write the twice-contracted Bianchi identity as $\nabla^\mu G_{\mu \nu} = 0$. The Einstein tensor of Eq. \ref{eq:einstens} appears in the first two terms of the EFE (Eq. \ref{eq:efe_si}), which makes clear its importance. Its full meaning will be explored in the next section.   

\subsection{Killing vectors}
\label{sec:kllvcts}

Symmetry is one of the most fundamental features of nature, and the mathematical formulation of physical theories tend to make use of them in different ways. In GR it is no different, as symmetries are of great importance in the solution of the EFE and in other contexts. However, the treatment of symmetries in a curved manifold deserves some caution.

Symmetries of the metric are called \emph{isometries}. A particularly useful kind of symmetry is that of translations, which are guaranteed by the independence of the metric with relation to the coordinate functions $x^\mu$, that is $\partial_\sigma g_{\mu \nu} = 0 \Rightarrow x^\sigma \rightarrow x^\sigma + a^\sigma$ is a symmetry, for fixed $a^\sigma$. Isometries of this kind are of great importance when applied to motion along geodesics, because they imply the conservation of the momentum component $p_\sigma$ of a particle in a timelike path $ \partial_\sigma g_{\mu \nu} = 0 \Rightarrow \dfrac{d p_\sigma}{d \tau} = 0$. These and other conserved quantities implied by isometries are quite useful in studying the motion through geodesics, as it will become clear.

There is, however, a more systematic way to characterize symmetries in spacetime. Let us consider the vector $K \equiv \partial_\sigma$, which is built as the partial derivative with respect to the coordinate $x^\sigma$, which $g_{\mu \nu}$ is independent of. In component notation, $K^\mu = (\partial_\sigma)^\mu = \delta^\mu_\sigma$. The vector $K^\mu$ is said to \emph{generate} the isometry, in the sense that a motion in the direction of $K^\mu$ is the infinitesimal expression of the transformation under which the geometry if invariant. That is, we can make the conserved quantity be written as $p_{\sigma} = K^\mu p_\mu$.

On the other hand, as already discussed, the fact that such a quantity is constant along the path is equivalent to the statement that its directional derivative through the geodesic is null, so that, if we write the geodesic equation as $p^\mu \nabla_\mu p^\nu = 0$, we can write $p^\mu \nabla_\mu(K_\nu p^\nu) = p^\mu p^\nu \nabla_{(\mu}K_{\nu)} = 0$. With this, we arrive at \emph{Killing's equation}, on the left side of the following implication,

\begin{equation}
 \nabla_{(\mu}K_{\nu)} = 0 \Rightarrow p^\mu \nabla_\mu(K_\nu p^\nu) = 0 \ .
 \label{eq:keq}
\end{equation}

What Eq. \ref{eq:keq} tell us is that any vector $K^\mu$ satisfying the Killing's equation will imply the conservation of $K_\mu p^\mu$ along a geodesic. In fact, every Killing vector implies the existence of constants of motion on geodesics. That makes sense, as, by construction, the metric is unchanging in the direction of the Killing vector. Also, it can also be shown that $K^\mu \nabla_\mu R = 0$, which formally expresses this idea.

This completes this mathematical appendix, which provides all the formalism necessary to make the main discussion self-contained and rigorously supported.

\chapter{Christoffel symbols and Riemann tensor for Schwarzschild metric}
\label{ap:cs_rt_sm}

Throughout this appendix, we will use the same index notation used in the text $(t, r, \theta, \varphi)$ instead of $(0, 1, 2, 3)$. Such notation provides a more natural and straightforward identification with the spherical coordinate system $\{ t, r, \theta, \varphi \}$ employed in the Schwarzschild solution. We will use Eqs. \ref{eq:christsymb} and \ref{eq:riemtens}, respectively for the Christoffel symbols and the Riemann tensor components,

\begin{equation}
\Gamma^\rho_{\mu \nu} = \frac{1}{2} g^{\rho \sigma} \left ( \partial_\mu g_{\nu \sigma} + \partial_\nu g_{\sigma \mu} - \partial_\sigma g_{\mu \nu} \right ) \ ;
\end{equation}

\vspace{-.4cm}

\begin{equation}
R\indices{^{\rho}_{\sigma \mu \nu}} = \partial_\mu \Gamma^\rho_{\nu \sigma} - \partial_\nu \Gamma^\rho_{\mu \sigma} + \Gamma^\rho_{\mu \lambda}\Gamma^\lambda_{\nu \sigma} - \Gamma^\rho_{\nu \lambda}\Gamma^\lambda_{\mu \sigma} \ .
\end{equation}

And for the Ricci tensor,

\begin{equation}
 R_{\mu \nu} = R\indices{^\lambda_{\mu \lambda \nu}} \ .
\end{equation}

First, we will present the Christoffel Symbols, the Riemann tensor components and the Ricci tensor components for the most generic metric written in terms of radial functions $\alpha(r)$ and $\beta(r)$,

\begin{equation}
 ds^2 = - e^{2\alpha(r)} \mathrm{d}t^2 + e^{2\beta(r)}\mathrm{d}r^2 + r^2\mathrm{d} \theta^2 + r^2\sin^2 \theta \mathrm{d} \varphi^2 \ .
\end{equation}

Thus, one has $g_{tt} = -e^{2\alpha(r)} \Rightarrow g^{tt} = -e^{-2\alpha(r)}$; $g_{rr} = e^{2\beta(r)} \Rightarrow g^{rr} = e^{-2\beta(r)}$; $g_{\theta \theta} = r^2 \Rightarrow g^{\theta \theta} = r^{-2} $ and $g_{\varphi \varphi} = r^2 \sin^2 \theta \Rightarrow g^{\varphi \varphi} = r^{-2} \sin^{-2} \theta$. As the metric coefficients depend only on $r$ and $\theta$, and we assumed a static solution, in everything that follows we will have $\partial_r(.) = \partial_\varphi(.) = \partial_t(.) =  0$, where $(.)$ denotes any general term. Also, throughout this appendix we will use partial $\partial_r$ to denote what is in fact a total derivative $\frac{d}{dr}$.

We will present the explicit calculation of a single component for each upper index $\rho$ of the Christoffel symbols, and then present the rest in matrix form,

\begin{equation}
\begin{gathered}
\Gamma^t_{tr} = \frac{1}{2} g^{t \sigma} ( \cancelto{^0}{\partial_t g_{r \sigma}} + \partial_r g_{\sigma t} - \partial_\sigma \cancelto{^0}{g_{t r}} )  
\\
\Rightarrow \Gamma^t_{tr} = \frac{1}{2} g^{tt} (\partial_r g_{tt}) = \frac{1}{2} (-e^{-2\alpha})\partial_r (-e^{2\alpha}) = \frac{1}{2} e^{-2\alpha}e^{2\alpha} (2\partial_r \alpha)  
\\
\Rightarrow \Gamma^t_{tr} = \partial_r \alpha \ ;
\end{gathered}
\end{equation}

\vspace{2mm}

\begin{equation}
\Gamma^t_{\mu \nu} = 
\begin{pmatrix}
0 & \partial_r \alpha & 0 & 0\\ 
\partial_r \alpha & 0 & 0 & 0 \\ 
0 & 0 & 0 & 0\\ 
0 & 0 & 0 & 0
\end{pmatrix}
\ ;
\end{equation}

\begin{equation}
\begin{gathered}
\Gamma^r_{tt} = \frac{1}{2} g^{r \sigma} ( \cancelto{^0}{\partial_t g_{t \sigma}} + \cancelto{^0}{\partial_t g_{\sigma t}} - \partial_\sigma g_{tt} ) 
\\
\Rightarrow \Gamma^r_{tt} = -\frac{1}{2} g^{rr} (\partial_r g_{tt}) = -\frac{1}{2} (e^{-2\beta}) \partial_r (-e^{2 \alpha}) = \frac{1}{2} e^{2\alpha}e^{-2\beta} (2\partial_r \alpha) 
\\
\Rightarrow \Gamma^r_{tt} = e^{2(\alpha - \beta)}(\partial_r \alpha) \ ;
\end{gathered}
\end{equation}

\vspace{2mm}

\begin{equation}
\Gamma^r_{\mu \nu} = 
\begin{pmatrix}
e^{2(\alpha - \beta)}(\partial_r \alpha) & 0 & 0 & 0\\ 
0 & \partial_r \beta & 0 & 0\\ 
0 & 0 & -r e^{-2\beta} & 0\\ 
0 & 0 & 0 & -r \sin^2 \theta e^{-2\beta} 
\end{pmatrix}
\ ;
\end{equation}

\begin{equation}
\begin{gathered}
\Gamma^\theta_{r \theta} = \frac{1}{2} g^{\theta \sigma} ( \partial_r g_{\theta \sigma} + \partial_\theta g_{\sigma r} - \partial_\sigma \cancelto{^0}{g_{r \theta}} ) 
\\
\Rightarrow \Gamma^\theta_{r \theta} = \frac{1}{2} g^{\theta \theta} (\partial_r g_{\theta \theta} + \partial_\theta \cancelto{^0}{g_{\theta r}})  = \frac{1}{2} \left( \frac{1}{r^2} \right) \partial_r (r^2) = \frac{1}{2} \frac{2r}{r^2} 
\\
\Rightarrow \Gamma^\theta_{r \theta} = \frac{1}{r} \ ;
\end{gathered}
\end{equation}

\vspace{2mm}

\begin{equation}
\Gamma^\theta_{\mu \nu} 
 \begin{pmatrix}
0 & 0 & 0 & 0\\ 
0 & 0 & \dfrac{1}{r} & 0\\ 
0 & \dfrac{1}{r} & 0 & 0\\ 
0 & 0 & 0 & \dfrac{-\sin(2\theta)}{2}
\end{pmatrix} 
\ ;
\end{equation}

\begin{equation}
\begin{gathered}
\Gamma^\varphi_{\theta \varphi} = \frac{1}{2} g^{\varphi \sigma} ( \partial_\theta g_{\varphi \sigma} + \partial_\varphi g_{\sigma \theta} - \partial_\sigma \cancelto{^0}{g_{\theta \varphi}} )  
\\
\Rightarrow \Gamma^\varphi_{\theta \varphi} = \frac{1}{2} g^{\varphi \varphi} (\partial_\theta g_{\varphi \varphi} + \partial_\varphi \cancelto{^0}{g_{\varphi \theta}}) = \frac{1}{2} \left ( \frac{1}{r^2 \sin^2(\theta)} \right) \partial_\theta (r^2 \sin^2(\theta)) = \frac{1}{2} \left ( \frac{2 \sin(\theta)\cos(\theta)}{ \sin^2 (\theta)} \right) 
\\
\Rightarrow \Gamma^\varphi_{\theta \varphi} = \mathrm{cot}(\theta) \ ;
\end{gathered}
\end{equation}

\begin{equation}
\Gamma^\varphi_{\mu \nu} = 
\begin{pmatrix}
0 & 0 & 0 & 0\\ 
0 & 0 & 0 & \dfrac{1}{r} \\ 
0 & 0 & 0 & \mathrm{cot}(\theta) \\ 
0 & \dfrac{1}{r} & \mathrm{cot}(\theta) & 0
\end{pmatrix}
\ .
\end{equation}

The non-vanishing and independent Riemann tensor components are,

\begin{equation}
\begin{aligned}
R\indices{^t_{r t r}} &= (\partial_r \alpha)(\partial_r \beta) - (\partial^2_{r^2} \alpha) - (\partial_r \alpha)^2 \ ; 
\\
R\indices{^t_{\theta t \theta}} &= -r e^{-2 \beta} \partial_r \alpha \ ; 
\\
R\indices{^t_{\varphi t \varphi}} &= -r e^{-2 \beta}  \sin^2(\theta) \partial_r \alpha \ ; 
\\
R\indices{^r_{\theta r \theta}} &= r e^{-2 \beta} \partial_r \beta \ ; 
\\
R\indices{^r_{\varphi r \varphi}} &=  r e^{-2 \beta} \sin^2(\theta)\partial_r \beta \ ; 
\\
R\indices{^\theta_{\varphi \theta \varphi}} &= (1 - e^{-2\beta}) \sin^2(\theta)\ .
\label{eq:riemcomp}
\end{aligned}
\end{equation}

Explicitly, the component $R\indices{^t_{\theta t \theta}}$ is calculated as,

\begin{equation}
\begin{gathered}
R\indices{^t_{\theta t \theta}} = \cancelto{^0}{\partial_t \Gamma^t_{\theta \theta}} - \partial_\theta \cancelto{^0}{\Gamma^t_{t \theta}} + \Gamma^t_{t \lambda}\Gamma^\lambda_{\theta \theta} - \cancelto{^0}{\Gamma^t_{\theta \lambda}}\Gamma^\lambda_{t \theta} \Rightarrow
\\
R\indices{^t_{\theta t \theta}} =  \Gamma^t_{t r}\Gamma^r_{\theta \theta} = (\partial_r \alpha) (-r e^{-2 \beta}) 
\\
\Rightarrow R\indices{^t_{\theta t \theta}} = -r e^{-2 \beta} \partial_r \alpha \ .
\end{gathered}
\end{equation}

And the non-vanishing Ricci tensor components are,

\begin{equation}
\begin{aligned}
  R_{tt} &= e^{2(\alpha - \beta)} \left (-(\partial_r \alpha) (\partial_r \beta) + \partial^2_{r^2} \alpha + (\partial_r \alpha)^2 + \frac{2}{r} \partial_r \alpha \right ) \ ;
  \\
  R_{rr} &= (\partial_r \alpha) (\partial_r \beta) -\partial^2_{r^2} \alpha - (\partial_r \alpha)^2 + \frac{2}{r} \partial_r \beta \ ;
  \\
  R_{\theta \theta} &= e^{-2\beta} (r\partial_r (\beta - \alpha) - 1) + 1 \ ;
  \\
  R_{\varphi \varphi} &= \sin^2 \theta (e^{-2\beta}(r \partial_r (\beta - \alpha) - 1) + 1) \ .
\end{aligned}
\end{equation}

Explicitly, the component $R_{tt}$ is obtained as follows,

\vspace{-5mm}

\begin{equation}
\begin{gathered}
R_{tt} = R\indices{^\lambda_{t \lambda t}} =  \cancelto{^0}{R\indices{^t_{t t t}}} +  R\indices{^r_{t r t}} +  R\indices{^\theta_{t \theta t}} +  R\indices{^\varphi_{t \varphi t}}   
\\
\Rightarrow R_{tt} = g^{rr}R_{rtrt} + g^{\theta \theta}R_{\theta t \theta t} + g^{\varphi \varphi}R_{ \varphi  t \varphi  t}  
\\
\Rightarrow R_{tt} = g^{rr}R_{trtr} + g^{\theta \theta}R_{t \theta t \theta} + g^{\varphi \varphi}R_{t \varphi  t \varphi  } 
\\
\Rightarrow R_{tt} = g^{rr} \left ( g_{t \lambda} R\indices{^\lambda_{rtr}} \right ) + g^{\theta \theta} \left( g_{t \lambda}R\indices{^\lambda_{\theta t \theta}} \right) + g^{\varphi \varphi} \left(g_{t \lambda}R\indices{^\lambda_{\varphi  t \varphi}} \right) 
\\
\Rightarrow R_{tt} = g^{rr} \left ( g_{tt} R\indices{^t_{rtr}} \right ) + g^{\theta \theta} ( g_{tt} R\indices{^t_{\theta t \theta}}) + g^{\varphi \varphi} \left(g_{t t}R\indices{^t_{\varphi  t \varphi}} \right) 
\\
\Rightarrow R_{tt} = (e^{-2 \beta})(-e^{2\alpha}) [ (\partial_r \alpha)(\partial_r \beta) - (\partial^2_{r^2} \alpha) - (\partial_r \alpha)^2] + \left(\frac{1}{r^2} \right)(-e^{2\alpha})(-re^{-2\beta} \partial_r \alpha) 
\\
+\left ( \frac{1}{r^2\sin^{2}(\theta)} \right)(-e^{2\alpha})(-r e^{-2 \beta}  \sin^2(\theta) \partial_r \alpha) 
\\
\Rightarrow R_{tt} = -e^{2 (\alpha -\beta)} [ (\partial_r \alpha)(\partial_r \beta) - (\partial^2_{r^2} \alpha) - (\partial_r \alpha)^2] + e^{2(\alpha - \beta)} \left ( \frac{1}{r} \right)(\partial_r \alpha) + e^{2(\alpha - \beta)} \left ( \frac{1}{r} \right)(\partial_r \alpha) 
\\
\Rightarrow R_{tt} = e^{2(\alpha - \beta)} \left (-(\partial_r \alpha) (\partial_r \beta) + \partial^2_{r^2} \alpha + (\partial_r \alpha)^2 + \frac{2}{r} \partial_r \alpha \right ) \ .
\end{gathered}
\end{equation}

Notice how, even though we did not present any component of the form $R\indices{^\lambda_{t \lambda t}}$ in Eq. \ref{eq:riemcomp}, it was possible to calculate the contraction $R_{tt}$, given the symmetries of the Riemann tensor. 

Now, for the solution Schwarzschild metric itself, given by Eq. \ref{eq:sm},

\begin{equation}
 ds^2 = - \left ( 1 - \frac{2M}{r} \right ) \mathrm{d}t^2 + \left ( 1 - \frac{2M}{r} \right )^{-1}\mathrm{d}r^2 +  r^2\mathrm{d} \theta^2 + r^2\sin^2 \theta \mathrm{d} \varphi^2 \ ,
\end{equation}

\noindent the Christoffel symbols are,

\begin{equation}
\Gamma^t_{\mu \nu} = 
\begin{pmatrix}
0 & \frac{M}{r(r-2M)} & 0 & 0\\ 
\frac{M}{r(r-2M)} & 0 & 0 & 0\\ 
0 & 0 & 0 & 0\\ 
0 & 0 & 0 & 0
\end{pmatrix}
\ ;
\end{equation}

\begin{equation}
\Gamma^r_{\mu \nu} = 
\begin{pmatrix}
\frac{M}{r^2}\left( 1- \frac{2M}{r} \right) & 0 & 0 & 0\\ 
0 & \frac{-M}{r(r-2M)} & 0 & 0\\ 
0 & 0 & 2M - r & 0\\ 
0 & 0 & 0  & (2M - r)\sin^2\theta
\end{pmatrix}
\ ;
\end{equation}

\begin{equation}
\Gamma^\theta_{\mu \nu} 
 \begin{pmatrix}
0 & 0 & 0 & 0\\ 
0 & 0 & \dfrac{1}{r} & 0\\ 
0 & \dfrac{1}{r} & 0 & 0\\ 
0 & 0 & 0 & \dfrac{-\sin(2\theta)}{2}
\end{pmatrix}
\ ;
\end{equation}

\begin{equation}
\Gamma^\varphi_{\mu \nu} = 
\begin{pmatrix}
0 & 0 & 0 & 0\\ 
0 & 0 & 0 & \dfrac{1}{r} \\ 
0 & 0 & 0 & \mathrm{cot}(\theta) \\ 
0 & \dfrac{1}{r} & \mathrm{cot}(\theta) & 0
\end{pmatrix}
\ .
\end{equation}

Notice that, since $\Gamma^\theta_{\mu \nu}$ and $\Gamma^\varphi_{\mu \nu}$ do not depend on $\alpha(r)$ and $\beta(r)$, they are the same as previously presented.

And the non-vanishing Riemann tensor components are,

\begin{equation}
\begin{aligned}
R\indices{^\theta_{t \theta t}} = R\indices{^\varphi_{t \varphi t}} &= \frac{-M (2M -r)}{r^4} \ ;
\\
R\indices{^r_{t r t}} &= \frac{2M (2M -r)}{r^4} \ ;
\\
R\indices{^\theta_{r \theta r}} = R\indices{^\varphi_{r \varphi r}} &= \frac{-M}{r^2(r-2M)} \ ;
\\
R\indices{^t_{r t r}} &= \frac{2M}{r^2(r-2M)} \ ; 
\\
R\indices{^r_{\theta r \theta}} = R\indices{^t_{\theta t \theta}} &= \frac{-M}{r} \ ; 
\\
R\indices{^\varphi_{\theta \varphi \theta}} &= \frac{2M}{r} \ ; 
\\
R\indices{^r_{\varphi r \varphi}} = R\indices{^t_{\varphi t \varphi}} &= \frac{-M \sin^2 \theta}{r} \ ; 
\\
R\indices{^\theta_{\varphi \theta \varphi}} &= \frac{2M \sin^2 \theta}{r}  \ .
\end{aligned}
\end{equation}

As it is easy to see from the Riemann tensor components above, one has, as expected for a vacuum solution, the Ricci tensor components $R_{tt} = R_{rr} = R_{\theta \theta} = R_{\varphi \varphi} = R_{\mu \nu} = 0$.

\chapter{Submanifolds and Hypersurfaces}
\label{ap:whisk}

In this appendix, which will closely follow  \cite{whisker}, we will present the basics of the mathematical formalism necessary to the study of the brane as a submanifold. The results here presented will be directly used in Sec. \ref{sec:effe}, although they may also appear throughout the text.

The brane is a 4-dimensional submanifold $\Sigma_4$ embedded in the 5-dimensional bulk $\mathcal{M}_5$. As the brane has one dimension less than the bulk, it is said to be a \emph{co-dimension one} hypersurface. $\Sigma_4$ can be described by a set of $5$ parametric equations which are functions of its internal coordinates $x^\mu$ $f^A = f^A(x^\mu)$, $\mu = (0, 1, 2, 3)$; $A=(0, 1, 2, 3, 4)$. 

Now, it is possible to construct a constraint equation defining the hypersurface $\Phi(f^A) = c$, where $c$ is a constant. That is, the hypersurface is defined as the set of points satisfying the constraint equation defined by the chosen constant. This then allows the construction of the unit vector normal to every point of $\Sigma_4$,

\begin{equation}
 n_A = \frac{\nabla_A \Phi}{\sqrt{\left | g^{AB}\nabla_B \Phi \nabla_A\Phi \right |}} \ ,
\end{equation}

\noindent where $g_{AB}$ is the metric of $\mathcal{M}_5$. From the definition of the normal vector, we define the \emph{induced metric} on $\Sigma_4$, $q_{AB}$, as

\begin{equation}
 q_{AB} = g_{AB} - n_A n_B \ .
\end{equation}

In fact, $q_{AB}$ is a tensor defined only on the hypersurface (the brane), and just like the metric $g_{AB}$, it encodes all the geometrical information of the submanifold. Therefore, naturally, all the objects constructed from the metric (like the Riemann and Ricci tensor) can be constructed from $q_{AB}$, and will be 4-dimensional, accordingly. 

The induced metric in the form $q\indices{^A_B}$ acts like a projection operator, from the tangent space $T_p$ of $\mathcal{M}_5$ to the tangent space $ \ ^{(4)}T_p$ of $\Sigma_4$, with $p \in \Sigma_4$. Generally, the projection of a tensor to $ \ ^{(4)}T_p$ is given by

\begin{equation}
 T\indices{^{A_1 \cdots A_k}_{B_1 \cdots B_l}} =  q\indices{^{A_1}_{C_1}} \cdots q\indices{^{A_k}_{C_k}} q\indices{_{B_1}^{D_1}} \cdots q\indices{_{B_l}^{D_l}} T\indices{^{C_1 \cdots C_k}_{D_1 \cdots D_l}} \ .
\end{equation}

An example is the projection of a vector $v^A$, decomposed into tangent and perpendicular parts to $\Sigma_4$ $v^A = v^A_{\bot} + v^A_{\parallel}$. Naturally, its projection to  $ \ ^{(4)}T_p$ is $q\indices{^A_B}v^B = v^A_{\parallel}$, which is intuitive and easily verifiable as true, since $q\indices{^A_B} = \delta^A_B - n^A n_B$. Also, we obviously have $q\indices{^A_B}n^B = 0$.

We can also define a covariant derivative on the brane, $D_A$, which is simply the projection of the covariant derivative of the bulk, $\nabla_A$, to the brane,

\begin{equation}
D_C T\indices{^{A_1 \cdots A_k}_{B_1 \cdots B_l}} =  q\indices{^{A_1}_{D_1}} \cdots q\indices{^{A_k}_{D_k}} q\indices{_{B_1}^{E_1}} \cdots q\indices{_{B_l}^{E_l}} \left (q\indices{_C^G}\nabla_G \right ) T\indices{^{D_1 \cdots D_k}_{E_1 \cdots E_l}} \ .
\end{equation}

This allows the definition of the \emph{extrinsic curvature} $K_{AB} = K_{(AB)}$,

\begin{equation}
 K_{AB} \equiv D_A n_B = q\indices{_A^C}\nabla_C n_B \ .
\end{equation}

Given these constructions, we can now introduce the Gauss-Codazzi equations,

\begin{equation}
\begin{gathered}
^{(4)}R\indices{_{ABC}^F} =  \  ^{(5)}R\indices{_{DGH}^E} q\indices{_E^F} q\indices{_A^D} q\indices{_B^G} q\indices{_C^H} + K_{CA}K\indices{_B^F} - K_{CB}K\indices{_A^F} \ ,
\end{gathered}
\end{equation}

\noindent and 

\begin{equation}
 D_A K\indices{^A_B} - D_B K = \  ^{(5)}R_{GH} n^H q\indices{^G_B} \ .
\end{equation}

A full derivation can be found on \cites{carroll, whisker, wald}. In the braneworld context, these equations are very important to construct 4-dimensional objects from 5-dimensional objects, through the mere projection of the bulk quantities into the brane and a correction given by the extrinsic curvature.

Now, through each point $p \in \Sigma_4$, there is an unique geodesic of tangent vector $n^A$. If we chose the chart $\{x^\mu\}$ for a portion of $\Sigma_4$ containing $p$, and label the points of the bulk with the parameter $y$ defined by the geodesic it lies, we construct the coordinate system $\{ x^\mu, y \}$, known as the \emph{Gaussian Normal coordinates (GNC)}, which is commonly used in braneworld scenarios. In this coordinates, the metric can be written as $ds^2 = q_{\mu \nu} \mathrm{d}x^\mu \otimes  \mathrm{d}x^\nu  + \mathrm{d}y^2$.

An important feature of the GNC system is that it allows the definition of the extrinsic curvature in a much more natural way $K_{AB} = \frac{1}{2}\partial_y q_{AB}$, which makes sense, since $y$ is defined tangentially to $\Sigma_4$.

\chapter{Calculation of the inverse boosted metric}
\label{ap:inverse}

In Eq. \ref{eq:boosted_metric_comps}, we have the components of the boosted metric,

\begin{equation}
\begin{gathered}
g_{tt} = \gamma^2 \left ( v^2 - \alpha^2 \left ( r - r_h \right ) \right ) \ ; \ 
g_{tr} = \gamma \alpha  \ ; \ 
g_{ti} = \frac{\gamma^2 \alpha^2}{r_c} \left ( r - r_c \right ) v_i \ ; \ 
\\
g_{rr} = 0; \ 
g_{ri} = -\frac{\gamma \alpha}{r_c}v_i \ ; \ 
g_{ij} = \delta_{ij} - \frac{\gamma^2 \alpha^2}{r_c^2}  \left (r -r_c \right ) v_i v_j \ .
\end{gathered}
\end{equation}

To calculate the inverse boosted metric components, we begin by the definition,

\begin{equation}
\begin{gathered}
g_{\mu \nu}g^{\nu \sigma} = \delta^\sigma_\mu 
\\
\Rightarrow g_{\mu t}g^{t \sigma} + g_{\mu r}g^{r \sigma} + g_{\mu j}g^{j \sigma} = \delta^\sigma_\mu \ .
\label{eq:inv_def}
\end{gathered}
\end{equation}

Fixing $\sigma = t, r$ or $i$ will give us three systems of three equations for the unknown inverse metric components. We shall start by fixing $\sigma = t$, which gives us,

\begin{empheq}[left=\empheqlbrace]{align}
g_{tt}g^{tt} + g_{tr}g^{rt} + g_{tj}g^{jt} = \delta^t_t = 1 \label{eq:t1} \\
g_{rt}g^{tt} + g_{rr}g^{rt} + g_{rj}g^{jt} = \delta^t_r = 0 \label{eq:t2} \\
g_{it}g^{tt} + g_{ir}g^{rt} + g_{ij}g^{jt} = \delta^t_i = 0 \label{eq:t3} 
\end{empheq}

Since $g_{rr}=0$, from Eq. \ref{eq:t2} we have

\begin{equation}
\begin{gathered}
\left(\gamma \alpha \right )g^{tt} + \left (-\frac{\gamma \alpha}{r_c} v_j \right )g^{jt} = 0 
\\
\Rightarrow g^{tt} = \frac{v_j}{r_c}g^{jt} \ .
\label{eq:t4}
\end{gathered}
\end{equation}

Multiplying Eq. \ref{eq:t1} by $-g_{ir}$ and Eq. \ref{eq:t3} by $g_{tr}$, and summing the resulting equations, yields,

\begin{equation}
\left (g_{tr}g_{it} - g_{ir}g_{tt} \right )g^{tt} + \left( g_{tr}g_{ij} - g_{ir}g_{tj}\right )g^{jt} = -g_{ir}
\label{eq:t5}
\end{equation}

Substituting Eq. \ref{eq:t4} in Eq. \ref{eq:t5} then allows us to find $g^{tj}$,

\begin{equation}
\begin{gathered}
\left (g_{tr}g_{it} - g_{ir}g_{tt} \right )\left (\frac{v_j}{r_c}g^{jt} \right ) + \left (g_{tr}g_{ij} - g_{ir}g_{tj} \right )g^{jt} = -g_{ir} 
\\
\Rightarrow \left ( g_{tr}g_{it}v_j - g_{ir}g_{tt}v_j + g_{tr}g_{ij}r_c - g_{ir}g_{tj}r_c \right )g^{jt} = -r_c g_{ir}  
\\
\Rightarrow \left [ g_{tr} \left ( g_{it}v_j + g_{ij} r_c \right ) - g_{ir} \left ( g_{tt} v_j + g_{tj}r_c \right ) \right ]g^{jt} = -r_c \left ( -\frac{\gamma \alpha}{r_c}v_i \right ) = \gamma \alpha v_i
\\
\Rightarrow \left (\gamma \alpha \right ) \left [ \left ( \frac{\gamma^2 \alpha^2}{r_c} \left (r - r_c \right ) v_i \right ) v_j + \left ( \delta_{ij} - \frac{\gamma^2 \alpha^2}{r_c^2}  \left (r -r_c \right ) v_i v_j  \right ) r_c \right ] g^{jt} 
\\
- \left ( -\frac{\gamma \alpha}{r_c}v_i \right ) \left [ \left (\gamma^2 \left ( v^2 - \alpha^2 \left ( r - r_h \right ) \right ) \right ) v_j + \left ( \frac{\gamma^2 \alpha^2}{r_c} \left ( r - r_c \right ) v_j \right ) r_c\right ] g^{jt} = \gamma \alpha v_i 
\\
\Rightarrow \left ( r_c \delta_{ij} + \frac{v_i}{r_c} \left [\gamma^2 v^2 v_j + \gamma^2 \alpha^2 v_j \left (r_h - r + r - r_c) \right ) \right ]\right ) g^{jt} = v_i  
\\
\Rightarrow \left ( r_c \delta_{ij} + \frac{\gamma^2}{r_c} \left [v^2 + \frac{r_c(r_h - r_c)}{r_c-r_h} \right ] v_i v_j\right ) g^{jt} = v_i  
\\
\Rightarrow \left ( r_c \delta_{ij} + {\gamma^2} \left [\frac{v^2}{r_c} - 1 \right ] v_i v_j\right ) g^{jt} = v_i  
\\
\Rightarrow \left ( r_c \delta_{ij} - v_i v_j\right ) g^{jt} = v_i  
\\
\Rightarrow \left ( r_c \delta_{ij}v^i - v_iv^i v_j\right ) g^{jt} = v_iv^i 
\\
\Rightarrow \left ( r_c v_j - v^2 v_j\right ) g^{jt} = v^2  
\\
\Rightarrow v_j g^{jt} = \frac{v^2}{r_c - v^2} = \frac{v^2}{r_c \left ( 1 - \frac{v^2}{r_c} \right ) } 
\\
\Rightarrow v_j g^{jt} = \frac{\gamma^2 v^2}{r_c} 
\\
\Rightarrow g^{jt} = \frac{\gamma^2}{r_c} v^j \ .
\end{gathered}
\end{equation}

From Eq. \ref{eq:t4} we then directly obtain $g^{tt}$,

\begin{equation}
\begin{gathered}
g^{tt} = \frac{v_j}{r_c}g^{jt} = \frac{v_j}{r_c} \left (\frac{\gamma^2}{r_c} v^j \right ) 
\\
\Rightarrow g^{tt} = \frac{\gamma^2 v^2}{r_c^2} \ .
\end{gathered}
\end{equation}

Now, using Eq. \ref{eq:t1} and substituting the components found above, we can determine the component $g^{tr}$,

\begin{equation}
\begin{gathered}
g_{tt}g^{tt} + g_{tr}g^{rt} + g_{tj}g^{jt} = \delta^t_t = 1 
\\
\Rightarrow \left [ \gamma^2 \left ( v^2 - \alpha^2 \left ( r - r_h \right ) \right ) \right ] \left ( \frac{\gamma^2 v^2}{r_c^2}\right ) + \left ( \gamma \alpha\right )g^{rt} + \left (\frac{\gamma^2 \alpha^2}{r_c} \left ( r - r_c \right ) v_j \right ) \left ( \frac{\gamma^2}{r_c} v^j\right ) = 1 
\\
\Rightarrow \frac{\gamma^2 v^2}{r_c^2} \left [\gamma^2v^2 - \gamma^2 \alpha^2 \left ( r - r_h - r + r_c\right ) \right ] + \gamma \alpha g^{rt} = 1 \Rightarrow
\\ 
\Rightarrow \gamma \alpha g^{rt} = 1 - \frac{\gamma^2 v^2}{r_c^2} \gamma^2 \left [v^2 - \frac{r_c \left (r_c - r_h \right )}{r_c-r_h} \right ] 
\\
\Rightarrow \gamma \alpha g^{rt} = 1 - \frac{\gamma^2 v^2}{r_c} \gamma^2 \left [\frac{v^2}{r_c} - 1 \right ] 
\\
\Rightarrow \gamma \alpha g^{rt} = 1 + \frac{\gamma^2v^2}{r_c} = 1 + \left ( \gamma^2 - 1 \right ) = \gamma^2 
\\
\Rightarrow g^{tr} = \frac{\gamma}{\alpha} \ ,
\end{gathered}
\end{equation}

\noindent where we used in the 6th line the identity of Eq. \ref{eq:id_gamma2-1}.

Now, fixing $\sigma=r$ in Eq. \ref{eq:inv_def} gives us a new system of three equations,

\begin{empheq}[left=\empheqlbrace]{align}
g_{tt}g^{tr} + g_{tr}g^{rr} + g_{tj}g^{jr} = \delta^r_t = 0 \label{eq:r1} \\
g_{rt}g^{tr} + g_{rr}g^{rr} + g_{rj}g^{jr} = \delta^r_r = 1 \label{eq:r2} \\
g_{it}g^{tr} + g_{ir}g^{rr} + g_{ij}g^{jr} = \delta^r_i = 0 \label{eq:r3} 
\end{empheq}

Eq. \ref{eq:r2} yields,

\begin{equation}
\begin{gathered}
g_{rt}g^{tr} + g_{rr}g^{rr} + g_{rj}g^{jr} = \left (\gamma \alpha \right ) \left ( \frac{\gamma}{\alpha}\right ) + \left (-\frac{\gamma\alpha}{r_c}v_j \right ) g^{jr} = 1 
\\
\Rightarrow v_j g^{jr} = \frac{r_c}{\gamma \alpha} \left ( \gamma^2 - 1 \right )  = \frac{r_c}{\gamma \alpha} \frac{\gamma^2 v^2}{r_c} 
\\
\Rightarrow v_j g^{jr} = \frac{\gamma v^2}{\alpha} 
\\
\Rightarrow g^{jr} = \frac{\gamma}{\alpha} v^j \ .
\end{gathered}
\end{equation}

On the other hand, from Eq. \ref{eq:r1} one gets,

\begin{equation}
\begin{gathered}
g_{tt}g^{tr} + g_{tr}g^{rr} + g_{tj}g^{jr} = 0 
\\
\Rightarrow \left [ \gamma^2 \left ( v^2 - \alpha^2 \left ( r - r_h \right ) \right ) \right ] \left ( \frac{\gamma}{\alpha} \right ) + \left ( \gamma \alpha \right ) g^{rr} +  \left ( \frac{\gamma^2 \alpha^2}{r_c} \left ( r - r_c \right ) v_j \right ) \left ( \frac{\gamma}{\alpha} v^j \right ) = 0 
\\
\Rightarrow \gamma \alpha g^{rr} = - \frac{\gamma^3}{\alpha} \left [ v^2 - \alpha^2 \left (r - r_h \right ) + \frac{\alpha^2 v^2}{r_c} \left (r -r_c \right ) \right ] 
\\
\Rightarrow g^{rr} = - \frac{\gamma^2}{\alpha^2} \left [ v^2 \left (1 + \frac{\alpha^2}{r_c} \left (r-r_c \right ) \right ) - \alpha^2 \left (r - r_h \right ) \right ] 
\\
\Rightarrow g^{rr} = - \frac{\gamma^2}{\alpha^2} \left [ v^2 \left (1 + \frac{r-r_c}{r_c-r_h} \right ) - \frac{r_c \left (r - r_h \right )}{r_c-r_h} \right ] 
\\
\Rightarrow g^{rr} = - \frac{\gamma^2}{\alpha^2} \left [ \frac{v^2 \left (r-r_h \right )}{r_c-r_h} - \frac{r_c \left (r - r_h \right )}{r_c-r_h} \right ] 
\\
\Rightarrow g^{rr} = -\frac{\gamma^2}{\alpha^2} \left (v^2 - r_c \right )  \left ( \frac{r - r_h}{r_c-r_h} \right )  
\\
\Rightarrow g^{rr} = \gamma^2 \left ( r_c - v^2 \right )  \left ( \frac{r - r_h}{r_c-r_h} \right ) \frac{1}{\alpha^2}  
\\
\Rightarrow g^{rr} = \gamma^2 r_c \left (1 - \frac{v^2}{r_c} \right ) \left ( \frac{r - r_h}{r_c-r_h} \right ) \frac{r_c - r_h}{r_c}  
\\
\Rightarrow g^{rr} = r-r_h \ .
\end{gathered}
\end{equation}

Notice that, as it should be, Eq. \ref{eq:r3} is satisfied,

\begin{equation}
\begin{gathered}
g_{it}g^{tr} + g_{ir}g^{rr} + g_{ij}g^{jr}  
\\
= \left ( \frac{\gamma^2 \alpha^2}{r_c} \left ( r - r_c \right ) v_i \right ) \left (\frac{\gamma}{\alpha} \right ) + \left ( -\frac{\gamma \alpha}{r_c}v_i \right ) \left (r-r_h \right ) + \left [ \delta_{ij} - \frac{\gamma^2 \alpha^2}{r_c^2}  \left (r -r_c \right ) v_i v_j \right ] \left (\frac{\gamma}{\alpha}v^j \right ) 
\\
= \left [ \frac{\gamma^3 \alpha}{r_c}\left (r -r_c \right ) \left (1 - \frac{v^2}{r_c} \right ) -\frac{\gamma \alpha}{r_c} \left (r-r_h \right ) + \frac{\gamma}{\alpha} \right ] v_i  
\\
= \left [ \frac{\alpha}{r_c} \left ( r - r_c - r + r_h \right ) + \frac{1}{\alpha} \right ] \gamma v_i  
\\
= \left [ \alpha \frac{r_h-r_c}{r_c} + \frac{1}{\alpha} \right ] \gamma v_i  
\\
= \left [ \alpha \left ( -\frac{1}{\alpha^2} \right ) + \frac{1}{\alpha} \right ] \gamma v_i = 0 
\\
\Rightarrow g_{it}g^{tr} + g_{ir}g^{rr} + g_{ij}g^{jr} = 0 \ .
\end{gathered}
\end{equation}

Finally, by fixing $\sigma = i$ on Eq. \ref{eq:inv_def}, we obtain,

\begin{empheq}[left=\empheqlbrace]{align}
g_{tt}g^{ti} + g_{tr}g^{ri} + g_{tj}g^{ji} = \delta^i_t = 0 \label{eq:i1} \\
g_{rt}g^{ti} + g_{rr}g^{ri} + g_{rj}g^{ji} = \delta^i_r = 0 \label{eq:i2} \\
g_{kt}g^{ti} + g_{kr}g^{ri} + g_{kj}g^{ji} = \delta^i_k \label{eq:i3} 
\end{empheq}

Eq. \ref{eq:i2} gives us directly,

\begin{equation}
\begin{gathered}
g_{rt}g^{ti} + g_{rr}g^{rr} + g_{rj}g^{ji} = \left (\gamma \alpha \right ) \left ( \frac{\gamma^2}{r_c}v^i \right ) + \left ( -\frac{\gamma \alpha}{r_c}v_j \right)g^{ji} = \delta^i_r = 0 
\\
\Rightarrow v_jg^{ji} = \gamma^2 v^i \ .
\label{eq:i4}
\end{gathered}
\end{equation}

We may have several expressions for $g^{ij}$ which solves the last line of Eq. \ref{eq:i4}. The most trivial ones would be $g^{ij} = \gamma^2 \delta^{ij}$ or $g^{ij} = \gamma^2 v^i v^j /v^2$. However, although both these solutions trivially satisfy Eq. \ref{eq:i2}, they do not satisfy Eq. \ref{eq:i3}, and therefore they cannot be right. In fact, the solution for $g^{ij}$ which satisfy all Eqs. \ref{eq:i1} to \ref{eq:i3} is,

\begin{equation}
 g^{ij} = \delta^{ij} + \frac{\gamma^2}{r_c} v^i v^j \ , 
\label{eq:i5}
\end{equation}

\noindent as one can explicitly verify,

\begin{equation}
\begin{gathered}
g_{tt}g^{ti} + g_{tr}g^{ri} + g_{tj}g^{ji}  
\\
= \left [ \gamma^2 \left ( v^2 - \alpha^2 \left ( r - r_h \right ) \right ) \right ] \left (\frac{\gamma^2}{r_c} v^i \right )  + \left (\gamma \alpha \right ) \left ( \frac{\gamma}{\alpha} v^i \right ) + \left [ \frac{\gamma^2 \alpha^2}{r_c} \left ( r - r_c \right ) v_j \right ] \left ( \delta^{ij} + \frac{\gamma^2}{r_c} v^i v^j \right ) 
\\ 
= \left [ \frac{\gamma^4}{r_c}\left ( v^2 - \alpha^2 \left (r-rh\right ) \right ) + \gamma^2 + \frac{\gamma^2 \alpha^2}{r_c} \left (r - r_c \right ) \left (1 + \frac{\gamma^2v^2}{r_c} \right )  \right ] v^i  
\\
= \frac{\gamma^4}{r_c} \left [ v^2 - \alpha^2 \left ( r - r_h \right ) + \frac{v^2 \alpha^2}{r_c} \left (r - r_c \right ) \right ]v^i + \gamma^2 \left [ 1 + \frac{\alpha^2 \left (r - r_c \right )}{r_c} \right ] v^i 
\\ 
= \frac{\gamma^4}{r_c} \left [ v^2 \left (  1 + \frac{\alpha^2 \left (r - r_c \right )}{r_c} \right ) - \alpha^2 \left ( r - r_h \right ) + \right ]v^i + \gamma^2 \left [ 1 + \frac{r - r_c}{r_c - r_h} \right ]v^i 
\\
= \frac{\gamma^4}{r_c} \left [ v^2 \left ( 1 + \frac{r - r_c}{r_c - r_h} \right ) - \alpha^2 \left ( r - r_h \right ) + \right ]v^i + \gamma^2 \left [\frac{r - r_h}{r_c - r_h} \right ]v^i 
\\
= \frac{\gamma^4}{r_c} \left [ v^2 \left ( \frac{r - r_h}{r_c - r_h} \right ) - \frac{r_c \left ( r - r_h \right )}{r_c-r_h} + \right ]v^i + \gamma^2 \left [\frac{r - r_h}{r_c - r_h} \right ]v^i 
\\
= \gamma^4 \left ( \frac{v^2 - r_c}{r_c} \right ) \left [ \frac{r - r_h}{r_c - r_h} \right ]v^i + \gamma^2 \left [\frac{r - r_h}{r_c - r_h} \right ]v^i 
\\
= \gamma^4 \left (- \frac{1}{\gamma^2} \right ) \left [ \frac{r - r_h}{r_c - r_h} \right ]v^i+ \gamma^2 \left [\frac{r - r_h}{r_c - r_h} \right ]v^i = 0 
\\
\Rightarrow g_{tt}g^{ti} + g_{tr}g^{ri} + g_{tj}g^{ji} = 0 \ ;
\end{gathered}
\end{equation}

\begin{equation}
\begin{gathered}
g_{rt}g^{ti} + g_{rr}g^{rr} + g_{rj}g^{ji}  
\\
= v_jg^{ji} - \gamma^2 v^i = v_j \left ( \delta^{ij} + \frac{\gamma^2}{r_c} v^i v^j \right ) - \gamma^2 v_i 
\\
= v_i + \frac{\gamma^2v^2}{r_c}v_i - \gamma^2v_i  
\\
= \left [ \gamma^2 \left (\frac{v^2}{r_c} - 1 \right ) + 1 \right ]v_i =  \left [ \gamma^2 \left (-\frac{1}{\gamma^2} \right ) + 1 \right ]v_i = 0 
\\
\Rightarrow g_{rt}g^{ti} + g_{rr}g^{rr} + g_{rj}g^{ji} = 0 \ ;
\end{gathered}
\end{equation}

\begin{equation}
\begin{gathered}
g_{kt}g^{ti} + g_{kr}g^{ri} + g_{kj}g^{ji}  
\\
= \left [ \frac{\gamma^2 \alpha^2}{r_c} \left ( r - r_c \right ) v_k \right ] \left ( \frac{\gamma^2}{r_c}v^i\right ) + \left ( -\frac{\gamma \alpha}{r_c}v_k \right ) \left ( \frac{\gamma}{\alpha}v^i \right ) + \left ( \delta_{kj} - \frac{\gamma^2 \alpha^2}{r_c^2}  \left (r -r_c \right ) v_k v_j \right ) \left ( \delta^{ji} + \frac{\gamma^2}{r_c} v^j v^i\right ) 
\\ 
= \frac{\gamma^4\alpha^2}{r_c^2} \left ( r - r_c \right ) v_k v^i - \frac{\gamma^2}{r_c} v_k v^i + \left [ \delta^i_k + \frac{\gamma^2}{r_c} v_k v^i - \frac{\gamma^2 \alpha^2}{r_c^2} \left (r - r_c \right ) \left (v_kv^i + \frac{\gamma^2v^2}{r_c}v_k v^i \right) \right ]  
\\
= \delta^i_k + \left [ \frac{\gamma^4\alpha^2}{r_c^2} \left ( r - r_c \right ) - \frac{\gamma^2\alpha^2}{r_c^2} \left (r - r_c \right ) \left (1 + \frac{\gamma^2v^2}{r_c} \right ) \right ] v_k v^i  
\\
= \delta^i_k + \left [ \frac{\gamma^4\alpha^2}{r_c^2} \left ( r - r_c \right ) - \frac{\gamma^2\alpha^2}{r_c^2} \left (r - r_c \right ) \left (1 + \left ( \gamma^2 - 1 \right ) \right) \right ] v_k v^i  
\\
= \delta^i_k + \left [ \frac{\gamma^4\alpha^2}{r_c^2} \left ( r - r_c \right ) - \frac{\gamma^4\alpha^2}{r_c^2} \left (r - r_c \right ) \right ] v_k v^i = \delta^i_k 
\\
\Rightarrow g_{kt}g^{ti} + g_{kr}g^{ri} + g_{kj}g^{ji} =  \delta^i_k \ .
\end{gathered}
\end{equation}

The solution in Eq. \ref{eq:i5} may be obtained directly from Eqs. \ref{eq:i1} or \ref{eq:i3}, although that is algebraically non trivial. One can then employ a trial-and-error approach, while imposing that Eqs. \ref{eq:i1} to \ref{eq:i3} must be solved. An alternative hint towards the determination of $g^{ij}$ comes from the direct inspection of Eq. \ref{eq:inducedup}. We know that, since $\gamma^{\mu \nu}$ is the inverse induced metric, we must necessarily have $\gamma^{i j} = \delta^{i j}$, which is equal to $g^{ij}-n^i n^j$ if and only we have $g^{ij}$ as in Eq. \ref{eq:i5}.

So, finally, we have the inverse boosted metric components,

\begin{equation}
\begin{gathered}
g^{tt} = \frac{\gamma^2 v^2}{r_c^2} \ ; \ 
g^{tr} = \frac{\gamma}{\alpha}  \ ; \ 
g^{ti} = \frac{\gamma^2}{r_c} v^i \ ; \ 
\\
g^{rr} = r-r_h; \ 
g^{ri} = \frac{\gamma}{\alpha}v^i \ ; \ 
g^{ij} = \delta^{ij} + \frac{\gamma^2}{r_c} v^i v^j \ .
\label{eq:boosted_metric_inv_comps_ap}
\end{gathered}
\end{equation}

As a final remark, notice that the calculation above does not change at all when we change the boundary conditions on Sec. \ref{sec:varying}, i.e., when we allow the spatial part of the induced metric to fluctuate. Indeed, all we have to do is the change $\delta_{ij} \mapsto \Omega_{ij} = \Omega_{ij}(x^a)$, and --- provided that we use $\Omega_{ij}$ and $\Omega^{ij}$ to respectively lower and raise the spatial indices, which is naturally the case we consider --- the inverse boosted metric components have the same form as in Eq. \ref{eq:boosted_metric_inv_comps_ap},

\begin{equation}
\begin{gathered}
g^{tt} = \frac{\gamma^2 v^2}{r_c^2} \ ; \ 
g^{tr} = \frac{\gamma}{\alpha}  \ ; \ 
g^{ti} = \frac{\gamma^2}{r_c} v^i \ ; \ 
\\
g^{rr} = r-r_h; \ 
g^{ri} = \frac{\gamma}{\alpha}v^i \ ; \ 
g^{ij} = \Omega^{ij} + \frac{\gamma^2}{r_c} v^i v^j \ .
\label{eq:boosted_metric_inv_comps_ap_bc2}
\end{gathered}
\end{equation}

\end{document}